\documentclass[12pt]{article}
\usepackage{amssymb}
\usepackage{amsmath}
\usepackage{bm}
\usepackage[dvips]{epsfig}
\usepackage{graphicx}

\begin{document}

\title{On the minimal set of conservation laws and the Hamiltonian
structure of the Whitham equations.}

\author{A.Ya. Maltsev.}

\date{
\centerline{\it L.D. Landau Institute for Theoretical Physics}
\centerline{\it 142432 Chernogolovka, pr. Ak. Semenova 1A,
maltsev@itp.ac.ru}}

\maketitle

\begin{abstract}
 We consider the questions connected with the Hamiltonian 
properties of the Whitham equations in case of several spatial 
dimensions. An essential point of our approach here is a connection 
of the Hamiltonian structure of the Whitham system with the 
finite-dimensional Poisson bracket defined on the space of
periodic or quasi-periodic solutions. From our point of view,
this approach gives a possibility to construct the Hamiltonian
structure of the Whitham equations under minimal requirements on 
the properties of the initial system. The Poisson bracket for
the Whitham system can be considered here as a deformation of
the finite-dimensional bracket with the aid of the 
Dubrovin - Novikov procedure of bracket averaging. At the end,
we consider the examples where the constructions of the paper
play an essential role for the construction of the Poisson
bracket for the Whitham system.
\end{abstract}

\section{Introduction.}

 In this article we review the questions related to the Hamiltonian
formulation of the Whitham averaging method. As is well known, the
Whitham method is connected with the slow modulations of periodic
or quasiperiodic solutions of partial differential equations (PDE's). 
Thus, we will consider here systems of PDE's having evolutionary form 
and the same will be assumed also about the corresponding Whitham system.
Besides that, we will assume that the initial evolutionary system
has a Hamiltonian structure given by a local field-theoretic
Poisson bracket with some local Hamiltonian functional. All our
considerations will be made for the $d$-dimensional space
$\mathbb{R}^{d}$ and one time variable $t$.

 For the systems described above we assume the existence
of finite-parametric families of $m$-phase periodic or quasiperiodic
solutions and consider slow modulations of parameters of these
solutions according to the Whitham approach. The main questions 
considered in the paper will be connected with the construction
of the Hamiltonian structure for the Whitham system under some
requirements of ``completeness'' and ``regularity'' of the
corresponding family of $m$-phase solutions.

 As an intermediate step, we will consider here also the 
finite-dimensional Poisson brackets, generated by the 
field-theoretic Hamiltonian structures on the corresponding 
families of $m$-phase solutions. The construction of these brackets
will be closely connected with the conservation laws of the initial
system, so we introduce here the requirement of existence of
a ``minimal'' set of the conservation laws for a regular family of 
$m$-phase solutions. As we will see, the same requirement is sufficient 
also for the construction of the Hamiltonian structure for the 
corresponding Whitham system.

 Our approach in many features will follow the scheme proposed
by B.A. Dubrovin and S.P. Novikov and based on the conservative 
form of the Whitham system. On the other hand, we would like to
make here a connection between the finite-dimensional Poisson
bracket, given by the restriction of the field-theoretic Poisson
bracket on the family of $m$-phase solutions, and the Hamiltonian 
structure of the Whitham system. As we will see, the Hamiltonian
structure for the Whitham system can be given as a deformation
of the finite-dimensional Poisson bracket with the aid of the
Dubrovin - Novikov procedure of averaging of a local Poisson 
bracket. Let us say that our considerations here will be based 
just on the minimal requirements on the Hamiltonian properties of 
the family of $m$-phase solutions.

 In Chapter 2 we consider in detail the Dirac restriction of the
field-theoretic Poisson bracket on the family of $m$-phase 
solutions of a Hamiltonian system. As we will see, many essential 
features of the construction can be demonstrated here on the
example of the KdV hierarchy, so we consider the KdV equation
as an illustration of our scheme at the end of the chapter. We give
here also some other examples, demonstrating some special cases
discussed in the chapter.

 In Chapter 3 we discuss the form of a regular Whitham system
for a complete Hamiltonian family of $m$-phase solutions and
the corresponding Hamiltonian structure on the space of slowly
modulated parameters.

 Chapter 4 is mainly technical in nature and devoted to 
justification of the construction of the Poisson bracket for
the Whitham system. 

 For the convenience, the technical chapters are provided with a 
brief description of the content in the beginning of the chapter.

 Finally, in Chapter 5 we consider the examples which can be
considered as a good illustration of the general scheme presented
in the paper. First, we continue here again with the KdV equation
and discuss the construction of the Hamiltonian structures for the
corresponding Whitham systems. Another example is connected with
the equation having just the minimal set of the conservation laws
according to our definition.

\section{Hamiltonian systems and the Poisson brackets on the
spaces of $m$-phase solutions.}
\setcounter{equation}{0}

 In this chapter we will consider the Dirac restriction of a 
field-theoretic Poisson bracket on a finite-parametric family of
$m$-phase solutions of a Hamiltonian system.

 In general, special features of the bracket restriction depend
strongly on the properties of the Hamiltonian operator on the
corresponding ``submanifold'' in the functional space. In our case,
the submanifolds of the $m$-phase solutions of infinite-dimensional
Hamiltonian systems will be characterized by the following common
features:

\vspace{1mm}

1) A part of the parameters on the family of $m$-phase solutions
represents the phase (angle) variables changing linearly with time 
according to the dynamics of the system. The other parameters remain 
unchanged according to the dynamics of the system.

\vspace{1mm}

2) There exists a set of commuting functionals 
$\, \{ I^{\gamma} \}, \,$
leaving invariant the submanifold of $m$-phase solutions according to
the infinite-dimensional Poisson structure and generating the linear 
shifts of the phase variables $\, \theta^{\alpha}_{0} \, $ on this
submanifold. (The values of the second part of parameters remain
unchanged).

\vspace{1mm}

 Here we will put some special requirements of ``regularity'' of
the submanifold of $m$-phase solutions and the existence of a 
``minimal'' set of commuting functionals $\, \{ I^{\gamma} \}$.
In particular, we will assume here that the values of the 
functionals $\, \{ I^{\gamma} \} \,$ on the family of $m$-phase 
solutions can be naturally chosen as a part of parameters on this 
family.

\vspace{1mm}

 As we will see, the main part of the Dirac restriction of the
infinite-dimensional Poisson bracket to the submanifold of $m$-phase 
solutions will be connected with the construction of the functionals,
representing the phase variables on the family of $m$-phase solutions
and leaving this family invariant according to the infinite-dimensional 
Poisson structure. This construction needs in general the resolvability
of nontrivial linear systems of PDE's on the space of $2\pi$-periodic
in each $\, \theta^{\alpha} \, $ functions 
$\, \beta_{i} \, (\theta^{1}, \dots, \theta^{m})$. The operators
of the corresponding systems are given by the pairwise Poisson brackets
of the constraints defining the $m$-phase solutions and are closely 
related with the Hamiltonian operator. As will be shown below, the
requirements of ``regularity'' of the family of $m$-phase solutions
and the existence of a ``minimal'' set of commuting functionals 
$\, \{ I^{\gamma} \} \,$ provide in fact the orthogonality of the 
right-hand parts of these systems to all the ``regular'' (left) 
eigen-vectors of the operators of the systems, corresponding to the zero 
eigen-value. As a consequence, we can claim in fact the resolvability of 
these systems in many simple cases, where the spectra of these operators 
have rather regular form. Thus, under the above requirements we can state 
in general the possibility of the regular Dirac restriction of an 
infinite-dimensional Poisson bracket on a submanifold of one-phase 
solutions of a Hamiltonian system, where the non-zero eigen-values of 
these operators are in common separated from zero. Other examples of
the ``regular'' situations are connected usually with rather simple
Hamiltonian operators, having simple spectral properties both in the
single-phase and the multi-phase situations.

 In case of possibility of regular Dirac restriction of a Poisson
bracket on the submanifold of $m$-phase solutions we can introduce
the restricted bracket on the space of parameters of the solutions
which can in fact be written in rather simple form. Thus, under the
requirements of ``regularity'' of the family of $m$-phase solutions
and the existence of a ``minimal'' set of commuting functionals
$\, \{ I^{\gamma} \} \,$ the total set of parameters on the family 
of $m$-phase solutions can be chosen in the form
$$\left( \theta_{0}^{1}, \dots , \, \theta_{0}^{m}, \,
{\bf k}_{1}, \dots, \, {\bf k}_{d} , \, U^{1}, \dots , \, U^{m+s}
\right) $$
where $\, {\bf k}_{q} \, = \, (k^{1}_{q}, \dots, \, k^{m}_{q}) \, $
represent the wave numbers of the solutions and the parameters
$\, U^{\gamma} \, $ coincide with the values of the functionals
$\, I^{\gamma} \, $ on the family. It can be stated then that after an 
appropriate choice of the initial phase shifts 
$\, \theta_{0}^{\alpha} \, $ the regular Dirac restriction of the
infinite-dimensional Poisson bracket on the family of $m$-phase 
solutions can be written in the form
\begin{equation}
\label{RestrictedBracket}
\begin{array}{c}
\left\{ \theta^{\alpha}_{0} \, , \, \theta^{\beta}_{0} \right\}
\,\, = \,\, 0
\,\,\, , \,\,\,\,\,
\left\{ \theta^{\alpha}_{0} \, , \, U^{\gamma} \right\} \,\, = \,\,
\omega^{\alpha\gamma}
\left( {\bf k}_{1}, \dots, {\bf k}_{d}, \, {\bf U} \right)
\,\,\, , \,\,\,\,\,
\left\{ \theta^{\alpha}_{0} \, , \, k^{\beta}_{p}
\right\} \,\, = \,\, 0  \,\,\, ,   \\   \\
\left\{ U^{\gamma} \, , \, U^{\rho} \right\} \,\, = \,\, 0
\,\,\, , \,\,\,\,\,
\left\{ U^{\gamma} \, , \, k^{\beta}_{p}
\right\} \,\, = \,\, 0  \,\,\, , \,\,\,\,\,
\left\{ k^{\alpha}_{q} \, , \, k^{\beta}_{p}
\right\} \,\, = \,\, 0
\end{array}
\end{equation}
where 
$\, \omega^{\alpha\gamma} \,
({\bf k}_{1}, \dots, {\bf k}_{d}, \, {\bf U}) \, $ represent the
frequencies, corresponding to the flows generated by the functionals
$\, I^{\gamma} \, $ on the family.

 Let us say, however, that the ``regular'' Dirac restriction of an
infinite-dimensional Poisson bracket on the full family of $m$-phase
solutions of a Hamiltonian system is in fact not possible in general
situation. Nevertheless, the bracket (\ref{RestrictedBracket}) can
still be associated with the infinite-dimensional Poisson bracket 
of a Hamiltonian system under some additional conditions. Namely,
under the same requirements of ``regularity'' of the family of 
$m$-phase solutions and the existence of a ``minimal'' set of commuting 
functionals (formulated below in detail) we have to require also the
possibility of the Dirac restriction of the infinite-dimensional 
bracket to some special submanifolds which form a dense set in the 
full family of $m$-phase solutions. Let us note here that this
requirement is usually rather weak and is satisfied in all examples
known to us. At the end of the chapter we give two rather simple
examples where two different situations with the Dirac restriction 
of an infinite-dimensional Poisson bracket to a submanifold 
of $m$-phase solutions naturally arise.

 Let us also note here, that in our scheme the bracket 
(\ref{RestrictedBracket}) can be considered in fact as an intermediate 
step to the bracket for the Whitham system, having more complicated
form. However, we must certainly say here that the theory of the
finite-dimensional brackets connected with the multi-phase solutions
of PDE's represents also one of the most interesting and important
branches of the theory of integrable systems 
(see \cite{VeselovNovikov1, VeselovNovikov2, KricheverCPAM, 
KricheverPhong}).

 We will start now more detailed consideration of the Dirac
restriction of an infinite-dimensional Poisson bracket on a
submanifold of $m$-phase solutions of a Hamiltonian system.

\vspace{0.2cm}

 In this paper we will consider quasiperiodic $m$-phase
solutions of Hamiltonian PDE's having the evolutionary form
\begin{equation}
\label{EvInSyst}
\varphi^{i}_{t} \,\, = \,\, F^{i} (\bm{\varphi}, \bm{\varphi}_{\bf x},
\bm{\varphi}_{\bf xx}, \dots ) \,\, \equiv \,\,
F^{i}(\bm{\varphi}, \bm{\varphi}_{x^{1}},
\dots, \bm{\varphi}_{x^{d}}, \dots )
\end{equation}
$i = 1, \dots, n$ ,
$\bm{\varphi} = (\varphi^{1}, \dots, \varphi^{n})$, with one time
and $d$ spatial dimensions.

 Let us say that we will everywhere define here a quasiperiodic
function $f ({\bf x})$ on $\mathbb{R}^{d}$ as a function coming
from a smooth periodic function of $m$ variables 
${\hat f} (\bm{\theta}) = {\hat f} (\theta^{1}, \dots, \theta^{m})$
according to the formula
$$f (x^{1}, \dots, \, x^{d}) \,\, = \,\, {\hat f}
({\bf k}_{1} x^{1} \, + \, \dots \, + \, {\bf k}_{d} x^{d} \, + \,
\bm{\theta}_{0} )$$

 The vectors ${\bf k}_{q} \, = \, (k^{1}_{q}, \dots, \, k^{m}_{q})$
will be called here the wave numbers of the function $f ({\bf x})$,
defining the corresponding mapping 
$\mathbb{R}^{d} \rightarrow \mathbb{T}^{m}$. Let us say, that for
our purposes here it will be not necessary to put additional 
requirements on the rational independence of the components of
$\, ({\bf k}_{1}, \dots, \, {\bf k}_{d})$, so our terminology here
does not in fact coincide completely with the standard one. Let us,
however, keep here this simplified terminology which will not play
important role in our considerations. For convenience we will
assume here that the function ${\hat f} (\bm{\theta})$ is always
$2\pi$-periodic with respect to each $\theta^{\alpha}$,
$\, \alpha = 1, \dots, m$.

 The $m$-phase solutions of system (\ref{EvInSyst}) have the form
$$\varphi^{i} ({\bf x}, t) \,\, = \,\, \Phi^{i} \left(
{\bf k}_{1}\, x^{1} \, + \, \dots \, + \,
{\bf k}_{d}\, x^{d}
\, + \, \bm{\omega}\, t \, + \, \bm{\theta}_{0} \right) $$
where the $2\pi$-periodic in each $\theta^{\alpha}$ functions
$\Phi^{i} (\bm{\theta})$ satisfy the system
\begin{equation}
\label{EvPhaseSyst}
\omega^{\alpha} \, \Phi^{i}_{\theta^{\alpha}} \,\, = \,\,
F^{i} \left( \bm{\Phi}, \,
k^{\beta_{1}}_{1} \, \bm{\Phi}_{\theta^{\beta_{1}}}, \dots,
k^{\beta_{d}}_{d} \, \bm{\Phi}_{\theta^{\beta_{d}}},
\dots \right)
\end{equation}

 We are going to consider smooth finite parametric 
families of the quasiperiodic solutions of (\ref{EvInSyst})
given by the formula 
\begin{equation}
\label{mphasesol}
\varphi^{i} ({\bf x}, t) \,\, = \,\, \Phi^{i} \left(
{\bf k}_{1}({\bf a})\, x^{1} \, + \, \dots \, + \,
{\bf k}_{d}({\bf a})\, x^{d}
\, + \, \bm{\omega}({\bf a})\, t \, + \, \bm{\theta}_{0}, \,
{\bf a} \right)
\end{equation}
with some smooth dependence of the functions $\Phi^{i}$ and
$({\bf k}_{q}, \, \bm{\omega})$ on the set of 
parameters  \linebreak
${\bf a} = (a^{1}, \dots, a^{N})$. We can see in fact that two
different types of parameters naturally arise in the definition
(\ref{mphasesol}). Thus, the parameters ${\bf a}$ define the 
``shape'' of the solutions $\varphi^{i} ({\bf x}, t)$ and the
corresponding ``frequencies'' and ``wave numbers''. At the 
same time, the parameters
$\bm{\theta}_{0} = (\theta^{1}_{0}, \dots, \theta^{m}_{0})$
represent just the initial phase shifts and take all possible
values on the full family of $m$-phase solutions. 

 Let us denote here by $\Lambda$ the corresponding set of the 
functions $\bm{\varphi} ({\bf x})$ in $\mathbb{R}^{d}$ given
by the formula 
\begin{equation}
\label{LambdaDef}
\varphi^{i} ({\bf x}) \,\, = \,\, \Phi^{i} \left(   
{\bf k}_{1}({\bf a})\, x^{1} \, + \, \dots \, + \,
{\bf k}_{d}({\bf a})\, x^{d}
\, + \, \bm{\theta}_{0}, \, {\bf a} \right) 
\end{equation}

 Now we don't put any special requirements on the parameters 
${\bf a}$. We can see according to (\ref{EvPhaseSyst}) that the family
$\Lambda$ is invariant with respect to the evolutionary system
(\ref{EvInSyst}). 

 To make a difference between the functions 
$\bm{\varphi} ({\bf x})$ and $\bm{\Phi} (\bm{\theta})$ let us
denote also by ${\hat \Lambda}$ the corresponding set of the
$2\pi$-periodic with respect to each $\theta^{\alpha}$ functions
$\, \bm{\Phi} (\bm{\theta} + \bm{\theta}_{0}, \, {\bf a})$,
smoothly depending on the parameters 
$\, {\bf a} = (a^{1}, \dots, a^{N})$. 
Rigorously speaking, we will
call here a smooth family of $m$-phase solutions of (\ref{EvInSyst})
a smooth family ${\hat \Lambda}$ of the functions
$\, \bm{\Phi} (\bm{\theta} + \bm{\theta}_{0}, \, {\bf a})$
with the smooth dependence $\, {\bf k}_{q} = {\bf k}_{q} ({\bf a})$,
$\, \bm{\omega} = \bm{\omega} ({\bf a})$, 
satisfying the corresponding system (\ref{EvPhaseSyst}).

 Let us say now that we will assume here also that system
(\ref{EvInSyst}) is Hamiltonian with respect to some local
field-theoretic Poisson bracket given in general by the 
expression
\begin{equation}
\label{MultDimPBr}
\{ \varphi^{i} ({\bf x}) \, , \, \varphi^{j} ({\bf y}) \} \,\, = \,\,
\sum_{l_{1},\dots,l_{d}} B^{ij}_{(l_{1},\dots,l_{d})}
(\bm{\varphi}, \bm{\varphi}_{\bf x}, \dots ) \,\,
\delta^{(l_{1})} (x^{1} - y^{1}) \, \dots \,
\delta^{(l_{d})} (x^{d} - y^{d})
\end{equation}
$(l_{1}, \dots, l_{d} \geq 0)$, and has a local Hamiltonian 
functional of the form
\begin{equation}
\label{MultDimHamFunc}
H \,\, = \,\, \int P_{H} \left(\bm{\varphi}, \bm{\varphi}_{\bf x},
\bm{\varphi}_{\bf xx}, \dots \right) \,\, d^{d} x
\end{equation}

 The Hamiltonian structure (\ref{MultDimPBr}) and the Hamiltonian
functional (\ref{MultDimHamFunc}) can be naturally considered on 
functional spaces of different types. Thus, the structure
(\ref{MultDimPBr}) is well defined on the space of rapidly
decreasing at infinity functions $\bm{\varphi} ({\bf x})$,
where (\ref{MultDimHamFunc}) represents a well defined 
translationally invariant functional under the appropriate
normalization of the density $P_{H}$:
$P_{H} (0, 0, \dots) = 0$. Other natural types of the functional
spaces can be represented by the spaces of the periodic or the
quasiperiodic functions $\bm{\varphi} ({\bf x})$. In this case
it is natural to define the functional $H$ in the form
$$H \,\, = \,\, \lim_{K\rightarrow\infty} \,\,\,
{1 \over (2K)^{d}} \,\, \int_{-K}^{K} \! \dots \int_{-K}^{K} \,
P_{H} \left( \bm{\varphi}, \, \bm{\varphi}_{\bf x}, \,
\bm{\varphi}_{\bf xx}, \dots \right) \, d^{d} x $$

 Let us note that in the last case we have to define the 
variation derivative of $H$ with respect to the variations 
of $\bm{\varphi} ({\bf x})$ having the same periodic or 
quasiperiodic properties as the original function. It's not
difficult to see that the standard Euler - Lagrange expressions
for the variation derivatives can be used also in this situation.

 Here we will write all the local translationally invariant
functionals in the general form 
\begin{equation}
\label{Integrals}
I \,\, = \,\, \int
P \left( \bm{\varphi}, \, \bm{\varphi}_{\bf x}, \,
\bm{\varphi}_{\bf xx},
\dots \right) \, d^{d} x  
\end{equation}
assuming an appropriate definition in every concrete
situation.

 Let us note also that the Hamiltonian structure (\ref{MultDimPBr})
and the functionals (\ref{Integrals}) can be also naturally 
considered on the spaces of quasiperiodic functions 
$\bm{\varphi} ({\bf x})$ with the fixed wave numbers
$({\bf k}_{1}, \dots, {\bf k}_{d})$, representing natural invariant
subspaces for the structure (\ref{MultDimPBr}).

\vspace{0.3cm}

 We are going to consider here ``maximal'' smooth families 
$\Lambda$ of $m$-phase solutions of system (\ref{EvInSyst})
in the sense which we will explain below. In particular, we will
always assume here that the values 
$({\bf k}_{1}, \dots, {\bf k}_{d})$ and $\bm{\omega}$ represent
independent parameters on the family $\Lambda$, such that the total
set of ${\bf a}$ includes $N = m (d + 1) + s$, $(s \geq 0)$
parameters $(a^{1}, \dots, a^{N})$. Thus, we assume in fact 
that the particular choice of ${\bf a}$ can be represented by the 
parameters
$({\bf k}_{1}, \dots, {\bf k}_{d}, \, \bm{\omega}, \, 
n^{1}, \dots, n^{s})$ where ${\bf k}_{q}$ and $\bm{\omega}$
are the wave numbers and the frequencies of the $m$-phase
solutions and ${\bf n} = (n^{1}, \dots, n^{s})$ are some
additional parameters (if any). Including the initial phase
shifts $\bm{\theta}_{0} = (\theta_{0}^{1}, \dots, \theta_{0}^{m})$
we then claim that the solutions from the family $\Lambda$ are
parametrized by the $m (d + 1) + s + m$ parameters
$({\bf k}_{1}, \dots, {\bf k}_{d}, \, \bm{\omega}, \, {\bf n}, \,
\bm{\theta}_{0})$.

 Let us introduce also the families
${\hat \Lambda}_{{\bf k}_{1}, \dots, {\bf k}_{d}} \subset 
{\hat \Lambda}$ consisting of the functions  \linebreak
$\bm{\Phi} (\bm{\theta} + \bm{\theta}_{0}, \,
{\bf k}_{1}, \dots, {\bf k}_{d}, \, \bm{\omega}, \, {\bf n}) 
\in {\hat \Lambda}$ with the fixed parameters 
$({\bf k}_{1}, \dots, {\bf k}_{d})$. The corresponding subsets
$\Lambda_{{\bf k}_{1}, \dots, {\bf k}_{d}} \subset \Lambda$
represent families of $m$-phase solutions of system 
(\ref{EvInSyst}) on the spaces of the quasiperiodic 
functions with the fixed wave numbers in the coordinate space. 
The full set of parameters on the families 
${\hat \Lambda}_{{\bf k}_{1}, \dots, {\bf k}_{d}}$
or $\Lambda_{{\bf k}_{1}, \dots, {\bf k}_{d}}$ can be represented 
by the values $(\bm{\omega}, \, {\bf n}, \, \bm{\theta}_{0})$.

 For convenience we will introduce here also the subset
${\cal M} \subset \{{\bf k}_{1}, \dots, {\bf k}_{d}\}$ in the
space of parameters $({\bf k}_{1}, \dots, {\bf k}_{d})$ given
by generic values of $({\bf k}_{1}, \dots, {\bf k}_{d})$, defined
by the requirement that the orbits of the group generated by the
set of constant vector fields $({\bf k}_{1}, \dots, {\bf k}_{d})$
are everywhere dense in $\mathbb{T}^{m}$. It is easy to see that 
the set ${\cal M}$ has the full measure in the space of the 
parameters $({\bf k}_{1}, \dots, {\bf k}_{d})$.

 Let us define the following quasiperiodic functions
\begin{multline*}
\bm{\varphi}_{\theta^{\alpha}} ({\bf x}) \,\, = \,\,
\bm{\Phi}_{\theta^{\alpha}} \left(
{\bf k}_{1} \, x^{1} \, + \, \dots \, + \, {\bf k}_{d} \, x^{d}
\, + \, \bm{\theta}_{0}, \, {\bf k}_{1}, \dots, {\bf k}_{d}, \,
\bm{\omega}, \, {\bf n} \right) \,\,\, ,  \\
\bm{\varphi}_{\omega^{\alpha}} ({\bf x}) \,\, = \,\,
\bm{\Phi}_{\omega^{\alpha}} \left(
{\bf k}_{1} \, x^{1} \, + \, \dots \, + \, {\bf k}_{d} \, x^{d}
\, + \, \bm{\theta}_{0}, \, {\bf k}_{1}, \dots, {\bf k}_{d}, \,
\bm{\omega}, \, {\bf n} \right) \,\,\, ,  \\
\bm{\varphi}_{n^{l}} ({\bf x}) \,\, = \,\,
\bm{\Phi}_{n^{l}} \left(
{\bf k}_{1} \, x^{1} \, + \, \dots \, + \, {\bf k}_{d} \, x^{d}
\, + \, \bm{\theta}_{0}, \, {\bf k}_{1}, \dots, {\bf k}_{d}, \,
\bm{\omega}, \, {\bf n} \right) \,\,\, ,
\end{multline*}
$\alpha = 1, \dots, m$, $\, l = 1, \dots, s$, on the family
$\Lambda$.

\vspace{0.2cm}

 We will say that the family 
$\Lambda_{{\bf k}_{1}, \dots, {\bf k}_{d}}$ represents a 
submanifold in the space of quasiperiodic functions with the
fixed wave numbers $({\bf k}_{1}, \dots, {\bf k}_{d})$ if the
functions 
$(\bm{\varphi}_{\theta^{\alpha}} ({\bf x}), \,
\bm{\varphi}_{\omega^{\alpha}} ({\bf x}), \,
\bm{\varphi}_{n^{l}} ({\bf x}))$ are linearly independent
for all values of $(\bm{\omega}, \, {\bf n}, \, \bm{\theta}_{0})$.

\vspace{0.3cm}

 It will be convenient here to assume all the properties formulated 
above whenever we mention the families $\, \Lambda \, $ or
$\, \Lambda_{{\bf k}_{1}, \dots, {\bf k}_{d}}$. So,
everywhere below we will assume that the submanifolds 
$\, \Lambda \, $ and 
$\, \Lambda_{{\bf k}_{1}, \dots, {\bf k}_{d}} \, $ represent 
the ``maximal'' families of $m$-phase solutions of system 
(\ref{EvInSyst}) in the above sense.

\vspace{0.3cm}

{\bf Definition 2.1.}

{\it We call the submanifold 
$\Lambda_{{\bf k}_{1}, \dots, {\bf k}_{d}}$ a regular Hamiltonian
submanifold in the space of quasiperiodic functions with the 
wave numbers $({\bf k}_{1}, \dots, {\bf k}_{d})$ if:

1) The bracket (\ref{MultDimPBr}) has on
$\Lambda_{{\bf k}_{1}, \dots, {\bf k}_{d}}$ 
constant number of ``annihilators'' defined by linearly
independent solutions
$\, {\bf v}^{(k)} ({\bf x}) \, = \, (v^{(k)}_{1}({\bf x}), \dots ,
v^{(k)}_{n}({\bf x}))$, $\,\, k = 1, \dots, s^{\prime}, \, $
of the equation
\begin{equation}
\label{AnnSolCoordRepr}
\sum_{l_{1},\dots,l_{d}} \left.
B^{ij}_{(l_{1},\dots,l_{d})} (\bm{\varphi},
\bm{\varphi}_{\bf x}, \dots)
\right|_{\Lambda} v^{(k)}_{j, \, l_{1} x^{1} \dots \, l_{d} x^{d}}
({\bf x}) \,\,\, = \,\,\, 0 
\end{equation}
on the space of quasiperiodic functions with the wave numbers
$({\bf k}_{1}, \dots, {\bf k}_{d})$;

2) We have: $m + s \geq s^{\prime}$, and:

$${\rm rank} \,
\begin{Vmatrix}
(\bm{\varphi}_{\omega^{\alpha}} \, \cdot \, 
{\bf v}^{(k)} )  \\
(\bm{\varphi}_{n^{l}} \, \cdot \,
{\bf v}^{(k)} )
\end{Vmatrix} \,\,\, = \,\,\, s^{\prime} $$
($\alpha = 1, \dots, m$, $\, l = 1, \dots, s$, 
$\, k = 1, \dots, s^{\prime}$), where
$$\left( \bm{\varphi}_{\omega^{\alpha}} \, \cdot \,
{\bf v}^{(k)} \right) \,\,\, \equiv \,\,\, \int
\varphi^{i}_{\omega^{\alpha}} ({\bf x}) \,\,
v^{(k)}_{i} ({\bf x}) \,\, d^{d} x \,\,\,\,\, , \,\,\,\,\,\,\,\,  
\left( \bm{\varphi}_{n^{l}} \, \cdot \,
{\bf v}^{(k)} \right) \,\,\, \equiv \,\,\, \int
\varphi^{i}_{n^{l}} ({\bf x}) \,\,
v^{(k)}_{i} ({\bf x}) \,\, d^{d} x $$
are the convolutions of the variation derivatives of
annihilators with the corresponding tangent vectors
$\bm{\varphi}_{\omega^{\alpha}}$, 
$\, \bm{\varphi}_{n^{l}}$.

\vspace{0.1cm}

 We call family $\Lambda$ a regular Hamiltonian submanifold
in the space of quasiperiodic functions if all the families
$\Lambda_{{\bf k}_{1}, \dots, {\bf k}_{d}}$ represent regular
Hamiltonian submanifolds in the spaces of quasiperiodic functions
with the wave numbers $({\bf k}_{1}, \dots, {\bf k}_{d})$
with the same number $s^{\prime}$.

}

\vspace{0.2cm}

 Let us note here that, according to our definition of a
quasiperiodic function, Definition 2.1 implies, in particular,
that for $({\bf k}_{1}, \dots, {\bf k}_{d}) \in {\cal M}$ the
number of smooth linearly independent  \linebreak
$2\pi$-periodic in each $\theta^{\alpha}$ solutions 
${\bf v}^{(k)} (\bm {\theta})$ of the equation

\begin{multline}
\label{BvAnnSyst}
\sum_{l_{1}, \dots, l_{d}}
B^{ij}_{(l_{1}, \dots, l_{d})}
\left( \bm{\Phi} (\bm{\theta} + \bm{\theta}_{0}, \, {\bf a}), \,
k^{\gamma_{1}}_{1} ({\bf a}) \, \bm{\Phi}_{\theta^{\gamma_{1}}},
\dots, \, k^{\gamma_{d}}_{d} ({\bf a}) \,
\bm{\Phi}_{\theta^{\gamma_{d}}}, \, \dots \right) \, \times \\
\times \,\, k^{\alpha^{1}_{1}}_{1} ({\bf a}) \dots
k^{\alpha^{1}_{l_{1}}}_{1} ({\bf a}) \, \dots \,
k^{\alpha^{d}_{1}}_{d} ({\bf a}) \dots
k^{\alpha^{d}_{l_{d}}}_{d} ({\bf a}) \,\,\,\,\,
v^{(k)}_{j, \, \theta^{\alpha^{1}_{1}} \dots 
\theta^{\alpha^{1}_{l_{1}}} \, \dots \, 
\theta^{\alpha^{d}_{1}} \dots
\theta^{\alpha^{d}_{l_{d}}}} \,\, = \,\, 0
\end{multline}
is exactly equal to $s^{\prime}$.

 Let us say also that the smooth periodic solutions 
of (\ref{BvAnnSyst}) define the annihilators with smooth 
variation derivatives of the Hamiltonian operator
\begin{multline}
\label{OperatorB}
{\hat B}^{ij}_{{\bf k}_{1}, \dots, {\bf k}_{d}} 
\,\, = \,\, \sum_{l_{1}, \dots, l_{d}}
B^{ij}_{(l_{1}, \dots, l_{d})}
\left( \bm{\varphi} (\bm{\theta}), \,
k^{\gamma_{1}}_{1} \, \bm{\varphi}_{\theta^{\gamma_{1}}},
\dots, \, k^{\gamma_{d}}_{d} \,
\bm{\varphi}_{\theta^{\gamma_{d}}}, \, \dots \right) \, \times \\
\times \,\, k^{\alpha^{1}_{1}}_{1} \dots
k^{\alpha^{1}_{l_{1}}}_{1}
\, \dots \,
k^{\alpha^{d}_{1}}_{d} \dots
k^{\alpha^{d}_{l_{d}}}_{d} \,\,\,\,\,
{\partial \over \partial \theta^{\alpha^{1}_{1}}} \dots
{\partial \over \partial \theta^{\alpha^{1}_{l_{1}}}} \, \dots \,
{\partial \over \partial \theta^{\alpha^{d}_{1}}} \dots
{\partial \over \partial \theta^{\alpha^{d}_{l_{d}}}} 
\end{multline}
defined on $\mathbb{T}^{m}$ for any set 
$({\bf k}_{1}, \dots, {\bf k}_{d})$.

\vspace{0.2cm}

{\bf Definition 2.2.}

{\it We say that a regular Hamiltonian submanifold
$\Lambda$ is equipped with a minimal
set of commuting integrals if there exist $m + s$ functionals
$I^{\gamma}$, $\, \gamma = 1, \dots, m + s$, having the form
\begin{equation}
\label{IgammaForm}
I^{\gamma} \,\, = \,\, \int
P^{\gamma} \left( \bm{\varphi}, \, \bm{\varphi}_{\bf x}, \,
\bm{\varphi}_{\bf xx}, \dots \right) \, d^{d} x 
\end{equation}
such that:

1) The functionals $I^{\gamma}$ commute with the Hamiltonian
(\ref{MultDimHamFunc}) and with each other with respect to the
bracket (\ref{MultDimPBr}):
\begin{equation}
\label{CommutativeSet}
\left\{ I^{\gamma} \, , \, H \right\} \,\, = \,\, 0 \,\,\,\,\, ,
\,\,\,\,\,\,\,\, \left\{ I^{\gamma} \, , \, I^{\rho} \right\}
\,\, = \,\, 0 \,\,\, , 
\end{equation}

2) The values $U^{\gamma}$ of the functionals $I^{\gamma}$ can be
used as the parameters $\, {\bf a}$ on every submanifold
$\Lambda_{{\bf k}_{1}, \dots, {\bf k}_{d}}$, such that the total
set of parameters on $\Lambda$ can be represented in the 
form \linebreak
$({\bf k}_{1}, \dots, {\bf k}_{d}, \, U^{1}, \dots, U^{m+s}, \,
\bm{\theta}_{0})$;

3) The Hamiltonian flows, generated by the functionals $I^{\gamma}$,
leave invariant the family $\Lambda$, generating the linear time
evolution of the phase shifts $\bm{\theta}_{0}$ with constant
frequencies 
$\bm{\omega}^{\gamma} = 
(\omega^{1\gamma}, \dots, \omega^{m\gamma})$,
such that
$${\rm rk} \,\, \left| \left| \,
\omega^{\alpha \gamma} ({\bf k}_{1}, \dots, {\bf k}_{d}, \, {\bf U})
\, \right| \right| \,\,\, = \,\,\, m $$ 
everywhere on $\Lambda$;

4) For every submanifold
$\Lambda_{{\bf k}_{1}, \dots, {\bf k}_{d}}$
the linear space generated by the variation derivatives
$\delta I^{\gamma} / \delta \varphi^{i} ({\bf x})$ contains
the variation derivatives of all the annihilators of
bracket (\ref{MultDimPBr}) on the corresponding space of
quasiperiodic functions. In other words,
on every submanifold $\Lambda_{{\bf k}_{1}, \dots, {\bf k}_{d}}$
we can write for some complete set $\{ {\bf v}^{(k)} ({\bf x}) \}$ 
of linearly independent quasiperiodic solutions of
(\ref{AnnSolCoordRepr}) the relations:
$$v^{(k)}_{i} ({\bf x}) \,\, = \,\,
\sum_{\gamma=1}^{m+s} \gamma^{k}_{\gamma} 
({\bf k}_{1}, \dots, {\bf k}_{d}, \, {\bf U}) \,\, \left.
{\delta I^{\gamma} \over \delta \varphi^{i} ({\bf x})} 
\right|_{\Lambda_{{\bf k}_{1}, \dots, {\bf k}_{d}}} $$
with some functions 
$\gamma^{k}_{\gamma} ({\bf k}_{1}, \dots, {\bf k}_{d}, \, {\bf U})$ 
on $\Lambda$.

}

\vspace{0.2cm}

 Let us also note that in Definition 2.2 we assume in particular
that the Jacobian of the coordinate transformation
$(\bm{\omega}, \, {\bf n}) \rightarrow 
(U^{1}, \dots, U^{m+s})$ is different from zero on every
$\Lambda_{{\bf k}_{1}, \dots, {\bf k}_{d}}$ whenever we introduce
the parameters $(U^{1}, \dots, U^{m+s})$ on $\Lambda$.

\vspace{0.2cm}

 We can see that for any regular Hamiltonian submanifold
$\Lambda$ equipped with a
minimal set of commuting integrals we must have the relation
$s = s^{\prime}$ connecting the number of annihilators of the bracket
(\ref{MultDimPBr}) on $\Lambda$ with the number of the additional
parameters $(n^{1}, \dots, n^{s})$. This requirement means in fact
that the solutions from $\Lambda$ are parametrized by the set of
parameters 
$({\bf k}_{1}, \dots, {\bf k}_{d}, \, \bm{\omega}, 
\, \bm{\theta}_{0})$ 
on every common level $N^{1} = {\rm const}$, $\dots$,
$N^{s^{\prime}} = {\rm const}$ of the annihilators of the bracket
(\ref{MultDimPBr}) on the space of quasiperiodic functions. Let us
say that this property presents in most of important examples
of $m$-phase solutions of Hamiltonian PDE's. So, we will always
in fact assume below that $s = s^{\prime}$, such that the
number of annihilators of the bracket (\ref{MultDimPBr}) is
exactly equal to the number of parameters
$(n^{1}, \dots, n^{s})$.

 It's not difficult to see, that we must have also the relations
\begin{equation}
\label{annomegarel}
\sum_{\gamma=1}^{m+s} \,\, \gamma^{k}_{\gamma}
\left( {\bf k}_{1}, \dots, {\bf k}_{d}, \, {\bf U} \right) \,\,
\omega^{\alpha\gamma} 
\left( {\bf k}_{1}, \dots, {\bf k}_{d}, \, {\bf U} \right)
\,\,\, \equiv \,\,\, 0
\end{equation}
for the functions $\gamma^{k}_{\gamma}$ and $\omega^{\alpha\gamma}$
on any regular Hamiltonian submanifold with a minimal set of
commuting integrals.

 In the full analogy with the Hamiltonian structure 
(\ref{OperatorB}) we can introduce also the functionals
\begin{equation}
\label{Jnu}
J^{\gamma} \,\, = \,\, 
\int_{0}^{2\pi}\!\!\!\!\!\dots\int_{0}^{2\pi}
P^{\gamma} \left( \bm{\varphi}, \,
k^{\beta_{1}}_{1} \, \bm{\varphi}_{\theta^{\beta_{1}}}, \dots,
k^{\beta_{d}}_{d} \, \bm{\varphi}_{\theta^{\beta_{d}}},
\dots \right) \, {d^{m} \theta \over (2\pi)^{m}} \,\,\,\,\, , 
\end{equation}
($\gamma = 1, \dots, m + s$),
on the space of $2\pi$-periodic in each
$\theta^{\alpha}$ functions.

 Let us note that for any submanifold
$\Lambda_{{\bf k}_{1}, \dots, {\bf k}_{d}}$ in the space of
quasiperiodic functions of ${\bf x}$ we can claim that the
corresponding functions
$\, \bm{\Phi}_{\theta^{\alpha}} 
(\bm{\theta} + \bm{\theta}_{0}, \, {\bf k}_{1}, \dots, {\bf k}_{d},
\, \bm{\omega}, \, {\bf n})$, 
$\, \bm{\Phi}_{\omega^{\alpha}}  
(\bm{\theta} + \bm{\theta}_{0}, \, {\bf k}_{1}, \dots, {\bf k}_{d},
\, \bm{\omega}, \, {\bf n})$, 
$\, \bm{\Phi}_{n^{l}}  
(\bm{\theta} + \bm{\theta}_{0}, \, {\bf k}_{1}, \dots, {\bf k}_{d},
\, \bm{\omega}, \, {\bf n}) $, 
$\,\, (\alpha = 1, \dots, m, \,\, l = 1, \dots, s )$, 
are also linearly independent on $\mathbb{T}^{m}$, being
linearly independent on the subset
$$\left. \left(
{\bf k}_{1} \, x^{1} \, + \, \dots \, + \, {\bf k}_{d} \, x^{d}
\, + \, \bm{\theta}_{0} \right) \, \right| \, {\rm mod} \,
(2\pi \, \mathbb{Z})^{m} 
\,\,\,\,\, \subset \,\,\,\,\, \mathbb{T}^{m} $$

 As a result, we can claim that for any submanifold
$\Lambda_{{\bf k}_{1}, \dots, {\bf k}_{d}}$ in the space of
quasiperiodic functions with the wave numbers 
$({\bf k}_{1}, \dots, {\bf k}_{d})$ the corresponding family
${\hat \Lambda}_{{\bf k}_{1}, \dots, {\bf k}_{d}}$ represents
a submanifold in the space of smooth functions in $\mathbb{T}^{m}$
in the same sense.

 In the generic case 
$({\bf k}_{1}, \dots, {\bf k}_{d}) \in {\cal M}$ the space of
quasiperiodic functions with the wave numbers
$({\bf k}_{1}, \dots, {\bf k}_{d})$ coincides with the space
of smooth functions on $\mathbb{T}^{m}$ according to our definition.
We can claim also, that the submanifold
${\hat \Lambda}_{{\bf k}_{1}, \dots, {\bf k}_{d}}$ corresponds
by definition to $\Lambda_{{\bf k}_{1}, \dots, {\bf k}_{d}}$
after the transition to the torus $\mathbb{T}^{m}$. The
corresponding Poisson structure (\ref{MultDimPBr}) will be
naturally written in this case as
\begin{multline}
\label{PBronTorus}
\left\{ \varphi^{i} (\bm{\theta}) \, , \,
\varphi^{j} (\bm{\theta}^{\prime}) \right\} \,\, = \,\, 
\sum_{l_{1}, \dots, l_{d}}
B^{ij}_{(l_{1}, \dots, l_{d})}
\left( \bm{\varphi} (\bm{\theta}), \,
k^{\gamma_{1}}_{1} \, \bm{\varphi}_{\theta^{\gamma_{1}}},
\dots, \, k^{\gamma_{d}}_{d} \,
\bm{\varphi}_{\theta^{\gamma_{d}}}, \, \dots \right) \, \times \\
\times \,\, k^{\alpha^{1}_{1}}_{1} \dots
k^{\alpha^{1}_{l_{1}}}_{1} 
\, \dots \,
k^{\alpha^{d}_{1}}_{d} \dots
k^{\alpha^{d}_{l_{d}}}_{d} \,\,\,\,\,
{\partial \over \partial \theta^{\alpha^{1}_{1}}} \dots
{\partial \over \partial \theta^{\alpha^{1}_{l_{1}}}} \, \dots \,
{\partial \over \partial \theta^{\alpha^{d}_{1}}} \dots
{\partial \over \partial \theta^{\alpha^{d}_{l_{d}}}} \,\,
\delta (\bm{\theta} - \bm{\theta}^{\prime})
\end{multline}
while the functionals $I^{\gamma}$ will be represented by
$J^{\gamma}$ after the transition to the functions on 
$\mathbb{T}^{m}$.

 For convenience, let us define here the delta-function
$\, \delta (\bm{\theta} - \bm{\theta}^{\prime})$
and its higher derivatives  \linebreak
$\delta_{\theta^{\alpha_{1}}\dots\theta^{\alpha_{s}}}
(\bm{\theta} - \bm{\theta}^{\prime})$
in the $\bm{\theta}$-space by the formula   
$$\int_{0}^{2\pi}\!\!\!\!\!\dots\int_{0}^{2\pi}
\delta_{\theta^{\alpha_{1}}\dots\theta^{\alpha_{s}}}
(\bm{\theta} - \bm{\theta}^{\prime}) \,\,
\psi (\bm{\theta}^{\prime}) \,\,
{d^{m} \theta^{\prime} \over (2\pi)^{m}} \,\,\, \equiv \,\,\,
\psi_{\theta^{\alpha_{1}}\dots\theta^{\alpha_{s}}} (\bm{\theta}) $$

 It will be also convenient here to define the variation
derivatives by the rule
$$\delta \, S \,\,\,\,\, \equiv \,\,\,\,\,
\int_{0}^{2\pi}\!\!\!\!\!\dots\int_{0}^{2\pi}
{\delta \, S \over \delta \varphi^{i} (\bm{\theta})} \,\,
\delta \varphi^{i} (\bm{\theta}) \,\,
{d^{m} \theta \over (2\pi)^{m}} $$
every time when the integration with respect to $\bm{\theta}$
is expected.

\vspace{0.2cm}

 For a regular Hamiltonian submanifold $\Lambda$ equipped with
a minimal set of commuting integrals
$(I^{1}, \dots, I^{m+s})$ we can claim then, that:

1) The functionals $J^{\gamma}$ commute with each other and
with the functional
$$J_{H} \,\, = \,\,
\int_{0}^{2\pi}\!\!\!\!\!\dots\int_{0}^{2\pi}
P_{H} \left( \bm{\varphi}, \,
k^{\beta_{1}}_{1} \, \bm{\varphi}_{\theta^{\beta_{1}}}, \dots,
k^{\beta_{d}}_{d} \, \bm{\varphi}_{\theta^{\beta_{d}}},
\dots \right) \, {d^{m} \theta \over (2\pi)^{m}} $$
according to the Hamiltonian structure (\ref{PBronTorus});

2) The values of the functionals $J^{\gamma}$ on 
${\hat \Lambda}_{{\bf k}_{1}, \dots, {\bf k}_{d}}$ coincide with
the corresponding values of $I^{\gamma}$ on the submanifolds
$\Lambda_{{\bf k}_{1}, \dots, {\bf k}_{d}}$;

3) The Hamiltonian flows, generated by the functionals
$J^{\gamma}$ on the space of smooth functions on $\mathbb{T}^{m}$
according to bracket (\ref{PBronTorus}), leave invariant the
submanifold ${\hat \Lambda}_{{\bf k}_{1}, \dots, {\bf k}_{d}}$,
generating the linear evolution of the phase shifts
$\bm{\theta}_{0}$ with the same frequencies
$\bm{\omega}^{\gamma} ({\bf k}_{1}, \dots, {\bf k}_{d}, {\bf U})$.

 Indeed, for $({\bf k}_{1}, \dots, {\bf k}_{d}) \in {\cal M}$
all the statements above just follow from the isomorphism of the
corresponding (Poisson) spaces of quasiperiodic functions to
the space of smooth functions on $\mathbb{T}^{m}$ with the
Poisson brackets (\ref{PBronTorus}). Just by continuity we also
obtain the same statements for arbitrary 
$({\bf k}_{1}, \dots, {\bf k}_{d})$. Let us note, however,
that in non-generic situation
$({\bf k}_{1}, \dots, {\bf k}_{d}) \notin {\cal M}$ the space
of quasiperiodic functions with the wave numbers
$({\bf k}_{1}, \dots, {\bf k}_{d})$, defined in our way, 
does not coincide actually with the space of the smooth functions 
on $\mathbb{T}^{m}$.

 The functions ${\bf v}^{(k)} (\bm{\theta})$, given by the
relations
\begin{equation}
\label{vkitheta}
v^{(k)}_{i} (\bm{\theta}) \,\,\, = \,\,\,
\sum_{\gamma=1}^{m+s} \gamma^{k}_{\gamma}
({\bf k}_{1}, \dots, {\bf k}_{d}, \, {\bf U}) \,\, \left.
{\delta J^{\gamma} \over \delta \varphi^{i} (\bm{\theta})}
\right|_{{\hat \Lambda}} 
\end{equation}
obviously define annihilators of the corresponding Hamiltonian
operators (\ref{OperatorB}) on $\mathbb{T}^{m}$. Moreover, for
$({\bf k}_{1}, \dots, {\bf k}_{d}) \in {\cal M}$ we can claim that
the functions (\ref{vkitheta}) define in fact the full set of 
annihilators of (\ref{OperatorB}) with smooth linearly
independent variation derivatives on $\mathbb{T}^{m}$.
However, it can be easily seen, that in non-generic situation
$({\bf k}_{1}, \dots, {\bf k}_{d}) \notin {\cal M}$ the number
of annihilators of the bracket (\ref{PBronTorus}) increases and in 
fact is equal to infinity. Thus, we can just claim here that for a
regular Hamiltonian submanifold $\Lambda$ equipped with a minimal
set of commuting integrals $(I^{1}, \dots, I^{m+s})$ the values
(\ref{vkitheta}) define the full set of the ``regular''
annihilators of the corresponding bracket (\ref{PBronTorus}),
smoothly depending on the parameters 
$({\bf k}_{1}, \dots, {\bf k}_{d}, \, {\bf U})$.

 Let us call here the bracket (\ref{PBronTorus}) the Poisson
bracket induced by the bracket (\ref{MultDimPBr}) and the set
$({\bf k}_{1}, \dots, {\bf k}_{d})$ on $\mathbb{T}^{m}$.

\vspace{2mm}

 We want to construct now on each submanifold
${\hat \Lambda}_{{\bf k}_{1}, \dots, {\bf k}_{d}}$ 
a Poisson bracket connected with
the Poisson bracket (\ref{PBronTorus}), defined on the space of
all smooth functions on $\mathbb{T}^{m}$. To construct the bracket
on ${\hat \Lambda}_{{\bf k}_{1}, \dots, {\bf k}_{d}}$ we are
going to use the Dirac procedure of restriction of a Poisson
bracket on a submanifold in a functional space. Thus, we will
try here to make the Dirac restriction of the bracket
(\ref{PBronTorus}) on the corresponding submanifold
${\hat \Lambda}_{{\bf k}_{1}, \dots, {\bf k}_{d}}$. As a result,
we have to obtain a Poisson bracket on the space of parameters
$(U^{1}, \dots, U^{m+s}, \, \theta_{0}^{1}, \dots, \theta_{0}^{m})$.

 Let us now discuss in detail the procedure of restriction of
bracket (\ref{PBronTorus}) on
${\hat \Lambda}_{{\bf k}_{1}, \dots, {\bf k}_{d}}$. From now on
we will assume here that the family $\Lambda$ represents a regular
Hamiltonian submanifold in the space of quasiperiodic functions
equipped with a minimal set of commuting integrals
$\{I^{1}, \dots, I^{m+s}\}$.

\vspace{0.2cm}

 For the Dirac restriction of the bracket (\ref{PBronTorus}) on
${\hat \Lambda}_{{\bf k}_{1}, \dots, {\bf k}_{d}}$ we need a set
of functionals  \linebreak
$\{G^{1}, \dots, G^{2m+s}\}$ possessing the
following properties:

 1) The values of the functionals $\{G^{1}, \dots, G^{2m+s}\}$
represent a coordinate system on
${\hat \Lambda}_{{\bf k}_{1}, \dots, {\bf k}_{d}}$;

 2) The Hamiltonian flows generated by the functionals
$\{G^{1}, \dots, G^{2m+s}\}$ according to bracket
(\ref{PBronTorus}) leave invariant the submanifold
${\hat \Lambda}_{{\bf k}_{1}, \dots, {\bf k}_{d}}$.

\vspace{0.2cm}

 The Dirac restriction of bracket (\ref{PBronTorus}) is given 
then in the coordinate system \linebreak
$(G^{1}, \dots, G^{2m+s})$ on
${\hat \Lambda}_{{\bf k}_{1}, \dots, {\bf k}_{d}}$ just by the
pairwise Poisson brackets of the functionals
$\{G^{1}, \dots, G^{2m+s}\}$, restricted on
${\hat \Lambda}_{{\bf k}_{1}, \dots, {\bf k}_{d}}$.

 It is easy to see that the functionals
$\{J^{1}, \dots, J^{m+s}\}$ can be used as a part of the set
$\{G^{1}, \dots, G^{2m+s}\}$ giving the coordinates
$(U^{1}, \dots, U^{m+s})$ on the submanifold
${\hat \Lambda}_{{\bf k}_{1}, \dots, {\bf k}_{d}}$. We need,
however, to construct also the other part of the set
$\{G^{1}, \dots, G^{2m+s}\}$, representing the coordinates
$(\theta_{0}^{1}, \dots, \theta_{0}^{m})$ on
${\hat \Lambda}_{{\bf k}_{1}, \dots, {\bf k}_{d}}$.

 We have to start with the remark that the values 
$\theta^{\alpha}_{0}$ represent the cyclic coordinates on
${\hat \Lambda}_{{\bf k}_{1}, \dots, {\bf k}_{d}}$ defined modulo
$2 \pi n^{\alpha}$, $\, n^{\alpha} \in \mathbb{Z}$. So, we have
to introduce in fact a set of local maps to define local
coordinates on the torus $\mathbb{T}^{m}$. Easy to see that we 
can cover $\mathbb{T}^{m}$ by a set of maps diffeomorphic to
$[0, 1]^{m} \subset \mathbb{R}^{m}$ and choose some definite 
values $\bm{\theta}_{0} \in \mathbb{R}^{m}$ in every map among
the discrete set defined by the cyclic coordinates on 
$\mathbb{T}^{m}$.

 On the next step we have to define functionals representing
the coordinates $\bm{\theta}_{0}$ on 
${\hat \Lambda}_{{\bf k}_{1}, \dots, {\bf k}_{d}}$. According to 
our remark above, we will actually define them locally in the 
vicinity of every point 
$({\bf U}, \, \bm{\theta}_{0})$ of the submanifold
${\hat \Lambda}_{{\bf k}_{1}, \dots, {\bf k}_{d}}$. Let us 
consider the functionals
$$\vartheta_{\alpha} \,\, = \,\, 
\int_{0}^{2\pi}\!\!\!\!\!\dots\int_{0}^{2\pi} \sum_{i=1}^{n} \,
\varphi^{i} (\bm{\theta}) \,\, 
\Phi^{i}_{\theta^{\alpha}} \left( \bm{\theta}, \,
{\bf k}_{1}, \dots, {\bf k}_{d}, \, J^{1}, \dots, J^{m+s} \right)
\,\, {d^{m} \theta \over (2\pi)^{m}} $$
$(\alpha = 1, \dots, m), \,$ on the space of smooth functions 
$\bm{\varphi} (\bm{\theta})$
on $\mathbb{T}^{m}$. 

 Substituting the functions $\bm{\varphi} (\bm{\theta})$
on $\mathbb{T}^{m}$ in the form
\begin{equation}
\label{varphirepr}
\varphi^{i} (\bm{\theta}) \,\, = \,\, \Phi^{i}
\left( \bm{\theta} + \bm{\theta}_{0}, \,
{\bf k}_{1}, \dots, {\bf k}_{d}, \, {\bf U} \right)
\end{equation}
we can see that the values of $\vartheta_{\alpha}$ on 
${\hat \Lambda}_{{\bf k}_{1}, \dots, {\bf k}_{d}}$ are equal to 
zero for 
$(\theta_{0}^{1}, \dots, \theta_{0}^{m}) = (0, \dots, 0)$
while the Jacobian 
$| \partial \vartheta_{\alpha} / \partial \theta_{0}^{\beta} |$
is given at 
$(\theta_{0}^{1}, \dots, \theta_{0}^{m}) = (0, \dots, 0)$
by the determinant ${\rm det} \, || M_{\alpha\beta} ||$, where
$$M_{\alpha\beta} \,\, = \,\,
\int_{0}^{2\pi}\!\!\!\!\!\dots\int_{0}^{2\pi} \sum_{i=1}^{n} \,
\Phi^{i}_{\theta^{\beta}} \left( \bm{\theta}, 
\, {\bf k}_{1}, \dots, {\bf k}_{d}, \, {\bf U} \right) \,\,
\Phi^{i}_{\theta^{\alpha}} \left( \bm{\theta}, 
\, {\bf k}_{1}, \dots, {\bf k}_{d}, \, {\bf U} \right) \,\,
{d^{m} \theta \over (2\pi)^{m}} $$

 According to the definition of the submanifold
${\hat \Lambda}_{{\bf k}_{1}, \dots, {\bf k}_{d}}$ we then have
\begin{equation}
\label{detMab}
{\rm det} \, || M_{\alpha\beta} || \,\,\, \neq \,\,\, 0 
\end{equation} 
so the mapping
\begin{equation}
\label{thetavarthetatrans}
(\theta_{0}^{1}, \dots, \theta_{0}^{m}) \,\,\, \rightarrow \,\,\,
\left( \left. \vartheta_{1} 
\right|_{{\hat \Lambda}_{{\bf k}_{1}, \dots, {\bf k}_{d}}}, 
\, \dots \, ,
\,\, \left. \vartheta_{m}
\right|_{{\hat \Lambda}_{{\bf k}_{1}, \dots, {\bf k}_{d}}} \right)
\end{equation}
is locally invertible.

 We can claim then that there exists a constant $K_{\bf U} > 0$
such that for the values $\theta^{\alpha}_{0}$, satisfying the
relation
\begin{equation}
\label{Smalltheta0}
- K_{\bf U} \,\, < \,\, \theta^{\alpha}_{0} \,\, < \,\, K_{\bf U}
\end{equation}
the transformations (\ref{thetavarthetatrans}) are invertible
in the neighborhood of the point ${\bf U}$ in the 
${\bf U}$-space, which we consider.

 Thus, we can locally write
$$\theta^{\alpha}_{0} \,\,\, = \,\,\, \tau^{\alpha}
\left( \left. \vartheta_{1}
\right|_{{\hat \Lambda}_{{\bf k}_{1}, \dots, {\bf k}_{d}}},
\, \dots \, ,
\,\, \left. \vartheta_{m}
\right|_{{\hat \Lambda}_{{\bf k}_{1}, \dots, {\bf k}_{d}}}, 
\,\, {\bf U} \right) $$
near the point $(U^{1}, \dots, U^{m+s}, \, 0, \dots, 0)$
of the submanifold
${\hat \Lambda}_{{\bf k}_{1}, \dots, {\bf k}_{d}}$.

 As a result, we can say, that the variables 
$\theta^{\alpha}_{0}$ on
${\hat \Lambda}_{{\bf k}_{1}, \dots, {\bf k}_{d}}$ can be
represented in the vicinity of the point
$(U^{1}, \dots, U^{m+s}, \, 0, \dots, 0)$ of the submanifold
${\hat \Lambda}_{{\bf k}_{1}, \dots, {\bf k}_{d}}$ as
the values of the functionals
$\tau^{\alpha} (\vartheta_{1}, \dots, \vartheta_{m}, \,
J^{1}, \dots, J^{m+s})$ on the functions (\ref{varphirepr}).

 In the same way, for any point of $\mathbb{T}^{m}$, having
coordinates
$(\theta_{0}^{1}, \dots, \theta_{0}^{m}) =
(\zeta^{1}, \dots, \zeta^{m})$ in some local map, we can
introduce the functionals
$$\vartheta_{\alpha}^{[\bm{\zeta}]} \,\, = \,\,
\int_{0}^{2\pi}\!\!\!\!\!\dots\int_{0}^{2\pi} \sum_{i=1}^{n} \,
\varphi^{i} (\bm{\theta}) \,\, 
\Phi^{i}_{\theta^{\alpha}} \left( \bm{\theta} + \bm{\zeta}, \,
{\bf k}_{1}, \dots, {\bf k}_{d}, \, J^{1}, \dots, J^{m+s} \right)
\,\, {d^{m} \theta \over (2\pi)^{m}} $$
$(\alpha = 1, \dots, m), \,$ and put
$$\theta^{\alpha}_{0} \,\,\, = \,\,\, \zeta^{\alpha} \,\, + \,\,
\tau^{\alpha} \left( \left. \vartheta_{1}^{[\bm{\zeta}]}
\right|_{{\hat \Lambda}_{{\bf k}_{1}, \dots, {\bf k}_{d}}},
\, \dots \, ,
\,\, \left. \vartheta_{m}^{[\bm{\zeta}]}
\right|_{{\hat \Lambda}_{{\bf k}_{1}, \dots, {\bf k}_{d}}},
\,\, {\bf U} \right) $$
(with the same functions $\, \tau^{\alpha}$) near the point 
$(U^{1}, \dots, U^{m+s}, \, \zeta^{1}, \dots, \zeta^{m})$
of the submanifold
${\hat \Lambda}_{{\bf k}_{1}, \dots, {\bf k}_{d}}$.
Including also the dependence on
$({\bf k}_{1}, \dots, {\bf k}_{d})$ as the parameters we can
locally represent the variables $\theta^{\alpha}_{0}$ as the
values of the functionals
$$\theta^{\alpha [\bm{\zeta}]}_{0} \,\,\, = \,\,\, 
\zeta^{\alpha} \,\, + \,\,
\tau^{\alpha} \left( \vartheta^{[\bm{\zeta}]}_{1}, 
\dots, \vartheta^{[\bm{\zeta}]}_{m}, \,
J^{1}, \dots, J^{m+s}, \, {\bf k}_{1}, \dots, {\bf k}_{d}
\right) $$
on the corresponding functions from 
${\hat \Lambda}_{{\bf k}_{1}, \dots, {\bf k}_{d}}$ near any
point \linebreak
$(U^{1}, \dots, U^{m+s}, \, \zeta^{1}, \dots, \zeta^{m}) \, $
of the submanifold
$\, {\hat \Lambda}_{{\bf k}_{1}, \dots, {\bf k}_{d}}$.

 The functionals $\, \theta^{\alpha [\bm{\zeta}]}_{0}$ give
(locally) the corresponding values of $\theta^{\alpha}_{0}$
on ${\hat \Lambda}_{{\bf k}_{1}, \dots, {\bf k}_{d}}$. In
particular, we can write for their Poisson brackets with
the functionals $J^{\gamma}$:
$$\left. \left\{ \theta^{\alpha [\bm{\zeta}]}_{0} \, , \,
J^{\gamma} \right\} 
\right|_{{\hat \Lambda}_{{\bf k}_{1}, \dots, {\bf k}_{d}}}
\,\,\, = \,\,\, \omega^{\alpha \gamma} ({\bf U}) $$
everywhere on 
${\hat \Lambda}_{{\bf k}_{1}, \dots, {\bf k}_{d}}$.

 Let us introduce the ``constraints'' 
$\, g^{i [\bm{\zeta}]} (\bm{\theta})$ defining (locally) the
submanifold ${\hat \Lambda}_{{\bf k}_{1}, \dots, {\bf k}_{d}}$.
We put
\begin{equation}
\label{gizeta}
g^{i [\bm{\zeta}]} (\bm{\theta}) \,\,\, = \,\,\,
\varphi^{i} (\bm{\theta}) \,\, - \,\, \Phi^{i} \left(
\bm{\theta} + \bm{\theta}_{0}^{[\bm{\zeta}]}, \,
{\bf k}_{1}, \dots, {\bf k}_{d}, \, J^{1}, \dots, J^{m+s}
\right) 
\end{equation}
near the point 
$(U^{1}, \dots, U^{m+s}, \, \zeta^{1}, \dots, \zeta^{m})$
of the submanifold
${\hat \Lambda}_{{\bf k}_{1}, \dots, {\bf k}_{d}}$.

 The functionals $\, g^{i [\bm{\zeta}]} (\bm{\theta})$ are
``numerated'' by the index $i = 1, \dots, n$ and the
``continuous index'' $\bm{\theta} \in \mathbb{T}^{m}$ and are
defined in the same region as the functionals
$\, \theta^{\alpha [\bm{\zeta}]}_{0}$ in the functional space.
The equations 
$$g^{i [\bm{\zeta}]} (\bm{\theta}) \,\,\, = \,\,\, 0 $$
define the submanifold 
$\, {\hat \Lambda}_{{\bf k}_{1}, \dots, {\bf k}_{d}}$ 
near its ``point'' with coordinates 
$(U^{1}, \dots, U^{m+s}, \, \zeta^{1}, \dots, \zeta^{m})$.
Let us denote here by 
$\, \Omega^{[\bm{\zeta}]} \, \subset \,
{\hat \Lambda}_{{\bf k}_{1}, \dots, {\bf k}_{d}}$
the corresponding part of the submanifold 
$\, {\hat \Lambda}_{{\bf k}_{1}, \dots, {\bf k}_{d}}$
containing the functions
$\, \bm{\Phi} ( \bm{\theta} + \bm{\theta}_{0}, 
{\bf k}_{1}, \dots, {\bf k}_{d}, {\bf U}) \, \in \,
{\hat \Lambda}_{{\bf k}_{1}, \dots, {\bf k}_{d}} \, $
close to the function
$\, \bm{\Phi} ( \bm{\theta} + \bm{\zeta},
{\bf k}_{1}, \dots, {\bf k}_{d}, {\bf U})$.

 The constraints (\ref{gizeta}) are obviously dependent since
the following identities take place for the ``gradients'' of
$\, g^{i [\bm{\zeta}]} (\bm{\theta})$ on
${\hat \Lambda}_{{\bf k}_{1}, \dots, {\bf k}_{d}}$:
\begin{equation}
\label{Jconstrdep}
\int_{0}^{2\pi}\!\!\!\!\!\dots\int_{0}^{2\pi}  \left. 
{\delta J^{\gamma} \over \delta \varphi^{i} (\bm{\theta})}
\right|_{{\hat \Lambda}_{{\bf k}_{1}, \dots, {\bf k}_{d}}}
\left. {\delta g^{i [\bm{\zeta}]} (\bm{\theta}) \over
\delta \varphi^{j} (\bm{\theta}^{\prime})} 
\right|_{{\hat \Lambda}_{{\bf k}_{1}, \dots, {\bf k}_{d}}}
\,\, {d^{m} \theta \over (2\pi)^{m}} \,\,\, \equiv \,\,\, 0
\,\,\, , \,\,\,\,\,\,\,\, \gamma = 1, \dots, m + s
\end{equation}
\begin{equation}
\label{thetaconstrdep}
\int_{0}^{2\pi}\!\!\!\!\!\dots\int_{0}^{2\pi}  \left.
{\delta \theta^{\alpha [\bm{\zeta}]}_{0}
\over \delta \varphi^{i} (\bm{\theta})}
\right|_{{\hat \Lambda}_{{\bf k}_{1}, \dots, {\bf k}_{d}}}
\left. {\delta g^{i [\bm{\zeta}]} (\bm{\theta}) \over
\delta \varphi^{j} (\bm{\theta}^{\prime})} 
\right|_{{\hat \Lambda}_{{\bf k}_{1}, \dots, {\bf k}_{d}}}
\,\, {d^{m} \theta \over (2\pi)^{m}} \,\,\, \equiv \,\,\, 0 
\,\,\, , \,\,\,\,\,\,\,\, \alpha = 1, \dots, m 
\end{equation}

 For our purposes we will not need in fact to construct an
independent system of constraints and will use system
(\ref{gizeta}) everywhere below.

\vspace{0.2cm}

 Let us make now one important remark. Namely, we can put in 
fact one more requirement on the functionals
$\, \theta^{\alpha [\bm{\zeta}]}_{0}$ giving the coordinates
$\theta^{\alpha}_{0}$
on ${\hat \Lambda}_{{\bf k}_{1}, \dots, {\bf k}_{d}}$.
Thus, we can actually define the functionals
$\, \theta^{\alpha [\bm{\zeta}]}_{0}$ in such a way, that
for the functions
$$h^{\alpha [\bm{\zeta}]}_{i} \left( \bm{\theta}; \,
{\bf U}, \bm{\theta}_{0} \right) \,\,\,\,\, \equiv \,\,\,\,\,
\left. {\delta \theta^{\alpha [\bm{\zeta}]}_{0}
\over \delta \varphi^{i} (\bm{\theta})}
\right|_{{\hat \Lambda}_{{\bf k}_{1}, \dots, {\bf k}_{d}}}
\,\,\, = \,\,\,\,\, 
\left. {\delta \theta^{\alpha [\bm{\zeta}]}_{0}
\over \delta \varphi^{i} (\bm{\theta})} 
\right|_{\bm{\varphi} = \bm{\Phi}_{[{\bf U}, \bm{\theta}_{0}]}}$$
we have the relations
$$h^{\alpha [\bm{\zeta}]}_{i} \left( \bm{\theta}; \,
{\bf U}, \bm{\zeta} + \Delta \bm{\theta}_{0} \right)
\,\,\,\,\, = \,\,\,\,\, h^{\alpha [\bm{\zeta}]}_{i} \left( 
\bm{\theta} + \Delta \bm{\theta}_{0} ; \,
{\bf U}, \bm{\zeta} \right) $$
provided that 
$\bm{\Phi}_{[{\bf U}, \bm{\zeta} + \Delta \bm{\theta}_{0}]} 
\, \in \, \Omega^{[\bm{\zeta}]}$.

 In other words, we can require the relations 
\begin{equation}
\label{thetagradinv}
\left. {\delta \theta^{\alpha [\bm{\zeta}]}_{0}
\over \delta \varphi^{i} (\bm{\theta} + \Delta \bm{\theta}_{0})}
\right|_{\bm{\varphi} = \bm{\Phi}_{[{\bf U}, \bm{\zeta}]}}
\,\,\, = \,\,\,\,\, \left. {\delta \theta^{\alpha [\bm{\zeta}]}_{0}
\over \delta \varphi^{i} (\bm{\theta})}
\right|_{\bm{\varphi} = \bm{\Phi}_{[{\bf U}, \bm{\zeta} 
+ \Delta \bm{\theta}_{0}]}} \,\,\, ,
\end{equation}
$\bm{\theta} \, \in \, [0, 2 \pi)$, 
$\, \bm{\Phi}_{[{\bf U}, \bm{\zeta} + \Delta \bm{\theta}_{0}]}
\, \in \, \Omega^{[\bm{\zeta}]}$, where
$\, \bm{\Phi}_{[{\bf U}, \bm{\theta}_{0}]} \, \in \,
{\hat \Lambda}_{{\bf k}_{1}, \dots, {\bf k}_{d}}$ represents the
corresponding ``point''
$$\bm{\varphi} (\bm{\theta}) \,\,\, = \,\,\,
\bm{\Phi} \left( \bm{\theta} + \bm{\theta}_{0}, \,
{\bf k}_{1}, \dots, {\bf k}_{d}, \, {\bf U} \right) $$
of the submanifold
${\hat \Lambda}_{{\bf k}_{1}, \dots, {\bf k}_{d}}$.

 Indeed, using the functions 
$\, h^{\alpha [\bm{\zeta}]}_{i} 
(\bm{\theta}; \, {\bf U}, \bm{\zeta})$
we can redefine the functionals 
$\, \theta^{\alpha [\bm{\zeta}]}_{0}$ putting
$${\tilde \theta}^{\alpha [\bm{\zeta}]}_{0} \,\,\, = \,\,\,
\theta^{\alpha [\bm{\zeta}]}_{0} \,\, + \,\,
\int_{0}^{2\pi}\!\!\!\!\!\dots\int_{0}^{2\pi}
h^{\alpha [\bm{\zeta}]}_{j} \left(
\bm{\theta}^{\prime} + \bm{\theta}_{0}^{[\bm{\zeta}]} -
\bm{\zeta}; \,\, {\bf J}, \bm{\zeta} \right) \,
g^{j [\bm{\zeta}]} (\bm{\theta}^{\prime}) \,\,
{d^{m} \theta^{\prime} \over (2\pi)^{m}} $$

 Using the relations
$${\delta g^{j [\bm{\zeta}]} (\bm{\theta}^{\prime}) \over
\delta \varphi^{i} (\bm{\theta})} \,\,\,\,\, = \,\,\,\,\,
\delta^{j}_{i} \,\, \delta (\bm{\theta}^{\prime} - \bm{\theta})
\,\,\, -  \,\,\, 
\Phi^{j}_{\theta^{\beta}} \left( \bm{\theta}^{\prime} + 
\bm{\theta}^{[\bm{\zeta}]}_{0}, \, {\bf J} \right) \,
{\delta \theta^{\beta [\bm{\zeta}]}_{0}
\over \delta \varphi^{i} (\bm{\theta})} \,\,\, -  \,\,\,
\Phi^{j}_{U^{\gamma}} \left( \bm{\theta}^{\prime} +  
\bm{\theta}^{[\bm{\zeta}]}_{0}, \, {\bf J} \right) \,
{\delta J^{\gamma} \over \delta \varphi^{i} (\bm{\theta})} $$
we can write
\begin{multline*}
\left. {\delta {\tilde \theta}^{\alpha [\bm{\zeta}]}_{0}
\over \delta \varphi^{i} (\bm{\theta})} 
\right|_{{\hat \Lambda}_{{\bf k}_{1}, \dots, {\bf k}_{d}}}
\,\,\,\,\, = \,\,\,\,\,
\left. {\delta \theta^{\alpha [\bm{\zeta}]}_{0}
\over \delta \varphi^{i} (\bm{\theta})}
\right|_{{\hat \Lambda}_{{\bf k}_{1}, \dots, {\bf k}_{d}}}
\,\,\,\,\, + \,\,\,\,\, h^{\alpha [\bm{\zeta}]}_{i} \left( 
\bm{\theta} \, + \, \bm{\theta}^{[\bm{\zeta}]}_{0} \, - \, \bm{\zeta}; 
\,\, {\bf J}, \, \bm{\zeta} \right) \,\, -  \\
- \,\, \int_{0}^{2\pi}\!\!\!\!\!\dots\int_{0}^{2\pi} 
h^{\alpha [\bm{\zeta}]}_{j} (\bm{\theta}^{\prime} + 
\bm{\theta}^{[\bm{\zeta}]}_{0} - \bm{\zeta}; \, {\bf J}, \bm{\zeta}) 
\,\, \Phi^{j}_{\theta^{\beta}} ( \bm{\theta}^{\prime} +  
\bm{\theta}^{[\bm{\zeta}]}_{0}, \, {\bf J} ) \,\,
{d^{m} \theta^{\prime} \over (2\pi)^{m}} \,\,\,
\left. {\delta \theta^{\beta [\bm{\zeta}]}_{0}
\over \delta \varphi^{i} (\bm{\theta})}
\right|_{{\hat \Lambda}_{{\bf k}_{1}, \dots, {\bf k}_{d}}}
\,\,\,\,\, - \\
- \,\, \int_{0}^{2\pi}\!\!\!\!\!\dots\int_{0}^{2\pi} 
h^{\alpha [\bm{\zeta}]}_{j} (\bm{\theta}^{\prime} +
\bm{\theta}^{[\bm{\zeta}]}_{0} - \bm{\zeta}; \, {\bf J}, \bm{\zeta}) 
\,\, \Phi^{j}_{U^{\gamma}} ( \bm{\theta}^{\prime} +
\bm{\theta}^{[\bm{\zeta}]}_{0}, \, {\bf J} ) \,\,
{d^{m} \theta^{\prime} \over (2\pi)^{m}} \,\,\,
\left. {\delta J^{\gamma} \over \delta \varphi^{i} (\bm{\theta})}
\right|_{{\hat \Lambda}_{{\bf k}_{1}, \dots, {\bf k}_{d}}}
\end{multline*}

 Using obvious relations
\begin{multline*}
\int_{0}^{2\pi}\!\!\!\!\!\dots\int_{0}^{2\pi} 
h^{\alpha [\bm{\zeta}]}_{j} (\bm{\theta}^{\prime} +
\bm{\theta}^{[\bm{\zeta}]}_{0} - \bm{\zeta}; \, {\bf J}, \bm{\zeta}) 
\,\, \Phi^{j}_{\theta^{\beta}} ( \bm{\theta}^{\prime} +
\bm{\theta}^{[\bm{\zeta}]}_{0}, \, {\bf J} )
\,\, {d^{m} \theta^{\prime} \over (2\pi)^{m}} \,\,\, = \\
= \,\,
\int_{0}^{2\pi}\!\!\!\!\!\dots\int_{0}^{2\pi}
h^{\alpha [\bm{\zeta}]}_{j} (\bm{\theta}^{\prime}; 
\, {\bf J}, \bm{\zeta}) \,\,
\Phi^{j}_{\theta^{\beta}} ( \bm{\theta}^{\prime} +
\bm{\zeta}, \, {\bf J} ) \,\,
{d^{m} \theta^{\prime} \over (2\pi)^{m}} \,\,\,\,\, \equiv
\,\,\,\,\, \delta^{\alpha}_{\beta}
\end{multline*}
\begin{multline*}
\int_{0}^{2\pi}\!\!\!\!\!\dots\int_{0}^{2\pi} 
h^{\alpha [\bm{\zeta}]}_{j} (\bm{\theta}^{\prime} +
\bm{\theta}^{[\bm{\zeta}]}_{0} - \bm{\zeta}; \, {\bf J}, \bm{\zeta}) 
\,\, \Phi^{j}_{U^{\gamma}} ( \bm{\theta}^{\prime} +
\bm{\theta}^{[\bm{\zeta}]}_{0}, \, {\bf J} ) \,\,
{d^{m} \theta^{\prime} \over (2\pi)^{m}} \,\,\, = \\
= \,\,
\int_{0}^{2\pi}\!\!\!\!\!\dots\int_{0}^{2\pi} 
h^{\alpha [\bm{\zeta}]}_{j} (\bm{\theta}^{\prime};
\, {\bf J}, \bm{\zeta}) \,\,
\Phi^{j}_{U^{\gamma}} ( \bm{\theta}^{\prime} +
\bm{\zeta}, \, {\bf J} ) \,\,
{d^{m} \theta^{\prime} \over (2\pi)^{m}} 
\,\,\, \equiv \,\,\, 0
\end{multline*}
we then easily get
$$\left. {\delta {\tilde \theta}^{\alpha [\bm{\zeta}]}_{0}
\over \delta \varphi^{i} (\bm{\theta})} 
\right|_{{\hat \Lambda}_{{\bf k}_{1}, \dots, {\bf k}_{d}}}
\,\,\,\,\, = \,\,\,\,\,
h^{\alpha [\bm{\zeta}]}_{i} (\bm{\theta} \, + \,
\bm{\theta}^{[\bm{\zeta}]}_{0} \, - \, \bm{\zeta}; 
\,\, {\bf J}, \, \bm{\zeta}) $$

 From the relations above it is obvious now that the functionals
$\, {\tilde \theta}^{\alpha [\bm{\zeta}]}_{0}$ satisfy the required 
conditions (\ref{thetagradinv}). Easy to see also, that the 
functionals $\, {\tilde \theta}^{\alpha [\bm{\zeta}]}_{0}$ have 
exactly the same values on 
${\hat \Lambda}_{{\bf k}_{1}, \dots, {\bf k}_{d}}$ as the
functionals $\, \theta^{\alpha [\bm{\zeta}]}_{0}$.

 Let us say, that the choice of the functionals
$\, \theta^{\alpha [\bm{\zeta}]}_{0}$ with the additional
invariance property (\ref{thetagradinv}) is in fact very
convenient in many cases. Thus, we will use this choice of
$\, \theta^{\alpha [\bm{\zeta}]}_{0}$ in Chapter 4,
where we are going to consider the averaging of the
Hamiltonian structures in the Whitham method.

 As a particular property of the functionals 
$\, \theta^{\alpha [\bm{\zeta}]}_{0}$, satisfying the
requirement (\ref{thetagradinv}), we can note the fact that
their pairwise Poisson brackets on
${\hat \Lambda}_{{\bf k}_{1}, \dots, {\bf k}_{d}}$  do not
depend actually on the coordinates $\bm{\theta}_{0}$. Indeed,
according to (\ref{thetagradinv}) the change of the
coordinates $\bm{\theta}_{0}$ on
${\hat \Lambda}_{{\bf k}_{1}, \dots, {\bf k}_{d}}$ produces
just the corresponding shift of $\bm{\theta}$ in the
``gradients'' of $\, \theta^{\alpha [\bm{\zeta}]}_{0}$.
Easy to see, that the same shift arises also in the 
coefficients of the operator 
${\hat B}^{ij}_{{\bf k}_{1}, \dots, {\bf k}_{d}}$, so it will
disappear after the integration w.r.t. $\bm{\theta}$.
Let us note also, that the functionals $J^{\gamma}$ satisfy
automatically the requirement
(\ref{thetagradinv}) view their invariance with respect to
the transformation $\bm{\theta} \, \rightarrow \, 
\bm{\theta} + \Delta \bm{\theta}$ on the functional space.

 In this chapter the requirement (\ref{thetagradinv}) will 
not actually play important role, so we do not impose it here.

\vspace{0.2cm}

 Finally, we have to construct the functionals for the
coordinates $\, \theta^{\alpha}_{0}$, possessing an additional 
property. Namely, we have to require that the Hamiltonian 
flows, generated by these functionals, leave invariant the 
submanifold ${\hat \Lambda}_{{\bf k}_{1}, \dots, {\bf k}_{d}}$.

\vspace{0.2cm}

 We have to modify now the functionals 
$\, \theta^{\alpha [\bm{\zeta}]}_{0}$ with the aid of the
constraints $\, g^{i [\bm{\zeta}]} (\bm{\theta})$ to get
the functionals commuting with all
$\, g^{i [\bm{\zeta}]} (\bm{\theta})$ on 
${\hat \Lambda}_{{\bf k}_{1}, \dots, {\bf k}_{d}}$.
Thus, we have to put now
\begin{equation}
\label{Invthetafunc}
\hat{\theta}^{\alpha [\bm{\zeta}]}_{0} \,\,\,\,\, = \,\,\,\,\,
\theta^{\alpha [\bm{\zeta}]}_{0} \,\,\, + \,\,\,
\int_{0}^{2\pi}\!\!\!\!\!\dots\int_{0}^{2\pi} 
g^{j [\bm{\zeta}]} (\bm{\theta}) \,\, 
\beta^{\alpha [\bm{\zeta}]}_{j} \left( \bm{\theta}; \,
{\bf J}, \bm{\theta}^{[\bm{\zeta}]}_{0} \right) \,
{d^{m} \theta \over (2\pi)^{m}} 
\end{equation}
where the functions 
$\beta^{\alpha [\bm{\zeta}]}_{i} (\bm{\theta}; \,
{\bf U}, \bm{\theta}_{0})$ are smooth
$2\pi$-periodic in each $\theta^{\alpha}$ functions defined by
the relations
$$\int_{0}^{2\pi}\!\!\!\!\!\dots\int_{0}^{2\pi}
\left. \left\{ g^{i [\bm{\zeta}]} (\bm{\theta}) \, , \,
g^{j [\bm{\zeta}]} (\bm{\theta}^{\prime}) \right\}
\right|_{{\hat \Lambda}_{{\bf k}_{1}, \dots, {\bf k}_{d}}} \,\,
\beta^{\alpha [\bm{\zeta}]}_{j} \left( \bm{\theta}^{\prime}; \,
{\bf U}, \bm{\theta}_{0} \right) \,\,
{d^{m} \theta^{\prime} \over (2\pi)^{m}} \,\,\,  = $$ 
\begin{equation}
\label{Systbeta}
= \,\,\, - \, 
\left. \left\{ g^{i [\bm{\zeta}]} (\bm{\theta}) \, , \,
\theta^{\alpha [\bm{\zeta}]}_{0} \right\}
\right|_{{\hat \Lambda}_{{\bf k}_{1}, \dots, {\bf k}_{d}}}
\end{equation}

 According to (\ref{Systbeta}) the functions
$\beta^{\alpha [\bm{\zeta}]}_{i} (\bm{\theta}; \,
{\bf U}, \bm{\theta}_{0})$ depend on the
coordinates $({\bf U}, \bm{\theta}_{0})$ on
${\hat \Lambda}_{{\bf k}_{1}, \dots, {\bf k}_{d}}$ in the vicinity
of the point 
$(U^{1}, \dots, U^{m+s}, \, \zeta^{1}, \dots, \zeta^{m})$
of the submanifold 
${\hat \Lambda}_{{\bf k}_{1}, \dots, {\bf k}_{d}}$.

 As we can see, the functions 
$\beta^{\alpha [\bm{\zeta}]}_{i} (\bm{\theta}; \,
{\bf U}, \bm{\theta}_{0})$ should satisfy at every
$({\bf U}, \bm{\theta}_{0})$ a linear system, where the ``matrix''
of the system is given by the pairwise Poisson brackets of the
constraints $g^{i [\bm{\zeta}]} (\bm{\theta})$ on
${\hat \Lambda}_{{\bf k}_{1}, \dots, {\bf k}_{d}}$. It's not 
difficult to check that the corresponding brackets can be 
represented in the form
\begin{multline}
\label{ConstrBr}
\left. \left\{ g^{i [\bm{\zeta}]} (\bm{\theta}) \, , \,
g^{j [\bm{\zeta}]} (\bm{\theta}^{\prime}) \right\}
\right|_{{\hat \Lambda}_{{\bf k}_{1}, \dots, {\bf k}_{d}}} 
\,\,\, = \,\,\,\,\,
\left. \left\{ \varphi^{i} (\bm{\theta}) \, , \,
\varphi^{j} (\bm{\theta}^{\prime}) \right\}
\right|_{{\hat \Lambda}_{{\bf k}_{1}, \dots, {\bf k}_{d}}} 
\,\,\, -   \\
- \,\, \Phi^{i}_{\theta^{\beta}} \left( 
\bm{\theta} + \bm{\theta}_{0}, \, {\bf k}_{1}, \dots, {\bf k}_{d},
{\bf U} \right) \, \left.
\left\{ \theta^{\beta [\bm{\zeta}]}_{0} \, , \,
\varphi^{j} (\bm{\theta}^{\prime}) \right\}
\right|_{{\hat \Lambda}_{{\bf k}_{1}, \dots, {\bf k}_{d}}}
\,\,\, -   \\
- \,\, \left. \left\{ \varphi^{i} (\bm{\theta}) \, , \,
\theta^{\gamma [\bm{\zeta}]}_{0} \right\}
\right|_{{\hat \Lambda}_{{\bf k}_{1}, \dots, {\bf k}_{d}}} \,
\Phi^{j}_{\theta^{\gamma}} \left(
\bm{\theta}^{\prime} + \bm{\theta}_{0}, \, 
{\bf k}_{1}, \dots, {\bf k}_{d},
{\bf U} \right) \,\,\, +  \\
+ \,\, \Phi^{i}_{\theta^{\beta}} \left(
\bm{\theta} + \bm{\theta}_{0}, \, {\bf k}_{1}, \dots, {\bf k}_{d},
{\bf U} \right) \, \left.
\left\{ \theta^{\beta [\bm{\zeta}]}_{0} \, , \,
\theta^{\gamma [\bm{\zeta}]}_{0} \right\}
\right|_{{\hat \Lambda}_{{\bf k}_{1}, \dots, {\bf k}_{d}}} \,
\Phi^{j}_{\theta^{\gamma}} \left(
\bm{\theta}^{\prime} + \bm{\theta}_{0}, \, 
{\bf k}_{1}, \dots, {\bf k}_{d},
{\bf U} \right) 
\end{multline}

 In the same way, the right-hand part of system (\ref{Systbeta})
can be written as:
\begin{multline*}
- \, \left. \left\{ g^{i [\bm{\zeta}]} (\bm{\theta}) \, , \,  
\theta^{\alpha [\bm{\zeta}]}_{0} \right\}
\right|_{{\hat \Lambda}_{{\bf k}_{1}, \dots, {\bf k}_{d}}}
\,\,\, = \,\,\,\,\,
- \, \left. \left\{ \varphi^{i} (\bm{\theta}) \, , \,
\theta^{\alpha [\bm{\zeta}]}_{0} \right\}
\right|_{{\hat \Lambda}_{{\bf k}_{1}, \dots, {\bf k}_{d}}} 
\,\,\, +  \\
+ \,\, \Phi^{i}_{U^{\gamma}} \left(
\bm{\theta} + \bm{\theta}_{0}, \, {\bf k}_{1}, \dots, {\bf k}_{d},
{\bf U} \right) \, \left. \left\{ J^{\gamma} \, , \,
\theta^{\alpha [\bm{\zeta}]}_{0} \right\}
\right|_{{\hat \Lambda}_{{\bf k}_{1}, \dots, {\bf k}_{d}}}   
\,\,\, +  \\
+ \,\, \Phi^{i}_{\theta^{\beta}} \left(
\bm{\theta} + \bm{\theta}_{0}, \, {\bf k}_{1}, \dots, {\bf k}_{d},
{\bf U} \right) \, \left.
\left\{ \theta^{\beta [\bm{\zeta}]}_{0} \, , \,
\theta^{\alpha [\bm{\zeta}]}_{0} \right\}
\right|_{{\hat \Lambda}_{{\bf k}_{1}, \dots, {\bf k}_{d}}}
\end{multline*}

 Let us say, however, that, due to the dependence of the
constraints $g^{i [\bm{\zeta}]} (\bm{\theta})$, system 
(\ref{Systbeta}) can be reduced in fact to a simpler form.
Let us prove here the following lemma:

\vspace{0.2cm}

{\bf Lemma 2.1.}

{\it Let the functions
$\beta^{\alpha [\bm{\zeta}]}_{i} (\bm{\theta}; \,
{\bf U}, \bm{\theta}_{0})$
satisfy the system
\begin{equation}
\label{ReducedSystbeta}
{\hat B}^{ij}_{{\bf k}_{1}, \dots, {\bf k}_{d}} \,\,\,
\beta^{\alpha [\bm{\zeta}]}_{j} (\bm{\theta}; \,
{\bf U}, \bm{\theta}_{0}) \,\,\,\,\, = \,\,\,\,\,
A^{i\alpha [\bm{\zeta}]} (\bm{\theta}; \, {\bf U}, \bm{\theta}_{0})
\end{equation}
where
\begin{multline*}
A^{i\alpha [\bm{\zeta}]} (\bm{\theta}; \, {\bf U}, \bm{\theta}_{0})
\,\,\, \equiv  \\
\equiv \,\,\, 
- \, \left. \left\{ \varphi^{i} (\bm{\theta}) \, , \,
\theta^{\alpha [\bm{\zeta}]}_{0} \right\}
\right|_{{\hat \Lambda}_{{\bf k}_{1}, \dots, {\bf k}_{d}}}
\,\,\, + \,\,\, \Phi^{i}_{U^{\gamma}} \left(
\bm{\theta} + \bm{\theta}_{0}, \, {\bf k}_{1}, \dots, {\bf k}_{d},
{\bf U} \right) \, \left. \left\{ J^{\gamma} \, , \,
\theta^{\alpha [\bm{\zeta}]}_{0} \right\}
\right|_{{\hat \Lambda}_{{\bf k}_{1}, \dots, {\bf k}_{d}}} 
\end{multline*}

 Then 
$\beta^{\alpha [\bm{\zeta}]}_{i} (\bm{\theta}; \,
{\bf U}, \bm{\theta}_{0})$ 
automatically satisfy system (\ref{Systbeta}).
}

\vspace{0.2cm}

 Proof.

 Let us first prove the following statement:

 Any 
$\beta^{\alpha [\bm{\zeta}]}_{i} (\bm{\theta}; \,
{\bf U}, \bm{\theta}_{0})$ satisfying system (\ref{ReducedSystbeta})
automatically satisfy the relations
\begin{equation}
\label{Betaortrel}
\int_{0}^{2\pi}\!\!\!\!\!\dots\int_{0}^{2\pi}
\, \Phi^{i}_{\theta^{\lambda}} \left(
\bm{\theta} + \bm{\theta}_{0}, \, {\bf k}_{1}, \dots, {\bf k}_{d},
{\bf U} \right) \, \beta^{\alpha [\bm{\zeta}]}_{i} (\bm{\theta}; \,
{\bf U}, \bm{\theta}_{0}) \,\, {d^{m} \theta \over (2\pi)^{m}}
\,\,\,\,\, \equiv \,\,\,\,\, 0
\end{equation}
$\lambda = 1, \dots, m$.

 Indeed, due to relations (\ref{Jconstrdep}) the right-hand part
of system (\ref{Systbeta}) is always orthogonal to the variation
derivatives $\delta J^{\gamma} / \delta \varphi^{i} (\bm{\theta})$,
$\gamma = 1, \dots, m + s$. Easy to see that the same property
then also takes place for the right-hand part of system
(\ref{ReducedSystbeta}). For any solution of (\ref{ReducedSystbeta})
this automatically implies the property
$$\omega^{\lambda \gamma} ({\bf U}) \,
\int_{0}^{2\pi}\!\!\!\!\!\dots\int_{0}^{2\pi}
\, \Phi^{i}_{\theta^{\lambda}} \left(
\bm{\theta} + \bm{\theta}_{0}, \, {\bf k}_{1}, \dots, {\bf k}_{d},
{\bf U} \right) \, \beta^{\alpha [\bm{\zeta}]}_{i} (\bm{\theta}; \,
{\bf U}, \bm{\theta}_{0}) \,\, {d^{m} \theta \over (2\pi)^{m}}
\,\,\,\,\, \equiv \,\,\,\,\, 0 $$
$\gamma = 1, \dots, m + s$. 

 From the part (3) of Definition 2.2 we then immediately get
relations (\ref{Betaortrel}). 

 Using relations (\ref{ConstrBr}) for the pairwise brackets of
constraints we can now claim that for any
$\beta^{\alpha [\bm{\zeta}]}_{i} (\bm{\theta}; \,
{\bf U}, \bm{\theta}_{0})$,
satisfying system (\ref{ReducedSystbeta}), the difference
between the left- and the right-hand parts of (\ref{Systbeta})
is given just by the expression
$$- \,\, \Phi^{i}_{\theta^{\beta}} \left(
\bm{\theta} + \bm{\theta}_{0}, \, {\bf k}_{1}, \dots, {\bf k}_{d},
{\bf U} \right) \, \int_{0}^{2\pi}\!\!\!\!\!\!\dots\!\int_{0}^{2\pi} 
\! \left. \left\{ \theta^{\beta [\bm{\zeta}]}_{0} \, , \,
\varphi^{j} (\bm{\theta}^{\prime}) \right\}
\right|_{{\hat \Lambda}_{{\bf k}_{1}, \dots, {\bf k}_{d}}} \,
\beta^{\alpha [\bm{\zeta}]}_{j} (\bm{\theta}^{\prime}; \,
{\bf U}, \bm{\theta}_{0}) \,\,
{d^{m} \theta^{\prime} \over (2\pi)^{m}} \,\, -  $$
$$- \,\,\, \Phi^{i}_{\theta^{\beta}} \left(
\bm{\theta} + \bm{\theta}_{0}, \, {\bf k}_{1}, \dots, {\bf k}_{d}, 
{\bf U} \right) \,\, \left. \left\{ \theta^{\beta [\bm{\zeta}]}_{0} 
\, , \, \theta^{\alpha [\bm{\zeta}]}_{0} \right\}
\right|_{{\hat \Lambda}_{{\bf k}_{1}, \dots, {\bf k}_{d}}} $$

 From the other hand, we know from (\ref{thetaconstrdep}) that
both the left- and the right-hand parts of (\ref{Systbeta}) are 
orthogonal to the variation derivatives
$\delta \theta^{\lambda [\bm{\zeta}]}_{0} / 
\delta \varphi^{i} (\bm{\theta})$,
($\lambda = 1, \dots, m$), for any
$\beta^{\alpha [\bm{\zeta}]}_{i} (\bm{\theta}; \,
{\bf U}, \bm{\theta}_{0})$. Applying this property to the above
expression we get immediately that the difference between the
left- and the right-hand part of system (\ref{Systbeta}) is
identically equal to zero for any
$\beta^{\alpha [\bm{\zeta}]}_{i} (\bm{\theta}; \,
{\bf U}, \bm{\theta}_{0})$
satisfying system (\ref{ReducedSystbeta}).

{\hfill Lemma 2.1 is proved.}

\vspace{0.2cm}

 Let us say that it will be rather convenient to us to choose
the solutions
$\beta^{\alpha [\bm{\zeta}]}_{i} (\bm{\theta}; \,
{\bf U}, \bm{\theta}_{0})$
of system (\ref{Systbeta}) to be also solutions of
(\ref{ReducedSystbeta}) and satisfy the additional property
(\ref{Betaortrel}).

 It is easy to see that system (\ref{ReducedSystbeta}) is
resolvable on the space of smooth $2\pi$-periodic in each
$\theta^{\alpha}$ functions if and only if the system
\begin{equation}
\label{betatildesyst}
{\hat B}^{ij}_{{\bf k}_{1}, \dots, {\bf k}_{d}} \,\,\,
{\tilde \beta}^{\alpha [\bm{\zeta}]}_{j} (\bm{\theta}; \,
{\bf U}, \bm{\theta}_{0}) \,\,\,\,\, = \,\,\,\,\,
\Phi^{i}_{U^{\gamma}} \left(
\bm{\theta} + \bm{\theta}_{0}, \, {\bf k}_{1}, \dots, {\bf k}_{d},
{\bf U} \right) \, \left. \left\{ J^{\gamma} \, , \,
\theta^{\alpha [\bm{\zeta}]}_{0} \right\}
\right|_{{\hat \Lambda}_{{\bf k}_{1}, \dots, {\bf k}_{d}}}
\end{equation}
is resolvable on the same space.

 Indeed, for any solution
${\tilde \beta}^{\alpha [\bm{\zeta}]}_{i} (\bm{\theta}; \,
{\bf U}, \bm{\theta}_{0})$ of system (\ref{betatildesyst}) we can 
just put
$$\beta^{\alpha [\bm{\zeta}]}_{i} (\bm{\theta}; \,
{\bf U}, \bm{\theta}_{0}) \,\,\,\,\, = \,\,\,\,\,
{\tilde \beta}^{\alpha [\bm{\zeta}]}_{i} (\bm{\theta}; \,
{\bf U}, \bm{\theta}_{0}) \,\,\, - \,\,\,
\left. {\delta \theta^{\alpha [\bm{\zeta}]}_{0}
\over \delta \varphi^{i} (\bm{\theta})}
\right|_{{\hat \Lambda}_{{\bf k}_{1}, \dots, {\bf k}_{d}}} $$
to get a solution of (\ref{ReducedSystbeta}). Using the
expression
$$\left. \left\{ J^{\gamma} \, , \,
\theta^{\alpha [\bm{\zeta}]}_{0} \right\}
\right|_{{\hat \Lambda}_{{\bf k}_{1}, \dots, {\bf k}_{d}}}
\,\,\, = \,\,\,\,\, - \,\, \omega^{\alpha\gamma} ({\bf U}) $$
everywhere on ${\hat \Lambda}_{{\bf k}_{1}, \dots, {\bf k}_{d}}$,
we can finally formulate the following statement:

\vspace{0.2cm}

 System (\ref{ReducedSystbeta}) is resolvable on the space of 
smooth $2\pi$-periodic in each $\theta^{\alpha}$ functions if 
and only if the following ``test'' system
\begin{equation}
\label{TestSystem}
{\hat B}^{ij}_{{\bf k}_{1}, \dots, {\bf k}_{d}} \,\,\,
{\tilde \beta}^{\alpha}_{j} (\bm{\theta}; \,
{\bf U}, \bm{\theta}_{0}) \,\,\,\,\, = \,\,\,\,\, - \,\,
\omega^{\alpha\gamma} ({\bf U}) \,\, \Phi^{i}_{U^{\gamma}} 
\left( \bm{\theta} + \bm{\theta}_{0}, \, 
{\bf k}_{1}, \dots, {\bf k}_{d}, {\bf U} \right)
\end{equation}
is resolvable on the same space.

\vspace{0.2cm}

 Let us note that system (\ref{TestSystem}) is well defined
globally on the whole submanifold 
${\hat \Lambda}_{{\bf k}_{1}, \dots, {\bf k}_{d}}$ and has
absolutely identical properties for all $\bm{\theta}_{0}$
under fixed values of ${\bf U}$. In particular, the spaces of 
solutions of (\ref{TestSystem}) are evidently isomorphic for
any two coordinate sets $({\bf U}, \bm{\theta}_{0})$ and
$({\bf U}, \bm{\theta}_{0}^{\prime})$.

 Let us give here the following definition:

\vspace{0.2cm}

{\bf Definition 2.3.}

{\it Let $\Lambda$ be a regular Hamiltonian submanifold in the
space of quasiperiodic functions, equipped with a minimal set
of commuting integrals $(I^{1}, \dots, I^{m+s})$. Let us fix 
some values $({\bf k}_{1}, \dots, {\bf k}_{d})$.

 We say that the submanifold
${\hat \Lambda}_{{\bf k}_{1}, \dots, {\bf k}_{d}}$ admits
regular Dirac restriction of bracket (\ref{PBronTorus}) if the
corresponding system (\ref{TestSystem}) is resolvable on the
space of smooth $2\pi$-periodic in each $\theta^{\alpha}$ 
functions for all values of ${\bf U}$ and has a smooth
$2\pi$-periodic in each $\theta^{\alpha}$ solution
${\tilde \beta}^{\alpha}_{i} (\bm{\theta} + \bm{\theta}_{0}, \,
{\bf U})$, smoothly depending on the parameters
${\bf U}$.
}

\vspace{0.2cm}

{\bf Lemma 2.2.}

{\it Let $\Lambda$ be a regular Hamiltonian submanifold in the
space of quasiperiodic functions, equipped with a minimal set
of commuting integrals $(I^{1}, \dots, I^{m+s})$. Let system
(\ref{TestSystem}) have near every value of $\, {\bf U}$ a smooth
$2\pi$-periodic in each $\theta^{\alpha}$ solution
${\tilde \beta}^{\alpha}_{i} (\bm{\theta} + \bm{\theta}_{0}, \,
{\bf U})$, smoothly depending on the parameters
${\bf U}$. Then we can construct a global Dirac restriction of
the bracket (\ref{PBronTorus}) on the corresponding submanifold
${\hat \Lambda}_{{\bf k}_{1}, \dots, {\bf k}_{d}}$, having the
form:
\begin{equation}
\label{GlDirRest}
\left\{ \theta^{\alpha}_{0} \, , \, \theta^{\beta}_{0} \right\}
\,\, = \,\, K^{\alpha\beta} ({\bf U}, \bm{\theta}_{0})
\,\,\, , \,\,\,\,\,
\left\{ \theta^{\alpha}_{0} \, , \, U^{\gamma} \right\} \,\, = \,\, 
\omega^{\alpha\gamma} ({\bf U}) \,\,\, , \,\,\,\,\,
\left\{ U^{\gamma} \, , \, U^{\lambda} \right\} \,\, = \,\, 0
\end{equation}
with some $2\pi$-periodic in each $\theta^{\alpha}$  functions
$K^{\alpha\beta} ({\bf U}, \bm{\theta}_{0})$.
}

\vspace{0.2cm}

Proof.

As we saw above, the existence of the solution
${\tilde \beta}^{\alpha}_{i} (\bm{\theta} + \bm{\theta}_{0}, \,
{\bf U})$ 
permits to construct the corresponding solutions
$\beta^{\alpha [\bm{\zeta}]}_{i} (\bm{\theta}; \,
{\bf U}, \bm{\theta}_{0})$ of system (\ref{Systbeta}) and then
to define the local functionals 
${\hat \theta}_{0}^{\alpha [\bm{\zeta}]}$ introduced in
(\ref{Invthetafunc}). The existence of the functionals
${\hat \theta}_{0}^{\alpha [\bm{\zeta}]}$, together with the
set $(J^{1}, \dots, J^{m+s})$, permits to define the Dirac
restriction of bracket (\ref{PBronTorus}) on
${\hat \Lambda}_{{\bf k}_{1}, \dots, {\bf k}_{d}}$, which can be
locally written in the form:

$$\left\{ \theta^{\alpha}_{0} \, , \, 
\theta^{\beta}_{0} \right\}^{[\bm{\zeta}]}
\,\,\, = \,\,\, 
\left. \left\{ {\hat \theta}_{0}^{\alpha [\bm{\zeta}]} \, , \,
{\hat \theta}_{0}^{\beta [\bm{\zeta}]} \right\}
\right|_{{\hat \Lambda}_{{\bf k}_{1}, \dots, {\bf k}_{d}}}
\,\,\, = \,\,\,\,\,
K^{\alpha\beta [\bm{\zeta}]} ({\bf U}, \bm{\theta}_{0}) $$
$$ \left\{ \theta^{\alpha}_{0} \, , \,
U^{\gamma} \right\}^{[\bm{\zeta}]} \,\,\, = \,\,\,
\omega^{\alpha\gamma} ({\bf U}) 
\,\,\,\,\, , \,\,\,\,\,\,\,\,
\left\{ U^{\gamma} \, , \, U^{\lambda} \right\}^{[\bm{\zeta}]}
\,\,\, = \,\,\, 0 $$

 What we actually have to prove is that any two Dirac restrictions
$\, \{ \dots , \dots \}_{D}^{[\bm{\zeta}]}$,
$\, \{ \dots , \dots \}_{D}^{[\bm{\zeta}^{\prime}]}$,
obtained with the aid of two different sets
$\{ {\hat \theta}_{0}^{\alpha [\bm{\zeta}]} \}$ and
$\{ {\hat \theta}_{0}^{\alpha [\bm{\zeta}^{\prime}]} \}$
in two local regions $\Omega^{[\bm{\zeta}]}$ and
$\Omega^{[\bm{\zeta}^{\prime}]}$ on
${\hat \Lambda}_{{\bf k}_{1}, \dots, {\bf k}_{d}}$, define in fact
the same Poisson bracket in the intersection
$\Omega^{[\bm{\zeta}, \bm{\zeta}^{\prime}]} \, = \,
\Omega^{[\bm{\zeta}]} \cap \Omega^{[\bm{\zeta}^{\prime}]}$.

 For the proof let us first note that the values of the
functionals 
${\hat \theta}_{0}^{\beta [\bm{\zeta}]}$ and
${\hat \theta}_{0}^{\beta [\bm{\zeta}^{\prime}]}$ on
${\hat \Lambda}_{{\bf k}_{1}, \dots, {\bf k}_{d}}$ by construction
can differ just by a constant $2 \pi n^{\beta}$, 
$n^{\beta} \in \mathbb{Z}$, in any connected part of the region
$\Omega^{[\bm{\zeta}, \bm{\zeta}^{\prime}]}$. To prove the Lemma
we have to prove then that the Hamiltonian flows, defined by
${\hat \theta}_{0}^{\beta [\bm{\zeta}]}$ and
${\hat \theta}_{0}^{\beta [\bm{\zeta}^{\prime}]}$ on
${\hat \Lambda}_{{\bf k}_{1}, \dots, {\bf k}_{d}}$, coincide in
$\Omega^{[\bm{\zeta}, \bm{\zeta}^{\prime}]}$.

 Let us note, that any functional
${\hat \theta}_{0}^{\beta [\bm{\zeta}]} \, - \,
{\hat \theta}_{0}^{\beta [\bm{\zeta}^{\prime}]}$ commutes on
${\hat \Lambda}_{{\bf k}_{1}, \dots, {\bf k}_{d}}$ with all the
functionals $J^{\gamma}$, $\, \gamma = 1, \dots, m + s$, and
$\,\,\, {\hat \theta}_{0}^{\alpha [\bm{\zeta}]}$,
$\, \alpha = 1, \dots, m$, in the region
$\Omega^{[\bm{\zeta}, \bm{\zeta}^{\prime}]}$.
Indeed, by construction, the Hamiltonian flows, generated by the
functionals $J^{\gamma}$ and 
${\hat \theta}_{0}^{\alpha [\bm{\zeta}]}$ leave invariant the
submanifold ${\hat \Lambda}_{{\bf k}_{1}, \dots, {\bf k}_{d}}$,
where the functional
${\hat \theta}_{0}^{\beta [\bm{\zeta}]} \, - \,
{\hat \theta}_{0}^{\beta [\bm{\zeta}^{\prime}]}$ has locally
constant values. As a result, we can claim, that all the brackets
$$\left. \left\{ J^{\gamma} \, , \,
{\hat \theta}_{0}^{\beta [\bm{\zeta}]} \right\}
\right|_{{\hat \Lambda}_{{\bf k}_{1}, \dots, {\bf k}_{d}}}
\,\,\,\,\, , \,\,\,\,\,\,\,\,
\left. \left\{ {\hat \theta}_{0}^{\alpha [\bm{\zeta}]} \, , \,
{\hat \theta}_{0}^{\beta [\bm{\zeta}]} \right\}
\right|_{{\hat \Lambda}_{{\bf k}_{1}, \dots, {\bf k}_{d}}} $$
($\gamma = 1, \dots, m + s$, $\, \alpha = 1, \dots, m$),
coincide with the corresponding brackets
$$\left. \left\{ J^{\gamma} \, , \,
{\hat \theta}_{0}^{\beta [\bm{\zeta}^{\prime}]} \right\} 
\right|_{{\hat \Lambda}_{{\bf k}_{1}, \dots, {\bf k}_{d}}}
\,\,\,\,\, , \,\,\,\,\,\,\,\,
\left. \left\{ {\hat \theta}_{0}^{\alpha [\bm{\zeta}]} 
\, , \, {\hat \theta}_{0}^{\beta [\bm{\zeta}^{\prime}]} \right\}
\right|_{{\hat \Lambda}_{{\bf k}_{1}, \dots, {\bf k}_{d}}} $$
in the region $\Omega^{[\bm{\zeta}, \bm{\zeta}^{\prime}]}$.
Since the values 
$(U^{1}, \dots, U^{m+s}, \theta_{0}^{1}, \dots, \theta_{0}^{m})$
represent a coordinate system on
${\hat \Lambda}_{{\bf k}_{1}, \dots, {\bf k}_{d}}$, we get 
immediately the required statement.

 Considering the values $(\theta_{0}^{1}, \dots, \theta_{0}^{m})$
as the cyclic coordinates on
${\hat \Lambda}_{{\bf k}_{1}, \dots, {\bf k}_{d}}$ we then
easily obtain the assertion of the Lemma.

{\hfill Lemma 2.2 is proved.}

\vspace{0.2cm}

 Let us give also the proof of the invariance of the restricted 
bracket with respect to the choice of the functionals
$\, (I^{1}, \dots, I^{m+s})$.

\vspace{0.2cm}

{\bf Lemma 2.3.}

{\it  Let $\Lambda$ be a regular Hamiltonian submanifold in the
space of quasiperiodic functions and there exist two different
sets of commuting integrals $\, (I^{1}, \dots, I^{m+s})$,
$\, (I^{\prime 1}, \dots, I^{\prime m+s})$ satisfying all the
requirements of Definition 2.2. Let a submanifold 
${\hat \Lambda}_{{\bf k}_{1}, \dots, {\bf k}_{d}}$ satisfy the
requirements of Lemma 2.2. Then the Dirac restrictions of bracket
(\ref{PBronTorus}) on 
${\hat \Lambda}_{{\bf k}_{1}, \dots, {\bf k}_{d}}$, obtained with 
the aid of the sets $\, (I^{1}, \dots, I^{m+s})$ and
$\, (I^{\prime 1}, \dots, I^{\prime m+s})$, coincide with each 
other.
}

\vspace{0.2cm}

Proof.

What we have to prove is that the brackets (\ref{GlDirRest}),
obtained with the aid of the sets $\, (I^{1}, \dots, I^{m+s})$ 
and $\, (I^{\prime 1}, \dots, I^{\prime m+s})$, transform into 
each other under the coordinate transformation
$$\Big( U^{1}, \dots, U^{m+s}, \,\, \theta_{0}^{1}, \dots, 
\theta_{0}^{m} \Big) \,\,\,\,\, \rightarrow \,\,\,\,\,
\Big( U^{\prime 1}, \dots, U^{\prime m+s}, \,\, \theta_{0}^{1}, 
\dots, \theta_{0}^{m} \Big) $$
where the coordinates $\, {\bf U}$ and $\, {\bf U}^{\prime}$
are given by the values of $\, {\bf I}$ and
$\, {\bf I}^{\prime}$ on the submanifold
${\hat \Lambda}_{{\bf k}_{1}, \dots, {\bf k}_{d}}$. Easy to see
that this requirement is given in our case by the relations
$$\omega^{\prime \alpha \gamma} \left( {\bf U} \right)
\,\,\,\,\, = \,\,\,\,\, 
{\partial U^{\prime \gamma} \over \partial U^{\rho}} \,\,\,
\omega^{\alpha \rho} \left( {\bf U} \right) $$
where $\, \omega^{\alpha \gamma} ({\bf U})$ and
$\, \omega^{\prime \alpha \gamma}({\bf U})$ are the frequencies,
corresponding to the sets $\, {\bf I}$ and $\, {\bf I}^{\prime}$
respectively.

 Consider the sets of the functionals $\, J^{\gamma}$ and
$\, J^{\prime \gamma}$, $\, \gamma = 1, \dots, m+s$,
introduced with the aid of the sets $\, {\bf I}$ and 
$\, {\bf I}^{\prime}$ according to formula (\ref{Jnu}). The
variation derivatives 
$\, \delta J^{\gamma} / \delta \varphi^{i} (\bm{\theta})$ and
$\, \delta J^{\prime \gamma} / \delta \varphi^{i} (\bm{\theta})$
represent regular covectors on the family ${\hat \Lambda}$,
smoothly depending on all the variables
$\, ({\bf k}_{1}, \dots, {\bf k}_{d}, \, {\bf U}, \, 
\bm{\theta}_{0})$. Besides that, the sets $\, {\bf J}$ and
$\, {\bf J}^{\prime}$ generate the same linear space of the 
vector fields at every point of ${\hat \Lambda}$ according to
bracket (\ref{PBronTorus}). Since the linear spaces, generated
by the gradients of $\, {\bf J}$ and $\, {\bf J}^{\prime}$ 
contain also all the regular annihilators of the bracket
(\ref{PBronTorus}) on ${\hat \Lambda}$, we can claim then,
that these spaces coincide with each other at every 
$\, ({\bf k}_{1}, \dots, {\bf k}_{d}, \, {\bf U}, \,
\bm{\theta}_{0})$. According to the translational invariance of
the functionals $\, {\bf J}$ and $\, {\bf J}^{\prime}$ we can
then write
$$\left. {\delta J^{\prime \gamma} \over 
\delta \varphi^{i} (\bm{\theta})} 
\right|_{{\hat \Lambda}_{{\bf k}_{1}, \dots, {\bf k}_{d}}}
\,\, = \,\,\,\,\, \lambda^{\gamma}_{\rho} \left(
{\bf k}_{1}, \dots, {\bf k}_{d}, \, {\bf U} \right) \,\,\,
\left. {\delta J^{\rho} \over
\delta \varphi^{i} (\bm{\theta})}
\right|_{{\hat \Lambda}_{{\bf k}_{1}, \dots, {\bf k}_{d}}} $$
on every submanifold 
${\hat \Lambda}_{{\bf k}_{1}, \dots, {\bf k}_{d}}$.

 Easy to see that we naturally have then the relations
$$\omega^{\prime \alpha \gamma} \left(
{\bf k}_{1}, \dots, {\bf k}_{d}, \, {\bf U} \right) 
\,\,\,\,\, = \,\,\,\,\, \lambda^{\gamma}_{\rho} \left(
{\bf k}_{1}, \dots, {\bf k}_{d}, \, {\bf U} \right) \,\,\,
\omega^{\alpha \rho} \left(
{\bf k}_{1}, \dots, {\bf k}_{d}, \, {\bf U} \right) $$
on every ${\hat \Lambda}_{{\bf k}_{1}, \dots, {\bf k}_{d}}$,
and also
$${\partial U^{\prime \gamma} \over \partial U^{\rho}} 
\,\,\,\,\, = \,\,\,\,\, \lambda^{\gamma}_{\rho} \left(
{\bf k}_{1}, \dots, {\bf k}_{d}, \, {\bf U} \right) $$
according to the definition of the coordinates
$\, {\bf U}$ and $\, {\bf U}^{\prime}$. Thus, we get now the
statement of the Lemma.

{\hfill Lemma 2.3 is proved.}

\vspace{0.2cm}

 Let us discuss now an analog of the action-angle variables for
the restricted bracket.

\vspace{0.2cm}

{\bf Theorem 2.1.}

{\it Let $\Lambda$ be a regular Hamiltonian submanifold in the
space of quasiperiodic functions, equipped with a minimal set
of commuting integrals $(I^{1}, \dots, I^{m+s})$. Let system
(\ref{TestSystem}) have near every value of ${\bf U}$ a smooth
$2\pi$-periodic in each $\theta^{\alpha}$ solution
${\tilde \beta}^{\alpha}_{i} (\bm{\theta} + \bm{\theta}_{0}, \,
{\bf U})$, smoothly depending on the parameters 
${\bf U}$. Then:

1) The Dirac restriction (\ref{GlDirRest}) of the bracket
(\ref{PBronTorus}) on 
${\hat \Lambda}_{{\bf k}_{1}, \dots, {\bf k}_{d}}$ has the form
\begin{equation}
\label{DiracRestr}
\left\{ \theta^{\alpha}_{0} \, , \, \theta^{\beta}_{0} \right\}
\,\, = \,\, K^{\alpha\beta} ({\bf U})
\,\,\, , \,\,\,\,\,
\left\{ \theta^{\alpha}_{0} \, , \, U^{\gamma} \right\} \,\, = \,\,
\omega^{\alpha\gamma} ({\bf U}) \,\,\, , \,\,\,\,\,
\left\{ U^{\gamma} \, , \, U^{\lambda} \right\} \,\, = \,\, 0
\end{equation}
with some skew-symmetric matrix $K^{\alpha\beta} ({\bf U})$, 
not depending on $\bm{\theta}_{0}$;

2) The relations
\begin{equation}
\label{omegaBracket}
\left\{ \theta^{\alpha}_{0} \, , \, \theta^{\beta}_{0} \right\}
\,\, = \,\, 0 
\,\,\, , \,\,\,\,\,
\left\{ \theta^{\alpha}_{0} \, , \, U^{\gamma} \right\} \,\, = \,\,
\omega^{\alpha\gamma} ({\bf U}) \,\,\, , \,\,\,\,\,
\left\{ U^{\gamma} \, , \, U^{\lambda} \right\} \,\, = \,\, 0
\end{equation}
define a Poisson bracket on the space 
$\, ({\bf U}, \bm{\theta}_{0})$.

3) In the ${\bf U}$-space there (locally) exists the
coordinate transformation
$$ Q_{\alpha} \,\,\, = \,\,\, Q_{\alpha} ({\bf U}) 
\,\,\,\,\, , \,\,\,\,\,\,\,\, 
N^{l} \,\,\, = \,\,\, N^{l} ({\bf U}) 
\,\,\,\,\, , \,\,\,\,\,\,\,\,
{\bar \theta}^{\alpha}_{0} \,\,\, = \,\,\, \theta^{\alpha}_{0} 
\,\, - \,\, q^{\alpha} ({\bf U}) $$
($\alpha = 1, \dots, m$, $\,\, l = 1, \dots, s$), such that the
bracket (\ref{DiracRestr}) takes the form:

\begin{equation}
\label{CanBracket}
\begin{array}{l}
\left\{ {\bar \theta}^{\alpha}_{0} \, , \,
{\bar \theta}^{\beta}_{0} \right\} \,\, = \,\, 0
\,\,\, , \,\,\,\,\,  
\left\{ {\bar \theta}^{\alpha}_{0} \, , \, Q_{\beta}
\right\} \,\, = \,\, \delta^{\alpha}_{\beta}
\,\,\, , \,\,\,\,\,
\left\{ {\bar \theta}^{\alpha}_{0} \, , \, N^{l} \right\}
\,\, = \,\, 0  \,\,\, ,   \\   \\
\left\{ Q_{\alpha} \, , \, Q_{\alpha} \right\} \,\, = \,\, 0
\,\,\, , \,\,\,\,\,
\left\{ Q_{\alpha} \, , \, N^{l} \right\} \,\, = \,\, 0
\,\,\, , \,\,\,\,\,
\left\{ N^{l} \, , \,  N^{p} \right\} \,\, = \,\, 0
\end{array}
\end{equation}

}

\vspace{0.2cm}

 Proof.

 Let us consider for bracket (\ref{GlDirRest}) the Jacobi
identities of the form
$$\left\{ \left\{ \theta^{\alpha}_{0} \, , \, \theta^{\beta}_{0}
\right\} \, , \, U^{\gamma} \right\} \,\, + \,\,
\left\{ \left\{ \theta^{\beta}_{0} \, , \, U^{\gamma} \right\}
\, , \, \theta^{\alpha}_{0} \right\} \,\, + \,\,
\left\{ \left\{ U^{\gamma} \, , \, \theta^{\alpha}_{0} 
\right\} \, , \, \theta^{\beta}_{0} \right\} 
\,\,\, \equiv \,\,\, 0  $$

 We immediately get the relations:
$${\partial K^{\alpha\beta} ({\bf U}, \bm{\theta}_{0}) \over
\partial \theta^{\lambda}_{0}} \,\, 
\omega^{\lambda \gamma} ({\bf U}) 
\,\,\,\,\, \equiv \,\,\,\,\,
\omega^{\alpha \mu} ({\bf U}) \, 
{\partial \omega^{\beta \gamma} ({\bf U}) \over
\partial  U^{\mu}} \,\,\, - \,\,\,
\omega^{\beta \mu} ({\bf U}) \,
{\partial \omega^{\alpha \gamma} ({\bf U}) \over
\partial  U^{\mu}} $$

 The right-hand part of the above identity obviously does not 
depend on $\bm{\theta}_{0}$, so we get the same for the 
left-hand part. Since the functions 
$K^{\alpha\beta} ({\bf U}, \bm{\theta}_{0})$ are periodic 
in each $\theta^{\alpha}$ we then actually get the relations
\begin{equation}
\label{kalphabetaomega0}
{\partial K^{\alpha\beta} ({\bf U}, \bm{\theta}_{0}) \over
\partial \theta^{\lambda}_{0}} \,\,
\omega^{\lambda \gamma} ({\bf U})
\,\,\,\,\, \equiv \,\,\,\,\, 0
\end{equation}
\begin{equation}
\label{commuteomega}
\omega^{\alpha \mu} ({\bf U}) \,
{\partial \omega^{\beta \gamma} ({\bf U}) \over
\partial  U^{\mu}} \,\,\, - \,\,\,
\omega^{\beta \mu} ({\bf U}) \,
{\partial \omega^{\alpha \gamma} ({\bf U}) \over
\partial  U^{\mu}} \,\,\,\,\, \equiv \,\,\,\,\, 0
\end{equation}
$\alpha, \beta = 1, \dots, m$, 
$\,\, \gamma = 1, \dots, m + s$.

 According to requirement (3) of Definition 2.2 we then get
immediately
$${\partial K^{\alpha\beta} ({\bf U}, \bm{\theta}_{0}) \over
\partial \theta^{\lambda}_{0}} \,\,\,\,\, \equiv \,\,\,\,\, 0 $$
which gives the first part of the Theorem.

 It's not difficult to check also that relations 
(\ref{commuteomega}) coincide with the Jacobi identity for the
bracket (\ref{omegaBracket}), so we get also the second part
of the Theorem.

 To prove the last part of the Theorem, let us note that
relations (\ref{commuteomega}) express in fact the 
commutativity of the vector fields
$${\vec \xi}_{(\alpha)} \,\,\, = \,\,\,
\left( \omega^{\alpha 1} ({\bf U}), \dots, \,
\omega^{\alpha \, m+s} ({\bf U}) \right)^{t} 
\,\,\,\,\, , \,\,\,\,\,\,\,\, \alpha = 1, \dots, m $$
on the ${\bf U}$-space. Since the vector fields
${\vec \xi}_{(\alpha)}$ are linearly independent, we can then 
claim that we can locally introduce a coordinate system
$(Q_{1}, \dots, Q_{m}, \, N^{1}, \dots, N^{s})$ in the
${\bf U}$-space in which the vector fields
${\vec \xi}_{(\alpha)}$ have the components:
\begin{equation}
\label{xicancomp}
{\vec \xi}_{(1)} \,\,\, = \,\,\, (1, 0, \dots, 0)^{t} \,\,\, ,
\,\,\,\,\, \dots \,\,\,\,\, , \,\,\,\,\, 
{\vec \xi}_{(m)} \,\,\, = \,\,\, 
(0, \dots, 0, 1, 0, \dots, 0)^{t}
\end{equation}

 Easy to see, that relations (\ref{xicancomp}) provide then the 
relations
$$\left\{ {\bar \theta}^{\alpha}_{0} \, , \, Q_{\beta}
\right\} \,\, = \,\, \delta^{\alpha}_{\beta}
\,\,\,\,\, , \,\,\,\,\,\,\,\,
\left\{ {\bar \theta}^{\alpha}_{0} \, , \, N^{l} \right\}
\,\, = \,\, 0 $$
for the Poisson bracket (\ref{DiracRestr}).

 From the Jacobi identities
$$\left\{ \left\{ \theta^{\alpha}_{0} \, , \,
\theta^{\beta}_{0} \right\} \, , \, \theta^{\gamma}_{0} \right\}
\,\,\,\,\, + \,\,\,\,\, {\rm c.p.} 
\,\,\,\,\,\,\,\, \equiv \,\,\,\,\,\,\,\, 0 $$
we have in the new coordinate system for the functions
$K^{\alpha\beta} ({\bf Q}, {\bf N})$:
$${\partial K^{\alpha\beta} ({\bf Q}, {\bf N}) \over
\partial Q_{\gamma}} \,\,\, + \,\,\,
{\partial K^{\beta\gamma} ({\bf Q}, {\bf N}) \over
\partial Q_{\alpha}} \,\,\, + \,\,\,
{\partial K^{\gamma\alpha} ({\bf Q}, {\bf N}) \over
\partial Q_{\beta}} \,\,\,\,\,\,\,\, \equiv \,\,\,\,\,\,\,\, 0 $$

 From the expression above it follows that the functions
$K^{\alpha\beta} ({\bf Q}, {\bf N})$ can be locally represented
in the form
$$K^{\alpha\beta} ({\bf Q}, {\bf N}) 
\,\,\,\,\, = \,\,\,\,\,
{\partial q^{\beta} ({\bf Q}, {\bf N}) \over \partial Q_{\alpha}}
\,\,\, - \,\,\, {\partial q^{\alpha} ({\bf Q}, {\bf N}) \over
\partial Q_{\beta}} $$
for some smooth functions $q^{\alpha} ({\bf Q}, {\bf N})$.
Putting then 
${\bar \theta}^{\alpha}_{0} \, = \,
\theta^{\alpha}_{0} \, - \, q^{\alpha} ({\bf Q}, {\bf N})$
we get immediately the relations
$\{{\bar \theta}^{\alpha}_{0} \, , \,{\bar \theta}^{\beta}_{0} \}
\, \equiv \, 0$ 
for the bracket (\ref{DiracRestr}).

{\hfill Theorem 2.1 is proved.}

\vspace{0.2cm}

 It is natural to call the variables $Q_{\alpha}$ the 
action-type variables and the variables 
${\bar \theta}^{\alpha}$ - the angle-type variables. The
variables $N^{l}$ represent the annihilators of the bracket
on ${\hat \Lambda}_{{\bf k}_{1}, \dots, {\bf k}_{d}}$.

\vspace{0.2cm}

 Let us discuss now the questions of the resolvability of the
test system (\ref{TestSystem}). First of all, we have to require
the orthogonality of the right-hand part of (\ref{TestSystem})
to the eigen-vectors of the operator 
${\hat B}^{ij}_{{\bf k}_{1}, \dots, {\bf k}_{d}}$, corresponding
to the zero eigen-value. 

 It's not difficult to see, that for generic case 
$({\bf k}_{1}, \dots, {\bf k}_{d}) \in {\cal M}$ this 
requirement is actually automatically satisfied for system
(\ref{TestSystem}). Indeed, the kernel vectors of the operator
${\hat B}^{ij}_{{\bf k}_{1}, \dots, {\bf k}_{d}}$ on the space
of smooth periodic functions are given in this case by the 
vectors (\ref{vkitheta}), such that we have
\begin{multline*}
\sum_{\gamma=1}^{m+s} \,\,
\omega^{\alpha\gamma} ({\bf k}_{1}, \dots, {\bf k}_{d}, {\bf U})
\,\,\,  \times  \\
\times \,\, \int_{0}^{2\pi}\!\!\!\!\!\dots\int_{0}^{2\pi} \!
v^{(k)}_{i} (\bm{\theta} + \bm{\theta}_{0}, \,
{\bf k}_{1}, \dots, {\bf k}_{d}, {\bf U}) \,\,
\Phi^{i}_{U^{\gamma}} (\bm{\theta} + \bm{\theta}_{0}, \,
{\bf k}_{1}, \dots, {\bf k}_{d}, {\bf U}) \,\,
{d^{m} \theta \over (2\pi)^{m}} \,\,\,\,\, \equiv   \\
\equiv \,\,\,\,\, \sum_{\gamma=1}^{m+s}
\gamma^{k}_{\gamma} ({\bf k}_{1}, \dots, {\bf k}_{d}, {\bf U})
\,\, \omega^{\alpha\gamma} 
({\bf k}_{1}, \dots, {\bf k}_{d}, {\bf U}) 
\end{multline*}
for 
$\, U^{\gamma} \, \equiv \, J^{\gamma} 
|_{{\hat \Lambda}_{{\bf k}_{1}, \dots, {\bf k}_{d}}}$

 The last expression coincides with the left-hand part of
(\ref{annomegarel}) and is identically equal to zero.

 On the other hand, we can see that the analogous property
can be definitely violated in the non-generic case
$({\bf k}_{1}, \dots, {\bf k}_{d}) \notin {\cal M}$ where
the number of annihilators of
${\hat B}^{ij}_{{\bf k}_{1}, \dots, {\bf k}_{d}}$ 
(if they exist) can be infinite and is not restricted by the 
set (\ref{vkitheta}).

 Easy to see that the simplest situation arises here in the
single-phase ($m = 1$) case where system (\ref{TestSystem}) 
is always resolvable. Indeed, all the annihilators
of ${\hat B}^{ij}_{{\bf k}_{1}, \dots, {\bf k}_{d}}$ are given
in this case by the set (\ref{vkitheta}), while the nonzero
eigen-values of 
${\hat B}^{ij}_{{\bf k}_{1}, \dots, {\bf k}_{d}}$ are separated
from zero. It's not difficult to see also that the corresponding
solutions of system (\ref{TestSystem}) or (\ref{Systbeta}) can
be chosen here as smooth functions of all the parameters,
including the wave numbers $({\bf k}_{1}, \dots, {\bf k}_{d})$.
We can then formulate here the following theorem:

\vspace{0.2cm}

 {\bf Theorem 2.2.}

{\it Let $\Lambda$ be a regular Hamiltonian submanifold in the
space of single-phase periodic functions in $\mathbb{R}^{d}$,
equipped with a minimal set of commuting integrals
$\, (I^{1}, \dots, I^{s+1})$. Then every submanifold
${\hat \Lambda}_{{\bf k}_{1}, \dots, {\bf k}_{d}}$ admits
regular Dirac restriction of the bracket (\ref{PBronTorus}),
smoothly depending on the parameters 
$\, ({\bf k}_{1}, \dots, {\bf k}_{d})$.
}

\vspace{0.2cm}

 The investigation of resolvability of system (\ref{TestSystem})
in the multi-phase ($m > 1$) situation is much more complicated 
in general. Thus, even under the requirement of orthogonality
of the right-hand part of (\ref{TestSystem}) to the kernel 
vectors of the operator
${\hat B}^{ij}_{{\bf k}_{1}, \dots, {\bf k}_{d}}$,
system (\ref{TestSystem}) can still be unresolvable on the space
of smooth $2\pi$-periodic in each $\theta^{\alpha}$ functions
if the eigen-values of
${\hat B}^{ij}_{{\bf k}_{1}, \dots, {\bf k}_{d}}$ are strongly
accumulated near the zero value. The properties of the
eigen-values of 
${\hat B}^{ij}_{{\bf k}_{1}, \dots, {\bf k}_{d}}$
strongly depend in fact on the properties of the numbers
$({\bf k}_{1}, \dots, {\bf k}_{d})$, so we have to find in
general a subset ${\cal S} \subset {\cal M}$ where the
corresponding systems (\ref{TestSystem}) are resolvable on the
space of smooth $2\pi$-periodic in each $\theta^{\alpha}$ 
functions. Easy to see that we can formulate the following
theorem:

\vspace{0.2cm}

{\bf Theorem 2.3.}

{\it Let $\Lambda$ be a regular Hamiltonian submanifold in the
space of quasiperiodic functions in $\mathbb{R}^{d}$, equipped
with a minimal set of commuting integrals 
$\, (I^{1}, \dots, I^{m+s})$. Let there exist a dense set
$\, {\cal S} \subset {\cal M}$ in the space
$\, ({\bf k}_{1}, \dots, {\bf k}_{d})$ on which the corresponding
systems (\ref{TestSystem}) satisfy the requirements of 
Theorem 2.1. Then the relations
\begin{equation}
\label{thetaUkBracket}
\begin{array}{c}
\left\{ \theta^{\alpha}_{0} \, , \, \theta^{\beta}_{0} \right\}
\,\, = \,\, 0
\,\,\, , \,\,\,\,\,
\left\{ \theta^{\alpha}_{0} \, , \, U^{\gamma} \right\} \,\, = \,\,
\omega^{\alpha\gamma} 
\left( {\bf k}_{1}, \dots, {\bf k}_{d}, \, {\bf U} \right) 
\,\,\, , \,\,\,\,\,
\left\{ \theta^{\alpha}_{0} \, , \, k^{\beta}_{p} 
\right\} \,\, = \,\, 0  \,\,\, ,   \\   \\
\left\{ U^{\gamma} \, , \, U^{\rho} \right\} \,\, = \,\, 0
\,\,\, , \,\,\,\,\,
\left\{ U^{\gamma} \, , \, k^{\beta}_{p}
\right\} \,\, = \,\, 0  \,\,\, , \,\,\,\,\,
\left\{ k^{\alpha}_{q} \, , \, k^{\beta}_{p}
\right\} \,\, = \,\, 0  
\end{array}  
\end{equation}
define a Poisson bracket on the space
$\, ({\bf k}_{1}, \dots, {\bf k}_{d}, \, {\bf U}, \, 
\bm{\theta}_{0})$.
}

\vspace{0.2cm}

 Indeed, it is not difficult to see that the Jacobi identity
for bracket (\ref{thetaUkBracket}) is given by relations
(\ref{commuteomega}) for the values
$$\omega^{\alpha\gamma} \,\,\, = \,\,\,
\omega^{\alpha\gamma} 
\left( {\bf k}_{1}, \dots, {\bf k}_{d}, \, {\bf U} \right) $$
which are obviously satisfied under the conditions of the
Theorem.

\vspace{0.2cm}

 Let us say, however, that the bracket (\ref{thetaUkBracket})
can not be obtained in general from the bracket 
(\ref{DiracRestr}) after a smooth coordinate transformation in 
the space 
$\, ({\bf k}_{1}, \dots, {\bf k}_{d}, \, {\bf U}, \,
\bm{\theta}_{0})$ since the dependence of the variables
$\bar{\bm{\theta}}_{0}$ on the values
$\, ({\bf k}_{1}, \dots, {\bf k}_{d})$ can have in general 
rather irregular form.

 Let us call here bracket (\ref{thetaUkBracket}) the bracket
associated with the bracket (\ref{MultDimPBr}) on the family
$\Lambda$.

\vspace{0.2cm}

 As an example, let us consider the Gardner - Zakharov - Faddeev 
bracket

\begin{equation}
\label{GardZakhFadd}
\{ \varphi (x) \, , \, \varphi (y) \} \,\,\, = \,\,\,
\delta^{\prime} (x - y)
\end{equation}
and the KdV equation
 
$$\varphi_{t} \,\, = \,\, \varphi \, \varphi_{x} \, - \,
\varphi_{xxx} $$
corresponding to the Hamiltonian functional 

$$H \,\,\, = \,\,\, \int \left( {\varphi^{3} \over 6} \, + \,  
{\varphi_{x}^{2} \over 2} \right) \, d x $$
 
 As it is well known (\cite{NovikovFuncAn}), the KdV equation
has a family of $m$-phase solutions for any $m \geq 0$, which can 
be represented as the set of extremals of the functionals given by 
all the linear combinations of the first $\, m + 2 \, $ integrals 
of KdV
\begin{equation}
\label{Extremals}
c_{1} \, \delta I^{1} \, + \, c_{2} \, \delta I^{2} \, + \, \dots
\, + \, c_{m+2} \, \delta I^{m+2} \,\, = \,\, 0
\end{equation}

 The first two integrals 
$$I^{1} \,\,\, = \,\,\, N \,\,\, = \,\,\,
\int \varphi \,\, d x   \,\,\,\,\,\,\,\, , \,\,\,\,\,\,\,\,\,\,
I^{2} \,\,\, = \,\,\, P \,\,\, = \,\,\,
\int \, {\varphi^{2} \over 2} \,\, d x $$
represent here the annihilator and the momentum functional of
the bracket (\ref{GardZakhFadd}). We have also $\, I^{3} = H$, 
and $\, I^{k}$, $\, k \geq 4$ represent the higher integrals of
the KdV equation.

  As was shown in \cite{NovikovFuncAn}, systems (\ref{Extremals}) 
represent completely integrable finite-dimensional 
systems having quasiperiodic solutions in the generic case.
According to \cite{NovikovFuncAn}, the parameters of the 
$m$-phase solutions are given by $2 m + 1$ real branching points
$\, (E_{1}, \dots , E_{2m+1})$, $\, E_{1} < E_{2} < \dots < E_{2m+1},$ 
of a hyperelliptic surface of genus $m$ and $m$ initial phases
$\, (\theta_{0}^{1}, \dots, \theta_{0}^{m})$. As it is also well-known,
the theory of the quasiperiodic solutions of KdV has a remarkable 
connection with the theory of theta-functions of Riemann surfaces
(\cite{DubrNovDokl, DubrNovZhETP, ItsMatveev1, ItsMatveev2,
Dubrovin1, DubrMatvNov, Dubrovin2}).

 It's not difficult to check that the families $\Lambda^{(m)}$ 
represent here regular Hamiltonian submanifolds in the space of
quasiperiodic functions, while the functionals
$\, (I^{1}, \dots, I^{m+1})$ give a minimal set of commuting
integrals for the family $\Lambda^{(m)}$.

 Easy to see that the operator ${\hat B}_{\bf k}$ has here the
form
$${\hat B}_{\bf k} \,\,\, = \,\,\, 
k^{1} \, {\partial \over \partial \theta^{1}} \,\, + \,\,
\dots \,\, + \,\,
k^{m} \, {\partial \over \partial \theta^{m}} $$

 Thus, we have to investigate the resolvability of the systems
\begin{equation}
\label{KdVTestSyst}
k^{1} \, {\tilde \beta}^{\alpha}_{\theta^{1}} \,\, + \,\,
\dots \,\, + \,\,  k^{m} \, {\tilde \beta}^{\alpha}_{\theta^{m}} 
\,\,\, = \,\,\, - \, \omega^{\alpha\gamma} \left( {\bf k}, {\bf U}
\right) \,\, \Phi_{U^{\gamma}} \left( \bm{\theta} + \bm{\theta}_{0},
\, {\bf k}, {\bf U} \right)
\end{equation}
to define the Dirac restriction of the bracket (\ref{GardZakhFadd})
on the submanifolds ${\hat \Lambda}_{\bf k}$.

 Let us write the Fourier expansion of the right-hand part of
(\ref{KdVTestSyst}) in the form:
\begin{equation}
\label{RHPFourExp}
- \, \omega^{\alpha\gamma} \left( {\bf k}, {\bf U}
\right) \, \Phi_{U^{\gamma}} \left( \bm{\theta},  
\, {\bf k}, {\bf U} \right) \,\,\, = 
\sum_{n_{1}, \dots, n_{m}} \!\! 
A_{n_{1} \dots n_{m}} \left( {\bf k}, {\bf U} \right)  \,\,
\exp \, ( i n_{1} \, \theta^{1} \, + \, \dots \, + \,
i n_{m} \, \theta^{m}) 
\end{equation}

 We have by definition $\, U^{\gamma} = \langle P^{\gamma} \rangle$,
$\, \gamma = 1, \dots, m + 1$, so we can write the following
relations
$$ \int_{0}^{2\pi}\!\!\!\!\!\dots\int_{0}^{2\pi}
\Phi_{U^{1}} (\bm{\theta}, \, {\bf k}, {\bf U}) \,\,
{d^{m} \theta \over (2\pi)^{m}} \,\, \equiv \,\, 1
\,\,\, , \,\,\,\,\,\,\,
\int_{0}^{2\pi}\!\!\!\!\!\dots\int_{0}^{2\pi}
\Phi_{U^{\gamma}} (\bm{\theta},  \, {\bf k}, {\bf U}) \,\,
{d^{m} \theta \over (2\pi)^{m}} \,\, \equiv \,\, 0
\,\,\, , \,\,\,\,\,\,\, \gamma \neq 1 $$
according to the definition of the functional $\, I^{1}$.
Using the relations 
$\, \omega^{\alpha 1} ({\bf k}, {\bf U}) \, \equiv \, 0$ we then
get immediately the relations 
$\, A_{0 \dots 0}  ({\bf k}, {\bf U}) \, \equiv \, 0$ for the
expansion (\ref{RHPFourExp}).

 From the theta-functional representation of the right-hand part
of (\ref{KdVTestSyst}) it is easy to get also that the values
$\, A_{0 \dots 0}  ({\bf k}, {\bf U})$ 
decay faster than any power of $|{\bf n}|$ at  
$|{\bf n}| \rightarrow \infty$, where

$$|{\bf n}| \,\, \equiv \,\,
\sqrt{ n_{1}^{2} \, + \, \dots \, + \, n_{m}^{2}} $$

 To investigate the resolvability of system (\ref{KdVTestSyst})
on the set $\, (k^{1}, \dots, k^{m}) \in {\cal M}$ let us define
the Diophantine conditions for the values 
$\, (k^{1}, \dots, k^{m})$.
Namely, the vector $\, (k^{1}, \dots, k^{m})$ represents a 
Diophantine vector with the index $\nu > 0$ and the coefficient 
$A > 0$, if
$$\left| n_{1} \, k^{1}({\bf U}) \, + \,\, \dots \,\, + \,
n_{m} \, k^{m}({\bf U}) \right| \,\,\, \geq \,\,\,
A \, |{\bf n}|^{-\nu} $$
for all $\, (n_{1}, \dots, n_{m}) \in \mathbb{Z}^{m}$
$\, ((n_{1}, \dots, n_{m}) \neq (0, \dots, 0))$.

 Let us denote here by ${\cal S}_{\nu}$ the set of all Diophantine 
vectors $\, (k^{1}, \dots, k^{m})$ with index $\nu$. The following
classical theorem (see e.g. \cite{Arnold, Shmidt}) can be formulated
about the space of $\, (k^{1}, \dots, k^{m})$:

\vspace{0.2cm}

{\it For any $\, \nu > m - 1$ the measure of the corresponding
set of non-Diophantine vectors 
$\, (k^{1}, \dots, k^{m})$ in $\mathbb{R}^{m}$ is equal 
to zero.}

\vspace{0.2cm}

 We can see now, that putting $\, {\cal S} = {\cal S}_{\nu}$ for
any $\, \nu > m - 1$ we get the dense set 
$\, {\cal S} \subset {\cal M}$ where system (\ref{KdVTestSyst})
is resolvable on the space of smooth $2\pi$-periodic in each
$\theta^{\alpha}$ functions.

 According to Theorem 2.3 we can claim now that the corresponding
relations (\ref{thetaUkBracket}) give a Poisson bracket on the
space of parameters $\, ({\bf k}, {\bf U}, \bm{\theta}_{0})$. 
It's not difficult to show also that bracket (\ref{thetaUkBracket}) 
gives one of the examples of the analytic Poisson brackets
compatible with the KdV theory, introduced by A.P. Veselov and 
S.P. Novikov (\cite{VeselovNovikov1, VeselovNovikov2}). Let us
say, that the methods demonstrated above are applicable in fact for
a wide class of Hamiltonian operators (\ref{MultDimPBr}).

\vspace{0.2cm}

 Let us note now, that in many examples the restriction of the
bracket (\ref{PBronTorus}) on the submanifolds
${\hat \Lambda}_{{\bf k}_{1}, \dots, {\bf k}_{d}}$ can in fact
be made in a simpler way than that described above. As an example,
let us consider the NLS equation
\begin{equation}
\label{NLS}
i \, \psi_{t} \,\, = \,\, \psi_{xx} \, + \, 
\kappa \, |\psi|^{2} \psi 
\end{equation}
and the Hamiltonian structure
\begin{equation}
\label{NLSBracket}
\{ \psi (x) \, , \, {\bar \psi} (y) \} \,\, = \,\, i \,
\delta (x - y) 
\end{equation}

 As it is well known, the equation (\ref{NLS}) has the families of
$m$-phase solutions for any $m$, given by the construction, 
analogous to the KdV case. The bracket (\ref{NLSBracket}) is
non-degenerate and it is easy to check that the full families
of $m$-phase solutions of NLS represent regular Hamiltonian
submanifolds in the space of quasiperiodic functions. Easy to see
also that the minimal sets of commuting integrals can be easily
constructed here with the aid of the higher integrals of the
NLS equation. The operator $\, {\hat B}^{ij}_{\bf k}$ has an
ultralocal form in this case and the system (\ref{TestSystem})
is trivially solvable for all $\, (k^{1}, \dots, k^{m})$. The Dirac
restriction of the bracket (\ref{PBronTorus}) on the submanifolds
${\hat \Lambda}_{\bf k}$ has in this case a regular character and
can be always written in the form (\ref{thetaUkBracket}) after some
smooth change of coordinates  \linebreak
$\, \theta_{0}^{\alpha} \, \rightarrow \, \theta_{0}^{\alpha} 
\, - \, q^{\alpha} ({\bf k}, {\bf U})$. The relations 
(\ref{thetaUkBracket}) define then a Poisson bracket on the space
$\, ({\bf k}, {\bf U}, \bm{\theta}_{0})$ which is always given here
by the Dirac restriction of bracket  (\ref{PBronTorus}) on the
submanifolds ${\hat \Lambda}_{\bf k}$ in the appropriate
coordinates.

 We have to say here also, that the Poisson and Symplectic structures,
defined on the spaces of $m$-phase solutions of different systems,
play extremely important role in many aspects of the theory of
integrable systems 
(see \cite{VeselovNovikov1, VeselovNovikov2, KricheverPhong}).

\vspace{0.2cm}

 As we will see, bracket (\ref{omegaBracket}) will play rather
important role in the Hamiltonian formulation of the Whitham
method, where the parameters 
$\, S^{\alpha} = \epsilon \, 
\theta_{0}^{\alpha} \, $ and $\, U^{\gamma}$
become slow functions of the spatial and time variables:
$\, S^{\alpha} \rightarrow S^{\alpha} ({\bf X}, T)$,
$\, U^{\gamma} \rightarrow  U^{\gamma} ({\bf X}, T)$,
$\, {\bf X} = \epsilon \, {\bf x}$, $\, T = \epsilon \, t$,
$\, \epsilon \rightarrow 0$. As we will show below, the first
two relations in (\ref{omegaBracket}) should be naturally
transformed to the relations
$$\left\{ S^{\alpha} ({\bf X}) \, , \, S^{\beta} ({\bf Y})
\right\} \,\,\, = \,\,\, 0 \,\,\, , $$
$$\left\{ S^{\alpha} ({\bf X}) \, , \, U^{\gamma}  ({\bf Y})
\right\} \,\,\, = \,\,\, \omega^{\alpha\gamma} \left(
{\bf S}_{X^{1}}, \dots, {\bf S}_{X^{d}}, {\bf U} ({\bf X}) 
\right) \,\,\, \delta ({\bf X} - {\bf Y}) $$
in this situation. On the other hand, the pairwise Poisson
brackets of the functionals $\, U^{\gamma} ({\bf X})$,
$\, U^{\rho} ({\bf Y})$ should be defined in this case by a
deformation of the bracket (\ref{omegaBracket}), given by the
Dubrovin - Novikov procedure of the bracket averaging. Let
us note also, that the values $\, k^{\alpha}_{q} ({\bf X})$
are given here by the relations
$\, k^{\alpha}_{q} ({\bf X}) = S^{\alpha}_{X^{q}}$ and do
not arise as additional parameters.

 Let us consider now the Hamiltonian formulation of the
Whitham method in more detail.

\section{Hamiltonian formulation of the Whitham method.}
\setcounter{equation}{0}

 In this chapter we will consider the formulation of the
multi-dimensional Whitham method from the Hamiltonian point of view.

\vspace{0.2cm}

 As it is well known, in the Whitham method 
(\cite{whith1, whith2, whith3}) we consider ``slow-modulated''
quasiperiodic solutions of PDE's, with parameters, slowly
depending on the coordinates ${\bf x}$ and $t$.

 More, precisely, we have to consider asymptotic solutions of
system (\ref{EvInSyst}) with the main term, having the form
\begin{equation}
\label{MainTermWhithSol}
\varphi^{i}_{(0)} ({\bf x}, t) \,\,\, = \,\,\,
\Phi^{i} \left( \bm{\theta} + 
{{\bf S} ({\bf X}, T) \over \epsilon} + 
\bm{\theta}_{0} ({\bf X}, T), \,\, {\bf S}_{X^{1}}, \dots,
{\bf S}_{X^{d}}, {\bf U} ({\bf X}, T) \right)
\end{equation}
where the functions ${\bf S} ({\bf X}, T)$ and
${\bf U} ({\bf X}, T)$ are functions of the ``slow'' variables
$\, X^{q} \, = \, \epsilon \, x^{q}$, 
$\, T \, = \, \epsilon \, t$, $\, \epsilon \rightarrow 0$.

 Following system (\ref{EvPhaseSyst}), it is easy to see that
solution (\ref{MainTermWhithSol}) satisfies system 
(\ref{EvInSyst}) in the main order of $\epsilon$ under the
additional requirement
\begin{equation}
\label{STomegaCond}
S^{\alpha}_{T} \,\,\, = \,\,\, \omega^{\alpha} \left(
{\bf S}_{X^{1}}, \dots, {\bf S}_{X^{d}}, {\bf U} ({\bf X}, T) 
\right)
\end{equation}

 However, to avoid growing secular terms in the next corrections
to (\ref{MainTermWhithSol}) we have to put additional 
requirements also to the functions ${\bf U} ({\bf X}, T)$. Thus,
using the substitution
\begin{equation}
\label{FirstCorrection}
\varphi^{i} ({\bf x}, t) \,\,\, \simeq \,\,\,
\varphi^{i}_{(0)} ({\bf x}, t) \,\, + \,\, \epsilon \,
\Psi^{i}_{(1)} \left( \bm{\theta} +
{{\bf S} ({\bf X}, T) \over \epsilon} +
\bm{\theta}_{0} ({\bf X}, T), \,\, {\bf X}, T \right)
\end{equation}
we get the linear system
\begin{equation}
\label{FirstApprSyst}
{\hat L}^{i}_{j} ({\bf X}, T) \,\, \Psi^{j}_{(1)} \left( 
\bm{\theta}, \, {\bf X}, T \right) \,\,\, = \,\,\,
f^{i}_{(1)} \left( \bm{\theta}, \, {\bf X}, T \right)
\end{equation}
where 
${\hat L}^{i}_{j} ({\bf X}, T) \, = \,
{\hat L}^{i}_{j [{\bf S}_{\bf X}, {\bf U} ({\bf X}, T)]}$
is the linear operator, given by linearization of system
(\ref{EvPhaseSyst}) on the function
$$\bm{\Psi}_{(0)} \left( \bm{\theta}, \, {\bf X}, T \right)
\,\,\, = \,\,\, \bm{\Phi} \left( \bm{\theta}, \,
{\bf S}_{X^{1}}, \dots, {\bf S}_{X^{d}}, {\bf U} ({\bf X}, T)
\right) $$
and $f^{i}_{(1)} ( \bm{\theta}, \, {\bf X}, T )$ is the first
``discrepancy'' defined from system (\ref{EvInSyst}).

 To study the resolvability of system (\ref{FirstApprSyst})
we have to investigate the properties of the operators
${\hat L}^{i}_{j [{\bf S}_{\bf X}, {\bf U} ({\bf X}, T)]}$
on the space of $2\pi$-periodic in each $\theta^{\alpha}$
functions. If we represent again the parameters $\, {\bf U}$
in the form:
$$\left( U^{1}, \dots, U^{m+s} \right) \,\,\, = \,\,\,
\left( \omega^{1}, \dots, \omega^{m}, \, n^{1}, \dots, n^{s}
\right) $$
we can see that the functions 
$$\bm{\Phi}_{\theta^{\alpha}} \left( \bm{\theta}, \,
{\bf S}_{X^{1}}, \dots, {\bf S}_{X^{d}}, \bm{\omega}, {\bf n}
\right) \,\,\, , \,\,\,\,\, \alpha = 1, \dots, m   \,\,\, , $$ 
$$\bm{\Phi}_{n^{l}} \left( \bm{\theta}, \,
{\bf S}_{X^{1}}, \dots, {\bf S}_{X^{d}}, \bm{\omega}, {\bf n}
\right) \,\,\, , \,\,\,\,\, l = 1, \dots, s  $$
belong to the kernel of the corresponding operator
${\hat L}^{i}_{j [{\bf S}_{\bf X}, {\bf U} ({\bf X}, T)]}$.
Let us say that for a good justification of the Whitham 
method we need in fact the requirement that the vectors
$\, \bm{\Phi}_{\theta^{\alpha}}$, $\, \bm{\Phi}_{n^{l}}$
represent the full set of linearly independent ``regular''
kernel vectors of the operators
${\hat L}^{i}_{j [{\bf S}_{\bf X}, {\bf U} ({\bf X}, T)]}$.
Besides that, we have to require also that the operator
${\hat L}^{i}_{j [{\bf S}_{\bf X}, {\bf U} ({\bf X}, T)]}$
has exactly $m + s$ regular left eigen-vectors 
(the eigen-vectors of the adjoint operator), corresponding 
to the zero eigen-value. More precisely, let us give here the 
definition of a complete regular family $\Lambda$ of 
$m$-phase solutions of system (\ref{EvInSyst}).

\vspace{0.2cm}

{\bf Definition 3.1.}

{\it  Let us call family $\Lambda$ a complete regular family
of $m$-phase solutions of system (\ref{EvInSyst}) if:

1) The total set of independent parameters on $\Lambda$ can
be represented by the values  \linebreak
$({\bf k}_{1}, \dots, {\bf k}_{d}, \, \bm{\omega}, \, {\bf n},
\bm{\theta}_{0})$;

2) The vectors
$\, \bm{\Phi}_{\theta^{\alpha}} (\bm{\theta}, \,
{\bf k}_{1}, \dots, {\bf k}_{d}, \bm{\omega}, {\bf n})$,
$\, \bm{\Phi}_{n^{l}} (\bm{\theta}, \,
{\bf k}_{1}, \dots, {\bf k}_{d}, \bm{\omega}, {\bf n}) \, $
are linearly independent and represent the maximal linearly 
independent set among the regular kernel vectors of the 
operator $\, {\hat L}^{i}_{j[{\bf k}_{1}, \dots, {\bf k}_{d}, 
\bm{\omega},{\bf n}]}$,
smoothly depending on the parameters
$\, ({\bf k}_{1}, \dots, {\bf k}_{d}, \, \bm{\omega}, \, {\bf n})$
on the whole set of parameters;

3) The operator
${\hat L}^{i}_{j[{\bf k}_{1}, \dots, {\bf k}_{d},
\bm{\omega},{\bf n}]}$
has exactly $m + s$ linearly independent regular left 
eigen-vectors 
$\bm{\kappa}^{(q)}_{[{\bf k}_{1}, \dots, {\bf k}_{d},
\bm{\omega}, {\bf n}]}
(\bm{\theta})$, $\,\, q = 1, \dots , m + s $, 
corresponding to the zero eigenvalue, 
smoothly depending on the parameters
$({\bf k}_{1}, \dots, {\bf k}_{d}, \, \bm{\omega}, \, {\bf n})$
on the whole set of parameters.
}

\vspace{0.2cm}

 Let us call here the regular Whitham system for a complete
regular family of $m$-phase solutions of (\ref{EvInSyst}) the
relations (\ref{STomegaCond}) together with the conditions of 
orthogonality of the discrepancy
${\bf f}_{(1)}(\bm{\theta}, {\bf X}, T)$ to the functions
$\bm{\kappa}^{(q)}_{[{\bf S}_{\bf X}, {\bf U}({\bf X},T)]}
(\bm{\theta})$:
\begin{equation}
\label{fkappaortcond}
\int_{0}^{2\pi}\!\!\!\!\!\dots\int_{0}^{2\pi}
\kappa^{(q)}_{i [{\bf S}_{\bf X}, {\bf U}({\bf X},T)]}
(\bm{\theta}) \,\,
f^{i}_{(1)} (\bm{\theta},{\bf X},T) \,\,
{d^{m} \theta \over (2\pi)^{m}} \,\,\, = \,\,\, 0
\,\,\,\,\,\,\,\,\,\, , \,\,\,\,\,\,\,\,\,\,
q = 1, \, \dots , \, m + s
\end{equation}

 In the single-phase case conditions (\ref{fkappaortcond})
provide the resolvability of system (\ref{FirstApprSyst}) at
every ${\bf X}$ and $T$ on the space of $2\pi$-periodic in
$\theta$ functions. Moreover, in the case $m = 1$ the
corresponding asymptotic solution 
$\bm{\varphi} (\theta, {\bf X}, T, \epsilon )$ of system
(\ref{EvInSyst}) can be usually represented in the form of the
regular asymptotic series:
$$\varphi^{i}(\theta, \, {\bf X}, T, \epsilon)
\,\,\, = \,\,\, \sum_{k\geq0} \, \epsilon^{k} \,
\Psi^{i}_{(k)} \left( {S ({\bf X},T) \over \epsilon} +
\theta, \, {\bf X}, \, T \right) $$
(see e.g. \cite{luke}).

 Certainly, the situation is much more complicated in the
multi-phase ($m > 1$) case, where the behavior of the 
eigen-values of ${\hat L}^{i}_{j} ({\bf X}, T)$ can be highly
nontrivial. As a rule, the study of the corrections to the main
approximation (\ref{MainTermWhithSol}) requires in this case
rather complicated mathematical methods. Let us give here the
references on the papers \cite{dobr1, dobr2, DobrKrichever} 
where the investigation of the multi-phase case can be found. 
As follows from the results of \cite{dobr1, dobr2, DobrKrichever}, 
the correction to the main term (\ref{MainTermWhithSol}) behaves
here in more complicated way than that represented in
(\ref{FirstCorrection}), however, it also vanishes at
$\epsilon \, \rightarrow \, 0$ under the fulfillment of
conditions (\ref{fkappaortcond}).

 In general, we can claim, that the regular Whitham system
(\ref{STomegaCond}), (\ref{fkappaortcond}) plays the central
role in the description of the slow modulations both in the
single-phase and the multi-phase situations.

 It is well known in the Whitham approach that the Whitham  
system does not give any restrictions on the functions
$\bm{\theta}_{0} ({\bf X}, T)$ and is connected just with
the functions
$\, {\bf k}_{q} ({\bf X}, T) \, = \, {\bf S}_{X^{q}}$ and
$\, {\bf U} ({\bf X}, T)$ (see e.g.
\cite{whith1, whith2, whith3, luke}). A simple proof of this
statement for system (\ref{fkappaortcond}) under the
assumptions, formulated above, can be found in 
\cite{Sigma, JMP2}. It can be also shown under some
assumptions that the phase shifts
$\theta^{\alpha}_{0}({\bf X},T)$ can in fact be 
absorbed by the functions
$S^{\alpha} ({\bf X},T)$ after an appropriate
correction of initial data (see e.g. 
\cite{Haberman1, Haberman2, MaltsevJMP, DobrMinenkov}).
At the same time, the corresponding initial phase shift can
still play rather important role in consideration of the slowly
modulated solutions in the so-called weakly nonlinear case
(\cite{Newell}, see also \cite{MaltsevAmerMath, DobrMinenkov}).
Let us represent here also just some incomplete list of the 
classical papers devoted to the foundations of the Whitham 
method:
\cite{AblBenny, dm, DobrMaslMFA, dobr1, dobr2, DobrKrichever, 
dn1, dn2, dn3, DubrovinCMP, ffm, Hayes, krichev1, KricheverCPAM,
luke, Newell, theorsol, Nov, whith1, whith2, whith3}. 

 In the rest of this paper we are going to consider the
Hamiltonian properties of the regular Whitham system, given
by (\ref{STomegaCond}) and (\ref{fkappaortcond}). As we will
see, the corresponding Hamiltonian structure will be connected
to some extend with the structures, considered in the previous
chapter.

 The Hamiltonian theory of the Whitham equations was started
in the pioneer works of B.A. Dubrovin and S.P. Novikov
(\cite{dn1, DubrNovDAN84, dn2, dn3}). In the approach of
B.A. Dubrovin and S.P. Novikov the Whitham system was 
considered as a system of Hydrodynamic Type. For the case of one
spatial dimension systems of this kind can be written in the
following general form:
\begin{equation}
\label{HTWhithSyst}
U^{\nu}_{T} \,\,\, = \,\,\, V^{\nu}_{\mu} ({\bf U}) \,\,
U^{\mu}_{X} \,\,\,\,\, , \,\,\,\,\,\,\,\,
\nu, \mu = 1, \dots, N
\end{equation}

 The general Dubrovin - Novikov bracket for system 
(\ref{HTWhithSyst}) can be written in the form
\begin{equation}
\label{DNBracket}
\{U^{\nu}(X), U^{\mu}(Y)\} \,\,\, = \,\,\, g^{\nu\mu}({\bf U}(X)) 
\,\, \delta^{\prime}(X-Y) \,\,\, + \,\,\,
b^{\nu\mu}_{\lambda}({\bf U}(X) ) \,\, U^{\lambda}_{X} \,\,\, 
\delta (X-Y) 
\end{equation}

 Theory of brackets (\ref{DNBracket}) is closely connected
with the Differential Geometry. Thus, expression 
(\ref{DNBracket}) with non-degenerate tensor 
$g^{\nu\mu}({\bf U})$ defines a Poisson bracket on the space of
fields ${\bf U} (X)$ if and only if
the values $g^{\nu\mu}({\bf U})$ represent a flat contravariant
pseudo-Riemannian metric on the space of 
$\, {\bf U}$, while the values
$\Gamma^{\mu}_{\nu\lambda} ({\bf U}) \, = \, - \,
g_{\nu\tau} ({\bf U}) \, b^{\tau\mu}_{\lambda} ({\bf U})$ 
coincide with the corresponding Christoffel symbols
($g_{\nu\tau} ({\bf U}) \, g^{\tau\mu}({\bf U}) \, \equiv \,
\delta^{\mu}_{\nu}$).

 The Hamiltonian properties of systems (\ref{HTWhithSyst}) are
also closely related with their integrability. Thus, according to
conjecture of S.P. Novikov, any diagonalizable system
(\ref{HTWhithSyst}), which is Hamiltonian with respect to some
bracket (\ref{DNBracket}), can be integrated. The Novikov 
conjecture was proved by S.P. Tsarev (\cite{tsarev, tsarev2}),
who suggested a method for solving diagonalizable Hamiltonian 
systems (\ref{HTWhithSyst}). In fact, the method of Tsarev is 
applicable to a wider class of diagonalizable systems 
(\ref{HTWhithSyst}). The corresponding class of systems 
(\ref{HTWhithSyst}) was called by S.P. Tsarev semi-Hamiltonian 
and includes also the systems, Hamiltonian with respect to the 
weakly nonlocal generalizations of the Dubrovin - Novikov bracket 
- the Mokhov - Ferapontov bracket (\cite{mohfer1}) and general 
Ferapontov brackets (\cite{fer1, fer2}).

 The Dubrovin - Novikov procedure of averaging of a Poisson
bracket is closely connected with the conservative form of the
Whitham system in the case of one spatial dimension ($d = 1$).
The initial system has in this case the form
\begin{equation}
\label{OneDimInSyst}
\varphi^{i}_{t} \,\,\, = \,\,\, F^{i} \left( \bm{\varphi}, 
\bm{\varphi}_{x}, \dots \right) 
\end{equation}
and is supposed to be Hamiltonian with respect to a local
Poisson bracket
$$\{\varphi^{i}(x) \, , \, \varphi^{j}(y)\} \,\,\, = \,\,\, 
\sum_{k \geq 0}
\, B^{ij}_{(k)} (\bm{\varphi}, \bm{\varphi}_{x}, \dots)
\,\, \delta^{(k)}(x-y) $$
with a local Hamiltonian functional
$$H \,\,\, = \,\,\, \int P_{H} ( \bm{\varphi}, 
\bm{\varphi}_{x}, \dots ) \,\, d x $$

 The Dubrovin - Novikov procedure is based on the
existence of $\, N \, = \, 2 m + s \, $ local integrals
$$I^{\nu} \,\,\, = \,\,\, \int
P^{\nu}(\bm{\varphi}, \bm{\varphi}_{x},\dots) \, dx $$
commuting with the Hamiltonian $H$ and with each other
$$\{I^{\nu} \, , \, H\} \, = \, 0 \,\,\,\,\, , \,\,\,\,\,\,\,\,
\{I^{\nu} \, , \, I^{\mu}\} \, = \, 0 $$

 For the time evolution of the densities $\, P^{\nu} (x)$ 
we can then write:
$$P^{\nu}_{t} (\bm{\varphi}, \bm{\varphi}_{x},\dots) 
\,\,\, \equiv \,\,\,
Q^{\nu}_{x} (\bm{\varphi}, \bm{\varphi}_{x},\dots) $$
for some functions 
$Q^{\nu} (\bm{\varphi}, \bm{\varphi}_{x}, \dots)$.

 In the same way, the pairwise Poisson brackets of the
densities $\, P^{\nu} (x)$, $\, P^{\mu} (y)$ can be
represented in the form:
$$\{P^{\nu}(x) \, , \, P^{\mu}(y)\} 
\,\,\, = \,\,\, \sum_{k\geq 0}
A^{\nu\mu}_{k}(\bm{\varphi}, \bm{\varphi}_{x}, \dots) \,\,
\delta^{(k)}(x-y) $$
where 
$$A^{\nu\mu}_{0}(\bm{\varphi}, \bm{\varphi}_{x}, \dots)
\,\, \equiv \,\,
\partial_{x} Q^{\nu\mu}(\bm{\varphi}, \bm{\varphi}_{x}, 
\dots) $$
for some functions 
$Q^{\nu\mu} (\bm{\varphi}, \bm{\varphi}_{x},\dots)$.

 The number of the functionals $I^{\nu}$ is equal here to the
number of the parameters  \linebreak 
$(k^{1}, \dots, k^{m}, \, \omega^{1}, \dots, \omega^{m}, \,
n^{1}, \dots, n^{s})$ and the full regular Whitham system can
be written in the conservative form
\begin{equation}
\label{OneDimWhithSyst}
\langle P^{\nu} \rangle_{T} \,\,\, = \,\,\,
\langle Q^{\nu} \rangle_{X} \,\,\,\,\, , \,\,\,\,\,\,\,\,
\nu = 1, \dots, N \,\,\,\,\, ,
\end{equation}
where the operation $\langle \dots \rangle$ means the averaging
on the family of $m$-phase solutions of (\ref{OneDimInSyst}):
$$\langle f \rangle \,\,\, = \,\,\,
\int_{0}^{2\pi}\!\!\!\!\!\dots\int_{0}^{2\pi}
f \left( \bm{\Phi}, \, k^{\alpha} \bm{\Phi}_{\theta^{\alpha}},
\dots \right) \,\, {d^{m} \theta \over (2\pi)^{m}} $$

 Using the parameters 
$\, U^{\nu} \, = \, \langle P^{\nu} \rangle$ instead of the
parameters $({\bf k}, \, \bm{\omega}, \, {\bf n})$ on $\Lambda$,
we can write the averaged Poisson bracket on the space of the
functions $\, {\bf U} ({\bf X})$ in the form
(\cite{dn1, DubrNovDAN84, dn2, dn3}):
\begin{equation}
\label{OneDimAvBr}
\{U^{\nu}(X) \, , \, U^{\mu}(Y)\} \,\,\, = \,\,\,
\langle A^{\nu\mu}_{1}\rangle
({\bf U}) \,\, \delta^{\prime}(X-Y) \,\, + \,\,
{\partial \langle Q^{\nu\mu} \rangle \over
\partial U^{\gamma}} \,\, U^{\gamma}_{X} \,\, \delta (X-Y)
\end{equation}

 At the same time, the averaging of the Hamiltonian functional
$H$ gives the Hamiltonian functional
$$H_{av} \,\,\, = \,\,\, \int_{-\infty}^{+\infty}
\langle P_{H} \rangle \left( {\bf U} (X) \right) \,\, d X $$ 
for the Whitham system (\ref{OneDimWhithSyst}).

 In paper \cite{izvestia} the proof of the Jacobi identity
for bracket (\ref{OneDimAvBr}) under certain regularity
assumptions about the family of $m$-phase solutions of system
(\ref{OneDimInSyst}) was suggested. Let us say also, that in
paper \cite{MalPav} the consistency of the bracket averaging
procedure with the averaging of the Lagrangian structure in 
the case when both the procedures are possible was also
established. The most detailed consideration of the 
justification of the averaging of local field-theoretic 
Hamiltonian structures in one-dimensional case can be found
in \cite{Sigma}, where the new results concerning the 
appearance of ``resonances'' on the space of $m$-phase 
solutions of (\ref{OneDimInSyst}) are also presented.
Let us also note, that the generalization of the Dubrovin - 
Novikov procedure for the weakly nonlocal Poisson brackets
was also suggested in \cite{malnloc2}.

 We have to say, however, that the form (\ref{HTWhithSyst})
of the Whitham system is not so convenient in the 
multi-dimensional situation because of the presence of
additional constraints in this case. Besides that, the 
requirement of the presence of a complete set of integrals
$\, I^{\nu}$, $\, \nu = 1, \dots, m (d + 1) + s$, seems to be
too strict in the case $d > 1$ and is not satisfied in a number 
of examples. Instead, it is more convenient to consider the
Whitham system in the form (\ref{STomegaCond}), 
(\ref{fkappaortcond}), where the parameters 
$\, ({\bf k}_{1}, \dots, {\bf k}_{d})$ are given by the derivatives
of the function $\, {\bf S} ({\bf X}, T)$. As a result, it is
also more natural to represent the corresponding Hamiltonian 
structure in the form, considered in the previous chapter.
Thus, we will separate here the ``phase'' variables
$\, (S^{1} ({\bf X}), \dots , S^{m} ({\bf X}))$ and the ``density''
variables $\, (U^{1} ({\bf X}), \dots , U^{m+s} ({\bf X}))$. 
As we will see, the procedure of the averaging of a Poisson
bracket can be made in this case just in presence of a minimal
set of commuting integrals $\, (I^{1}, \dots , I^{m+s})$ on the
family of $m$-phase solutions of (\ref{EvInSyst}). Unfortunately,
the procedure described in the paper still can not be considered
as a general method for the multi-dimensional ($d \geq 2$) integrable
PDE's since both the Hamiltonian structures and the conservation
laws usually have there more complicated form. However, the scheme
we present here seems to be rather useful in many interesting
examples.

 Let us note also here that in paper \cite{PseudoPhaseAv} the
possibility of reduction of the necessary number of commuting 
integrals was considered from another point of view, using
the concept of ``pseudo-phases'', first introduced by Whitham
(\cite{whith3}). The approach, used in \cite{PseudoPhaseAv}
is applicable mostly to special physical systems, having some
additional symmetry properties. 

 Thus, we put again now $d \geq 1$ and consider $m$-phase
solutions of system (\ref{EvInSyst}), which is supposed to be
Hamiltonian with respect to the multi-dimensional Poisson bracket
(\ref{MultDimPBr}) with the Hamiltonian functional 
(\ref{MultDimHamFunc}).

 We will also assume here that the family $\Lambda$ represents
a complete regular family of $m$-phase solutions of system
(\ref{EvInSyst}) and there exist $m + s$ functionals 
(\ref{IgammaForm}), such that their values can be used as a full
set of parameters $\, {\bf U}$ at every fixed values of
$\, ({\bf k}_{1}, \dots, {\bf k}_{d})$.

 From the conditions
$${\rm rk} \,\, \left| \left|
{\partial U^{\gamma} \over \partial \omega^{\alpha}} \quad
{\partial U^{\gamma} \over \partial n^{l}} \right| \right|
\,\,\, = \,\,\, m \, + \, s $$
we easily get the following statement:

1) The vectors
$$\left\{ \bm{\Phi}_{\omega^{\alpha}}
(\bm{\theta}, \, {\bf k}_{1}, \dots , {\bf k}_{d}, 
\, \bm{\omega}, \, {\bf n}) \, ,
\,\,\,
\bm{\Phi}_{n^{l}}
(\bm{\theta}, \, {\bf k}_{1}, \dots , {\bf k}_{d},
\, \bm{\omega}, \, {\bf n}) 
\right\} $$
($\alpha = 1, \dots, m, \,\, l = 1, \dots, s$),
are linearly independent at every
$\, ({\bf k}_{1}, \dots, {\bf k}_{d}, \, \bm{\omega}, \, {\bf n})$;

2) The covectors
\begin{equation}
\label{zetaVarDer}
\zeta^{(\gamma)}_{i [{\bf k}_{1}, \dots, {\bf k}_{d}, \bm{\omega}, 
{\bf n}]} (\bm{\theta}) \,\,\,\,\, = \,\,\,\,\, \left.
{\delta J^{\gamma} \over \delta \varphi^{i} (\bm{\theta})}
\right|_{\bm{\varphi}(\bm{\theta}) \, = \, \bm{\Phi}
(\bm{\theta}, {\bf k}_{1}, \dots, {\bf k}_{d}, \bm{\omega}, {\bf n})}
\end{equation}
($\gamma = 1, \dots, m + s$), are linearly independent at every 
$\, ({\bf k}_{1}, \dots, {\bf k}_{d}, \, \bm{\omega}, \, {\bf n})$;

3) 
$${\rm rk} \,\, \left| \left| \left( \bm{\zeta}^{(\gamma)}
\, \cdot \, \bm{\Phi}_{\omega^{\alpha}} \right) \quad
\left( \bm{\zeta}^{(\gamma)} \, \cdot \, \bm{\Phi}_{n^{l}} \right) 
\right| \right|
\,\,\, = \,\,\, m \, + \, s $$
at every
$\, ({\bf k}_{1}, \dots, {\bf k}_{d}, \, \bm{\omega}, \, {\bf n})$,
where $\, ( \quad \cdot \quad )$ means the standard convolution
$$\left( \bm{\zeta} \, \cdot \, \bm{\xi} \right) 
\,\,\, \equiv \,\,\, \int_{0}^{2\pi}\!\!\!\!\!\dots\int_{0}^{2\pi}
\zeta_{i} (\bm{\theta}) \,\, \xi^{i} (\bm{\theta}) \,\, 
{d^{m} \theta \over (2\pi)^{m}} $$
on the space of $2\pi$-periodic in each $\theta^{\alpha}$
functions.

 It is not difficult to check also, that for any translationally
invariant first integral $\, I^{\gamma}$ of system (\ref{EvInSyst}),
having the form (\ref{IgammaForm}), the corresponding covector
$\bm{\zeta}^{(\gamma)}_{[{\bf k}_{1}, \dots, {\bf k}_{d}, {\bf U}]} 
(\bm{\theta})$ represents a regular left eigen-vector of the
operator 
${\hat L}^{i}_{j [{\bf k}_{1}, \dots, {\bf k}_{d}, {\bf U}]}$,
corresponding to the zero eigen-value. Thus, we can formulate here
the following proposition:

\vspace{0.2cm}

{\bf Proposition 3.1.}

{\it Let the family $\Lambda$ represent
a complete regular family of $m$-phase solutions of system
(\ref{EvInSyst}) and there exist $m + s$ first integrals
(\ref{IgammaForm}), such that their values can be used as a full
set of parameters $\, {\bf U}$ at every fixed values of   
$\, ({\bf k}_{1}, \dots, {\bf k}_{d})$. Then the corresponding
covectors 
$\bm{\zeta}^{(\gamma)}_{[{\bf k}_{1}, \dots, {\bf k}_{d}, {\bf U}]}
(\bm{\theta})$, defined by (\ref{zetaVarDer}), generate the full
space of the regular left eigen-vectors of the operators
${\hat L}^{i}_{j [{\bf k}_{1}, \dots, {\bf k}_{d}, {\bf U}]}$,
corresponding to the zero eigen-value. 
}

\vspace{0.2cm}

 For consideration of the Hamiltonian structure it will be 
convenient to write now the regular Whitham system in a slightly
different form.

\vspace{0.2cm}

{\bf Lemma 3.1.}

{\it Let the family $\Lambda$ represent
a complete regular family of $m$-phase solutions of system
(\ref{EvInSyst}) and there exist $m + s$ first integrals
(\ref{IgammaForm}), such that their values can be used as a full
set of parameters $\, {\bf U}$ at every fixed values of
$\, ({\bf k}_{1}, \dots, {\bf k}_{d})$. Let the time evolution of 
the corresponding densities $\, P^{\gamma} ({\bf x})$ according
to system (\ref{EvInSyst}) have the form
$$P^{\gamma}_{t} \left( \bm{\varphi}, \, \bm{\varphi}_{\bf x}, \,
\bm{\varphi}_{\bf xx}, \dots \right) \,\,\, = \,\,\,   
Q^{\gamma 1}_{x^{1}} \left( \bm{\varphi}, \,
\bm{\varphi}_{\bf x}, \, \bm{\varphi}_{\bf xx}, \dots \right) \, + \,
\dots \,\, + \,\, Q^{\gamma d}_{x^{d}} \left( \bm{\varphi}, \,
\bm{\varphi}_{\bf x}, \, \bm{\varphi}_{\bf xx}, \dots \right) $$
with some functions 
$(Q^{\gamma 1} (\bm{\varphi}, \dots ), \dots ,
Q^{\gamma d} (\bm{\varphi}, \dots ))$.

 Then the system
\begin{equation}
\label{MultDimConsWhithSyst}
\begin{array}{c}
S^{\alpha}_{T} \,\,\, = \,\,\, \omega^{\alpha} \left(
{\bf S}_{X^{1}}, \dots, {\bf S}_{X^{d}}, {\bf U} 
\right) \,\,\, , \,\,\,\,\,\,\,\, 
\alpha \, = \, 1, \dots, m \, ,  \\  \\
U^{\gamma}_{T} \,\, = \,\,
\langle Q^{\gamma 1} \rangle_{X^{1}} \, + \, \dots \, + \,
\langle Q^{\gamma d} \rangle_{X^{d}} \,\,\, , \,\,\,\,\,\,\,\, 
\gamma \, = \, 1, \dots , m + s
\end{array}
\end{equation}
($\, U^{\gamma} \, \equiv \, \langle P^{\gamma} \rangle$),
is equivalent to system (\ref{STomegaCond}), (\ref{fkappaortcond}).
}

\vspace{0.2cm}

 Proof.

 We have actually to prove the equivalence of the second part of
system (\ref{MultDimConsWhithSyst}) to relations 
(\ref{fkappaortcond}) under conditions (\ref{STomegaCond}).

 For convenience, let us introduce the functions
$$\Pi^{\gamma (l_{1} \dots l_{d})}_{i}
(\bm{\varphi}, \bm{\varphi}_{\bf x}, \dots )
\,\, \equiv \,\,
{\partial P^{\gamma} (\bm{\varphi}, \bm{\varphi}_{\bf x}, \dots )   
\over \partial \varphi^{i}_{l_{1} x^{1} \dots l_{d} x^{d}} }
\,\,\,\,\, , \,\,\,\,\,\,\,\, l_{1}, \dots , l_{d} \, \geq \, 0 $$

 From the expression for the evolution of the densities
$P^{\gamma} (\bm{\varphi}, \epsilon \bm{\varphi}_{\bf X}, \dots )$
we then get the following identities:
$$\sum_{l_{1}, \dots , l_{d}}
\epsilon^{l_{1} + \dots + l_{d}} \,\,
\Pi^{\nu (l_{1} \dots l_{d})}_{i}
(\bm{\varphi}, \, \epsilon \, \bm{\varphi}_{\bf X}, \dots ) \,
\left( F^{i} (\bm{\varphi}, \,
\epsilon \, \bm{\varphi}_{\bf X}, \dots )
\right)_{l_{1} X^{1} \dots l_{d} X^{d}} \,\,\, \equiv  $$
\begin{equation}
\label{PnuQnuIdentity}
\equiv \,\,\, \epsilon \, Q^{\nu 1}_{X^{1}}
(\bm{\varphi}, \, \epsilon \, \bm{\varphi}_{\bf X}, \dots )
\, + \, \dots \, + \, \epsilon \, Q^{\nu d}_{X^{d}}
(\bm{\varphi}, \, \epsilon \, \bm{\varphi}_{\bf X}, \dots )
\end{equation}  

 Let us substitute now the functions
\begin{equation}
\label{VarphiSubs}
\varphi^{i} (\bm{\theta}, {\bf X}) \,\, = \,\, \Phi^{i}
\left( \bm{\theta} + {{\bf S} ({\bf X}) \over \epsilon}, \, 
{\bf S}_{X^{1}}, \dots, {\bf S}_{X^{d}}, {\bf U}({\bf X}) \right)
\end{equation}
in the identity (\ref{PnuQnuIdentity}). Easy to see that every
operator $\, \epsilon \, \partial / \partial X^{p}$ acting on 
functions (\ref{VarphiSubs}) can be naturally represented as
the sum of the term 
$\, S^{\alpha}_{X^{p}} \, \partial / \partial \theta^{\alpha}$
and the terms, proportional to $\epsilon$. So, for any function
$f (\bm{\varphi}, \epsilon \, \bm{\varphi}_{\bf X}, \dots )$
on the submanifold (\ref{VarphiSubs}) we can write in fact the 
expansion
$$f (\bm{\varphi}, \epsilon \, \bm{\varphi}_{\bf X}, \dots ) 
\,\,\, = \,\,\, \sum_{l \geq 0} \, \epsilon^{l} \, 
f_{(l)} \left( \bm{\theta} + {{\bf S} ({\bf X}) \over \epsilon}, 
\,\, {\bf X} \right) $$
where all $f_{(l)} (\bm{\theta}, {\bf X})$ represent some local
functions of 
$\, ({\bf S}_{X^{1}}, \dots, {\bf S}_{X^{d}}, {\bf U} ({\bf X}))$
and their derivatives.

 It's not difficult to see also, that we can write:
$$\langle Q^{\gamma 1} \rangle_{X^{1}} 
\, + \, \dots \, + \,
\langle Q^{\gamma d} \rangle_{X^{d}} \,\, = \,\,
\int_{0}^{2\pi}\!\!\!\!\!\!\dots\!\int_{0}^{2\pi} 
\left( Q^{\gamma 1}_{X^{1} (1)} \, + \, \dots \, + \,
Q^{\gamma d}_{X^{d} (1)} \right)
\, {d^{m} \theta \over (2\pi)^{m}} \,\, = $$  
$$= \,\, \int_{0}^{2\pi}\!\!\!\!\!\!\dots\!\int_{0}^{2\pi}
\sum_{l_{1}, \dots, l_{d}}
\left( \Pi^{\gamma (l_{1} \dots l_{d})}_{i \, (0)} \,
F^{i}_{l_{1} X^{1} \dots l_{d} X^{d} \, (1)} \, + \,
\Pi^{\gamma (l_{1} \dots l_{d})}_{i \, (1)} \,
F^{i}_{l_{1} X^{1} \dots l_{d} X^{d} \, (0)} \right)
{d^{m} \theta \over (2\pi)^{m}} \,\, = $$
$$= \,\, \int_{0}^{2\pi}\!\!\!\!\!\!\dots\!\int_{0}^{2\pi}  
\sum_{l_{1}, \dots, l_{d}} \left(
\Pi^{\gamma (l_{1} \dots l_{d})}_{i \, (0)} \,\,
S^{\gamma^{1}_{1}}_{X^{1}} \dots 
S^{\gamma^{1}_{l_{1}}}_{X^{1}} \,\, \dots \,\,
S^{\gamma^{d}_{1}}_{X^{d}} \dots 
S^{\gamma^{d}_{l_{d}}}_{X^{d}} \,\,\,
F^{i}_{(1) \, \theta^{\gamma^{1}_{1}}\dots\theta^{\gamma^{1}_{l_{1}}}
\, \dots \, \theta^{\gamma^{d}_{1}}\dots\theta^{\gamma^{d}_{l_{d}}}}
\,\,\, +  \right. $$
$$+ \,\,\, \left.
\Pi^{\gamma (l_{1} \dots l_{d})}_{i \, (0)} \,
\left( \omega^{\beta} \, \Phi^{i}_{\theta^{\beta}}
\right)_{l_{1} X^{1} \dots l_{d} X^{d} \, (1)} \,\, + \,\,
\Pi^{\gamma (l_{1} \dots l_{d})}_{i \, (1)} \,
\left( \omega^{\beta} \, \Phi^{i}_{\theta^{\beta}}
\right)_{l_{1} X^{1} \dots l_{d} X^{d} \, (0)} \right)
\, {d^{m} \theta \over (2\pi)^{m}} \,\,\, =$$

$$=  \int_{0}^{2\pi}\!\!\!\!\!\!\dots\!\int_{0}^{2\pi}
\!\!\! \sum_{l_{1}, \dots, l_{d}} \!\! \left( \!
S^{\gamma^{1}_{1}}_{X^{1}} \dots 
S^{\gamma^{1}_{l_{1}}}_{X^{1}} \, \dots \,
S^{\gamma^{d}_{1}}_{X^{d}} \dots 
S^{\gamma^{d}_{l_{d}}}_{X^{d}} \,\,\,
(-1)^{l_{1} + \dots + l_{d}} \,\,
\Pi^{\gamma (l_{1} \dots l_{d})}_{i \, (0) \,
\theta^{\gamma^{1}_{1}}\dots\theta^{\gamma^{1}_{l_{1}}}
\, \dots \, \theta^{\gamma^{d}_{1}}\dots\theta^{\gamma^{d}_{l_{d}}}}
\,\, F^{i}_{(1)} \, +  \right. $$
$$+ \,\,\, \omega^{\beta}_{X^{1}} \,\,   
\Pi^{\gamma (l_{1} \dots l_{d})}_{i(0)} \,\, l_{1} \,\,
\Phi^{i}_{\theta^{\beta} \, (l_{1}-1) X^{1} \dots l_{d} X^{d} \, (0)}
\,\,\, + \, \dots \, + \,\,\,
\omega^{\beta}_{X^{d}} \,\,
\Pi^{\gamma (l_{1} \dots l_{d})}_{i(0)} \,\, l_{d} \,\,
\Phi^{i}_{\theta^{\beta} \, l_{1} X^{1} \dots (l_{d}-1) X^{d} \, (0)}
\,\,\, +$$
$$\left. + \,\,\, \omega^{\beta} \,\,
\Pi^{\gamma (l_{1} \dots l_{d})}_{i(0)} \,\,
\Phi^{i}_{\theta^{\beta} \, l_{1} X^{1} \dots l_{d} X^{d} \, (1)}
\,\,\, + \,\,\, \omega^{\beta} \,\,
\Pi^{\gamma (l_{1} \dots l_{d})}_{i(1)} \,\,
\Phi^{i}_{\theta^{\beta} \, l_{1} X^{1} \dots l_{d} X^{d} \, (0)}
\, \right) \, {d^{m} \theta \over (2\pi)^{m}} $$  

 Let us note now that the last two terms in the expression above
represent the integral of the value
$$\omega^{\beta} \, \sum_{l_{1}, \dots, l_{d}}
\left( \Pi^{\gamma (l_{1} \dots l_{d})}_{i} \,
\Phi^{i}_{\theta^{\beta} \, l_{1} X^{1} \dots l_{d} X^{d}}  
\right)_{(1)} \,\,\, \equiv \,\,\, \omega^{\beta} \, 
\partial P^{\gamma}_{(1)} / \partial \theta^{\beta}$$
and disappear after integration w.r.t. $\bm{\theta}$.
We can assume also, that the unessential
phase shift $\, {\bf S} ({\bf X}) / \epsilon $ in the integrands
above is omitted after taking all the differentiations w.r.t. 
${\bf X}$.

 From the other hand, we can write for the time derivative 
of the average $\langle P^{\gamma} \rangle$:
$$\langle P^{\gamma} \rangle_{T} \,\,\, \equiv   
\, \int_{0}^{2\pi}\!\!\!\!\!\!\dots\!\int_{0}^{2\pi} \!\!
\sum_{l_{1}, \dots, l_{d}}
\Pi^{\gamma (l_{1} \dots l_{d})}_{i \, (0)} 
\left(
k^{\gamma^{1}_{1}}_{1} \dots k^{\gamma^{1}_{l_{1}}}_{1} \, \dots \,
k^{\gamma^{d}_{1}}_{d} \dots k^{\gamma^{d}_{l_{d}}}_{d} \,\,
\Phi^{i}_{\theta^{\gamma^{1}_{1}}\dots\theta^{\gamma^{1}_{l_{1}}}
\, \dots \, \theta^{\gamma^{d}_{1}}\dots\theta^{\gamma^{d}_{l_{d}}}}
\right)_{T}  {d^{m} \theta \over (2\pi)^{m}} $$

 Putting $\, k^{\alpha}_{p} \, = \, S^{\alpha}_{X^{p}}$ and using
relations (\ref{STomegaCond}), we can see then after the 
integration by parts that the relations
$$\,\, \langle P^{\gamma} \rangle_{T} \,\, - \,\,
\langle Q^{\gamma 1} \rangle_{X^{1}}
- \dots - \langle Q^{\gamma d} \rangle_{X^{d}} \,\,\, = \,\,\, 0 $$
can be actually written in the form:
$$\int_{0}^{2\pi}\!\!\!\!\!\!\dots\!\int_{0}^{2\pi}
\zeta^{(\gamma)}_{i[{\bf S}_{\bf X}, {\bf U}({\bf X}, T)]} 
(\bm{\theta}) \,\, \Big( \Phi^{i}_{T} 
( \bm{\theta}, \, {\bf S}_{\bf X}, {\bf U}({\bf X}, T)) \, - \,
F^{i}_{[1]} (\bm{\theta}, {\bf X}, T) \Big) \,\,
{d^{m} \theta \over (2\pi)^{m}} \,\, = \,\, 0 $$

 We can see now, that the expressions above represent the
orthogonality of the covectors  \linebreak
$\, \bm{\zeta}^{(\gamma)}_{[{\bf S}_{\bf X}, {\bf U}({\bf X}, T)]}
(\bm{\theta})$ to the discrepancy
${\bf f}_{(1)} (\bm{\theta}, {\bf X}, T)$, introduced in
(\ref{FirstApprSyst}). From Proposition 3.1 we get then the
statement of the Lemma.

{\hfill Lemma 3.1 is proved.}

\vspace{0.2cm}

 For our further considerations we will need also the definitions,
characterizing the Hamiltonian properties of a complete regular
family of $m$-phase solutions of (\ref{EvInSyst}).

\vspace{0.2cm}

{\bf Definition 3.2.}

{\it We call a complete regular family $\Lambda$ a complete
Hamiltonian family of $m$-phase solutions of (\ref{EvInSyst})
if it represents a regular Hamiltonian submanifold in the space
of quasiperiodic functions according to Definition 2.1.
}

\vspace{0.2cm}

{\bf Definition 3.3.}

{\it We say that a complete Hamiltonian family $\Lambda$ of 
$m$-phase solutions of system (\ref{EvInSyst}) is equipped with 
a minimal set of commuting integrals if it represents a regular 
Hamiltonian submanifold equipped with a minimal set of commuting 
integrals according to Definition 2.2.
}

\vspace{0.2cm}

 We want to suggest now the Hamiltonian structure for the
Whitham system (\ref{MultDimConsWhithSyst}) under the requirement
that the family $\Lambda$ represents a complete Hamiltonian family 
of $m$-phase solutions of system (\ref{EvInSyst}) equipped with a
minimal set of commuting integrals $\, (I^{1}, \dots, I^{m+s})$.
As we will see, the Hamiltonian structures for system
(\ref{MultDimConsWhithSyst}) can be considered as a deformation
of the structures, arising in Chapter 2, with the aid of the
Dubrovin - Novikov procedure. Let us describe now the
corresponding scheme.

 As we said already, we assume that the integrals $\, I^{\gamma}$
have the form (\ref{IgammaForm}) with some densities
$P^{\gamma} (\bm{\varphi}, \bm{\varphi}_{\bf x}, \dots )$.
In complete analogy with the Dubrovin - Novikov scheme, we can
represent the pairwise Poisson brackets of the densities
$P^{\gamma} ({\bf x})$, $P^{\rho} ({\bf y})$ in the form
$$\{ P^{\gamma} ({\bf x}) \, , \, P^{\rho} ({\bf y}) \} \,\, = \,\,
\sum_{l_{1},\dots,l_{d}} A^{\gamma\rho}_{l_{1} \dots l_{d}}
(\bm{\varphi}, \bm{\varphi}_{\bf x}, \dots ) \,\,
\delta^{(l_{1})} (x^{1} - y^{1}) \, \dots \,
\delta^{(l_{d})} (x^{d} - y^{d}) $$
($l_{1}, \dots, l_{d} \geq 0$).

 View relations (\ref{CommutativeSet}) we can also write here
\begin{equation}
\label{A0QDerRel}
A^{\gamma\rho}_{0 \dots 0} (\bm{\varphi}, \bm{\varphi}_{\bf x}, \dots )
\,\,\, \equiv \,\,\, \partial_{x^{1}} \, Q^{\gamma\rho 1}
(\bm{\varphi}, \bm{\varphi}_{\bf x}, \dots ) \, + \, \dots \, + \,
\partial_{x^{d}} \, Q^{\gamma\rho d}
(\bm{\varphi}, \bm{\varphi}_{\bf x}, \dots ) 
\end{equation}
for some functions 
$(Q^{\gamma\rho 1} (\bm{\varphi}, \bm{\varphi}_{\bf x}, \dots ), 
\dots, Q^{\gamma\rho d} (\bm{\varphi}, \bm{\varphi}_{\bf x}, \dots ))$.

 Let us define now the averaged Poisson bracket 
$\{ \dots , \dots \}_{\rm AV}$ on the space of fields
$\, ({\bf S} ({\bf X}), {\bf U} ({\bf X}))$ by the formula
\begin{equation}
\label{AveragedBracket}
\begin{array}{c}
\left\{ S^{\alpha} ({\bf X})  \, , \,
S^{\beta} ({\bf Y}) \right\}_{\rm AV}
\,\,\, =  \,\,\, 0
\,\,\,\,\,\,\,\, , \,\,\,\,\,\,\,\,\,\,\,\,\,\,\,
\alpha, \beta \, = \, 1, \dots , m \, ,  \\  \\
\left\{ S^{\alpha} ({\bf X})  \, , \,
U^{\gamma} ({\bf Y}) \right\}_{\rm AV} \,\,\, = \,\,\,
\omega^{\alpha\gamma}
\left({\bf S}_{X^{1}}, \dots, {\bf S}_{X^{d}},
{\bf U} ({\bf X}) \! \right)
\, \delta ({\bf X} - {\bf Y})  \,\,\,\,\, ,  \\  \\
\left\{ U^{\gamma} ({\bf X})\, , \,
U^{\rho} ({\bf Y}) \right\}_{\rm AV}
\,\,\, = \,\,\, \langle A^{\gamma\rho}_{10\dots0} \rangle
\left({\bf S}_{X^{1}}, \dots, {\bf S}_{X^{d}},
{\bf U} ({\bf X}) \! \right) \,\,
\delta_{X^{1}} ({\bf X} - {\bf Y})
\,\,\, + \,\, \dots \,\, +   \\  \\
+ \,\,\, \langle A^{\gamma\rho}_{0\dots01} \rangle
\left({\bf S}_{X^{1}}, \dots, {\bf S}_{X^{d}},
{\bf U} ({\bf X}) \! \right) \,\,\,
\delta_{X^{d}} ({\bf X} - {\bf Y}) \,\,\, +   \\  \\
+ \,\,\, \left[ \langle Q^{\gamma\rho \, p} \rangle
\left({\bf S}_{X^{1}}, \dots, {\bf S}_{X^{d}},
{\bf U} ({\bf X}) \! \right)
\right]_{X^{p}} \,\,\, \delta ({\bf X} - {\bf Y})
\,\,\,\,\,\,\,\, , \,\,\,\,\,\,\,\,\,\,\,\,\,\,\,
\gamma, \rho \, = \, 1, \dots , m + s
\end{array}
\end{equation}

 In the next chapter we will prove the Jacobi identity for the
bracket (\ref{AveragedBracket}) and show that system
(\ref{MultDimConsWhithSyst}) is Hamiltonian with respect to
bracket (\ref{AveragedBracket}) with the Hamiltonian functional
\begin{equation}
\label{HamFunctional}
H_{av} \,\,\, = \,\,\, \int \langle P_{H} \rangle \,
\left({\bf S}_{\bf X}, \, U^{1}({\bf X}), \dots, U^{m+s}({\bf X})
\right) \,\, d^{d} X 
\end{equation}

 In the main features, the next chapter will be of a technical 
nature.

 Let us note now, that the construction of the Poisson bracket
(\ref{AveragedBracket}) can be also justified in terms
of the Dubrovin - Novikov scheme in the presence of a 
``complete'' set of commuting integrals 
$\, (I^{1}, \dots, I^{N})$, $\, N = m (d + 1) + s$ (see \cite{JMP2}).
However, as we said already, the presence of a complete set of
commuting integrals can be in fact too strict in the multi-dimensional
situation. All the considerations here will involve
just the presence of a minimal set of commuting integrals
$\, (I^{1}, \dots, I^{m+s})$, which is a much weaker requirement
in general case.

\section{The averaging of a local Poisson bracket.}
\setcounter{equation}{0}

 In this chapter we will consider in detail the proof of the Jacobi
identity for the averaged Poisson bracket. The proof will follow the
general idea of the Dirac restriction of a Poisson bracket on a
submanifold and have some common features with the considerations
of Chapter 2. Let us note also that the considerations of this chapter
will contain quite a lot of technical calculations. 

 First of all, we will introduce here a special submanifold
${\cal K}$ in the space of functions 
$\, \bm{\varphi} (\bm{\theta}, {\bf X}) \, $ consisting of the
functions
$$\varphi^{i} (\bm{\theta}, {\bf X}) \,\,\, = \,\,\,
\Phi^{i} \left( \bm{\theta} +
{{\bf S} ({\bf X}) \over \epsilon}, \, {\bf S}_{X^{1}},
\dots, {\bf S}_{X^{d}}, {\bf U} ({\bf X}) \right) \,\, , $$   
such that the functions 
$\, \bm{\varphi} (\bm{\theta}, {\bf X}) \, \in \, {\cal K} \, $ 
represent the functions from the family $\, \Lambda \, $
at every fixed $\, {\bf X}$. As we can see, the functions
$\, {\bf S} ({\bf X}) \, = \, 
(S^{1} ({\bf X}), \dots, S^{m} ({\bf X})) \, $ and
$\, {\bf U} ({\bf X}) \, = \,
(U^{1} ({\bf X}), \dots, U^{m+s} ({\bf X})) \, $ represent a
special ``coordinate system'' on the submanifold ${\cal K}$, 
depending on the parameter $\, \epsilon$.

 Like in Chapter 2, we will introduce here special functionals
$\, ({\bf S}^{[\bm{\zeta}]} ({\bf X}), \,
{\bf U}^{[\bm{\zeta}]} ({\bf X})) \, $ (for any fixed values
$\, \bm{\zeta} ({\bf X}) \, = \, 
(\zeta^{1} ({\bf X}), \dots, \zeta^{m} ({\bf X})) \, $
representing the coordinates
$\, ({\bf S} ({\bf X}), \, {\bf U} ({\bf X})) \, $
after the restriction on ${\cal K}$ and defined near the ``points''
$$\varphi^{i} (\bm{\theta}, {\bf X}) \,\,\, = \,\,\,
\Phi^{i} \left( \bm{\theta} +
{\bm{\zeta} ({\bf X}) \over \epsilon}, \, \bm{\zeta}_{X^{1}},
\dots, \bm{\zeta}_{X^{d}}, {\bf U} ({\bf X}) \right) 
\,\,\,\,\, \in \,\,\,\,\, {\cal K} $$
in the functional space.

 As will be shown below, for any regular Hamiltonian family 
$\, \Lambda $, equipped with a minimal set of commuting integrals
$\, (I^{1}, \dots, I^{m+s})$, the pairwise Poisson brackets of the 
functionals  \linebreak
$\, ({\bf S}^{[\bm{\zeta}]} ({\bf X}), \,
{\bf U}^{[\bm{\zeta}]} ({\bf X})) \, $ on ${\cal K}$ coincide in the
main order of $\, \epsilon \, $ with the relations 
(\ref{AveragedBracket}). Besides that, it can be shown, that for the
functionals $\, ({\bf S}^{[\bm{\zeta}]} ({\bf X}), \,
{\bf U}^{[\bm{\zeta}]} ({\bf X})) \, $ defined in the right way
the higher terms in $\, \epsilon \, $ of their pairwise Poisson 
brackets demonstrate necessary regularity properties in the vicinity
of the submanifold ${\cal K}$. The proof of the corresponding statements 
is based on some technical calculations and is represented in the first
part of the chapter.

 The proof of the Jacobi identity for the bracket 
(\ref{AveragedBracket}) is based on the idea of the Dirac restriction 
of a Poisson bracket on a submanifold and requires the resolvability
of some special linear systems of PDE's on the submanifold ${\cal K}$.
The corresponding systems are connected with the functionals
$\, ({\bf S}^{[\bm{\zeta}]} ({\bf X}), \,
{\bf U}^{[\bm{\zeta}]} ({\bf X})) \, $ on the submanifold ${\cal K}$
and represent systems of PDE's on the space of $2\pi$-periodic
in each $\, \theta^{\alpha} \, $ functions 
$\, \beta_{i} (\theta^{1}, \dots, \theta^{m}) \, $ 
at every fixed $\, {\bf X}$. The operators of the systems are given 
now by the main terms of the pairwise Poisson brackets of constraints
$\, g^{i} (\bm{\theta}, {\bf X}) \, $ on the submanifold ${\cal K}$
and we have actually here the situation similar to that considered in 
Chapter 2. It can be shown again, that under the assumptions of our
scheme the right-hand parts of these systems are automatically 
orthogonal to all the ``regular'' (left) eigen-vectors of the system
operators corresponding to the zero eigen-value, which provides
in fact the resolvability of these systems in a large set of
``regular'' examples. In particular, we can claim the ``regular''
behavior of the solutions of these systems in the single-phase case
and also for a special class of rather simple Poisson brackets
(\ref{MultDimPBr}).

 However, in the general multi-phase case the situation is more
complicated and the systems mentioned above can be actually
unresolvable at some ``points'' of the submanifold ${\cal K}$ because
of rather irregular behavior of the spectra of the system operators.
Fortunately, for the proof of the Jacobi identity for the bracket
(\ref{AveragedBracket}) it is enough to require again the solvability
of the mentioned systems just on some dense set of parameters on the 
family $\, \Lambda$. Theorem 4.1 is devoted to the proof of the Jacobi 
identity for the bracket (\ref{AveragedBracket}) under the
corresponding assumptions. Let us say here also, that the assumptions
of Theorem 4.1 seem in fact to be rather natural in the multi-phase
situation and are satisfied in the known examples.

 Theorem 4.2 is devoted to the proof of the invariance of the
procedure of bracket averaging with respect to the choice of the set
of functionals $\, \{ I^{\gamma} \}$. 

 Finally, in Theorem 4.3 we prove that the averaged bracket 
(\ref{AveragedBracket}) gives the Hamiltonian structure for the system
(\ref{MultDimConsWhithSyst}) with the Hamiltonian functional
(\ref{HamFunctional}).

\vspace{0.2cm}

 Let us start now the detailed consideration of the stated scheme. 
First, we have to introduce the extended space of fields making the 
change
$$\bm{\varphi}({\bf x}) \,\, \rightarrow \,\,
\bm{\varphi}(\bm{\theta}, {\bf x}) $$
where the new functions $\bm{\varphi}(\bm{\theta}, {\bf x})$
are $2\pi$-periodic with respect to each $\theta^{\alpha}$
at every fixed values of ${\bf X}$.

 Easy to see that we can define the Poisson bracket on the extended 
field space by expression
$$\{\varphi^{i}(\bm{\theta}, {\bf x}) \, , \,
\varphi^{j}(\bm{\theta}^{\prime}, {\bf y})\}
\,\,\, =  \,\,\, \sum_{l_{1},\dots,l_{d}}
B^{ij}_{(l_{1},\dots,l_{d})}
(\bm{\varphi}, \bm{\varphi}_{\bf x}, \dots)
\,\, \delta^{(l_{1})}(x^{1} - y^{1}) \dots
\delta^{(l_{d})}(x^{d} - y^{d})
\,\, \delta (\bm{\theta} - \bm{\theta}^{\prime}) $$

 After the replacement 
${\bf x} \rightarrow {\bf X} = \epsilon {\bf x}$ we can then
introduce the Poisson bracket
\begin{multline}
\label{EpsExtBracket}
\{\varphi^{i}(\bm{\theta}, {\bf X}) \, , \,
\varphi^{j}(\bm{\theta}^{\prime}, {\bf Y})\}
\, = \\
= \, \sum_{l_{1},\dots,l_{d}}  \!
\epsilon^{l_{1} + \dots + l_{d}} \,
B^{ij}_{(l_{1},\dots,l_{d})}
(\bm{\varphi}, \, \epsilon\, \bm{\varphi}_{\bf X}, \dots)
\,\, \delta^{(l_{1})}(X^{1} - Y^{1})  \dots
\delta^{(l_{d})}(X^{d} - Y^{d})
\, \delta (\bm{\theta} - \bm{\theta}^{\prime})
\end{multline}
on the space of fields $\bm{\varphi} (\bm{\theta}, {\bf X})$.

 Let us describe now the submanifold ${\cal K}$ in the extended 
field space connected with the slow modulated $m$-phase solutions
of system (\ref{EvInSyst}). Namely, we will assume that the functions
$\bm{\varphi}(\bm{\theta}, {\bf X}) \in {\cal K}$ represent functions
from the family ${\hat \Lambda}$ at any fixed value of ${\bf X}$.
Besides that, we set, that the functions
$${\bf S} ({\bf X}) \,\, = \,\,
\left( S^{1} ({\bf X}), \dots, S^{m} ({\bf X}) \right) \,\,\, ,
\,\,\,\,\, {\bf U} ({\bf X}) \,\, = \,\,
\left( U^{1} ({\bf X}), \dots, U^{m+s} ({\bf X}) \right) $$
give a ``coordinate system'' on the submanifold ${\cal K}$ such that
the functions $\bm{\varphi}(\bm{\theta}, {\bf X}) \in {\cal K}$
have the form
\begin{equation}
\label{FormaPhinaK}
\varphi^{i} (\bm{\theta}, {\bf X}) \,\, = \,\, \Phi^{i} \left(
{{\bf S} ({\bf X}) \over \epsilon}
\, + \, \bm{\theta}, \, {\bf S}_{X^{1}}, \dots , {\bf S}_{X^{d}}, \,
{\bf U} ({\bf X}) \right)
\end{equation}

 We can see that, according to our definition, the parameters
$({\bf k}_{1}, \dots, {\bf k}_{d})$ on ${\hat \Lambda}$ are 
directly connected here with the derivatives
$({\bf S}_{X^{1}}, \dots, {\bf S}_{X^{d}})$ of the functions
$(S^{1} ({\bf X}), \dots, S^{m} ({\bf X}))$.

 Like in Chapter 2, we will need to introduce here the
functionals on the extended field space giving the values of the
parameters ${\bf S} ({\bf X})$ and ${\bf U} ({\bf X})$ on the
submanifold ${\cal K}$.

 Let us define the following functionals on the space of
functions $\bm{\varphi} (\bm{\theta}, {\bf X})$:
\begin{equation}
\label{FuncJnuX}
J^{\gamma}({\bf X}) \,\, = \,\, 
\int_{0}^{2\pi}\!\!\!\!\!\dots\int_{0}^{2\pi}
P^{\gamma} (\bm{\varphi}, \, \epsilon \bm{\varphi}_{\bf X}, \,
\epsilon^{2} \bm{\varphi}_{\bf XX}, \dots ) \,\,
{d^{m} \theta \over (2\pi)^{m}} \,\,\,\,\,\,\,\, , \,\,\,\,\, 
\gamma = 1, \dots, m + s
\end{equation}

 It's not difficult to see that we can write for the values of
$J^{\gamma}({\bf X})$ on the submanifold ${\cal K}$:
\begin{equation}
\label{JUtransform}
J^{\gamma}({\bf X}) \,\, = \,\, 
U^{\gamma}({\bf X}) \, + \, \sum_{l\geq1} \epsilon^{l} \, 
J^{\gamma}_{(l)} \left( {\bf X}, [{\bf S}_{\bf X}, {\bf U}]
\right) 
\,\,\,\,\,\,\,\, , \,\,\,\,\,
\gamma = 1, \dots, m + s
\end{equation}
where $J^{\gamma}_{(l)}({\bf X}, [{\bf S}_{\bf X}, {\bf U}])$ 
- are polynomials in the 
derivatives $({\bf U}_{\bf X}, {\bf U}_{\bf XX}, \dots)$
and \linebreak
$({\bf S}_{\bf XX}, {\bf S}_{\bf XXX}, \dots)$
with coefficients depending on the values ${\bf U} ({\bf X})$
and ${\bf k}_{q} ({\bf X}) = {\bf S}_{X^{q}}$ on ${\cal K}$.
Easy to see also that the right-hand part of expression
(\ref{JUtransform}) does not contain an explicit dependence on
the values ${\bf S} ({\bf X}) / \epsilon$ view the invariance 
of the functionals $J^{\gamma}({\bf X})$ w.r.t. the 
transformation
$${\bf S} ({\bf X}) \,\,\, \rightarrow \,\,\, {\bf S} ({\bf X})
\,\, + \,\, {\rm const} $$

 Let us introduce here the following gradation rule. Namely, we
prescribe gradation degree $0$ to the functions 
${\bf U} ({\bf X})$ and ${\bf k}_{q} ({\bf X}) = {\bf S}_{X^{q}}$
on ${\cal K}$ while every differentiation w.r.t. ${\bf X}$
increases the gradation degree by $1$. The expression
(\ref{JUtransform}) represents then the graded expansion of the
values of $J^{\gamma}({\bf X})$ on ${\cal K}$. Let us note also
that the functions $S^{\alpha} ({\bf X})$ have gradation degree
$-1$ in this scheme.

 It is easy to see that the transformations (\ref{JUtransform})
can be inverted as a formal series in $\epsilon$ at every fixed
${\bf S} ({\bf X})$, such that we can write
\begin{equation} 
\label{UJtransform}
U^{\gamma}({\bf X}) \,\, = \,\, 
J^{\gamma}({\bf X}) \, + \, \sum_{l\geq1} \epsilon^{l} \, 
U^{\gamma}_{(l)} \left({\bf X}, [{\bf S}_{\bf X}, {\bf J}] \right) 
\,\,\,\,\,\,\,\, , \,\,\,\,\,
\gamma = 1, \dots, m + s
\end{equation}
on the functions of the submanifold ${\cal K}$.

 The functions 
$\, U^{\gamma}_{(l)} ({\bf X}, [{\bf S}_{\bf X}, {\bf J}]) \, $ 
are given here by polynomials in 
$\, ({\bf J}_{\bf X}, {\bf J}_{\bf XX}, \dots) \, $
and \linebreak
$\, ({\bf S}_{\bf XX}, {\bf S}_{\bf XXX}, \dots) \, $
with coefficients depending on the values of ${\bf J}$ and
${\bf S}_{\bf X}$ at the point ${\bf X}$. Now the functions
$U^{\gamma}_{(l)}$ have degree $l$ according to analogous rule
applied to the functions ${\bf J}({\bf X})$ and 
${\bf S}_{\bf X}$. Let us note that this gradation rule does
not coincide exactly with the rule formulated above view the
nontrivial connection between the values of ${\bf U}({\bf X})$
and ${\bf J}({\bf X})$ on ${\cal K}$.

 Let us now fix the values 
$\, S^{\alpha} ({\bf X}) = \zeta^{\alpha} ({\bf X})$ on
the submanifold ${\cal K}$. Like in Chapter 2, we 
are going to define the local functionals
$S^{\alpha [\bm{\zeta}]} ({\bf X})$ in the region of the
extended field space near the point
$({\bf U}({\bf X}), \bm{\zeta} ({\bf X}))$ of the submanifold
${\cal K}$.

 Let us introduce here the functionals
$$\vartheta^{[\bm{\zeta}]}_{\alpha} ({\bf X}) \, = \,
\int_{0}^{2\pi}\!\!\!\!\!\dots\int_{0}^{2\pi}
\sum_{i=1}^{n} \, \varphi^{i} (\bm{\theta}, {\bf X}) \,\,
\Phi^{i}_{\theta^{\alpha}} \left( \bm{\theta} + 
{\bm{\zeta} ({\bf X}) \over \epsilon}, \, \bm{\zeta}_{X^{1}},
\dots, \bm{\zeta}_{X^{d}}, {\bf J} ({\bf X}) \right) 
{d^{m} \theta \over (2\pi)^{m}} $$

 For the functions
$$\varphi^{i} (\bm{\theta}, {\bf X}) \,\,\, = \,\,\,
\Phi^{i} \left( \bm{\theta} +
{{\bf S} ({\bf X}) \over \epsilon}, \, {\bf S}_{X^{1}},
\dots, {\bf S}_{X^{d}}, {\bf U} ({\bf X}) \right) \,\, , $$
where 
$\, {\bf S}({\bf X}) \, = \, \bm{\zeta} ({\bf X}) + 
\epsilon \, \Delta \bm{\theta}_{0} ({\bf X})$, we then have
$$\vartheta^{[\bm{\zeta}]}_{\alpha} ({\bf X}) \, = \,
\int_{0}^{2\pi}\!\!\!\!\!\!\dots\!\int_{0}^{2\pi} \sum_{i=1}^{n} 
\Phi^{i} \left( \bm{\theta} + 
\Delta \bm{\theta}_{0} ({\bf X}), 
\bm{\zeta}_{\bf X} + 
\epsilon \Delta \bm{\theta}_{0 {\bf X}}, 
{\bf U}({\bf X}) \right) \,
\Phi^{i}_{\theta^{\alpha}} \left( \bm{\theta},
\bm{\zeta}_{\bf X}, {\bf J}({\bf X}) |_{\cal K} \right)
{d^{m} \theta \over (2\pi)^{m}} $$

 View relations (\ref{UJtransform}) and (\ref{detMab})
on ${\cal K}$ we can see that the mapping
$$\Delta \bm{\theta}_{0} ({\bf X}) 
\,\,\,\,\, \rightarrow \,\,\,\,\,
\bm{\vartheta}^{[\bm{\zeta}]} ({\bf X}) |_{\cal K} $$
is locally invertible in the main order of $\epsilon$ at any 
fixed values of ${\bf J}({\bf X})$. As a result, we can locally
represent the values of $\Delta \bm{\theta}_{0} ({\bf X})$ 
on ${\cal K}$ in the form of the asymptotic expansions:
$$\Delta \bm{\theta}_{0} ({\bf X}) \,\, = \,\,
\tau^{\alpha} \left( 
\vartheta^{[\bm{\zeta}]}_{1} ({\bf X}) |_{\cal K},
\dots, \vartheta^{[\bm{\zeta}]}_{m} ({\bf X}) |_{\cal K}, \,
J^{1} ({\bf X}) |_{\cal K}, \dots, J^{m+s} ({\bf X}) |_{\cal K}
\right) \,\, +   $$
$$+ \,\, \sum_{l\geq1} \, \epsilon^{l} \,\,
\Delta \theta^{\alpha}_{0(l)} \left( {\bf X}, \,
\left[ \bm{\vartheta}^{[\bm{\zeta}]} ({\bf X}) |_{\cal K}, \,
{\bf J}({\bf X}) |_{\cal K}, \,
\bm{\zeta}_{\bf X} \right] \right) $$
where $\Delta \theta^{\alpha}_{0(l)} ({\bf X})$ are some 
local functions of their arguments and their derivatives,
polynomial in the derivatives and
having gradation degree $l$ in terms of the total number
of differentiations of the functions 
$\, \bm{\vartheta}^{[\bm{\zeta}]} ({\bf X}) |_{\cal K}$,
$\, {\bf J}({\bf X}) |_{\cal K}$, and
$\, \bm{\zeta}_{X^{q}}$ w.r.t. ${\bf X}$.

 Let us define now the functionals
${\tilde S}^{\alpha [\bm{\zeta}]} ({\bf X})$ 
by the asymptotic series
$${\tilde S}^{\alpha [\bm{\zeta}]} ({\bf X}) \,\, = \,\,
\bm{\zeta} ({\bf X}) \, + \, 
\epsilon \, \tau^{\alpha} \left( 
\vartheta^{[\bm{\zeta}]}_{1} ({\bf X}),
\dots, \vartheta^{[\bm{\zeta}]}_{m} ({\bf X}), \,
J^{1} ({\bf X}), \dots, J^{m+s} ({\bf X}) 
\right) \,\, +   $$
\begin{equation}
\label{SalphazetaDef}
+ \,\, \epsilon \, \sum_{l\geq1} \, \epsilon^{l} \,\,
\Delta \theta^{\alpha}_{0(l)} \left( {\bf X}, \,
\left[ \bm{\vartheta}^{[\bm{\zeta}]} ({\bf X}), 
\, {\bf J}({\bf X}), \,
\bm{\zeta}_{\bf X} \right] \right) 
\end{equation}

 Substituting the functionals 
${\tilde {\bf S}}^{[\bm{\zeta}]} ({\bf X})$ 
in (\ref{UJtransform}) we
can then also define the functionals
\begin{equation} 
\label{UgammazetaDef}
{\tilde U}^{\gamma [\bm{\zeta}]}({\bf X}) \,\,\, = \,\,\,
J^{\gamma}({\bf X}) \,\, + \,\, \sum_{l\geq1} \epsilon^{l} \,
U^{\gamma}_{(l)} 
\left({\bf X}, 
[{\tilde {\bf S}}^{[\bm{\zeta}]}_{\bf X}, {\bf J}] \right)
\,\,\,\,\, , \,\,\,\,\,
\gamma = 1, \dots, m + s 
\end{equation}
in the neighborhood of the ``points'' 
$({\bf U}({\bf X}), \bm{\zeta} ({\bf X}))$ of the submanifold
${\cal K}$.

 Like in Chapter 2, we can introduce here also the
constraints
$${\tilde g}^{i [\bm{\zeta}]} (\bm{\theta}, {\bf X}) 
\,\,\, = \,\,\, \varphi^{i} (\bm{\theta}, {\bf X})
\,\, - \,\, \Phi^{i} \left( \bm{\theta} \, + \,
{{\tilde {\bf S}}^{[\bm{\zeta}]} ({\bf X}) \over \epsilon}\, , \,\, 
{\tilde {\bf S}}^{[\bm{\zeta}]}_{X^{1}}, \dots, 
{\tilde {\bf S}}^{[\bm{\zeta}]}_{X^{d}}, 
{\tilde {\bf U}}^{[\bm{\zeta}]} ({\bf X}) \right) $$

 The constraints 
${\tilde g}^{i [\bm{\zeta}]} (\bm{\theta}, {\bf X})$ are
``numerated'' now by the index $\, i = 1, \dots, n \,$ and the
``continuous indices'' $\, \theta^{\alpha} \in [0, 2\pi )$,
$\, X^{q} \in (-\infty , +\infty)$.

 Let us note here that we will mainly consider the ``regularized''
functionals
$$J_{[{\bf a}]} \,\,\, = \,\,\, \int J^{\gamma} ({\bf X}) \,\,
a_{\gamma} ({\bf X}) \,\, d^{d} X \,\,\,\,\, , \,\,\,\,\,\,\,\,
{\tilde U}^{[\bm{\zeta}]}_{[{\bf a}]} \,\,\, = \,\,\, 
\int {\tilde U}^{\gamma [\bm{\zeta}]} ({\bf X}) \,\,
a_{\gamma} ({\bf X}) \,\, d^{d} X \,\,\,\,\, , $$
$${\tilde S}^{[\bm{\zeta}]}_{[{\bf f}]} \,\,\, = \,\,\, \int
{\tilde S}^{\alpha [\bm{\zeta}]} ({\bf X}) \,\,
f_{\alpha} ({\bf X}) \,\, d^{d} X $$
with smooth compactly supported functions $\, {\bf a} ({\bf X})$,
$\, {\bf f} ({\bf X})$.

 We can see that the ``gradients'' of the functionals
$\, {\tilde S}^{[\bm{\zeta}]}_{[{\bf f}]}$ have the order
$O (\epsilon)$ on ${\cal K}$ according to the definition
of the functionals 
$\, {\tilde S}^{\alpha [\bm{\zeta}]} ({\bf X})$.

\vspace{0.2cm}

 Easy to see that the functionals $J_{[{\bf a}]}$ are
invariant w.r.t. the transformations
\begin{equation}
\label{thetashift}
\bm{\varphi} \, (\bm{\theta}, \, {\bf X}) 
\,\,\, \rightarrow \,\,\, 
\bm{\varphi} \, (\bm{\theta} + \Delta \bm{\theta}, \, {\bf X})
\end{equation}

 As a corollary, we can write for the ``gradients'' of
$J_{[{\bf a}]}$:
$${\cal L}_{\alpha} \left[
{\delta J_{[{\bf a}]} \over 
\delta \varphi^{i} (\bm{\theta}, {\bf X})} \right] 
\,\,\, \equiv \,\,\, 0 \,\,\,\,\, , \,\,\,\,\,\,\,\,
\alpha = 1, \dots, m $$
where ${\cal L}_{\alpha}$ denote the Lie derivatives along
the vector fields
$$\bm{\nu}_{\alpha} \,\,\, = \,\,\, \int
{d^{m} \theta \over (2\pi)^{m}} \,\, d^{d} X \,\,\,
\varphi^{i}_{\theta^{\alpha}} (\bm{\theta}, {\bf X}) \,\,
{\delta \over \delta \varphi^{i} (\bm{\theta}, {\bf X})} $$
corresponding to transformations (\ref{thetashift}).

 Let us note that the functionals
$\, {\tilde U}^{[\bm{\zeta}]}_{[{\bf a}]}$ 
do not posses in general 
the analogous property being defined with the aid of the 
functionals ${\tilde {\bf S}}^{[\bm{\zeta}]} ({\bf X})$.

 Now, let us redefine the functionals
$\, {\tilde S}^{\alpha [\bm{\zeta}]} ({\bf X})$
putting
\begin{equation}
\label{NewSDefinition}
S^{\alpha [\bm{\zeta}]} ({\bf X}) \,\,\,\,\, = \,\,\,\,\,
{\tilde S}^{\alpha [\bm{\zeta}]} ({\bf X}) \,\,\, + 
\end{equation}
$$+ \,\,\, \epsilon \int \,
{\tilde h}^{\alpha [\bm{\zeta}]}_{j} \left(
{\bf X}, \, \bm{\theta} +
{{\tilde {\bf S}}^{[\bm{\zeta}]} ({\bf Y}) \over \epsilon}
- {\bm{\zeta} ({\bf Y}) \over \epsilon} ,
\, {\bf Y}; \,\, \left[ {\tilde {\bf U}}^{[\bm{\zeta}]}, 
\bm{\zeta}, \epsilon \right] \right) \,\,
{\tilde g}^{j [\bm{\zeta}]} (\bm{\theta}, {\bf Y}) \,\,\,
{d^{m} \theta \over (2\pi)^{m}} \,\, d^{d} Y $$
where the distribution 
${\tilde h}^{\alpha [\bm{\zeta}]}_{j} (
{\bf X}, \, \bm{\theta}, \, {\bf Y}; \,\, \left[ {\bf U},
{\bf S}, \epsilon \right] )$ is defined by the formula
$$\epsilon \,\, {\tilde h}^{\alpha [\bm{\zeta}]}_{j} \left(
{\bf X}, \, \bm{\theta}, \, {\bf Y}; \,\, \left[ {\bf U}, 
{\bf S}, \epsilon \right] \right) 
\,\,\,\,\, \equiv \,\,\,\,\,  
\left. {\delta {\tilde S}^{\alpha [\bm{\zeta}]} ({\bf X}) 
\over \delta \varphi^{j} (\bm{\theta}, {\bf Y})}
\right|_{\bm{\varphi} = \bm{\Phi}_{[{\bf U}, {\bf S}]}} $$

 The last notation means here that the variation derivative
of $\, {\tilde S}^{\alpha [\bm{\zeta}]} ({\bf X})$ is taken on
the function
$$\bm{\varphi} (\bm{\theta}, {\bf Y}) \,\,\, = \,\,\,
\bm{\Phi} \left( \bm{\theta} +
{{\bf S} ({\bf Y}) \over \epsilon},
\,\, {\bf S}_{Y^{1}}, \dots, {\bf S}_{Y^{d}},
{\bf U} ({\bf Y}) \right) $$
from the submanifold ${\cal K}$.

 According to their definition, the functionals 
$\, S^{\alpha [\bm{\zeta}]} ({\bf X})$ take the same values 
on the submanifold ${\cal K}$ as the functionals
$\, {\tilde S}^{\alpha [\bm{\zeta}]} ({\bf X})$.

 Let us use now the smooth functions
$${\tilde h}^{[\bm{\zeta}]}_{j [{\bf f}]}
\left( \bm{\theta}, {\bf Y}; \, \left[ {\bf U},
{\bf S}, \epsilon \right] \right) 
\,\,\,\,\, \equiv \,\,\,\,\, \int \,
{\tilde h}^{\alpha [\bm{\zeta}]}_{j} \left(
{\bf X}, \, \bm{\theta}, {\bf Y};
\, \left[ {\bf U}, {\bf S}, \epsilon \right] \right) \,\,
f_{\alpha} ({\bf X}) \,\,\, d ^{d} X $$
with compactly supported $f_{\alpha} ({\bf X})$, corresponding to
the regularized functionals 
${\tilde S}^{[\bm{\zeta}]}_{[{\bf f}]}$. From the definition of
the functionals $\, {\tilde S}^{\alpha [\bm{\zeta}]} ({\bf X})$
it's not difficult to get the following relations for
$\, {\bf S} ({\bf X}) \equiv \bm{\zeta} ({\bf X})$:
$${\tilde h}^{[\bm{\zeta}]}_{j [{\bf f}]}
\left( \bm{\theta}, {\bf Y}; \, \left[ {\bf U},
\bm{\zeta}, \epsilon \right] \right) \,\,\,\,\, = \,\,\,\,\, 
\sum_{l\geq0} \, \epsilon^{l} \, 
{\tilde h}^{[\bm{\zeta}]}_{j [l] [{\bf f}]} \left(
\bm{\theta} + {\bm{\zeta} ({\bf Y}) \over \epsilon}, \, {\bf Y}; \,
\left[ {\bf U}, \bm{\zeta}_{\bf Y} \right] \right) $$
where all ${\tilde h}^{[\bm{\zeta}]}_{j [l] [{\bf f}]}$ are local 
functions of $\, ({\bf U} ({\bf Y}), \, \bm{\zeta}_{\bf Y})$ and
their derivatives, polynomial in the derivatives
$\, ({\bf U}_{\bf Y}, \bm{\zeta}_{\bf YY},
{\bf U}_{\bf YY}, \bm{\zeta}_{\bf YYY})$ and having degree $l$ 
according to our gradation rule.

 We can write on ${\cal K}$:
$$\left. {\delta S^{[\bm{\zeta}]}_{[{\bf f}]} \over
\delta \varphi^{i} (\bm{\theta}, {\bf X})} 
\right|_{\bm{\varphi} = \bm{\Phi}_{[{\bf U}, {\bf S}]}}
\, = \,\,
\left. {\delta {\tilde S}^{[\bm{\zeta}]}_{[{\bf f}]} \over
\delta \varphi^{i} (\bm{\theta}, {\bf X})}
\right|_{\bm{\varphi} = \bm{\Phi}_{[{\bf U}, {\bf S}]}}
\, + \,\, \epsilon \, {\tilde h}^{[\bm{\zeta}]}_{i [{\bf f}]} 
\left( \bm{\theta} + {{\bf S} ({\bf X}) \over \epsilon}
- {\bm{\zeta} ({\bf X}) \over \epsilon}, \,
{\bf X}; \,\, \left[ {\bf U}, \bm{\zeta}, \epsilon \right] 
\right) \, - $$
\begin{multline*}
- \,\, \epsilon \int \, {\tilde h}^{[\bm{\zeta}]}_{j [{\bf f}]} 
\left( \bm{\theta} + {{\bf S} ({\bf Y}) \over \epsilon}
- {\bm{\zeta} ({\bf Y}) \over \epsilon}, \,
{\bf Y}; \,\, \left[ {\bf U}, \bm{\zeta}, \epsilon \right]
\right) \,\, \Phi^{j}_{\theta^{\alpha}} \left( \bm{\theta} + 
{{\bf S} ({\bf Y}) \over \epsilon}, \,\, 
{\bf S}_{\bf Y}, {\bf U} ({\bf Y}) \right) \, \times  \\
\times \, {1 \over \epsilon} \, \left.
{\delta {\tilde S}^{\alpha [\bm{\zeta}]} ({\bf Y}) \over
\delta \varphi^{i} (\bm{\theta}, {\bf X})}
\right|_{\bm{\varphi} = \bm{\Phi}_{[{\bf U}, {\bf S}]}} \,
{d^{m} \theta \over (2\pi)^{m}} \,\, d^{d} Y \,\, -
\end{multline*}
\begin{multline*}
- \,\, \epsilon \int \, {\tilde h}^{[\bm{\zeta}]}_{j [{\bf f}]}
\left( \bm{\theta} + {{\bf S} ({\bf Y}) \over \epsilon}
- {\bm{\zeta} ({\bf Y}) \over \epsilon}, \,
{\bf Y}; \,\, \left[ {\bf U}, \bm{\zeta}, \epsilon \right]
\right) \,\, \Phi^{j}_{k^{\alpha}_{q}} \left( \bm{\theta} +
{{\bf S} ({\bf Y}) \over \epsilon}, \,\,
{\bf S}_{\bf Y}, {\bf U} ({\bf Y}) \right) \, \times  \\
\times \,  \left. 
{\delta {\tilde S}^{\alpha [\bm{\zeta}]}_{Y^{q}} ({\bf Y}) 
\over \delta \varphi^{i} (\bm{\theta}, {\bf X})} 
\right|_{\bm{\varphi} = \bm{\Phi}_{[{\bf U}, {\bf S}]}} \,
{d^{m} \theta \over (2\pi)^{m}} \,\, d^{d} Y \,\, -
\end{multline*}
\begin{multline*}
- \,\, \epsilon \int \, {\tilde h}^{[\bm{\zeta}]}_{j [{\bf f}]}
\left( \bm{\theta} + {{\bf S} ({\bf Y}) \over \epsilon}
- {\bm{\zeta} ({\bf Y}) \over \epsilon}, \,
{\bf Y}; \,\, \left[ {\bf U}, \bm{\zeta}, \epsilon \right]
\right) \,\, \Phi^{j}_{U^{\gamma}} \left( \bm{\theta} +
{{\bf S} ({\bf Y}) \over \epsilon}, \,\,
{\bf S}_{\bf Y}, {\bf U} ({\bf Y}) \right) \, \times  \\
\times \, \left.
{\delta {\tilde U}^{\gamma [\bm{\zeta}]} ({\bf Y}) \over
\delta \varphi^{i} (\bm{\theta}, {\bf X})}
\right|_{\bm{\varphi} = \bm{\Phi}_{[{\bf U}, {\bf S}]}} \,
{d^{m} \theta \over (2\pi)^{m}} \,\, d^{d} Y 
\end{multline*}

 In the case when the functions
$\bm{\zeta} ({\bf X})$ and ${\bf S} ({\bf X})$ belong to the
same orbit of the group (\ref{thetashift})
(i.e. $\, \bm{\zeta}_{X^{q}} \, \equiv \, {\bf S}_{X^{q}}$,
$\, q = 1, \dots, d$), we can write
$$\left. {\delta S^{[\bm{\zeta}]}_{[{\bf f}]} \over
\delta \varphi^{i} (\bm{\theta}, {\bf X})}
\right|_{\bm{\varphi} = \bm{\Phi}_{[{\bf U}, {\bf S}]}}
\, = \,\,\,
\left. {\delta {\tilde S}^{[\bm{\zeta}]}_{[{\bf f}]} \over
\delta \varphi^{i} (\bm{\theta}, {\bf X})}
\right|_{\bm{\varphi} = \bm{\Phi}_{[{\bf U}, {\bf S}]}}
\,\, + \,\,\, \epsilon \, {\tilde h}^{[\bm{\zeta}]}_{i [{\bf f}]}
\left( \bm{\theta} + {{\bf S} ({\bf X}) \over \epsilon}
- {\bm{\zeta} ({\bf X}) \over \epsilon}, \,
{\bf X}; \,\, \left[ {\bf U}, \bm{\zeta}, \epsilon \right]
\right) \, - $$
\begin{multline*}
- \,\,\, \epsilon \int \, {\tilde h}^{[\bm{\zeta}]}_{j [{\bf f}]}
\left( \bm{\theta}, \,
{\bf Y}; \,\, \left[ {\bf U}, \bm{\zeta}, \epsilon \right]
\right) \,\, \Phi^{j}_{\theta^{\alpha}} \left( \bm{\theta} +
{\bm{\zeta} ({\bf Y}) \over \epsilon}, \,\,
\bm{\zeta}_{\bf Y}, {\bf U} ({\bf Y}) \right) 
\,\, \times  \\
\times \,\, {1 \over \epsilon} \, \left.  
{\delta {\tilde S}^{\alpha [\bm{\zeta}]} ({\bf Y}) \over
\delta \varphi^{i} (\bm{\theta}, {\bf X})}
\right|_{\bm{\varphi} = \bm{\Phi}_{[{\bf U}, {\bf S}]}} \,
{d^{m} \theta \over (2\pi)^{m}} \,\, d^{d} Y \,\, - 
\end{multline*}
\begin{multline*}
- \,\,\, \epsilon \int \, {\tilde h}^{[\bm{\zeta}]}_{j [{\bf f}]}
\left( \bm{\theta}, \,
{\bf Y}; \,\, \left[ {\bf U}, \bm{\zeta}, \epsilon \right]
\right) \,\, \Phi^{j}_{k^{\alpha}_{q}} \left( \bm{\theta} + 
{\bm{\zeta} ({\bf Y}) \over \epsilon}, \,\,  
\bm{\zeta}_{\bf Y}, {\bf U} ({\bf Y}) \right) \,\, \times  \\
\times \,\, \left. 
{\delta {\tilde S}^{\alpha [\bm{\zeta}]}_{Y^{q}} ({\bf Y})
\over \delta \varphi^{i} (\bm{\theta}, {\bf X})}
\right|_{\bm{\varphi} = \bm{\Phi}_{[{\bf U}, {\bf S}]}} \,
{d^{m} \theta \over (2\pi)^{m}} \,\, d^{d} Y \,\, - 
\end{multline*}
$$- \,\, \epsilon \int  {\tilde h}^{[\bm{\zeta}]}_{j [{\bf f}]}
\left( \bm{\theta}, \,
{\bf Y}; \,\, \left[ {\bf U}, \bm{\zeta}, \epsilon \right]
\right) \,\, \Phi^{j}_{U^{\gamma}} \! \left( \bm{\theta} +   
{\bm{\zeta} ({\bf Y}) \over \epsilon}, \,\,
\bm{\zeta}_{\bf Y}, {\bf U} ({\bf Y}) \right) \, \left.
{\delta {\tilde U}^{\gamma [\bm{\zeta}]} ({\bf Y}) \over
\delta \varphi^{i} (\bm{\theta}, {\bf X})} 
\right|_{\bm{\varphi} = \bm{\Phi}_{[{\bf U}, {\bf S}]}} \,
{d^{m} \theta \over (2\pi)^{m}} \, d^{d} Y $$

 Using the convolution with arbitrary variation
$\, \delta \varphi^{i} (\bm{\theta}, {\bf X})$ we can see then 
that the last term in the above expression is identically equal 
to zero, while the first term is canceled with the third and 
the forth terms according to the definition of the functions
${\tilde h}^{[\bm{\zeta}]}_{j [{\bf f}]} (\bm{\theta}, 
\, {\bf Y}; \,\, [ {\bf U}, \bm{\zeta}, \epsilon ] ) $. 
So, we can write for 
$\, {\bf S}_{X^{q}} \, \equiv \, \bm{\zeta}_{X^{q}}$:
$$\left. {\delta S^{[\bm{\zeta}]}_{[{\bf f}]} \over
\delta \varphi^{i} (\bm{\theta}, {\bf X})}
\right|_{\bm{\varphi} = \bm{\Phi}_{[{\bf U}, {\bf S}]}} 
\, = \,\,\, {\tilde h}^{[\bm{\zeta}]}_{i [{\bf f}]}
\left( \bm{\theta} + {{\bf S} ({\bf X}) \over \epsilon}
- {\bm{\zeta} ({\bf X}) \over \epsilon}, \,
{\bf X}; \,\, \left[ {\bf U}, \bm{\zeta}, \epsilon \right]
\right) $$
in the full analogy with Chapter 2.

 Easy to see that the 1-form
$\, \delta S^{[\bm{\zeta}]}_{[{\bf f}]} /
\delta \varphi^{i} (\bm{\theta}, {\bf X})$ is invariant in this
case with respect to the transformations (\ref{thetashift}),
so we can write
\begin{equation}
\label{LieDerSf}
{\cal L}_{\alpha} \left. \left[
{\delta S^{[\bm{\zeta}]}_{[{\bf f}]} \over
\delta \varphi^{i} (\bm{\theta}, {\bf X})} \right]
\right|_{\bm{\varphi} = \bm{\Phi}_{[{\bf U}, \bm{\zeta}]}}
\,\,\, \equiv \,\,\, 0
\end{equation}

 In general, we can write:
\begin{equation}
\label{NewSVarDer1}
\left. {\delta S^{[\bm{\zeta}]}_{[{\bf f}]} \over
\delta \varphi^{i} (\bm{\theta}, {\bf X})}
\right|_{\bm{\varphi} = \bm{\Phi}_{[{\bf U}, {\bf S}]}}
\, = \,\,\, {\tilde h}^{[\bm{\zeta}]}_{i [{\bf f}]}
\left( \bm{\theta} + {{\bf S} ({\bf X}) \over \epsilon}
- {\bm{\zeta} ({\bf X}) \over \epsilon}, \,
{\bf X}; \,\, \left[ {\bf U}, \bm{\zeta}, \epsilon \right]
\right) \,\,\, +
\end{equation}
\begin{multline*}
+ \, \epsilon \int {\tilde h}^{[\bm{\zeta}]}_{j [{\bf f}]}
\left( \bm{\theta}, \,
{\bf Y}; \,\, \left[ {\bf U}, \bm{\zeta}, \epsilon \right]
\right) \,\, \times  \\
\times \,\, \left( \Phi^{j}_{\theta^{\alpha}} 
\left( \bm{\theta} + {\bm{\zeta} ({\bf Y}) \over \epsilon}, \,\,
{\bf S}_{\bf Y}, {\bf U} ({\bf Y}) \right) \,\, - \,\,
\Phi^{j}_{\theta^{\alpha}} \left( \bm{\theta} +
{\bm{\zeta} ({\bf Y}) \over \epsilon}, \,\,
\bm{\zeta}_{\bf Y}, {\bf U} ({\bf Y}) \right) \right)
\times  \\
\times \, {1 \over \epsilon} \, \left.
{\delta {\tilde S}^{\alpha [\bm{\zeta}]} ({\bf Y}) \over
\delta \varphi^{i} (\bm{\theta}, {\bf X})}
\right|_{\bm{\varphi} = \bm{\Phi}_{[{\bf U}, {\bf S}]}} \,
{d^{m} \theta \over (2\pi)^{m}} \,\, d^{d} Y \,\, +
\end{multline*}
\begin{multline*}
+ \, \epsilon \int {\tilde h}^{[\bm{\zeta}]}_{j [{\bf f}]}
\left( \bm{\theta}, \,
{\bf Y}; \,\, \left[ {\bf U}, \bm{\zeta}, \epsilon \right]     
\right) \,\, \times  \\
\times \,\, \left( \Phi^{j}_{k^{\alpha}_{q}} \left( \bm{\theta} 
+ {\bm{\zeta} ({\bf Y}) \over \epsilon}, \,\,
{\bf S}_{\bf Y}, {\bf U} ({\bf Y}) \right) \,\, - \,\,
\Phi^{j}_{k^{\alpha}_{q}} \left( \bm{\theta} +
{\bm{\zeta} ({\bf Y}) \over \epsilon}, \,\,
\bm{\zeta}_{\bf Y}, {\bf U} ({\bf Y}) \right) \right)
\times  \\
\times \,  \left.   
{\delta {\tilde S}^{\alpha [\bm{\zeta}]}_{Y^{q}} ({\bf Y})
\over \delta \varphi^{i} (\bm{\theta}, {\bf X})}   
\right|_{\bm{\varphi} = \bm{\Phi}_{[{\bf U}, {\bf S}]}} \,
{d^{m} \theta \over (2\pi)^{m}} \,\, d^{d} Y \,\, +
\end{multline*}
\begin{multline*}
+ \, \epsilon \int \, {\tilde h}^{[\bm{\zeta}]}_{j [{\bf f}]}
\left( \bm{\theta}, \,
{\bf Y}; \,\, \left[ {\bf U}, \bm{\zeta}, \epsilon \right]
\right) \,\, \times  \\ 
\times \,\, \left( \Phi^{j}_{U^{\gamma}} \left( \bm{\theta} 
+ {\bm{\zeta} ({\bf Y}) \over \epsilon}, \,\,
{\bf S}_{\bf Y}, {\bf U} ({\bf Y}) \right) \,\, - \,\,
\Phi^{j}_{U^{\gamma}} \left( \bm{\theta} +
{\bm{\zeta} ({\bf Y}) \over \epsilon}, \,\,
\bm{\zeta}_{\bf Y}, {\bf U} ({\bf Y}) \right) \right)
\times  \\
\times \, \left.
{\delta {\tilde U}^{\gamma [\bm{\zeta}]} ({\bf Y}) \over
\delta \varphi^{i} (\bm{\theta}, {\bf X})}
\right|_{\bm{\varphi} = \bm{\Phi}_{[{\bf U}, {\bf S}]}} \,
{d^{m} \theta \over (2\pi)^{m}} \,\, d^{d} Y
\end{multline*}
or, for close values of $\, \bm{\zeta}_{\bf Y}$ and
$\, {\bf S}_{\bf Y}$:
\begin{equation}
\label{NewSVarDer2}
\left. {\delta S^{[\bm{\zeta}]}_{[{\bf f}]} \over
\delta \varphi^{i} (\bm{\theta}, {\bf X})}
\right|_{\bm{\varphi} = \bm{\Phi}_{[{\bf U}, {\bf S}]}}
\, = \,\,\, {\tilde h}^{[\bm{\zeta}]}_{i [{\bf f}]}
\left( \bm{\theta} + {{\bf S} ({\bf X}) \over \epsilon}
- {\bm{\zeta} ({\bf X}) \over \epsilon}, \,
{\bf X}; \,\, \left[ {\bf U}, \bm{\zeta}, \epsilon \right]
\right) \,\,\, +
\end{equation}
\begin{multline*}
+ \,\,\, \epsilon \int \, 
\left( S^{\beta}_{Y^{p}} \, - \, \zeta^{\beta}_{Y^{p}} \right)
\,\, {\tilde h}^{[\bm{\zeta}]}_{j [{\bf f}]}
\left( \bm{\theta}, \,
{\bf Y}; \,\, \left[ {\bf U}, \bm{\zeta}, \epsilon \right] \right) 
\,\, \Phi^{j}_{\theta^{\alpha} k^{\beta}_{p}} \left( \bm{\theta} +
{\bm{\zeta} ({\bf Y}) \over \epsilon}, \,\,
\bm{\zeta}_{\bf Y}, {\bf U} ({\bf Y}) \right) \, \times  \\
\times \, {1 \over \epsilon} \, \left.
{\delta {\tilde S}^{\alpha [\bm{\zeta}]} ({\bf Y}) \over
\delta \varphi^{i} (\bm{\theta}, {\bf X})}
\right|_{\bm{\varphi} = \bm{\Phi}_{[{\bf U}, {\bf S}]}} \,
{d^{m} \theta \over (2\pi)^{m}} \,\, d^{d} Y \,\,\, +
\end{multline*}
\begin{multline*}
+ \,\,\, \epsilon \int \, 
\left( S^{\beta}_{Y^{p}} \, - \, \zeta^{\beta}_{Y^{p}} \right) 
\,\, {\tilde h}^{[\bm{\zeta}]}_{j [{\bf f}]}
\left( \bm{\theta}, \, {\bf Y}; 
\,\, \left[ {\bf U}, \bm{\zeta}, \epsilon \right] \right) \,\, 
\Phi^{j}_{k^{\alpha}_{q} k^{\beta}_{p}} \left( \bm{\theta} + 
{\bm{\zeta} ({\bf Y}) \over \epsilon}, \,\,
\bm{\zeta}_{\bf Y}, {\bf U} ({\bf Y}) \right) \, \times  \\
\times \,\, \left.   
{\delta {\tilde S}^{\alpha [\bm{\zeta}]}_{Y^{q}} ({\bf Y})
\over \delta \varphi^{i} (\bm{\theta}, {\bf X})}   
\right|_{\bm{\varphi} = \bm{\Phi}_{[{\bf U}, {\bf S}]}} \,
{d^{m} \theta \over (2\pi)^{m}} \,\, d^{d} Y \,\,\, +
\end{multline*}
\begin{multline*}
+ \,\,\, \epsilon \int \, 
\left( S^{\beta}_{Y^{p}} \, - \, \zeta^{\beta}_{Y^{p}} \right)
\,\, {\tilde h}^{[\bm{\zeta}]}_{j [{\bf f}]}
\left( \bm{\theta}, \,
{\bf Y}; \,\, \left[ {\bf U}, \bm{\zeta}, \epsilon \right] \right) 
\,\, \Phi^{j}_{U^{\gamma} k^{\beta}_{p}} \left( \bm{\theta} +
{\bm{\zeta} ({\bf Y}) \over \epsilon}, \,\,
\bm{\zeta}_{\bf Y}, {\bf U} ({\bf Y}) \right) \, \times  \\
\times \, \left.
{\delta {\tilde U}^{\gamma [\bm{\zeta}]} ({\bf Y}) \over
\delta \varphi^{i} (\bm{\theta}, {\bf X})}
\right|_{\bm{\varphi} = \bm{\Phi}_{[{\bf U}, {\bf S}]}} \,  
{d^{m} \theta \over (2\pi)^{m}} \,\, d^{d} Y \,\,\, +
\end{multline*} 
$$+ \,\,\,\,\, O \left( \, \left| 
{\bf S}_{\bf Y} - \bm{\zeta}_{\bf Y} \right|^{2} \, \right) $$

 Using the functionals $\, S^{\alpha [\bm{\zeta}]} ({\bf X})$
we can introduce also the new functionals
\begin{equation}
\label{NewFuncU}
U^{\gamma [\bm{\zeta}]}({\bf X}) \,\,\, = \,\,\,
J^{\gamma}({\bf X}) \,\, + \,\, \sum_{l\geq1} \epsilon^{l} \,
U^{\gamma}_{(l)}
\left({\bf X},
[{\bf S}^{[\bm{\zeta}]}_{\bf X}, {\bf J}] \right)
\,\,\,\,\, , \,\,\,\,\,
\gamma = 1, \dots, m + s
\end{equation}
and the new constraints
\begin{equation}
\label{gconstr}
g^{i [\bm{\zeta}]} (\bm{\theta}, {\bf X}) \,\,\, = \,\,\,
\varphi^{i} (\bm{\theta}, {\bf X})
\,\, - \,\, \Phi^{i} \left( \bm{\theta} \, + \,
{{\bf S}^{[\bm{\zeta}]} ({\bf X}) \over \epsilon}\, , \,\,
{\bf S}^{[\bm{\zeta}]}_{X^{1}}, \dots, 
{\bf S}^{[\bm{\zeta}]}_{X^{d}},
{\bf U}^{[\bm{\zeta}]} ({\bf X}) \right)
\end{equation}

 The equations
$$g^{i [\bm{\zeta}]} (\bm{\theta}, {\bf X}) \,\,\, = \,\,\, 0$$
define the submanifold ${\cal K}$ near the points
$({\bf U} ({\bf X}), \bm{\zeta} ({\bf X}))$.

 Like in Chapter 2, the constraints 
$g^{i [\bm{\zeta}]} (\bm{\theta}, {\bf X})$ are also not 
independent here. Thus, in the same way, we can write here the
following set of relations
\begin{equation}
\label{dependence}
\int \!\! \int_{0}^{2\pi}\!\!\!\!\!\dots\int_{0}^{2\pi}
\left. {\delta S^{\alpha [\bm{\zeta}]} ({\bf Z}) \over
\delta \varphi^{i} (\bm{\theta}, {\bf X})}\right|_{\cal K}
\,\, \left. {\delta g^{i [\bm{\zeta}]} (\bm{\theta}, {\bf X}) \over
\delta \varphi^{j} (\bm{\theta}^{\prime}, {\bf Y})}\right|_{\cal K}
\,\, {d^{m} \theta \over (2\pi)^{m}} \,\, d^{d} X 
\,\, \equiv \,\, 0
\,\,\,\,\, , \,\,\,\,\, \alpha = 1, \dots, m
\end{equation}
\begin{equation}
\label{dependence2}
\int \!\! \int_{0}^{2\pi}\!\!\!\!\!\dots\int_{0}^{2\pi}
\left. {\delta U^{\gamma [\bm{\zeta}]} ({\bf Z}) \over
\delta \varphi^{i} (\bm{\theta}, {\bf X})}\right|_{\cal K}
\,\, \left. {\delta g^{i [\bm{\zeta}]} (\bm{\theta}, {\bf X}) \over
\delta \varphi^{j} (\bm{\theta}^{\prime}, {\bf Y})}\right|_{\cal K}
\,\, {d^{m} \theta \over (2\pi)^{m}} \,\, d^{d} X \,\, \equiv \,\, 0
\,\,\,\,\, , \,\,\,\,\, \gamma = 1, \dots, m + s
\end{equation}
which take place identically for their ``gradients'' on the 
submanifold ${\cal K}$.

 As we can see, the functionals ${\bf U}^{[\bm{\zeta}]} ({\bf X})$,
${\bf S}^{[\bm{\zeta}]} ({\bf X})$, and
${\bf g}^{[\bm{\zeta}]} (\bm{\theta}, {\bf X})$ are introduced just
as formal series in $\epsilon$ near the points of the submanifold
${\cal K}$ with the set of coordinates
$({\bf U}({\bf Z}), \bm{\zeta} ({\bf Z}))$. 
We can see also, that the values of
$S^{\alpha [\bm{\zeta}]} ({\bf X})$ are restricted by the
conditions 
$S^{\alpha} ({\bf X}) \, = \, \zeta^{\alpha} ({\bf X})
+ O (\epsilon)$, while their gradients have the order
$O (\epsilon)$ on ${\cal K}$. This definition will be 
sufficient for us here, since we are going to consider actually
just the pairwise Poisson brackets of these functionals
and their gradients at the points of the submanifold
${\cal K}$ with coordinates
$({\bf U}({\bf Z}), \bm{\zeta} ({\bf Z}))$.

 The functionals 
$({\bf U}^{[\bm{\zeta}]} ({\bf X}), \,
{\bf S}^{[\bm{\zeta}]} ({\bf X}), \,
{\bf g}^{[\bm{\zeta}]} (\bm{\theta}, {\bf X}))$ can be considered
as an ``overdetermined'' coordinate system near the points
$({\bf U}({\bf Z}), \bm{\zeta} ({\bf Z}))$ of the submanifold
${\cal K}$, since we always have relations
(\ref{dependence}) - (\ref{dependence2}) for the gradients of the
coordinates ${\bf g}^{[\bm{\zeta}]} (\bm{\theta}, {\bf X})$.

 Let us define regularized functionals
$g^{[\bm{\zeta}]}_{[{\bf Q}]}$ by the formula
$$g^{[\bm{\zeta}]}_{[{\bf Q}]} \,\, = \,\, \int
\int_{0}^{2\pi}\!\!\!\!\!\dots\int_{0}^{2\pi}
g^{i [\bm{\zeta}]} (\bm{\theta}, {\bf X}) \,\,
Q_{i} \left( {{\bf S}^{[\bm{\zeta}]}({\bf X}) \over \epsilon}
+ \bm{\theta}, {\bf X} \right) \,\,
{d^{m} \theta \over (2\pi)^{m}} \, d^{d} X $$
with smooth, compactly supported in ${\bf X}$ and $2\pi$-periodic
in each $\theta^{\alpha}$ functions $Q_{i} (\bm{\theta}, {\bf X})$.
In fact, it will be convenient to put some additional requirements
on the functions $Q_{i} (\bm{\theta}, {\bf X})$. Namely, let us
define for arbitrary smooth, compactly supported in ${\bf X}$ and
$2\pi$-periodic in each $\theta^{\alpha}$ functions
${\tilde {\bf Q}} (\bm{\theta}, {\bf X}) =
({\tilde Q}_{1}(\bm{\theta}, {\bf X}), \dots ,
{\tilde Q}_{n}(\bm{\theta}, {\bf X}))$
the functionals
\begin{multline*}
Q_{i}(\bm{\theta}, {\bf X}) \,\,\, = \,\,\,
{\tilde Q}_{i}(\bm{\theta}, {\bf X}) \,\,\, - \,\,\,
\Phi^{i}_{\theta^{\beta}}
(\bm{\theta}, {\bf S}^{[\bm{\zeta}]}_{\bf X},
{\bf U}^{[\bm{\zeta}]} ({\bf X}))
\,\, M^{\beta\gamma} ({\bf S}^{[\bm{\zeta}]}_{\bf X},
{\bf U}^{[\bm{\zeta}]} ({\bf X}))
\, \times   \\
\times \, \int_{0}^{2\pi}\!\!\!\!\!\dots\int_{0}^{2\pi}
{\tilde Q}_{j}(\bm{\theta}^{\prime}, {\bf X}) \,\,
\Phi^{j}_{\theta^{\prime\gamma}}
(\bm{\theta}^{\prime}, {\bf S}^{[\bm{\zeta}]}_{\bf X},
{\bf U}^{[\bm{\zeta}]} ({\bf X})) \,\,
{d^{m} \theta^{\prime} \over (2\pi)^{m}}
\end{multline*}

 Here the matrix
$M^{\beta\gamma} ({\bf S}_{\bf X}, {\bf U}({\bf X}))$
represents the inverse of the matrix
$$M_{\beta\gamma} ({\bf S}_{\bf X}, {\bf U}({\bf X})) \,\, = \,\,
\int_{0}^{2\pi}\!\!\!\!\!\dots\int_{0}^{2\pi} \sum_{i=1}^{n} \,
\Phi^{i}_{\theta^{\beta}}
(\bm{\theta}, {\bf S}_{\bf X}, {\bf U}({\bf X})) \,\,
\Phi^{i}_{\theta^{\gamma}}
(\bm{\theta}, {\bf S}_{\bf X}, {\bf U}({\bf X})) \,\,
{d^{m} \theta \over (2\pi)^{m}} $$
on a complete regular family $\Lambda$.

 By definition, the functionals $Q_{i}(\bm{\theta}, {\bf X})$
represent local functions of
$({\bf S}^{[\bm{\zeta}]}_{\bf X},
{\bf U}^{[\bm{\zeta}]} ({\bf X}))$
$$Q_{i}(\bm{\theta}, {\bf X}) \,\,\, \equiv \,\,\,
Q_{i}(\bm{\theta}, {\bf X}, \, {\bf S}^{[\bm{\zeta}]}_{\bf X},
{\bf U}^{[\bm{\zeta}]}({\bf X})) $$
and the arbitrary fixed functions
${\tilde {\bf Q}} (\bm{\theta}, {\bf X})$.
We will assume everywhere below that
${\bf Q} (\bm{\theta}, {\bf X})$
represents a functional of this type.

 We can see that for fixed values of the functionals
$( {\bf S}^{[\bm{\zeta}]} ({\bf Z}), 
{\bf U}^{[\bm{\zeta}]} ({\bf Z}))$
the values of $Q_{i}(\bm{\theta}, {\bf X})$
with arbitrary ${\tilde {\bf Q}} (\bm{\theta}, {\bf X})$
represent all possible smooth compactly
supported in ${\bf X}$ and $2\pi$-periodic in each $\theta^{\alpha}$
functions satisfying the restrictions
\begin{equation}
\label{OgranQ}
\int_{0}^{2\pi}\!\!\!\!\!\dots\int_{0}^{2\pi}
Q_{i} (\bm{\theta}, {\bf X}) \,\, \Phi^{i}_{\theta^{\alpha}}
(\bm{\theta}, {\bf S}^{[\bm{\zeta}]}_{\bf X},
{\bf U}^{[\bm{\zeta}]} ({\bf X})) 
\,\, {d^{m} \theta \over (2\pi)^{m}}
\,\, = \,\, 0 \,\,\,\,\,\,\, , \,\,\, \forall \, {\bf X}   
\, , \,\,\,\,\, \alpha = 1, \dots , m
\end{equation}

 It's not difficult to see also that for the functionals
(\ref{NewFuncU}) - (\ref{gconstr}), defined with the aid of
${\bf S}^{[\bm{\zeta}]} ({\bf X})$ and ${\bf J} ({\bf X})$,
we can write the relations, analogous to (\ref{LieDerSf}), i.e.
\begin{equation}   
\label{UgVDerInv}
{\cal L}_{\alpha} \left. \left[
{\delta U^{[\bm{\zeta}]}_{[{\bf q}]} \over
\delta \varphi^{i} (\bm{\theta}, {\bf X})} \right]
\right|_{\bm{\varphi} = \bm{\Phi}_{[{\bf U}, \bm{\zeta}]}}
\,\,\, \equiv \,\,\, 0  \,\,\,\,\, , \,\,\,\,\,\,\,\,
{\cal L}_{\alpha} \left. \left[
{\delta g^{[\bm{\zeta}]}_{[{\bf Q}]} \over
\delta \varphi^{i} (\bm{\theta}, {\bf X})} \right]
\right|_{\bm{\varphi} = \bm{\Phi}_{[{\bf U}, \bm{\zeta}]}}
\,\,\, \equiv \,\,\, 0
\end{equation}

 Using the evident relation
\begin{equation}
\label{StandardExp}
\delta \, {\cal F} \,\,\, = \,\,\, \int
{\delta {\cal F} \over \delta \varphi^{i} (\bm{\theta}, {\bf X})}
\,\, \delta \varphi^{i} (\bm{\theta}, {\bf X}) \,\,
{d^{m} \theta \over (2\pi)^{m}} \,\, d^{d} X 
\end{equation}
and the specific definition of the constraints
$g^{i [\bm{\zeta}]} (\bm{\theta}, {\bf X})$, it is easy to obtain
the following relation at the points of the submanifold
${\cal K}$:
\begin{multline}
\label{SpecialExp}
\delta \, {\cal F} \,\,\, = \,\,\, \int \left.
{\delta {\cal F} \over \delta \varphi^{i} (\bm{\theta}, {\bf X})}
\right|_{\cal K} \, 
\delta g^{i [\bm{\zeta}]} (\bm{\theta}, {\bf X}) \,\,
{d^{m} \theta \over (2\pi)^{m}} \,\, d^{d} X \,\,\, +   \\
+ \,\,\, \int \left[ \delta {\cal F} |_{\cal K} /
\delta U^{\gamma} ({\bf X}) \right] \,\,\,
\delta U^{\gamma [\bm{\zeta}]} ({\bf X}) \,\,\, d^{d} X 
\,\,\,\,\, + \,\,\, \int \left[ \delta  {\cal F} |_{\cal K} /
\delta S^{\alpha} ({\bf X}) \right] \,\,\,
\delta S^{\alpha [\bm{\zeta}]} ({\bf X}) \,\,\, d^{d} X
\end{multline}
for any functional ${\cal F}$. Let us note that here and below
we always assume that the derivatives
$\, \delta / \delta U^{\gamma} ({\bf X})$ and
$\, \delta / \delta S^{\alpha} ({\bf X})$ are taken under the
conditions $\, {\bf g} (\bm{\theta}, {\bf W}) \, = \, 0$
(all $\bm{\theta}$ and ${\bf W}$).

 We need to discuss now the pairwise Poisson brackets of the
functionals ${\bf J} ({\bf X})$,
${\bf U}^{[\bm{\zeta}]} ({\bf X})$,
${\bf S}^{[\bm{\zeta}]} ({\bf X})$, and
${\bf g}^{[\bm{\zeta}]} (\bm{\theta}, {\bf X})$ on the 
submanifold ${\cal K}$.

 Let us note that, according to (\ref{A0QDerRel}), we can write 
here
\begin{equation}
\label{epsAQRel}   
A^{\gamma\rho}_{0 \dots 0}
(\bm{\varphi}, \epsilon \bm{\varphi}_{\bf X}, \dots)
\,\, \equiv \,\, \epsilon \, \partial_{X^{1}} \,
Q^{\gamma\rho 1}
(\bm{\varphi}, \epsilon \bm{\varphi}_{\bf X}, \dots)
\, + \, \dots \, + \,
\epsilon \, \partial_{X^{d}} \,
Q^{\gamma\rho d} (\bm{\varphi}, \epsilon \bm{\varphi}_{\bf X}, 
\dots) 
\end{equation}

 Using the definition of the functionals $J^{\gamma} ({\bf X})$
and relations (\ref{epsAQRel}) we can then write for the functionals
$$J_{[{\bf a}]} \, = \,
\int J^{\gamma} ({\bf X}) \,\, a_{\gamma} ({\bf X}) \,\, d^{d} X
\,\,\,\,\,\,\,\, , \,\,\,\,\,\,\,\,
J_{[{\bf b}]} \, = \,
\int J^{\gamma} ({\bf X}) \,\, b_{\gamma} ({\bf X}) \,\, d^{d} X $$
the relations
\begin{equation}
\label{JaJbBr}
\left. \left\{ J_{[{\bf a}]} \, , \, J_{[{\bf b}]}
\right\}\right|_{\cal K} \,\, = \,\, \epsilon \,
\left. \left\{ J_{[{\bf a}]} \, , \, J_{[{\bf b}]}
\right\}\right|_{{\cal K}[1]} \,\, + \,\, \epsilon^{2} \,
\left. \left\{ J_{[{\bf a}]} \, , \, J_{[{\bf b}]}
\right\}\right|_{{\cal K}[2]} \,\, + \,\, \dots
\end{equation}

 The functionals
$\{ J_{[{\bf a}]} \, , \, J_{[{\bf b}]} \} |_{{\cal K}[l]}$
depend on the coordinates 
$[{\bf S} ({\bf X}), \, {\bf U} ({\bf X})]$ on ${\cal K}$
and are invariant under the shift  
\begin{equation}
\label{ShiftSTransform}
S^{\alpha} ({\bf X}) \,\, \rightarrow \,\, S^{\alpha} ({\bf X})
\,\, + \,\, {\rm const}
\end{equation}

 Let us say also, that all the statements above
are also valid for the functions  \linebreak  
$\{ J^{\gamma} ({\bf X})  \, , \, J_{[{\bf b}]} \}|_{{\cal K}[l]}$  
and the distributions
$\{ J^{\gamma} ({\bf X})  \, , \, J^{\rho} ({\bf Y})
\}|_{{\cal K}[l]}$ on ${\cal K}$.

 Using the expressions
\begin{multline}
\label{JqJpVarDeriv}
\left. {\delta \{ J_{[{\bf a}]} \, , \, J_{[{\bf b}]} \} \over
\delta \varphi^{k} (\bm{\theta}, {\bf W}) }\right|_{\cal K}
\,\,\, \equiv  \\
\equiv  \,\, \sum_{l_{1}, \dots, l_{d}}
\int \!\! \left. \left[
{\delta \over \delta \varphi^{k} (\bm{\theta}, {\bf W})}   
\int_{0}^{2\pi}\!\!\!\!\!\dots\int_{0}^{2\pi} \!\!
\epsilon^{l_{1} + \dots + l_{d}} \,
A^{\gamma\rho}_{l_{1} \dots l_{d}}
\left( \bm{\varphi} (\bm{\theta}^{\prime}, {\bf X}),
\dots \right) \, {d^{m} \theta^{\prime} \over (2\pi)^{m}} \right]
\! \right|_{\cal K} \,\, \times  \\
\times  \,\, a_{\gamma}({\bf X}) \,
b_{\rho, \, l_{1} X^{1} \dots l_{d} X^{d}} ({\bf X}) \,\, d^{d} X 
\end{multline}
we can represent the values (\ref{JqJpVarDeriv}) as the graded
decompositions on ${\cal K}$ depending on the values
$k^{\alpha}_{q} ({\bf X}) = S^{\alpha}_{X^{q}}$,
$U^{\gamma} ({\bf X})$ and their derivatives at the corresponding
point ${\bf W}$.

 According to (\ref{epsAQRel}), we can claim again that the 
corresponding expansion of (\ref{JqJpVarDeriv}) starts with 
the first degree of $\epsilon$:
\begin{equation}
\label{JqJpSkobRazl}
\left. {\delta \{ J_{[{\bf a}]} \, , \, J_{[{\bf b}]} \} \over
\delta \varphi^{k} (\bm{\theta}, {\bf W}) }\right|_{\cal K}
\,\,\, = \,\,\,
\epsilon \, \left. {\delta \{ J_{[{\bf a}]} \, , \, J_{[{\bf b}]} \}
\over \delta
\varphi^{k} (\bm{\theta}, {\bf W}) }\right|_{{\cal K}[1]} \,\,
+ \,\, \epsilon^{2} \,
\, \left. {\delta \{ J_{[{\bf a}]} \, , \, J_{[{\bf b}]} \}
\over \delta
\varphi^{k} (\bm{\theta}, {\bf W}) }\right|_{{\cal K}[2]} \,\,
+ \,\,\, \dots
\end{equation}

 Every term in decomposition (\ref{JqJpSkobRazl}) represents
a local function of $k^{\alpha}_{q} ({\bf W}) = S^{\alpha}_{W^{q}}$,
$U^{\gamma} ({\bf W})$, $a_{\gamma} ({\bf W})$,
$b_{\gamma} ({\bf W})$ and their derivatives w.r.t. ${\bf W}$.
All the terms in (\ref{JqJpSkobRazl}) are polynomial in the
derivatives of 
$({\bf k}_{q} ({\bf W}), {\bf U} ({\bf W}), {\bf a} ({\bf W}),  
{\bf b} ({\bf W}))$ and have degree $l$ given by the total number
of differentiations of these functions w.r.t. ${\bf W}$.
At the same time the
dependence of the terms of (\ref{JqJpSkobRazl}) on the variables
$\bm{\theta}$ appears with the common shift
${\bf S} ({\bf W}) / \epsilon$.

 Easy to see that we have the relations
\begin{equation}
\label{VarDerJaJbSU}
{\delta \{ J_{[{\bf a}]} , J_{[{\bf b}]} \}|_{\cal K} \over
\delta S^{\alpha} ({\bf W})} \,\,\, = \,\,\,
O (\epsilon) \,\,\,\,\,\,\,\, , \,\,\,\,\,\,\,\,
{\delta \{ J_{[{\bf a}]} , J_{[{\bf b}]} \}|_{\cal K} \over 
\delta U^{\gamma} ({\bf W})} \,\,\, = \,\,\,
O (\epsilon) 
\end{equation}

 Besides that, we can write
\begin{equation}
\label{VarDerJaJbSUComp}
{\delta \{ J_{[{\bf a}]} , J_{[{\bf b}]} \}|_{\cal K} \over
\delta S^{\alpha} ({\bf W})} \, _{[1]} \,\, = \,\,
{\delta \{ J_{[{\bf a}]} , J_{[{\bf b}]}
|_{{\cal K}[1]} \over \delta S^{\alpha} ({\bf W})} 
\,\,\,\,\, ,  \,\,\,\,\,\,\,\,
{\delta \{ J_{[{\bf a}]} , J_{[{\bf b}]} \}|_{\cal K} \over
\delta U^{\gamma} ({\bf W})} \, _{[1]} \,\, = \,\,
{\delta \{ J_{[{\bf a}]} , J_{[{\bf b}]} \}
|_{{\cal K}[1]} \over \delta U^{\gamma} ({\bf W})}
\end{equation}
on the submanifold ${\cal K}$.

 In particular, all the relations
(\ref{JqJpSkobRazl}) - (\ref{VarDerJaJbSUComp}) can be used
at the ``points''
$$\bm{\varphi} \left(\bm{\theta}, {\bf X} \right) \,\,\, = \,\,\,
\bm{\Phi} \left( \bm{\theta} + 
{\bm{\zeta}({\bf X}) \over \epsilon}, \,\, \bm{\zeta}_{\bf X},
\, {\bf U} ({\bf X}) \right) $$
of the submanifold ${\cal K}$.

 From the definition of the functionals 
$\, {\tilde {\bf S}}^{[\bm{\zeta}]} ({\bf X})$ and
$\, {\bf S}^{[\bm{\zeta}]} ({\bf X})$ 
it's not difficult to get
the following relations
\begin{equation}
\label{SJSSK1}
\left. \left\{ S^{[\bm{\zeta}]}_{[{\bf f}]} , J_{[{\bf b}]} 
\right\}\right|_{\bm{\varphi} = \bm{\Phi}_{[{\bf U}, \bm{\zeta}]}}
= \,\,\, \epsilon
\left. \left\{ S^{[\bm{\zeta}]}_{[{\bf f}]} , J_{[{\bf b}]}
\right\}
\right|_{\bm{\varphi} = \bm{\Phi}_{[{\bf U}, \bm{\zeta}]} \, [0]}
+ \,\, \epsilon^{2}
\left. \left\{ S^{[\bm{\zeta}]}_{[{\bf f}]} , J_{[{\bf b}]}
\right\}
\right|_{\bm{\varphi} = \bm{\Phi}_{[{\bf U}, \bm{\zeta}]} \, [1]}
+ \,\, \dots   
\end{equation}
\begin{equation}
\label{SJSSK2}
\left. \left\{ S^{[\bm{\zeta}]}_{[{\bf f}]} \, , \,
S^{[\bm{\zeta}]}_{[{\bf h}]} \right\}
\right|_{\bm{\varphi} = \bm{\Phi}_{[{\bf U}, \bm{\zeta}]}}
= \,\,\, \epsilon^{2} \,
\left. \left\{ S^{[\bm{\zeta}]}_{[{\bf f}]} \, , \,
S^{[\bm{\zeta}]}_{[{\bf h}]} \right\}
\right|_{\bm{\varphi} = \bm{\Phi}_{[{\bf U}, \bm{\zeta}]} \, [0]}
+ \,\, \dots 
\end{equation}
where all the terms represent local graded expressions of
$\bm{\zeta}_{\bf X}$, ${\bf U} ({\bf X})$, ${\bf f} ({\bf X})$,
${\bf h} ({\bf X})$, ${\bf b} ({\bf X})$ and their derivatives
integrated over $\mathbb{R}^{d}$. In the same way, it is not 
difficult to get also at 
$\, {\bf S} ({\bf X}) \equiv \bm{\zeta} ({\bf X})$:
\begin{equation}
\label{VarDerSfJbphi}
\left. {\delta \{ S^{[\bm{\zeta}]}_{[{\bf f}]} ,
J_{[{\bf b}]} \} \over
\delta \varphi^{k} (\bm{\theta}, {\bf W}) }
\right|_{\bm{\varphi} = \bm{\Phi}_{[{\bf U}, \bm{\zeta}]}}
\!\! = \,\, \epsilon 
\left. {\delta \{ S^{[\bm{\zeta}]}_{[{\bf f}]} ,
J_{[{\bf b}]} \} \over
\delta \varphi^{k} (\bm{\theta}, {\bf W}) }
\right|_{\bm{\varphi} = \bm{\Phi}_{[{\bf U}, \bm{\zeta}]} \, [0]}
\!\! + \,\, \epsilon^{2} 
\left. {\delta \{ S^{[\bm{\zeta}]}_{[{\bf f}]} ,
J_{[{\bf b}]} \} \over
\delta \varphi^{k} (\bm{\theta}, {\bf W}) }
\right|_{\bm{\varphi} = \bm{\Phi}_{[{\bf U}, \bm{\zeta}]} \, [1]}
\!\! + \,\, \dots
\end{equation}
\begin{equation}
\label{VarDerSfShphi}
\left. {\delta \{ S^{[\bm{\zeta}]}_{[{\bf f}]} ,
S^{[\bm{\zeta}]}_{[{\bf f}]} \} \over
\delta \varphi^{k} (\bm{\theta}, {\bf W}) }
\right|_{\bm{\varphi} = \bm{\Phi}_{[{\bf U}, \bm{\zeta}]}}
= \,\,\, \epsilon^{2} \,
\left. {\delta \{ S^{[\bm{\zeta}]}_{[{\bf f}]} ,
S^{[\bm{\zeta}]}_{[{\bf h}]} \} \over
\delta \varphi^{k} (\bm{\theta}, {\bf W}) }
\right|_{\bm{\varphi} = \bm{\Phi}_{[{\bf U}, \bm{\zeta}]} \, [0]}
\, + \,\, \dots
\end{equation}

 Every term in (\ref{VarDerSfJbphi}) - (\ref{VarDerSfShphi})
represents a local function of 
$k^{\alpha}_{q} ({\bf W}) = \zeta^{\alpha}_{W^{q}}$,
$U^{\gamma} ({\bf W})$, $b_{\gamma} ({\bf W})$,
$f_{\alpha} ({\bf W})$, $h_{\alpha} ({\bf W})$ 
and their derivatives w.r.t. ${\bf W}$. All the terms again 
are polynomial in the derivatives of
$({\bf k}_{q} ({\bf W}), {\bf U} ({\bf W}), {\bf b} ({\bf W}),
{\bf f} ({\bf W}), {\bf h} ({\bf W}))$ with the corresponding 
gradation degree. The dependence of (\ref{VarDerSfJbphi}) and
(\ref{VarDerSfShphi}) on the variables
$\bm{\theta}$ appears with the common shift
$\bm{\zeta} ({\bf W}) / \epsilon$.

 Let us say now that the Poisson brackets containing the
functionals $\, {\tilde {\bf S}}^{[\bm{\zeta}]} ({\bf X})$ 
and $\, {\bf S}^{[\bm{\zeta}]} ({\bf X})$
have in fact a little bit more complicated structure at the 
points of ${\cal K}$ for general values of the coordinates
$\, ({\bf S} ({\bf X})$, ${\bf U} ({\bf X}))$. 
This circumstance is caused by the fact that,
according to the definition, the densities of the functionals
$\bm{\vartheta}^{[\bm{\zeta}]} ({\bf X})$ 
contain an explicit dependence
on the variables $\bm{\theta}$. Calculating the pairwise brackets
of $\vartheta^{[\bm{\zeta}]}_{\alpha} ({\bf X})$ 
with $J_{[{\bf b}]}$ and
$\vartheta^{[\bm{\zeta}]}_{\beta} ({\bf Y})$ 
everywhere on ${\cal K}$ we can represent them in the form:
\begin{equation}
\label{varthetaJbExp}
\epsilon \,
\left. \left\{ \vartheta^{[\bm{\zeta}]}_{\alpha} ({\bf X}) 
\, , \, J_{[{\bf b}]}
\right\}\right|_{\cal K} \,\, = \,\,\, \epsilon \,
\left. \left\{ \vartheta^{[\bm{\zeta}]}_{\alpha} ({\bf X}) 
\, , \, J_{[{\bf b}]}
\right\}\right|_{{\cal K}[0]} \,\, + \,\, \epsilon^{2} \,
\left. \left\{ \vartheta^{[\bm{\zeta}]}_{\alpha} ({\bf X}) 
\, , \, J_{[{\bf b}]}
\right\}\right|_{{\cal K}[1]} \,\, + \,\, \dots   
\end{equation}
\begin{equation}
\label{varthetavarthetaExp}
\epsilon^{2} \,
\left. \left\{ \vartheta^{[\bm{\zeta}]}_{\alpha} ({\bf X})
\, , \, \vartheta^{[\bm{\zeta}]}_{\beta} ({\bf Y})
\right\}\right|_{\cal K} \,\, = \,\,\, \epsilon^{2} \,
\left. \left\{ \vartheta^{[\bm{\zeta}]}_{\alpha} ({\bf X}) 
\, , \, \vartheta^{[\bm{\zeta}]}_{\beta} ({\bf Y})
\right\}\right|_{{\cal K}[0]} \,\, + \,\, \dots
\end{equation}

 The terms in (\ref{varthetaJbExp}) - (\ref{varthetavarthetaExp})
have the decreasing orders in $\epsilon$ like in (\ref{JaJbBr}), 
with the analogous dependence on the derivatives of the functions
$\, {\bf k}_{q} ({\bf X}) = {\bf S}_{X^{q}}$,
$\, {\bf U} ({\bf X})$ and $\, \bm{\zeta}_{\bf X}$. 
However, the dependence on the values ${\bf S} ({\bf X})$ 
on ${\cal K}$ here is more complicated. Namely, in addition to 
the dependence on the values 
$({\bf S}_{\bf X}, {\bf U} ({\bf X}), \bm{\zeta}_{\bf X})$
and their derivatives, the terms in 
(\ref{varthetaJbExp}) - (\ref{varthetavarthetaExp}) 
can contain now also an explicit dependence on 
$\Delta \bm{\theta}_{0} ({\bf X}) =  
({\bf S} ({\bf X}) - \bm{\zeta} ({\bf X})) / \epsilon$
due to the definition of the functionals
$\vartheta^{[\bm{\zeta}]}_{\alpha} ({\bf X})$. 
The corresponding dependence has an oscillating 
character and is periodic with the period
$$S^{\alpha} ({\bf X}) \,\, \rightarrow \,\, 
S^{\alpha} ({\bf X}) \,\, + \,\, 2 \pi \epsilon $$
for each function $S^{\alpha} ({\bf X})$.

 According to relations (\ref{SalphazetaDef}) and
(\ref{NewSVarDer1}), the expansions
(\ref{varthetaJbExp}) - (\ref{varthetavarthetaExp})
naturally generate the expansions depending on
$\Delta \bm{\theta}_{0} ({\bf X})$
for the brackets of the functionals
$\, {\tilde S}^{\alpha [\bm{\zeta}]} ({\bf X})$ and 
$\, S^{\alpha [\bm{\zeta}]} ({\bf Y})$ on ${\cal K}$, 
which can be used in the region of definition 
of these functionals
($|{\bf S} ({\bf X}) - \bm{\zeta} ({\bf X})| = O (\epsilon)$).
Thus, we can write:

$$ \left. \left\{ S^{\alpha [\bm{\zeta}]} ({\bf X})
\, , \, J_{[{\bf b}]}
\right\}\right|_{\cal K} \,\, = \,\,\, \epsilon \,
\left. \left\{ S^{\alpha [\bm{\zeta}]} ({\bf X})
\, , \, J_{[{\bf b}]}
\right\}\right|_{{\cal K}[0]} \,\, + \,\, \epsilon^{2} \,
\left. \left\{ S^{\alpha [\bm{\zeta}]} ({\bf X})
\, , \, J_{[{\bf b}]}
\right\}\right|_{{\cal K}[1]} \,\, + \,\, \dots $$
$$ \left. \left\{ S^{\alpha [\bm{\zeta}]} ({\bf X})
\, , \, S^{\beta [\bm{\zeta}]} ({\bf Y})
\right\}\right|_{\cal K} \,\, = \,\,\, \epsilon^{2} \,
\left. \left\{ S^{\alpha [\bm{\zeta}]} ({\bf X})
\, , \, S^{\beta [\bm{\zeta}]} ({\bf Y})  
\right\}\right|_{{\cal K}[0]} \,\, + \,\, \dots $$
or
\begin{equation}
\label{SfJbBronKGen}
\left. \left\{ S^{[\bm{\zeta}]}_{[{\bf f}]} , J_{[{\bf b}]}
\right\}\right|_{\cal K} = \,\,\, \epsilon 
\left. \left\{ S^{[\bm{\zeta}]}_{[{\bf f}]} , J_{[{\bf b}]}
\right\} \right|_{{\cal K} \, [0]}
+ \,\, \epsilon^{2} 
\left. \left\{ S^{[\bm{\zeta}]}_{[{\bf f}]} , J_{[{\bf b}]}
\right\}
\right|_{{\cal K} \, [1]}
+ \,\, \dots 
\end{equation}
\begin{equation}
\label{SfShBronKGen}
\left. \left\{ S^{[\bm{\zeta}]}_{[{\bf f}]} \, , \,
S^{[\bm{\zeta}]}_{[{\bf h}]} \right\}
\right|_{\cal K}
= \,\,\, \epsilon^{2} \,
\left. \left\{ S^{[\bm{\zeta}]}_{[{\bf f}]} \, , \,
S^{[\bm{\zeta}]}_{[{\bf h}]} \right\}
\right|_{{\cal K} \, [0]} 
+ \,\, \dots 
\end{equation}

 All the terms above represent local expressions of
${\bf S} _{\bf X}$, $\bm{\zeta}_{\bf X}$, ${\bf U} ({\bf X})$, 
${\bf f} ({\bf X})$, ${\bf h} ({\bf X})$, ${\bf b} ({\bf X})$ 
and their derivatives with the additional dependence on
$\Delta \bm{\theta}_{0} ({\bf X})$, integrated over $\mathbb{R}^{d}$.
Let us note also that the indices $[k]$ mean here just the total
number of differentiations of the functions
${\bf S} _{\bf X}$, $\bm{\zeta}_{\bf X}$, ${\bf U} ({\bf X})$,
${\bf f} ({\bf X})$, ${\bf h} ({\bf X})$, ${\bf b} ({\bf X})$ w.r.t.
${\bf X}$. For $\, {\bf S} ({\bf X}) \equiv \bm{\zeta} ({\bf X})$
the relations above represent just the regular graded expansions
in our previous sense.

 The operator $\, \delta / \delta S^{\alpha} ({\bf X})$ can be
written here as the operator
\begin{equation}
\label{SVarDerRepr}
{\delta \over \delta S^{\alpha} ({\bf X})} \,\,\,\,\, = \,\,\,\,\,
{1 \over \epsilon} \, 
{\partial \over \partial \Delta \theta^{\alpha}_{0}} \,\,\, + \,\,\,
{\partial \over \partial X^{q}} \,
{\partial \over \partial S^{\alpha}_{X^{q}}} \,\,\, + \,\,\,
{\partial^{2} \over \partial X^{q} \partial X^{p}} \,\,
{\partial \over \partial S^{\alpha}_{X^{q} X^{p}}} \,\,\, + \,\,\, 
\dots
\end{equation}
applied to the integrands in 
(\ref{SfJbBronKGen}) - (\ref{SfShBronKGen}).
According to relations (\ref{NewSVarDer2}) we can see that the
first term in (\ref{SVarDerRepr}) can be actually omitted at
the ``points'' of ${\cal K}$ with
$\, {\bf S} ({\bf X}) \equiv \bm{\zeta} ({\bf X})$.
As a result, we can actually claim that the values 
$$\left. {\delta \{ S^{[\bm{\zeta}]}_{[{\bf f}]} ,
J_{[{\bf b}]} \} \over \delta S^{\alpha} ({\bf X})}
\right|_{\bm{\varphi} = \bm{\Phi}_{[{\bf U}, \bm{\zeta}]}}
\,\,\,\,\,\,\,\, , \,\,\,\,\,\,\,\,\,\,
\left. {\delta \{ S^{[\bm{\zeta}]}_{[{\bf f}]} ,
S^{[\bm{\zeta}]}_{[{\bf h}]} \} \over \delta S^{\alpha} ({\bf X})}
\right|_{\bm{\varphi} = \bm{\Phi}_{[{\bf U}, \bm{\zeta}]}} $$
are still represented by the regular graded expansions, and, 
besides that, we can write
$$\left. {\delta \{ S^{[\bm{\zeta}]}_{[{\bf f}]} ,
J_{[{\bf b}]} \} \over \delta S^{\alpha} ({\bf X})}
\right|_{\bm{\varphi} = \bm{\Phi}_{[{\bf U}, \bm{\zeta}]}}
= \,\,\, \epsilon \left. \left[ \left. \delta \left\{
S^{[\bm{\zeta}]}_{[{\bf f}]} , J_{[{\bf b}]} \right\}
\right|_{{\cal K} [0]} \, \big/ \, \delta S^{\alpha} ({\bf X})
\right] \right|_{\bm{\varphi} = \bm{\Phi}_{[{\bf U}, \bm{\zeta}]}}
\,\,\,\,\, + \,\,\,\,\, O (\epsilon^{2}) $$
$$\left. {\delta \{ S^{[\bm{\zeta}]}_{[{\bf f}]} ,
J_{[{\bf b}]} \} \over \delta U^{\gamma} ({\bf X})}
\right|_{\bm{\varphi} = \bm{\Phi}_{[{\bf U}, \bm{\zeta}]}}
= \,\,\, \epsilon \left. \left[ \left. \delta \left\{
S^{[\bm{\zeta}]}_{[{\bf f}]} , J_{[{\bf b}]} \right\}
\right|_{{\cal K} [0]} \, \big/ \, \delta U^{\gamma} ({\bf X})
\right] \right|_{\bm{\varphi} = \bm{\Phi}_{[{\bf U}, \bm{\zeta}]}}
\,\,\,\,\, + \,\,\,\,\, O (\epsilon^{2}) $$
$$\left. {\delta \{ S^{[\bm{\zeta}]}_{[{\bf f}]} , 
S^{[\bm{\zeta}]}_{[{\bf h}]} \} \over \delta S^{\alpha} ({\bf X})}
\right|_{\bm{\varphi} = \bm{\Phi}_{[{\bf U}, \bm{\zeta}]}}
= \,\,\, O (\epsilon^{2}) \,\,\,\,\,\,\,\, , \,\,\,\,\,\,\,\,\,\,
\left. {\delta \{ S^{[\bm{\zeta}]}_{[{\bf f}]} , 
S^{[\bm{\zeta}]}_{[{\bf h}]} \} \over \delta U^{\gamma} ({\bf X})}
\right|_{\bm{\varphi} = \bm{\Phi}_{[{\bf U}, \bm{\zeta}]}}
= \,\,\, O (\epsilon^{2}) $$
at the points $({\bf U} ({\bf X}), \bm{\zeta} ({\bf X}))$ of the
submanifold ${\cal K}$.

 Using the same arguments we can claim also the analogous
statements about the functionals, defined with the aid of
the functionals $S^{\alpha [\bm{\zeta}]} ({\bf X})$.
Thus, using the definition (\ref{NewFuncU}), we can also
write
$$\left. {\delta \{ S^{[\bm{\zeta}]}_{[{\bf f}]} ,
U^{[\bm{\zeta}]}_{[{\bf b}]} \} \over \delta S^{\alpha} ({\bf X})}
\right|_{\bm{\varphi} = \bm{\Phi}_{[{\bf U}, \bm{\zeta}]}}
= \,\,\, \epsilon \left. \left[ \left. \delta \left\{
S^{[\bm{\zeta}]}_{[{\bf f}]} , 
U^{[\bm{\zeta}]}_{[{\bf b}]} \right\}   
\right|_{{\cal K} [0]} \, \big/ \, \delta S^{\alpha} ({\bf X})
\right] \right|_{\bm{\varphi} = \bm{\Phi}_{[{\bf U}, \bm{\zeta}]}}
\,\,\,\,\, + \,\,\,\,\, O (\epsilon^{2}) $$
$$\left. {\delta \{ S^{[\bm{\zeta}]}_{[{\bf f}]} ,  
U^{[\bm{\zeta}]}_{[{\bf b}]} \} \over \delta U^{\gamma} ({\bf X})}
\right|_{\bm{\varphi} = \bm{\Phi}_{[{\bf U}, \bm{\zeta}]}}
= \,\,\, \epsilon \left. \left[ \left. \delta \left\{
S^{[\bm{\zeta}]}_{[{\bf f}]} , 
U^{[\bm{\zeta}]}_{[{\bf b}]} \right\}
\right|_{{\cal K} [0]} \, \big/ \, \delta U^{\gamma} ({\bf X})
\right] \right|_{\bm{\varphi} = \bm{\Phi}_{[{\bf U}, \bm{\zeta}]}}
\,\,\,\,\, + \,\,\,\,\, O (\epsilon^{2}) $$
at the points
$({\bf U} ({\bf X}), \bm{\zeta} ({\bf X}))$ on ${\cal K}$.

 Consider the Hamiltonian flow, generated by the functional
$J_{[{\bf b}]}$, at the points of the submanifold
${\cal K}$:
$$\bm{\varphi} (\bm{\theta}, {\bf X}) \,\,\, = \,\,\,
\bm{\Phi} \left( \bm{\theta} +
{{\bf S} ({\bf X}) \over \epsilon}, \,\, {\bf S}_{\bf X}, \,
{\bf U} ({\bf X}) \right) $$

 It's not difficult to see, that in the main order of
$\epsilon$ the flow leaves invariant the submanifold
${\cal K}$, generating the linear evolution of the phase   
shifts $\theta_{0}^{\alpha} ({\bf X})$ with the frequencies 
$\, \omega^{\alpha\gamma} ({\bf X}) \, b_{\gamma} ({\bf X})$
such that we have
$$\left. \left\{ \varphi^{i} (\bm{\theta}, {\bf X}) \, , \,
J_{[{\bf b}]} \right\}\right|_{\cal K} \,\, = \,\,\,
\Phi^{i}_{\theta^{\alpha}} \left( \! \bm{\theta} +
{{\bf S} ({\bf X}) \over \epsilon} , \,
{\bf S}_{\bf X}, {\bf U} ({\bf X}) \! \right) \,
\omega^{\alpha\gamma} ({\bf X}) \, b_{\gamma} ({\bf X})
\,\,\, + \,\,\, O (\epsilon) $$

 So, using the values of the functionals
${\bf S}^{[\bm{\zeta}]} ({\bf X})$ on ${\cal K}$, we can
write:
$$\left. \left\{ S^{\alpha [\bm{\zeta}]} ({\bf X})
\, , \, J_{[{\bf b}]}
\right\}\right|_{\cal K} \,\, = \,\,\, \epsilon \,\,
\omega^{\alpha\gamma} \left( {\bf S}_{\bf X},
{\bf U} ({\bf X}) \right) \, b_{\gamma} ({\bf X}) \,\,\, +
\,\,\, \dots $$
where the next corrections are represented by the terms
containing the higher derivatives of the functions
${\bf S} _{\bf X}$, $\bm{\zeta}_{\bf X}$, ${\bf U} ({\bf X})$,
${\bf f} ({\bf X})$, ${\bf b} ({\bf X})$. We then immediately
get:
$$\left. \left\{ S^{[\bm{\zeta}]}_{[{\bf f}]} , J_{[{\bf b}]}
\right\} \right|_{{\cal K} \, [0]} \,\,\, = \,\,\, \int
\omega^{\alpha\gamma} \left( {\bf S}_{\bf X},
{\bf U} ({\bf X}) \right) \, f_{\alpha} ({\bf X})
\, b_{\gamma} ({\bf X}) \,\, d^{d} X $$

 According to definition (\ref{NewFuncU}), we have also
\begin{equation}
\label{varphiUzetabr}
\left. \left\{ \varphi^{i} (\bm{\theta}, {\bf X}) \, , \,  
U^{[\bm{\zeta}]}_{[{\bf b}]} \right\}
\right|_{\bm{\varphi} = \bm{\Phi}_{[{\bf U}, \bm{\zeta}]}}
 = \,\,
\Phi^{i}_{\theta^{\alpha}} \! \left( \! \bm{\theta} +
{{\bf S} ({\bf X}) \over \epsilon} , \,
{\bf S}_{\bf X}, {\bf U} ({\bf X}) \! \right) \,
\omega^{\alpha\gamma} ({\bf X}) \, b_{\gamma} ({\bf X})
\,\,\, + \,\,\, O (\epsilon) 
\end{equation}
$$\left. \left\{ S^{\alpha [\bm{\zeta}]} ({\bf X})
\, , \, U^{[\bm{\zeta}]}_{[{\bf b}]}
\right\}\right|_{\cal K} \,\, = \,\,\, \epsilon \,\,
\omega^{\alpha\gamma} \left( {\bf S}_{\bf X},
{\bf U} ({\bf X}) \right) \, b_{\gamma} ({\bf X}) \,\,\, +
\,\,\, \dots $$

 At the same time we can write
\begin{equation}
\label{varphiSzetabr}
\left. \left\{ \varphi^{i} (\bm{\theta}, {\bf X}) \, , \,
S^{[\bm{\zeta}]}_{[{\bf h}]} \right\}
\right|_{\bm{\varphi} = \bm{\Phi}_{[{\bf U}, \bm{\zeta}]}}
\, = \,\,\, O (\epsilon)
\end{equation}

 Let us introduce now the unified set of functionals 
$\, G^{\nu [\bm{\zeta}]} ({\bf X})$, $\, \nu = 1, \dots, 2m + s$, 
putting
$$\left( G^{1 [\bm{\zeta}]} ({\bf X}), \, \dots, \, 
G^{2m+s \, [\bm{\zeta}]} ({\bf X}) \right)
\,\,\, =  \,\,\, 
\left( S^{1 [\bm{\zeta}]} ({\bf X}), \, \dots, \, 
S^{m [\bm{\zeta}]} ({\bf X}), \, 
U^{1 [\bm{\zeta}]} ({\bf X}), \, \dots, \,
U^{m+s \, [\bm{\zeta}]} ({\bf X}) \right) $$
and the unified notations for the coordinates 
$\, ({\bf U} ({\bf X}), {\bf S} ({\bf X}))$ on ${\cal K}$:
$$\left( G^{1} ({\bf X}), \, \dots, \,
G^{2m+s} ({\bf X}) \right)
\,\,\,= \,\,\, \left( S^{1} ({\bf X}), \, \dots, \,
S^{m} ({\bf X}), \, U^{1} ({\bf X}), \, \dots, \,
U^{m+s} ({\bf X}) \right) $$

 We define also the regularized functionals
$$G^{[\bm{\zeta}]}_{[{\bf q}]} \,\, = \,\, \int
G^{\nu [\bm{\zeta}]} ({\bf X}) \,\, 
q_{\nu} ({\bf X}) \,\, d^{d} X $$
(summation in $\, \nu = 1, \dots, 2 m + s$) for smooth compactly 
supported vector-valued functions  \linebreak
$\, {\bf q} ({\bf X}) = (q_{1}({\bf X}), \dots , 
q_{2 m + s}({\bf X}))$.

 According to our considerations above, we have then in general
case:
$$\left. \left\{ G^{[\bm{\zeta}]}_{[{\bf q}]} \, , \, 
G^{[\bm{\zeta}]}_{[{\bf p}]} \right\}
\right|_{\bm{\varphi} = \bm{\Phi}_{[{\bf U}, \bm{\zeta}]}}
\, = \,\,\, O (\epsilon)\,\,\,\,\,\,\,\, , \,\,\,\,\,\,\,\,\,\,
\left. {\delta \{ G^{[\bm{\zeta}]}_{[{\bf q}]} , 
G^{[\bm{\zeta}]}_{[{\bf p}]} \} 
\over \delta G^{\nu} ({\bf X})}
\right|_{\bm{\varphi} = \bm{\Phi}_{[{\bf U}, \bm{\zeta}]}}
\, = \,\,\, O (\epsilon) $$

 Besides that, we can write
\begin{equation}
\label{AvBrRel1}
\left. \left\{ G^{[\bm{\zeta}]}_{[{\bf q}]} \, , \, 
G^{[\bm{\zeta}]}_{[{\bf p}]} \right\}
\right|_{\bm{\varphi} = \bm{\Phi}_{[{\bf U}, \bm{\zeta}]} \, (1)}
\, = \,\,\, \left. \left\{ G_{[{\bf q}]} \, , \, G_{[{\bf p}]} 
\right\}_{\rm AV} \right|_{[{\bf U}, \bm{\zeta}]} 
\end{equation}
\begin{equation}
\label{AvBrRel2}
\left. {\delta \{ G^{[\bm{\zeta}]}_{[{\bf q}]} , 
G^{[\bm{\zeta}]}_{[{\bf p}]} \}
\over \delta G^{\nu} ({\bf X})}
\right|_{\bm{\varphi} = \bm{\Phi}_{[{\bf U}, \bm{\zeta}]} \, (1)}
\, = \,\,\, \left. {\delta \{ G_{[{\bf q}]} , G_{[{\bf p}]} 
\}_{\rm AV} \over \delta G^{\nu} ({\bf X})}
\right|_{[{\bf U}, \bm{\zeta}]} 
\end{equation}
where the form $\{ \dots , \dots \}_{\rm AV}$ on the
space $\, ({\bf S} ({\bf X}), {\bf U} ({\bf X}))$ is defined
by relations (\ref{AveragedBracket}).

 Let us note that we use here the notation $(k)$ to designate
just the corresponding order of $\epsilon$. As we saw above, this 
order can be different from the gradation degree $[k]$ for
expressions containing the functionals 
${\bf S}^{[\bm{\zeta}]} ({\bf X})$.

 According to the considerations above we can also write
\begin{equation}
\label{GGVarDerphiOrd}
\left. {\delta \{ G^{[\bm{\zeta}]}_{[{\bf q}]} ,
G^{[\bm{\zeta}]}_{[{\bf p}]} \}
\over \delta \varphi^{k} (\bm{\theta}, {\bf X})}
\right|_{\bm{\varphi} = \bm{\Phi}_{[{\bf U}, \bm{\zeta}]}}
\, = \,\,\, O (\epsilon) 
\end{equation}
where all the expressions above are given by the regular graded
series at the point $({\bf U} ({\bf X}), \bm{\zeta} ({\bf X}))$.

 Using our remark about the flow, generated by the functional
$J_{[{\bf b}]}$, we can easily get the relations
$$\left. \left\{ 
g^{i [\bm{\zeta}]} (\bm{\theta}, {\bf X}) \, , \,
J_{[{\bf b}]} \right\} 
\right|_{\bm{\varphi} = \bm{\Phi}_{[{\bf U}, \bm{\zeta}]}}
\,\,\, = \,\,\, O (\epsilon) \,\,\,\,\, , \,\,\,\,\,\,\,\,
\left. \left\{ 
g^{[\bm{\zeta}]}_{[{\bf Q}]} \, , \, J_{[{\bf b}]} \right\} 
\right|_{\bm{\varphi} = \bm{\Phi}_{[{\bf U}, \bm{\zeta}]}}
\,\,\, = \,\,\, O (\epsilon) $$
and also
$$\left. \left\{
g^{i [\bm{\zeta}]} (\bm{\theta}, {\bf X}) \, , \,
U^{[\bm{\zeta}]}_{[{\bf b}]} \right\}
\right|_{\bm{\varphi} = \bm{\Phi}_{[{\bf U}, \bm{\zeta}]}}
\,\,\, = \,\,\, O (\epsilon) \,\,\,\,\, , \,\,\,\,\,\,\,\,
\left. \left\{
g^{[\bm{\zeta}]}_{[{\bf Q}]} \, , \, 
U^{[\bm{\zeta}]}_{[{\bf b}]} \right\} 
\right|_{\bm{\varphi} = \bm{\Phi}_{[{\bf U}, \bm{\zeta}]}}
\,\,\, = \,\,\, O (\epsilon) $$

 However, we should write in general
$$\left. { \delta \{ g^{[\bm{\zeta}]}_{[{\bf Q}]}  , 
J_{[{\bf b}]} \} \over \delta \varphi^{k} (\bm{\theta}, {\bf X})}
\right|_{\bm{\varphi} = \bm{\Phi}_{[{\bf U}, \bm{\zeta}]}}
\,\,\, = \,\,\, O (1) \,\,\,\,\, , \,\,\,\,\,\,\,\,
\left. { \delta \{ g^{[\bm{\zeta}]}_{[{\bf Q}]}  ,
U^{[\bm{\zeta}]}_{[{\bf b}]} \} \over 
\delta \varphi^{k} (\bm{\theta}, {\bf X})}
\right|_{\bm{\varphi} = \bm{\Phi}_{[{\bf U}, \bm{\zeta}]}}    
\,\,\, = \,\,\, O (1) $$

 For the functionals $S^{[\bm{\zeta}]}_{[{\bf f}]}$ it's not
difficult to get the relations
$$\left. \left\{
g^{[\bm{\zeta}]}_{[{\bf Q}]} \, , \, 
S^{[\bm{\zeta}]}_{[{\bf f}]} \right\}
\right|_{\bm{\varphi} = \bm{\Phi}_{[{\bf U}, \bm{\zeta}]}}
\,\,\, = \,\,\, O (\epsilon) \,\,\,\,\, , \,\,\,\,\,\,\,\,
\left. { \delta \{ g^{[\bm{\zeta}]}_{[{\bf Q}]}  ,
S^{[\bm{\zeta}]}_{[{\bf f}]} \} 
\over \delta \varphi^{k} (\bm{\theta}, {\bf X})}
\right|_{\bm{\varphi} = \bm{\Phi}_{[{\bf U}, \bm{\zeta}]}}
\,\,\, = \,\,\, O (\epsilon) $$

 In general, we have to write for the regularized functionals
$G^{[\bm{\zeta}]}_{[{\bf q}]}$:
\begin{equation}
\label{gGOrd}
\left. \left\{
g^{i [\bm{\zeta}]} (\bm{\theta}, {\bf X}) \, , \,
G^{[\bm{\zeta}]}_{[{\bf q}]} \right\}
\right|_{\bm{\varphi} = \bm{\Phi}_{[{\bf U}, \bm{\zeta}]}}
\,\,\, = \,\,\, O (\epsilon) \,\,\,\,\, , \,\,\,\,\,\,\,\,
\left. \left\{
g^{[\bm{\zeta}]}_{[{\bf Q}]} \, , \,
G^{[\bm{\zeta}]}_{[{\bf q}]} \right\}
\right|_{\bm{\varphi} = \bm{\Phi}_{[{\bf U}, \bm{\zeta}]}}
\,\,\, = \,\,\, O (\epsilon) 
\end{equation}
\begin{equation}
\label{gGVarDerphiOrd}
\left. { \delta \{ g^{[\bm{\zeta}]}_{[{\bf Q}]}  ,
G^{[\bm{\zeta}]}_{[{\bf q}]} \}
\over \delta \varphi^{k} (\bm{\theta}, {\bf X})}
\right|_{\bm{\varphi} = \bm{\Phi}_{[{\bf U}, \bm{\zeta}]}}
\,\,\, = \,\,\, O (1) 
\end{equation}

 In the same way as before, we can state also for our 
functionals $\, {\bf S}^{[\bm{\zeta}]} ({\bf X})$,
$\, {\bf U}^{[\bm{\zeta}]} ({\bf X})$,
$\, {\bf g}^{[\bm{\zeta}]} (\bm{\theta}, {\bf X})$,
the relations:
$$\left. {\delta \{ g^{[\bm{\zeta}]}_{[{\bf Q}]} , 
G^{[\bm{\zeta}]}_{[{\bf p}]} \}
\over \delta S^{\alpha} ({\bf X})}
\right|_{\bm{\varphi} = \bm{\Phi}_{[{\bf U}, \bm{\zeta}]}}
\, = \,\,\, O (\epsilon) \,\,\,\,\, , \,\,\,\,\,\,\,\,
\left. {\delta \{ g^{[\bm{\zeta}]}_{[{\bf Q}]} ,
G^{[\bm{\zeta}]}_{[{\bf p}]} \}
\over \delta U^{\gamma} ({\bf X})}
\right|_{\bm{\varphi} = \bm{\Phi}_{[{\bf U}, \bm{\zeta}]}}
\, = \,\,\, O (\epsilon)$$
or, in the ``unified'' form:
\begin{equation}
\label{gGVarDerGOrd}
\left. {\delta \{ g^{[\bm{\zeta}]}_{[{\bf Q}]} ,
G^{[\bm{\zeta}]}_{[{\bf p}]} \}
\over \delta G^{\nu} ({\bf X})}
\right|_{\bm{\varphi} = \bm{\Phi}_{[{\bf U}, \bm{\zeta}]}}
\, = \,\,\, O (\epsilon) 
\end{equation}

 The pairwise Poisson brackets of the functionals
$g^{i [\bm{\zeta}]} (\bm{\theta}, {\bf X})$ can be represented 
by the following a little bit bulky expression
$$\left\{ g^{i [\bm{\zeta}]} (\bm{\theta}, {\bf X}) \, , \,
g^{j [\bm{\zeta}]} (\bm{\theta}^{\prime}, {\bf Y}) \right\} 
\,\, = \,\, \left\{ \varphi^{i} (\bm{\theta}, {\bf X}) \, , \,
\varphi^{j} (\bm{\theta}^{\prime}, {\bf Y}) \right\}  \,\, - $$
$$- \,\, \left\{ \varphi^{i} (\bm{\theta}, {\bf X}) \, , \,
U^{\lambda [\bm{\zeta}]}({\bf Y}) \right\} \,\,\,
\Phi^{j}_{U^{\lambda}} 
\left( {{\bf S}^{[\bm{\zeta}]}({\bf Y}) \over \epsilon} + 
\bm{\theta}^{\prime} , {\bf Y} \right)  \, - $$
$$- \,\, \left\{ \varphi^{i} (\bm{\theta}, {\bf X}) \, , \,
S^{\beta [\bm{\zeta}]}_{Y^{p}} \right\} \,\,\,
\Phi^{j}_{k^{\beta}_{p}}
\left( {{\bf S}^{[\bm{\zeta}]}({\bf Y}) \over \epsilon} +
\bm{\theta}^{\prime} , {\bf Y} \right)  \, - $$
$$- \, \Phi^{i}_{U^{\nu}} 
\left( {{\bf S}^{[\bm{\zeta}]}({\bf X}) \over \epsilon} +
\bm{\theta} , {\bf X} \right) \, 
\left\{ U^{\nu [\bm{\zeta}]}({\bf X}) 
\, , \, \varphi^{j} (\bm{\theta}^{\prime}, {\bf Y}) 
\right\}  \,\,\, - $$
$$- \, \Phi^{i}_{k^{\alpha}_{q}}
\left( {{\bf S}^{[\bm{\zeta}]}({\bf X}) \over \epsilon} +
\bm{\theta} , {\bf X} \right) \, 
\left\{ S^{\alpha [\bm{\zeta}]}_{X^{q}}
\, , \, \varphi^{j} (\bm{\theta}^{\prime}, {\bf Y}) 
\right\}  \,\,\, + $$
$$+ \, \Phi^{i}_{U^{\nu}} 
\left( {{\bf S}^{[\bm{\zeta}]}({\bf X}) \over \epsilon} +
\bm{\theta} , {\bf X} \right) \,
\left\{ U^{\nu [\bm{\zeta}]}({\bf X}) \, , \,
U^{\lambda [\bm{\zeta}]}({\bf Y}) \right\} \,\,\,
\Phi^{j}_{U^{\lambda}} 
\left( {{\bf S}^{[\bm{\zeta}]}({\bf Y}) \over \epsilon} +
\bm{\theta}^{\prime} , {\bf Y} \right) \, + $$
$$+ \, \Phi^{i}_{k^{\alpha}_{q}}
\left( {{\bf S}^{[\bm{\zeta}]}({\bf X}) \over \epsilon} +
\bm{\theta} , {\bf X} \right) \,
\left\{ S^{\alpha [\bm{\zeta}]}_{X^{q}} \, , \,
U^{\lambda [\bm{\zeta}]}({\bf Y}) \right\} \,\,\,
\Phi^{j}_{U^{\lambda}}
\left( {{\bf S}^{[\bm{\zeta}]}({\bf Y}) \over \epsilon} +
\bm{\theta}^{\prime} , {\bf Y} \right) \, + $$
$$+ \, \Phi^{i}_{U^{\nu}}
\left( {{\bf S}^{[\bm{\zeta}]}({\bf X}) \over \epsilon} +
\bm{\theta} , {\bf X} \right) \,
\left\{ U^{\nu [\bm{\zeta}]}({\bf X}) \, , \,
S^{\beta [\bm{\zeta}]}_{Y^{p}} \right\} \,\,\,
\Phi^{j}_{k^{\beta}_{p}}
\left( {{\bf S}^{[\bm{\zeta}]}({\bf Y}) \over \epsilon} +
\bm{\theta}^{\prime} , {\bf Y} \right) \, + $$
$$+ \, \Phi^{i}_{k^{\alpha}_{q}}
\left( {{\bf S}^{[\bm{\zeta}]}({\bf X}) \over \epsilon} +
\bm{\theta} , {\bf X} \right) \,
\left\{ S^{\alpha [\bm{\zeta}]}_{X^{q}} \, , \,
S^{\beta [\bm{\zeta}]}_{Y^{p}} \right\} \,\,\,
\Phi^{j}_{k^{\beta}_{p}}
\left( {{\bf S}^{[\bm{\zeta}]}({\bf Y}) \over \epsilon} +
\bm{\theta}^{\prime} , {\bf Y} \right) \, - $$
$$- \,\, {1 \over \epsilon} \,\,
\left\{ \varphi^{i} (\bm{\theta}, {\bf X}) \, , \,
S^{\beta [\bm{\zeta}]}({\bf Y}) \right\} \,\,\,
\Phi^{j}_{\theta^{\prime\beta}}   
\left( {{\bf S}^{[\bm{\zeta}]}({\bf Y}) \over \epsilon} +
\bm{\theta}^{\prime} , {\bf Y} \right)  \,\, - $$
$$- \,\, {1 \over \epsilon} \,\, \Phi^{i}_{\theta^{\alpha}}
\left( {{\bf S}^{[\bm{\zeta}]}({\bf X}) \over \epsilon} +
\bm{\theta} , {\bf X} \right) \,
\left\{  S^{\alpha [\bm{\zeta}]}({\bf X}) \, , \,
\varphi^{j} (\bm{\theta}^{\prime}, {\bf Y}) \right\}
\,\,\, + $$
$$+ \,\,\, {1 \over \epsilon} \,\,
\Phi^{i}_{\theta^{\alpha}} \left(
{{\bf S}^{[\bm{\zeta}]}({\bf X}) \over \epsilon} + \bm{\theta} ,
{\bf X} \right) \,
\left\{  S^{\alpha [\bm{\zeta}]} ({\bf X}) \, , \, 
U^{\lambda [\bm{\zeta}]}({\bf Y}) \right\}
\,\,\, \Phi^{j}_{U^{\lambda}}
\left( {{\bf S}^{[\bm{\zeta}]}({\bf Y}) \over \epsilon} +
\bm{\theta}^{\prime} , {\bf Y} \right)  \,\,\, + $$
$$+ \,\,\, {1 \over \epsilon} \,\,
\Phi^{i}_{\theta^{\alpha}} \left( 
{{\bf S}^{[\bm{\zeta}]}({\bf X}) \over \epsilon} + \bm{\theta} ,
{\bf X} \right) \,
\left\{  S^{\alpha [\bm{\zeta}]} ({\bf X}) \, , \,
S^{\beta [\bm{\zeta}]}_{Y^{p}} \right\} \,\,\,
\Phi^{j}_{k^{\beta}_{p}}
\left( {{\bf S}^{[\bm{\zeta}]}({\bf Y}) \over \epsilon} +
\bm{\theta}^{\prime} , {\bf Y} \right) \, + $$
$$+ \,\,\, {1 \over \epsilon} \,\,
\Phi^{i}_{U^{\nu}} \left( 
{{\bf S}^{[\bm{\zeta}]}({\bf X}) \over \epsilon} +
\bm{\theta} , {\bf X} \right)
\left\{ U^{\nu [\bm{\zeta}]}({\bf X}) \, , \,
S^{\beta [\bm{\zeta}]}({\bf Y}) \right\} \,\,\,
\Phi^{j}_{\theta^{\prime\beta}} 
\left( {{\bf S}^{[\bm{\zeta}]}({\bf Y}) \over \epsilon}
+ \bm{\theta}^{\prime} , {\bf Y} \right)  \,\,\, + $$
$$+ \,\, {1 \over \epsilon} \,\, \Phi^{i}_{k^{\alpha}_{q}}
\left( {{\bf S}^{[\bm{\zeta}]}({\bf X}) \over \epsilon} +
\bm{\theta} , {\bf X} \right) \,
\left\{ S^{\alpha [\bm{\zeta}]}_{X^{q}} \, , \,
S^{\beta [\bm{\zeta}]}({\bf Y}) \right\} \,\,\,
\Phi^{j}_{\theta^{\prime\beta}} 
\left( {{\bf S}^{[\bm{\zeta}]}({\bf Y}) \over \epsilon}
+ \bm{\theta}^{\prime} , {\bf Y} \right)  \,\,\, + $$
$$+ \,\, {1 \over \epsilon^{2}} \,\,
\Phi^{i}_{\theta^{\alpha}}  \left(
{{\bf S}^{[\bm{\zeta}]}({\bf X}) \over \epsilon} + \bm{\theta} ,
{\bf X} \right) \,
\left\{  S^{\alpha [\bm{\zeta}]}({\bf X}) \, , \, 
S^{\beta [\bm{\zeta}]}({\bf Y}) \right\}
\,\,\, \Phi^{j}_{\theta^{\prime\beta}}   
\left( {{\bf S}^{[\bm{\zeta}]}({\bf Y}) \over \epsilon} + 
\bm{\theta}^{\prime} , {\bf Y} \right) $$
where we put
$\bm{\Phi} (\bm{\theta}, {\bf X}) \, \equiv \,
\bm{\Phi} (\bm{\theta}, \, {\bf S}^{[\bm{\zeta}]}_{X^{1}},
\dots , {\bf S}^{[\bm{\zeta}]}_{X^{d}}, 
{\bf U}^{[\bm{\zeta}]} ({\bf X}))$.

 However, using relations (\ref{varphiUzetabr}) and 
(\ref{OgranQ}), it's not difficult to check that
$$\left. \left\{ g^{[\bm{\zeta}]}_{[{\bf Q}]} \, , \,
g^{[\bm{\zeta}]}_{[{\bf P}]} \right\}
\right|_{\bm{\varphi} = \bm{\Phi}_{[{\bf U}, \bm{\zeta}]} \, (0)}
\,\,\, = \,\,\, \int  Q_{i} (\bm{\theta}, {\bf X}) \,\,
{\hat B}^{ij}_{[0]} ({\bf X}) \,\,
P_{j} (\bm{\theta}, {\bf X}) \,\,
{d^{m} \theta \over (2\pi)^{m}} \, d^{d} X $$
where
\begin{multline*}
{\hat B}^{ij}_{[0]} ({\bf X}) \,\, = \,\, \sum_{l_{1}, \dots, l_{d}}
B^{ij}_{(l_{1}, \dots , l_{d})}
\left( \bm{\Phi} (\bm{\theta}, \, \bm{\zeta}_{\bf X},
{\bf U}({\bf X})), \,
\zeta^{\gamma_{1}}_{X^{1}} \, 
\bm{\Phi}_{\theta^{\gamma_{1}}},
\dots, \, \zeta^{\gamma_{d}}_{X^{d}} \,
\bm{\Phi}_{\theta^{\gamma_{d}}}, \, \dots \right) \, \times \\
\times \,\, \zeta^{\alpha^{1}_{1}}_{X^{1}} \dots
\zeta^{\alpha^{1}_{l_{1}}}_{X^{1}}
\, \dots \,
\zeta^{\alpha^{d}_{1}}_{X^{d}} \dots
\zeta^{\alpha^{d}_{l_{d}}}_{X^{d}} \,\,\,\,\,
{\partial \over \partial \theta^{\alpha^{1}_{1}}} \dots
{\partial \over \partial \theta^{\alpha^{1}_{l_{1}}}} \, \dots \,
{\partial \over \partial \theta^{\alpha^{d}_{1}}} \dots
{\partial \over \partial \theta^{\alpha^{d}_{l_{d}}}}
\end{multline*}

 It's not difficult to check also the relations:
\begin{equation}
\label{ggVarDerOrd}
\left. { \delta \{ g^{[\bm{\zeta}]}_{[{\bf Q}]}  ,
g^{[\bm{\zeta}]}_{[{\bf P}]} \} 
\over \delta G^{\nu} ({\bf X})}
\right|_{\bm{\varphi} = \bm{\Phi}_{[{\bf U}, \bm{\zeta}]}}
\,\,\, = \,\,\, O (1) \,\,\,\,\, , \,\,\,\,\,\,\,\,
\left. { \delta \{ g^{[\bm{\zeta}]}_{[{\bf Q}]}  ,
g^{[\bm{\zeta}]}_{[{\bf P}]} \}
\over \delta \varphi^{k} (\bm{\theta}, {\bf X})}
\right|_{\bm{\varphi} = \bm{\Phi}_{[{\bf U}, \bm{\zeta}]}}
\,\,\, = \,\,\, O (1)
\end{equation}

 Using relations (\ref{LieDerSf}), (\ref{UgVDerInv}) and the 
invariance of the bracket (\ref{EpsExtBracket}) w.r.t. the
transformations 
$\, \theta^{\alpha} \, \rightarrow \, 
\theta^{\alpha} \, + \, {\rm const}$, 
we can also write for all the brackets
$\, \{ G^{[\bm{\zeta}]}_{[{\bf q}]} ,
G^{[\bm{\zeta}]}_{[{\bf p}]} \}$,
$\, \{ g^{[\bm{\zeta}]}_{[{\bf Q}]} ,
G^{[\bm{\zeta}]}_{[{\bf p}]} \}$,
$\, \{ g^{[\bm{\zeta}]}_{[{\bf Q}]} ,
g^{[\bm{\zeta}]}_{[{\bf P}]} \}$ the relations:
$$\int \left. {\delta \{ G^{[\bm{\zeta}]}_{[{\bf q}]} ,
G^{[\bm{\zeta}]}_{[{\bf p}]} \}
\over \delta \varphi^{i} (\bm{\theta}, {\bf X})}
\right|_{\bm{\varphi} = \bm{\Phi}_{[{\bf U}, \bm{\zeta}]}} \,\,\,
\Phi^{i}_{\theta^{\alpha}} \left( \bm{\theta} +
{\bm{\zeta} ({\bf X}) \over \epsilon} \, ,
\,\, \bm{\zeta}_{\bf X} \, , {\bf U} ({\bf X}) \right) \,\,
{d^{m} \theta \over (2\pi)^{m}} \, d^{d} X
\,\,\, \equiv \,\,\, 0 $$
(etc. $\dots$).

 The relations above take place in all orders of $\epsilon$
and can in fact be strengthened in the main order. Indeed, we can
see that the arbitrary functions $q_{\nu} ({\bf X})$ appear in
the main order of $\epsilon$ just as local factors in the
integrands of the above expressions. In the same way,
making the change
$\, {\tilde Q}_{i}(\bm{\theta}, {\bf W}) \rightarrow
{\tilde Q}_{i}(\bm{\theta}, {\bf W}) \, \mu_{i}({\bf W})$
with arbitrary smooth functions $\mu_{i}({\bf W})$,
we get the same change
$\, Q_{i}(\bm{\theta}, {\bf W}) \rightarrow
Q_{i}(\bm{\theta}, {\bf W}) \, \mu_{i}({\bf W})$
for the functionals $\, Q_{i}(\bm{\theta}, {\bf W})$.
Here again the functions $\mu_{i}({\bf W})$ 
appear just as local factors $\mu_{i}({\bf X})$ in the main order 
of $\epsilon$. We can see then, that we can omit the integration
w.r.t. ${\bf X}$ in the main order of $\epsilon$ and write for
any ${\bf X}$:
\begin{equation}
\label{GGOrtCond}
\int_{0}^{2\pi}\!\!\!\!\!\dots\int_{0}^{2\pi}
\left. {\delta \{ G^{[\bm{\zeta}]}_{[{\bf q}]} ,
G^{[\bm{\zeta}]}_{[{\bf p}]} \}
\over \delta \varphi^{i} (\bm{\theta}, {\bf X})}
\right|_{\bm{\varphi} = \bm{\Phi}_{[{\bf U}, \bm{\zeta}]} \, (1)}
\,\,\, \Phi^{i}_{\theta^{\alpha}} \left( \bm{\theta} +
{\bm{\zeta} ({\bf X}) \over \epsilon} \, ,
\,\, \bm{\zeta}_{\bf X} \, , {\bf U} ({\bf X}) \right) \,\,
{d^{m} \theta \over (2\pi)^{m}}
\,\,\, \equiv \,\,\, 0
\end{equation}
\begin{equation}
\label{gGOrtCond}
\int_{0}^{2\pi}\!\!\!\!\!\dots\int_{0}^{2\pi}
\left. {\delta \{ g^{[\bm{\zeta}]}_{[{\bf Q}]} ,
G^{[\bm{\zeta}]}_{[{\bf p}]} \}
\over \delta \varphi^{i} (\bm{\theta}, {\bf X})}
\right|_{\bm{\varphi} = \bm{\Phi}_{[{\bf U}, \bm{\zeta}]} \, (0)}
\,\,\, \Phi^{i}_{\theta^{\alpha}} \left( \bm{\theta} +
{\bm{\zeta} ({\bf X}) \over \epsilon} \, ,
\,\, \bm{\zeta}_{\bf X} \, , {\bf U} ({\bf X}) \right) \,\,
{d^{m} \theta \over (2\pi)^{m}}
\,\,\, \equiv \,\,\, 0
\end{equation}
\begin{equation}
\label{ggOrtCond}
\int_{0}^{2\pi}\!\!\!\!\!\dots\int_{0}^{2\pi}
\left. {\delta \{ g^{[\bm{\zeta}]}_{[{\bf Q}]} ,
g^{[\bm{\zeta}]}_{[{\bf P}]} \}
\over \delta \varphi^{i} (\bm{\theta}, {\bf X})}
\right|_{\bm{\varphi} = \bm{\Phi}_{[{\bf U}, \bm{\zeta}]} \, (0)}
\,\,\, \Phi^{i}_{\theta^{\alpha}} \left( \bm{\theta} +
{\bm{\zeta} ({\bf X}) \over \epsilon} \, ,
\,\, \bm{\zeta}_{\bf X} \, , {\bf U} ({\bf X}) \right) \,\,
{d^{m} \theta \over (2\pi)^{m}}
\,\,\, \equiv \,\,\, 0
\end{equation}

\vspace{0.2cm}

 One of the main goals of this chapter is to prove that the form
(\ref{AveragedBracket}) gives in fact a Poisson bracket on the 
space of fields $\, ({\bf S} ({\bf X}), {\bf U} ({\bf X}))$.
For the justification of this fact the resolvability of the
following systems
\begin{equation}
\label{BgSsystem}
{\hat B}^{ij}_{[0] [\bm{\zeta}]} ({\bf X}) \,\,
B^{\rm I}_{j [{\bf f}]} \left( \bm{\theta} +
{\bm{\zeta} ({\bf X}) \over \epsilon}, \, {\bf X} \right)
\,\,\, = \,\,\, - \, \left. \left\{ 
g^{i [\bm{\zeta}]} (\bm{\theta}, {\bf X})
\, , \, S^{[\bm{\zeta}]}_{[{\bf f}]} \right\}
\right|_{\bm{\varphi} = \bm{\Phi}_{[{\bf U}, \bm{\zeta}]} \, (1)}
\end{equation}
\begin{equation}
\label{BgUsystem}
{\hat B}^{ij}_{[0] [\bm{\zeta}]} ({\bf X}) \,\,
B^{\rm II}_{j [{\bf b}]} \left( \bm{\theta} +
{\bm{\zeta} ({\bf X}) \over \epsilon}, \, {\bf X} \right)
\,\,\, = \,\,\, - \, \left. \left\{ 
g^{i [\bm{\zeta}]} (\bm{\theta}, {\bf X})
\, , \, U^{[\bm{\zeta}]}_{[{\bf b}]} \right\}
\right|_{\bm{\varphi} = \bm{\Phi}_{[{\bf U}, \bm{\zeta}]} \, (1)}
\end{equation}
where
\begin{multline*}
{\hat B}^{ij}_{[0] [\bm{\zeta}]} ({\bf X}) \,\, \equiv \,\,
\sum_{l_{1}, \dots, l_{d}} B^{ij}_{(l_{1}, \dots , l_{d})}
\left( \bm{\Phi} \Big( \bm{\theta} + 
{\bm{\zeta} ({\bf X}) \over \epsilon}, 
\,\, \bm{\zeta}_{\bf X},
{\bf U}({\bf X}) \Big), \, \dots \right) \,\, \times  \\  
\times \,\, \zeta^{\alpha^{1}_{1}}_{X^{1}} \dots
\zeta^{\alpha^{1}_{l_{1}}}_{X^{1}}
\, \dots \, 
\zeta^{\alpha^{d}_{1}}_{X^{d}} \dots
\zeta^{\alpha^{d}_{l_{d}}}_{X^{d}} \,\,\,\,\,
{\partial \over \partial \theta^{\alpha^{1}_{1}}} \dots
{\partial \over \partial \theta^{\alpha^{1}_{l_{1}}}} \, \dots \,
{\partial \over \partial \theta^{\alpha^{d}_{1}}} \dots
{\partial \over \partial \theta^{\alpha^{d}_{l_{d}}}}
\end{multline*}
on the space of $2\pi$-periodic in each $\theta^{\alpha}$
functions will play important role.

 Let us consider systems (\ref{BgSsystem}) - (\ref{BgUsystem}) 
in more detail under the assumptions that the family $\Lambda$ 
represents a regular Hamiltonian submanifold equipped with a
minimal set of commuting integrals $(I^{1}, \dots, I^{m+s})$.
Let us note, that systems (\ref{BgSsystem}) - (\ref{BgUsystem})
are in fact analogous to system (\ref{ReducedSystbeta})
considered in Chapter 2. Easy to see that systems
(\ref{BgSsystem}) - (\ref{BgUsystem}) represent linear
non-homogeneous differential systems in $\bm{\theta}$ with
periodic coefficients at every fixed value of ${\bf X}$.
Systems (\ref{BgSsystem}) - (\ref{BgUsystem}) are independent
for different ${\bf X}$, while the operators
$\, {\hat B}^{ij}_{[0] [\bm{\zeta}]} ({\bf X})$ coincide
with the corresponding operators
$\, {\hat B}^{ij}_{{\bf k}_{1}, \dots, {\bf k}_{d}}$ after
the trivial shift:
$\, \bm{\theta} \,\, \rightarrow \,\,
\bm{\theta} \, + \, \bm{\zeta} ({\bf X}) / \epsilon $.
We can also claim here, that all the ``regular'' kernel vectors
of the operators 
$\, {\hat B}^{ij}_{[0] [\bm{\zeta}]} ({\bf X})$ on the space
of $2\pi$-periodic in each $\theta^{\alpha}$ functions are
given by the values
$$v^{(k)}_{i [\bm{\zeta}_{\bf X}, {\bf U}({\bf X})]} 
\left( \bm{\theta} + {\bm{\zeta} ({\bf X}) \over \epsilon} 
\right) \,\,\, = \,\,\, 
\sum_{\gamma=1}^{m+s} \gamma^{k}_{\gamma}
\left( \bm{\zeta}_{X^{1}}, \dots, \bm{\zeta}_{X^{d}}, 
{\bf U} ({\bf X}) \right) \,\, 
\zeta^{(\gamma)}_{i [\bm{\zeta}_{\bf X}, {\bf U}({\bf X})]}
\left( \bm{\theta} + {\bm{\zeta} ({\bf X}) \over \epsilon}
\right) $$
where the functions
$\zeta^{(\gamma)}_{i [{\bf k}_{1}, \dots, {\bf k}_{d},
{\bf U}]}(\bm{\theta})$ are defined by relations 
(\ref{zetaVarDer}).

 Like in the case of system (\ref{ReducedSystbeta}), we can
formulate here the following lemma:

\vspace{0.2cm}

{\bf Lemma 4.1.}
 
{\it Let the family $\Lambda$ represent a regular Hamiltonian
submanifold equipped with a minimal set of commuting integrals
$\, (I^{1}, \dots, I^{m+s})$. Then the right-hand parts of 
systems (\ref{BgSsystem}) - (\ref{BgUsystem}) are automatically 
orthogonal to the regular kernel vectors of the corresponding
operators $\, {\hat B}^{ij}_{[0] [\bm{\zeta}]} ({\bf X})$.
}

\vspace{0.2cm}

 Proof.

 Indeed, view the relations (\ref{dependence2}) the convolution 
of the variation derivatives \linebreak
$\delta U^{\gamma [\bm{\zeta}]}({\bf Z}) / 
\delta \varphi^{i} (\bm{\theta}, {\bf X})$
with the brackets
$\{ g^{i [\bm{\zeta}]} (\bm{\theta}, {\bf X}) \, , \, 
S^{[\bm{\zeta}]}_{[{\bf f}]} \}|_{\cal K}$ or
$\{ g^{i [\bm{\zeta}]} (\bm{\theta}, {\bf X}) \, , \, 
U^{[\bm{\zeta}]}_{[{\bf b}]} \}|_{\cal K}$
is identically equal to zero on 
${\cal K}$. From the relations (\ref{NewFuncU})
we can see then that the values 
$\delta U^{\gamma [\bm{\zeta}]}({\bf Z}) / 
\delta \varphi^{i} (\bm{\theta}, {\bf X})$ 
can be replaced in the leading order of $\epsilon$ by the values
$\delta J^{\gamma}({\bf Z}) / 
\delta \varphi^{i} (\bm{\theta}, {\bf X})$ 
at $\bm{\varphi} = \bm{\Phi}_{[{\bf U}, \bm{\zeta}]}$.
Using the explicit expressions for the derivatives
$\delta J^{\gamma}({\bf Z}) / 
\delta \varphi^{i} (\bm{\theta}, {\bf X})$
at $\bm{\varphi} = \bm{\Phi}_{[{\bf U}, \bm{\zeta}]}$:
$$\left. {\delta J^{\gamma}({\bf Z}) \over 
\delta \varphi^{i} (\bm{\theta}, {\bf X})}
\right|_{\bm{\varphi} = \bm{\Phi}_{[{\bf U}, \bm{\zeta}]}} 
\,\,\, =  \,\, \sum_{l_{1},\dots,l_{d}}  
\Pi^{\gamma (l_{1} \dots l_{d})}_{i}
\left( \! \bm{\Phi} \Big( 
\bm{\theta} + {\bm{\zeta}({\bf Z}) \over \epsilon} , \,
{\bf Z} \Big), \dots \! \right) \, 
\epsilon^{l_{1} + \dots + l_{d}} \,\, 
\delta_{l_{1} Z^{1} \dots l_{d} Z^{d}} ({\bf Z} - {\bf X}) $$
we can then write the corresponding convolutions as the action 
of the operator
$$\int_{0}^{2\pi}\!\!\!\!\!\!\dots \! \int_{0}^{2\pi} \!
{d^{m} \theta \over (2\pi)^{m}} \,
\sum_{l_{1},\dots,l_{d}} \Pi^{\gamma (l_{1} \dots l_{d})}_{i}
\left( \bm{\Phi} \Big(
\bm{\theta} + {\bm{\zeta}({\bf Z}) \over \epsilon} , \,
{\bf Z} \Big), \dots \! \right) \, 
\epsilon^{l_{1} + \dots + l_{d}} \,   
{d^{l_{1}} \over d Z^{1 \, l_{1}}} \, \dots \,
{d^{l_{d}} \over d Z^{d \, l_{d}}} $$
on the  functions
$\, \{ g^{i [\bm{\zeta}]} (\bm{\theta}, {\bf Z}) \, , \,   
S^{[\bm{\zeta}]}_{[{\bf f}]} \}
|_{\bm{\varphi} = \bm{\Phi}_{[{\bf U}, \bm{\zeta}]}}$,
$\, \{ g^{i [\bm{\zeta}]} (\bm{\theta}, {\bf Z}) \, , \,   
U^{[\bm{\zeta}]}_{[{\bf b}]} \}
|_{\bm{\varphi} = \bm{\Phi}_{[{\bf U}, \bm{\zeta}]}}$.
Easy to see that in the leading order of $\epsilon$ the
corresponding action is given by the action of the operator
\begin{multline}
\label{MainOrderOpAction}
\int_{0}^{2\pi}\!\!\!\!\!\!\dots \! \int_{0}^{2\pi} \!
{d^{m} \theta \over (2\pi)^{m}} \,
\sum_{l_{1},\dots,l_{d}} \Pi^{\gamma (l_{1} \dots l_{d})}_{i}
\left( \bm{\Phi} \Big(
\bm{\theta} + {\bm{\zeta}({\bf Z}) \over \epsilon} , \,
{\bf Z} \Big), \dots \! \right) \,\, \times   \\
\times \,\, \zeta^{\alpha^{1}_{1}}_{Z^{1}} \dots
\zeta^{\alpha^{1}_{l_{1}}}_{Z^{1}}
\, \dots \,
\zeta^{\alpha^{d}_{1}}_{Z^{d}} \dots
\zeta^{\alpha^{d}_{l_{d}}}_{Z^{d}} \,\,\,\,\,
{\partial \over \partial \theta^{\alpha^{1}_{1}}} \dots
{\partial \over \partial \theta^{\alpha^{1}_{l_{1}}}} \, \dots \,
{\partial \over \partial \theta^{\alpha^{d}_{1}}} \dots  
{\partial \over \partial \theta^{\alpha^{d}_{l_{d}}}}
\end{multline}
on the same functions.

 Since the right-hand parts of systems
(\ref{BgSsystem}) - (\ref{BgUsystem}) represent the leading 
order of the corresponding brackets
at $\bm{\varphi} = \bm{\Phi}_{[{\bf U}, \bm{\zeta}]}$, we get
that they should vanish under the action of the operator
(\ref{MainOrderOpAction}). After the integration by parts
we then get immediately the orthogonality of the right-hand 
parts of (\ref{BgSsystem}) - (\ref{BgUsystem}) to the values
$\zeta^{(\gamma)}_{i [\bm{\zeta}_{\bf X}, {\bf U}({\bf X})]}
(\bm{\theta} + \bm{\zeta}({\bf X})/\epsilon)$.

{\hfill Lemma 4.1 is proved.}

\vspace{0.2cm}

 In general, we can write
\begin{equation}
\label{zetarhpOrt}
\int_{0}^{2\pi}\!\!\!\!\!\dots\int_{0}^{2\pi}
\zeta^{(\gamma)}_{i [\bm{\zeta}_{\bf X}, {\bf U}({\bf X})]}
\left( \bm{\theta} + {\bm{\zeta} ({\bf X}) \over \epsilon} \right)
\,\, \left. \left\{ g^{i [\bm{\zeta}]} (\bm{\theta}, {\bf X}) 
\, , \, G^{[\bm{\zeta}]}_{[{\bf q}]} \right\} 
\right|_{\bm{\varphi} = \bm{\Phi}_{[{\bf U}, \bm{\zeta}]} \, (1)}
\,\, {d^{m} \theta \over (2\pi)^{m}} \,\,\, = \,\,\, 0
\end{equation}

 Like in the case of system (\ref{ReducedSystbeta}), we can
also claim here that systems 
(\ref{BgSsystem}) - (\ref{BgUsystem}) are resolvable on the
space of $2\pi$-periodic functions of $\theta$ in the
single-phase situation. However, the investigation of 
resolvability of systems (\ref{BgSsystem}) - (\ref{BgUsystem}) 
is much more complicated in the general multi-phase case, where 
the behavior of eigen-values of the operators
$\, {\hat B}^{ij}_{[0] [\bm{\zeta}]} ({\bf X})$ can be rather
nontrivial. In general, we should expect the resolvability
of systems (\ref{BgSsystem}) - (\ref{BgUsystem}) just on the
subset ${\cal S}^{\prime}$ in the space of parameters
$$\big( {\bf k}_{1} ({\bf X}) , \dots, 
{\bf k}_{d} ({\bf X}) , {\bf U} ({\bf X})
\big) \,\,\, = \,\,\, \big( \bm{\zeta}_{X^{1}}, \dots,
\bm{\zeta}_{X^{d}}, {\bf U} ({\bf X}) \big) $$

 For the corresponding values of 
$(\bm{\zeta}_{X^{1}}, \dots, \bm{\zeta}_{X^{d}}, 
{\bf U} ({\bf X}))$
we can then write in the general form
$$\left. \left\{ G^{[\bm{\zeta}]}_{[{\bf q}]} \, , \,
g^{i [\bm{\zeta}]} (\bm{\theta}, {\bf X}) \right\} 
\right|_{\bm{\varphi} = \bm{\Phi}_{[{\bf U}, \bm{\zeta}]} \, (1)}
\,\,\, = \,\,\, {\hat B}^{ij}_{[0][\bm{\zeta}]} ({\bf X}) \,\,
B_{j [{\bf q}]} \left( \bm{\theta} +
{\bm{\zeta} ({\bf X}) \over \epsilon}, \, {\bf X} \right) $$

 In the full analogy with (\ref{Betaortrel}) we can then also 
write here the relations
\begin{equation}
\label{BiOrtCond}
\int_{0}^{2\pi}\!\!\!\!\!\dots\int_{0}^{2\pi} 
\Phi^{i}_{\theta^{\alpha}} \left( \bm{\theta} +
{\bm{\zeta} ({\bf X}) \over \epsilon}, \,\,
\bm{\zeta}_{\bf X}, {\bf U} ({\bf X}) \right) \, 
B_{i [{\bf q}]} \left( \bm{\theta} +
{\bm{\zeta} ({\bf X}) \over \epsilon}, \, {\bf X} \right) \,
{d^{m} \theta \over (2\pi)^{m}} \,\,\, = \,\,\, 0
\end{equation}
for the corresponding solutions of 
(\ref{BgSsystem}) - (\ref{BgUsystem}).

 Let us denote here by $\, {\cal M}^{\prime}$ the subset in the
space of parameters 
$\, ({\bf k}_{1}, \dots, {\bf k}_{d}, \, {\bf U})$, such that
$\, ({\bf k}_{1}, \dots, {\bf k}_{d}) \, \in \, {\cal M}$. Easy
to see that the set $\, {\cal M}^{\prime}$ has the full measure
in the space $\, ({\bf k}_{1}, \dots, {\bf k}_{d}, \, {\bf U})$.

 Let us recall also that the notation $\Lambda$ denotes the family
of $m$-phase solutions of (\ref{EvInSyst}) given by formula
(\ref{LambdaDef}). As before, the notation ${\hat \Lambda}$
represents the corresponding set of $2\pi$-periodic in each
$\theta^{\alpha}$ functions
$\, \bm{\Phi} (\bm{\theta} \, + \, \bm{\theta}_{0}, \, 
{\bf k}_{1}, \dots, {\bf k}_{d}, \, {\bf U})$, 
satisfying system (\ref{EvPhaseSyst}).

\vspace{0.2cm}

{\bf Theorem 4.1.}

{\it Let the family $\Lambda$ represent a regular Hamiltonian 
submanifold equipped with a minimal set of commuting integrals
$\, (I^{1}, \dots, I^{m+s})$. Let the parameter space
$\, ({\bf k}_{1}, \dots, {\bf k}_{d}, \, {\bf U})$ of the family
$\Lambda$ have a dense set 
${\cal S}^{\prime} \, \subset \, {\cal M}^{\prime}$
on which systems (\ref{BgSsystem}) - (\ref{BgUsystem}) are
resolvable on the space of smooth  \linebreak
$2\pi$-periodic in each $\theta^{\alpha}$ functions. Then the 
relations (\ref{AveragedBracket}) define a Poisson bracket on the 
space of fields $\, ({\bf S}({\bf X}), \, {\bf U}({\bf X}))$.
}

\vspace{0.2cm}

Proof.

 The skew-symmetry of the form (\ref{AveragedBracket}) easily
follows from the skew-symmetry of the bracket (\ref{EpsExtBracket}).
Let us prove the Jacobi identity for the form 
(\ref{AveragedBracket}). Without loss of generality, we will give
here the proof for the point 
${\bf G} ({\bf X}) \, = \, (\bm{\zeta} ({\bf X}), 
{\bf U} ({\bf X}))$ of the submanifold ${\cal K}$.

 Let us fix now the functions 
${\bf q} ({\bf X})$, ${\bf p} ({\bf X})$
and the functional ${\bf Q} (\bm{\theta}, {\bf X})$
and consider the Jacobi identity of the form
\begin{equation}
\label{JacobigJJ}
\left\{ g^{[\bm{\zeta}]}_{[{\bf Q}]} \, , \, 
\left\{ G^{[\bm{\zeta}]}_{[{\bf q}]} \, , \, 
G^{[\bm{\zeta}]}_{[{\bf p}]} \right\} \right\} \, + \,
\left\{ G^{[\bm{\zeta}]}_{[{\bf p}]} \, , \, 
\left\{g^{[\bm{\zeta}]}_{[{\bf Q}]} \, , \, 
G^{[\bm{\zeta}]}_{[{\bf q}]}
\right\} \right\} \, + \,
\left\{ G^{[\bm{\zeta}]}_{[{\bf q}]} \, , \, 
\left\{ G^{[\bm{\zeta}]}_{[{\bf p}]} \, , \, 
g^{[\bm{\zeta}]}_{[{\bf Q}]}
\right\} \right\} \,\, \equiv \,\, 0
\end{equation}

 Using expansions (\ref{StandardExp}) - (\ref{SpecialExp}) and
relations (\ref{GGVarDerphiOrd}), 
(\ref{gGOrd}) - (\ref{gGVarDerGOrd}), 
it is not difficult to see that the leading term  ($\sim \epsilon$) 
of relation (\ref{JacobigJJ}) at
$\bm{\varphi} = \bm{\Phi}_{[{\bf U}, \bm{\zeta}]}$
can be written as 
\begin{multline*}
\int \left. \left\{ g^{[\bm{\zeta}]}_{[{\bf Q}]} \, , \, 
\varphi^{k} (\bm{\theta}, {\bf X}) \right\} 
\right|_{\bm{\varphi} = \bm{\Phi}_{[{\bf U}, \bm{\zeta}]} \, (0)}
\,\, \left. {\delta \{ G^{[\bm{\zeta}]}_{[{\bf q}]} , 
G^{[\bm{\zeta}]}_{[{\bf p}]} \} 
\over \delta \varphi^{k} (\bm{\theta}, {\bf X}) } 
\right|_{\bm{\varphi} = \bm{\Phi}_{[{\bf U}, \bm{\zeta}]} \, (1)}
\,\, {d^{m} \theta \over (2\pi)^{m}} \,\, d^{d} X \,\, + \\   
+ \, \int \left. \left\{ G^{[\bm{\zeta}]}_{[{\bf p}]} \, , \,  
g^{k [\bm{\zeta}]} (\bm{\theta}, {\bf X}) \right\} 
\right|_{\bm{\varphi} = \bm{\Phi}_{[{\bf U}, \bm{\zeta}]} \, (1)} 
\,\, \left. {\delta \{ g^{[\bm{\zeta}]}_{[{\bf Q}]} , 
G^{[\bm{\zeta}]}_{[{\bf q}]} \} 
\over  \delta \varphi^{k} (\bm{\theta}, {\bf X}) } 
\right|_{\bm{\varphi} = \bm{\Phi}_{[{\bf U}, \bm{\zeta}]} \, (0)}
\,\, {d^{m} \theta \over (2\pi)^{m}} \,\, d^{d} X \,\, - \\
- \, \int \left. \left\{ G^{[\bm{\zeta}]}_{[{\bf q}]} \, , \,
g^{k [\bm{\zeta}]} (\bm{\theta}, {\bf X}) \right\} 
\right|_{\bm{\varphi} = \bm{\Phi}_{[{\bf U}, \bm{\zeta}]} \, (1)}
\,\, \left. {\delta \{ g^{[\bm{\zeta}]}_{[{\bf Q}]} , 
G^{[\bm{\zeta}]}_{[{\bf p}]} \} 
\over \delta \varphi^{k} (\bm{\theta}, {\bf X}) } 
\right|_{\bm{\varphi} = \bm{\Phi}_{[{\bf U}, \bm{\zeta}]} \, (0)} 
\,\, {d^{m} \theta \over (2\pi)^{m}} \,\, d^{d} X 
\,\, \equiv \,\, 0
\end{multline*}

 We can write again the above identity in a stronger form.
Thus, making again the change
$\, {\tilde Q}_{i}(\bm{\theta}, {\bf X}) \rightarrow
{\tilde Q}_{i}(\bm{\theta}, {\bf X}) \, \mu_{i}({\bf X})$,
we can change again the functionals 
$\, Q_{i}(\bm{\theta}, {\bf X})$:
$Q_{i}(\bm{\theta}, {\bf X}) \rightarrow
Q_{i}(\bm{\theta}, {\bf X}) \, \mu_{i}({\bf X})$
by arbitrary smooth multipliers $\mu_{i}({\bf X})$.
Easy to see also that the functions $\mu_{i}({\bf X})$ appear then
just as simple local multipliers in the integrands in the zero
order of $\epsilon$. By the arbitrariness of $\mu_{i}({\bf X})$,     
we can then write again for every ${\bf X}$:
\begin{multline*}
\int_{0}^{2\pi}\!\!\!\!\!\!\dots \! \int_{0}^{2\pi}
\left. \left\{ g^{[\bm{\zeta}]}_{[{\bf Q}]} \, , \, 
\varphi^{k} (\bm{\theta}, {\bf X}) \right\} 
\right|_{\bm{\varphi} = \bm{\Phi}_{[{\bf U}, \bm{\zeta}]} \, (0)}
\,\, \left.{\delta \{ G^{[\bm{\zeta}]}_{[{\bf q}]} , 
G^{[\bm{\zeta}]}_{[{\bf p}]} \} \over
\delta \varphi^{k} (\bm{\theta}, {\bf X}) } 
\right|_{\bm{\varphi} = \bm{\Phi}_{[{\bf U}, \bm{\zeta}]} \, (1)}
\,\, {d^{m} \theta \over (2\pi)^{m}} \,\, + \\
+ \, \int_{0}^{2\pi}\!\!\!\!\!\! \dots \! \int_{0}^{2\pi}
\left. \left\{ G^{[\bm{\zeta}]}_{[{\bf p}]} \, , \,
g^{k [\bm{\zeta}]} (\bm{\theta}, {\bf X}) \right\} 
\right|_{\bm{\varphi} = \bm{\Phi}_{[{\bf U}, \bm{\zeta}]} \, (1)}
\,\, \left. {\delta \{ g^{[\bm{\zeta}]}_{[{\bf Q}]} , 
G^{[\bm{\zeta}]}_{[{\bf q}]} \} 
\over \delta \varphi^{k} (\bm{\theta}, {\bf X}) } 
\right|_{\bm{\varphi} = \bm{\Phi}_{[{\bf U}, \bm{\zeta}]} \, (0)}
\,\, {d^{m} \theta \over (2\pi)^{m}} \,\, - \\
- \, \int_{0}^{2\pi}\!\!\!\!\!\!\dots \! \int_{0}^{2\pi}
\left. \left\{ G^{[\bm{\zeta}]}_{[{\bf q}]} \, , \,
g^{k [\bm{\zeta}]} (\bm{\theta}, {\bf X}) \right\} 
\right|_{\bm{\varphi} = \bm{\Phi}_{[{\bf U}, \bm{\zeta}]} \, (1)}
\,\, \left. {\delta \{ g^{[\bm{\zeta}]}_{[{\bf Q}]} , 
G^{[\bm{\zeta}]}_{[{\bf p}]} \} 
\over \delta \varphi^{k} (\bm{\theta}, {\bf X}) } 
\right|_{\bm{\varphi} = \bm{\Phi}_{[{\bf U}, \bm{\zeta}]} \, (0)}
\,\, {d^{m} \theta \over (2\pi)^{m}} \,\, \equiv \,\, 0
\end{multline*}

 Using relations (\ref{OgranQ}) we can write
$$\left. \left\{ g^{[\bm{\zeta}]}_{[{\bf Q}]} \, , \,  
\varphi^{k} (\bm{\theta}, {\bf X}) \right\}
\right|_{\bm{\varphi} = \bm{\Phi}_{[{\bf U}, \bm{\zeta}]}}
\,\,\, =  $$
$$= \, \int Q_{j} \left( \bm{\theta}^{\prime} +
{\bm{\zeta} ({\bf W}) \over \epsilon}, \, {\bf W} \right) \,
\left. \left\{ g^{j [\bm{\zeta}]} (\bm{\theta}^{\prime}, {\bf W}) 
\, , \, \varphi^{k} (\bm{\theta}, {\bf X}) \right\}
\right|_{\bm{\varphi} = \bm{\Phi}_{[{\bf U}, \bm{\zeta}]}}
\,\, {d^{m} \theta^{\prime} \over (2\pi)^{m}} \, d^{d} W 
\,\,\, =  $$
$$= \, \int Q_{j} \left( \bm{\theta}^{\prime} +
{\bm{\zeta} ({\bf W}) \over \epsilon}, \, {\bf W} \right) \,
\left. \left\{ \varphi^{j} (\bm{\theta}^{\prime}, {\bf W})
\, , \, \varphi^{k} (\bm{\theta}, {\bf X}) \right\}
\right|_{\bm{\varphi} = \bm{\Phi}_{[{\bf U}, \bm{\zeta}]}}
\,\, {d^{m} \theta^{\prime} \over (2\pi)^{m}} \, d^{d} W
\,\,\, -  $$
\begin{multline*}
- \, \int Q_{j} \left( \bm{\theta}^{\prime} +
{\bm{\zeta} ({\bf W}) \over \epsilon}, \, {\bf W} \right) \,
\Phi^{j}_{k^{\alpha}_{q}} \left( \bm{\theta}^{\prime} +
{\bm{\zeta} ({\bf W}) \over \epsilon}, \, {\bf W} \right) 
\,\, \times  \\
\times  \,\, \left. \left\{ S^{\alpha [\bm{\zeta}]}_{W^{q}} 
\, , \, \varphi^{k} (\bm{\theta}, {\bf X}) \right\}
\right|_{\bm{\varphi} = \bm{\Phi}_{[{\bf U}, \bm{\zeta}]}}
\,\, {d^{m} \theta^{\prime} \over (2\pi)^{m}} \, d^{d} W
\,\,\, - 
\end{multline*}
\begin{multline*}  
- \, \int Q_{j} \left( \bm{\theta}^{\prime} +
{\bm{\zeta} ({\bf W}) \over \epsilon}, \, {\bf W} \right) \,
\Phi^{j}_{U^{\gamma}} \left( \bm{\theta}^{\prime} +
{\bm{\zeta} ({\bf W}) \over \epsilon}, \, {\bf W} \right) 
\,\, \times  \\
\times  \,\, \left. \left\{ U^{\gamma [\bm{\zeta}]} ({\bf W}) 
\, , \, \varphi^{k} (\bm{\theta}, {\bf X}) \right\}
\right|_{\bm{\varphi} = \bm{\Phi}_{[{\bf U}, \bm{\zeta}]}}
\,\, {d^{m} \theta^{\prime} \over (2\pi)^{m}} \, d^{d} W 
\end{multline*}

 Thus, we have in the main order of $\epsilon$:
\begin{multline*}
\left. \left\{ g^{[\bm{\zeta}]}_{[{\bf Q}]} \, , \,
\varphi^{k} (\bm{\theta}, {\bf X}) \right\}
\right|_{\bm{\varphi} = \bm{\Phi}_{[{\bf U}, \bm{\zeta}]} \, (0)}
\,\,\,\,\, =  \,\,\, - \, 
{\hat B}^{kj}_{[0] [\bm{\zeta}]} ({\bf X}) \,\,
Q_{j} \left( \bm{\theta} + {\bm{\zeta} ({\bf X}) \over \epsilon}, 
\, {\bf X} \right) \,\,\, -   \\
- \, \int_{0}^{2\pi}\!\!\!\!\!\!\dots \! \int_{0}^{2\pi}
Q_{j} \left( \bm{\theta}^{\prime}, \, {\bf X} \right) \,\,
\Phi^{j}_{U^{\gamma}} 
\left( \bm{\theta}^{\prime}, \, {\bf X} \right) \,\,
{d^{m} \theta^{\prime} \over (2\pi)^{m}} \,\,\,\, \cdot \,\,\,\,
\omega^{\alpha\gamma} ({\bf X}) \,\,
\Phi^{k}_{\theta^{\alpha}} \left( \bm{\theta} + 
{\bm{\zeta} ({\bf X}) \over \epsilon}, \, {\bf X} \right) 
\end{multline*}
(view the skew-symmetry of the operator
${\hat B}^{kj}_{[0] [\bm{\zeta}]} ({\bf X})$).

 Using relations (\ref{GGOrtCond}), we have then
\begin{multline*}
\int_{0}^{2\pi}\!\!\!\!\!\! \dots \! \int_{0}^{2\pi}
\left. \left\{ g^{[\bm{\zeta}]}_{[{\bf Q}]} \, , \,
\varphi^{k} (\bm{\theta}, {\bf X}) \right\}
\right|_{\bm{\varphi} = \bm{\Phi}_{[{\bf U}, \bm{\zeta}]} \, (0)}
\,\, \left.{\delta \{ G^{[\bm{\zeta}]}_{[{\bf q}]} , 
G^{[\bm{\zeta}]}_{[{\bf p}]} \} 
\over \delta \varphi^{k} (\bm{\theta}, {\bf X}) }
\right|_{\bm{\varphi} = \bm{\Phi}_{[{\bf U}, \bm{\zeta}]} \, (1)}
\,\, {d^{m} \theta \over (2\pi)^{m}} \,\, =   \\
= \, -  \int_{0}^{2\pi}\!\!\!\!\!\! \dots \! \int_{0}^{2\pi}
\left[ {\hat B}^{kj}_{[0] [\bm{\zeta}]} ({\bf X}) \,\,
Q_{j} \! \left( \bm{\theta} +
{\bm{\zeta} ({\bf X}) \over \epsilon}, \, {\bf X} \right) \right]
\, \left.{\delta \{ G^{[\bm{\zeta}]}_{[{\bf q}]} , 
G^{[\bm{\zeta}]}_{[{\bf p}]} \} 
\over \delta \varphi^{k} (\bm{\theta}, {\bf X}) }
\right|_{\bm{\varphi} = \bm{\Phi}_{[{\bf U}, \bm{\zeta}]} \, (1)}
\,\, {d^{m} \theta \over (2\pi)^{m}}
\end{multline*}

 Now assume that 
$(\bm{\zeta}_{\bf X}, {\bf U}({\bf X})) \in {\cal S}^{\prime}$,
so we can write the relations
\begin{equation}
\label{ABBPredstav}
\left. \left\{ G^{[\bm{\zeta}]}_{[{\bf q}]} \, , \,
g^{k [\bm{\zeta}]} (\bm{\theta}, {\bf X}) \right\} 
\right|_{\bm{\varphi} = \bm{\Phi}_{[{\bf U}, \bm{\zeta}]} \, (1)}
\,\,\, = \,\,\, {\hat B}^{kj}_{[0] [\bm{\zeta}]} ({\bf X}) \,\,
B_{j[{\bf q}]} \left( \bm{\theta} +
{\bm{\zeta} ({\bf X}) \over \epsilon}, \, {\bf X} \right) 
\end{equation}
where $B_{j[{\bf q}]} (\bm{\theta}, {\bf X})$ represent some
smooth $2\pi$-periodic in each $\theta^{\alpha}$ functions of the
variables $\bm{\theta}$.

 We can write then at the corresponding 
$\, {\bf X} \in \mathbb{R}^{d}$: 
\begin{multline}
\label{TozhdgQJqJp}
\int_{0}^{2\pi}\!\!\!\!\!\!\dots \! \int_{0}^{2\pi} \!
\left[ {\hat B}^{kj}_{[0] [\bm{\zeta}]} ({\bf X}) \,\,
Q_{j} \! \left( \bm{\theta} +
{\bm{\zeta} ({\bf X}) \over \epsilon}, \, {\bf X} \right) \right]
\left.{\delta \{ G^{[\bm{\zeta}]}_{[{\bf q}]} , 
G^{[\bm{\zeta}]}_{[{\bf p}]} \} 
\over \delta \varphi^{k} (\bm{\theta}, {\bf X}) }
\right|_{\bm{\varphi} = \bm{\Phi}_{[{\bf U}, \bm{\zeta}]} \, (1)}
\,\, {d^{m} \theta \over (2\pi)^{m}} \,\, -   \\
- \, \int_{0}^{2\pi}\!\!\!\!\!\!\dots \! \int_{0}^{2\pi} \!
\left[ {\hat B}^{kj}_{[0] [\bm{\zeta}]} ({\bf X}) \,\,
B_{j[{\bf p}]} \! \left( \bm{\theta} +
{\bm{\zeta} ({\bf X}) \over \epsilon}, \, {\bf X} \right)
\right] \, \left.
{\delta \{ g^{[\bm{\zeta}]}_{[{\bf Q}]} , 
G^{[\bm{\zeta}]}_{[{\bf q}]} \} 
\over \delta \varphi^{k} (\bm{\theta}, {\bf X}) } 
\right|_{\bm{\varphi} = \bm{\Phi}_{[{\bf U}, \bm{\zeta}]} \, (0)}
\, {d^{m} \theta \over (2\pi)^{m}} \,\, +   \\
+ \, \int_{0}^{2\pi}\!\!\!\!\!\!\dots \! \int_{0}^{2\pi} \!
\left[ {\hat B}^{kj}_{[0] [\bm{\zeta}]} ({\bf X}) \,\,
B_{j[{\bf q}]} \! \left( \bm{\theta} +
{\bm{\zeta} ({\bf X}) \over \epsilon}, \, {\bf X} \right)
\right] \, \left.
{\delta \{ g^{[\bm{\zeta}]}_{[{\bf Q}]} , 
G^{[\bm{\zeta}]}_{[{\bf p}]} \} 
\over \delta \varphi^{k} (\bm{\theta}, {\bf X}) } 
\right|_{\bm{\varphi} = \bm{\Phi}_{[{\bf U}, \bm{\zeta}]} \, (0)}
\, {d^{m} \theta \over (2\pi)^{m}} \,\, \equiv \,\, 0 
\end{multline}

 Let us consider now the Jacobi identity of the form
\begin{equation}
\label{JacobiggJ}
\left\{ g^{[\bm{\zeta}]}_{[{\bf P}]} \, , \, 
\left\{ g^{[\bm{\zeta}]}_{[{\bf Q}]} \, , \,
G^{[\bm{\zeta}]}_{[{\bf q}]} \right\} \right\} \, + \,
\left\{ g^{[\bm{\zeta}]}_{[{\bf Q}]} \, , \, 
\left\{  G^{[\bm{\zeta}]}_{[{\bf q}]} \, , \,
g^{[\bm{\zeta}]}_{[{\bf P}]} \right\} \right\} \, + \,
\left\{  G^{[\bm{\zeta}]}_{[{\bf q}]} \, , \, 
\left\{ g^{[\bm{\zeta}]}_{[{\bf P}]} \, , \,
g^{[\bm{\zeta}]}_{[{\bf Q}]} \right\} \right\} 
\,\, \equiv \,\, 0
\end{equation}
where the functionals
$\, {\bf P}(\bm{\theta}, {\bf X})$ and 
$\, {\bf Q}(\bm{\theta}, {\bf X})$
are defined with the aid of arbitrary functions
$\, {\tilde {\bf P}}(\bm{\theta}, {\bf X})$,
$\, {\tilde {\bf Q}}(\bm{\theta}, {\bf X})$.

 According to relations (\ref{varphiUzetabr}), 
(\ref{varphiSzetabr}) and (\ref{ggOrtCond}), we can write then
in the main order of $\epsilon$:
\begin{multline*}
\int \left. \left\{ g^{[\bm{\zeta}]}_{[{\bf P}]} 
\, , \, \varphi^{k} (\bm{\theta}, {\bf X})
\right\} 
\right|_{\bm{\varphi} = \bm{\Phi}_{[{\bf U}, \bm{\zeta}]} \, (0)}
\,\, \left. {\delta \{ g^{[\bm{\zeta}]}_{[{\bf Q}]} \, , \, 
G^{[\bm{\zeta}]}_{[{\bf q}]} \} 
\over \delta \varphi^{k} (\bm{\theta}, {\bf X}) } 
\right|_{\bm{\varphi} = \bm{\Phi}_{[{\bf U}, \bm{\zeta}]} \, (0)}
\,\, {d^{m} \theta \over (2\pi)^{m}} \, d^{d} X \,\,\, -   \\
- \, \int \left. \left\{ g^{[\bm{\zeta}]}_{[{\bf Q}]} \, , \,
\varphi^{k} (\bm{\theta}, {\bf X}) \right\} 
\right|_{\bm{\varphi} = \bm{\Phi}_{[{\bf U}, \bm{\zeta}]} \, (0)}
\,\, \left. {\delta \{ g^{[\bm{\zeta}]}_{[{\bf P}]} \, , \, 
G^{[\bm{\zeta}]}_{[{\bf q}]} \} \over
\delta \varphi^{k} (\bm{\theta}, {\bf X}) } 
\right|_{\bm{\varphi} = \bm{\Phi}_{[{\bf U}, \bm{\zeta}]} \, (0)}
\,\, {d^{m} \theta \over (2\pi)^{m}} \, d^{d} X 
\,\,\, \equiv \,\,\, 0  
\end{multline*}

 Having the functions $q_{\nu}({\bf X})$ just as simple local
factors in the integrands, we can write again the above relation
in the stronger form:
\begin{multline*}
\int_{0}^{2\pi}\!\!\!\!\!\! \dots \! \int_{0}^{2\pi} 
\left. \left\{ g^{[\bm{\zeta}]}_{[{\bf P}]}
\, , \, \varphi^{k} (\bm{\theta}, {\bf X})
\right\}
\right|_{\bm{\varphi} = \bm{\Phi}_{[{\bf U}, \bm{\zeta}]} \, (0)}
\,\, \left. {\delta \{ g^{[\bm{\zeta}]}_{[{\bf Q}]} \, , \,
G^{[\bm{\zeta}]}_{[{\bf q}]} \}
\over \delta \varphi^{k} (\bm{\theta}, {\bf X}) }
\right|_{\bm{\varphi} = \bm{\Phi}_{[{\bf U}, \bm{\zeta}]} \, (0)}
\,\, {d^{m} \theta \over (2\pi)^{m}} \,\,\, -   \\
- \, \int_{0}^{2\pi}\!\!\!\!\!\! \dots \! \int_{0}^{2\pi}
\left. \left\{ g^{[\bm{\zeta}]}_{[{\bf Q}]} \, , \,
\varphi^{k} (\bm{\theta}, {\bf X}) \right\}  
\right|_{\bm{\varphi} = \bm{\Phi}_{[{\bf U}, \bm{\zeta}]} \, (0)}
\,\, \left. {\delta \{ g^{[\bm{\zeta}]}_{[{\bf P}]} \, , \,
G^{[\bm{\zeta}]}_{[{\bf q}]} \} \over
\delta \varphi^{k} (\bm{\theta}, {\bf X}) }   
\right|_{\bm{\varphi} = \bm{\Phi}_{[{\bf U}, \bm{\zeta}]} \, (0)}
\,\, {d^{m} \theta \over (2\pi)^{m}} \,\,\, \equiv \,\,\, 0
\end{multline*}

 Using the same calculations as before with the relations
(\ref{gGOrtCond}), we can rewrite the above relations in
the form:
\begin{multline*}
\int_{0}^{2\pi}\!\!\!\!\!\!\dots \! \int_{0}^{2\pi} \!
\left[ {\hat B}^{kj}_{[0] [\bm{\zeta}]} ({\bf X}) \,\,
P_{j} \! \left( \bm{\theta} +
{\bm{\zeta} ({\bf X}) \over \epsilon}, \, {\bf X} \right) \right]
\,\, \left. {\delta \{ g^{[\bm{\zeta}]}_{[{\bf Q}]} \, , \,
G^{[\bm{\zeta}]}_{[{\bf q}]} \}
\over \delta \varphi^{k} (\bm{\theta}, {\bf X}) }
\right|_{\bm{\varphi} = \bm{\Phi}_{[{\bf U}, \bm{\zeta}]} \, (0)}
\,\, {d^{m} \theta \over (2\pi)^{m}} \,\,\, -   \\
- \, \int_{0}^{2\pi}\!\!\!\!\!\!\dots \! \int_{0}^{2\pi} \!
\left[ {\hat B}^{kj}_{[0] [\bm{\zeta}]} ({\bf X}) \,\,
Q_{j} \! \left( \bm{\theta} +
{\bm{\zeta} ({\bf X}) \over \epsilon}, \, {\bf X} \right) \right]
\,\, \left. {\delta \{ g^{[\bm{\zeta}]}_{[{\bf P}]} \, , \,
G^{[\bm{\zeta}]}_{[{\bf q}]} \}
\over \delta \varphi^{k} (\bm{\theta}, {\bf X}) }
\right|_{\bm{\varphi} = \bm{\Phi}_{[{\bf U}, \bm{\zeta}]} \, (0)}
\,\, {d^{m} \theta \over (2\pi)^{m}} \,\,\, \equiv \,\,\, 0
\end{multline*}

 It is easy to see that the values
$$\left. {\delta \{ g^{[\bm{\zeta}]}_{[{\bf Q}]} \, , \,
G^{[\bm{\zeta}]}_{[{\bf q}]} \}
\over \delta \varphi^{k} (\bm{\theta}, {\bf X}) }
\right|_{\bm{\varphi} = \bm{\Phi}_{[{\bf U}, \bm{\zeta}]} \, (0)} $$
can be represented in the form:
$$\left. {\delta \{ g^{[\bm{\zeta}]}_{[{\bf Q}]} \, , \,
G^{[\bm{\zeta}]}_{[{\bf q}]} \}
\over \delta \varphi^{k} (\bm{\theta}, {\bf X}) }
\right|_{\bm{\varphi} = \bm{\Phi}_{[{\bf U}, \bm{\zeta}]} \, (0)}
\, = \,\,\,
q_{\nu} ({\bf X}) \,\, 
{\hat K}^{\nu i}_{k [{\bf U}, \bm{\zeta}]} ({\bf X}) \,\,
Q_{i} \left( \bm{\theta} +
{\bm{\zeta} ({\bf X}) \over \epsilon}, \, {\bf X} \right) $$
where $\, {\hat K}^{\nu i}_{k [{\bf U}, \bm{\zeta}]} ({\bf X})$
is a linear operator on the space of smooth periodic functions of
$\bm{\theta}$.

 The above relations can then be finally written as:
\begin{multline*}
\int_{0}^{2\pi}\!\!\!\!\!\!\dots \! \int_{0}^{2\pi} \!
\left[ {\hat B}^{kj}_{[0] [\bm{\zeta}]} ({\bf X}) \,\,
P_{j} \! \left( \bm{\theta} +
{\bm{\zeta} ({\bf X}) \over \epsilon}, \, {\bf X} \right) \right]
\,\, \times   \\
\times \, \left[ q_{\nu} ({\bf X}) \,\,
{\hat K}^{\nu i}_{k [{\bf U}, \bm{\zeta}]} ({\bf X}) \,\,
Q_{i} \left( \bm{\theta} +
{\bm{\zeta} ({\bf X}) \over \epsilon}, \, {\bf X} \right) \right]
\,\, {d^{m} \theta \over (2\pi)^{m}} \,\,\, -   
\end{multline*}
\begin{multline}
\label{ggJRasp}
- \, \int_{0}^{2\pi}\!\!\!\!\!\!\dots \! \int_{0}^{2\pi} \!
\left[ {\hat B}^{kj}_{[0] [\bm{\zeta}]} ({\bf X}) \,\,
Q_{j} \! \left( \bm{\theta} +
{\bm{\zeta} ({\bf X}) \over \epsilon}, \, {\bf X} \right) \right]
\,\, \times   \\
\times \, \left[ q_{\nu} ({\bf X}) \,\,
{\hat K}^{\nu i}_{k [{\bf U}, \bm{\zeta}]} ({\bf X}) \,\,
P_{i} \left( \bm{\theta} +
{\bm{\zeta} ({\bf X}) \over \epsilon}, \, {\bf X} \right) \right]
\,\, {d^{m} \theta \over (2\pi)^{m}} \,\,\, \equiv \,\,\, 0
\end{multline}

 Let us note now that the values 
${\bf Q} (\bm{\theta}, {\bf X})$ and
${\bf P} (\bm{\theta}, {\bf X})$ represent here 
arbitrary smooth $2\pi$-periodic in each $\theta^{\alpha}$
functions, satisfying conditions (\ref{OgranQ}).
In particular, for
$(\bm{\zeta}_{\bf X}, {\bf U}({\bf X})) \in {\cal S}^{\prime}$
we can substitute the values
\begin{equation}
\label{PBPB}
{\bf P} (\bm{\theta}, {\bf X}) \,\, = \,\,
{\bf B}_{[{\bf p}]} (\bm{\theta}, {\bf X}) \,\,\,\,\,\,\,\,
\text{or}
\,\,\,\,\,\,\,\, {\bf P} (\bm{\theta}, {\bf X}) \,\, = \,\,
{\bf B}_{[{\bf q}]} (\bm{\theta}, {\bf X})
\end{equation}
in the identity (\ref{ggJRasp}).

 As a result, for
$(\bm{\zeta}_{\bf X}, {\bf U}({\bf X})) \in {\cal S}^{\prime}$
we can rewrite relations (\ref{TozhdgQJqJp}) in 
the form:
\begin{multline*}
\int_{0}^{2\pi}\!\! \!\!\!\! \dots \! \int_{0}^{2\pi}
{d^{m} \theta \over (2\pi)^{m}} \,
\left[ {\hat B}^{kj}_{[0][\bm{\zeta}]} ({\bf X}) \,\,
Q_{j} \left(\bm{\theta} + {\bm{\zeta}({\bf X}) \over \epsilon},
{\bf X} \right) \right] \,\, \times  \\
\times \,\, \left( \left.
{\delta \{ G^{[\bm{\zeta}]}_{[{\bf q}]} ,  
G^{[\bm{\zeta}]}_{[{\bf p}]} \} \over
\delta \varphi^{k} (\bm{\theta}, {\bf X}) } 
\right|_{\bm{\varphi} = \bm{\Phi}_{[{\bf U}, \bm{\zeta}]} \, (1)}
\, - \,\,\, q_{\nu} ({\bf X}) \,\,
{\hat K}^{\nu i}_{k [{\bf U}, \bm{\zeta}]} ({\bf X}) \,\,
B_{i [{\bf p}]} \left( \bm{\theta} +
{\bm{\zeta} ({\bf X}) \over \epsilon}, \, {\bf X} \right)
\,\, +   \right. \\
\left. + \,\,\, p_{\nu} ({\bf X}) \,\,
{\hat K}^{\nu i}_{k [{\bf U}, \bm{\zeta}]} ({\bf X}) \,\,
B_{i [{\bf q}]} \left( \bm{\theta} +
{\bm{\zeta} ({\bf X}) \over \epsilon}, \, {\bf X} \right)
\right) \,\,\, \equiv \,\,\, 0 
\end{multline*}

 Using the skew-symmetry of the operator  
${\hat B}^{kj}_{[0][\bm{\zeta}]} ({\bf X})$
we can then write under the same conditions
\begin{multline*}
\int_{0}^{2\pi}\!\! \!\!\!\! \dots \! \int_{0}^{2\pi}
{d^{m} \theta \over (2\pi)^{m}} \,\,
Q_{j} \left(\bm{\theta} + {\bm{\zeta}({\bf X}) \over \epsilon},
{\bf X} \right) \,\, \times   \\
\times \,\, {\hat B}^{kj}_{[0][\bm{\zeta}]} ({\bf X})
\left( \left.
{\delta \{ G^{[\bm{\zeta}]}_{[{\bf q}]} , 
G^{[\bm{\zeta}]}_{[{\bf p}]} \} \over
\delta \varphi^{k} (\bm{\theta}, {\bf X}) }
\right|_{\bm{\varphi} = \bm{\Phi}_{[{\bf U}, \bm{\zeta}]} \, (1)}
\, - \,\,\, q_{\nu} ({\bf X}) \,\,
{\hat K}^{\nu i}_{k [{\bf U}, \bm{\zeta}]} ({\bf X}) \,\,
B_{i [{\bf p}]} \left( \bm{\theta} +
{\bm{\zeta} ({\bf X}) \over \epsilon}, \, {\bf X} \right)
\,\, +   \right. \\
\left. + \,\,\, p_{\nu} ({\bf X}) \,\,
{\hat K}^{\nu i}_{k [{\bf U}, \bm{\zeta}]} ({\bf X}) \,\,
B_{i [{\bf q}]} \left( \bm{\theta} +
{\bm{\zeta} ({\bf X}) \over \epsilon}, \, {\bf X} \right)
\right) \,\,\, \equiv \,\,\, 0
\end{multline*}

 The values $Q_{j} (\bm{\theta}, {\bf X} )$ represent here
arbitrary smooth $2\pi$-periodic in each $\theta^{\alpha}$
functions satisfying the restriction (\ref{OgranQ}). 
At the same time the values in the brackets represent some 
smooth $2\pi$-periodic in each $\theta^{\alpha}$
functions of $\bm{\theta}$ for 
$(\bm{\zeta}_{\bf X}, {\bf U}({\bf X})) \in {\cal S}^{\prime}$.
As a corollary, we can then write for some values
$a^{\alpha}_{[{\bf q}, {\bf p}]} ({\bf X})$:
\begin{multline*}
{\hat B}^{kj}_{[0][\bm{\zeta}]} ({\bf X}) 
\left( 
{\delta \{ G^{[\bm{\zeta}]}_{[{\bf q}]}  , 
G^{[\bm{\zeta}]}_{[{\bf p}]} \} \over
\delta \varphi^{k} (\bm{\theta}, {\bf X}) }
\right|_{\bm{\varphi} = \bm{\Phi}_{[{\bf U}, \bm{\zeta}]} \, (1)}
- \,\, q_{\nu} ({\bf X}) \,
{\hat K}^{\nu i}_{k [{\bf U}, \bm{\zeta}]} ({\bf X}) \,
B_{i [{\bf p}]} \! \left( \bm{\theta} +
{\bm{\zeta} ({\bf X}) \over \epsilon}, {\bf X} \right)
\, +    \\
\left. + \,\, p_{\nu} ({\bf X}) \,
{\hat K}^{\nu i}_{k [{\bf U}, \bm{\zeta}]} ({\bf X}) \,
B_{i [{\bf q}]} \! \left( \bm{\theta} +
{\bm{\zeta} ({\bf X}) \over \epsilon}, {\bf X} \right) \!
\right) \,\,\, \equiv \,\, \sum_{\alpha=1}^{m}
a^{\alpha}_{[{\bf q}, {\bf p}]}
({\bf X}) \,\, \Phi^{j}_{\theta^{\alpha}} \!
\left( \bm{\theta} +
{\bm{\zeta} ({\bf X}) \over \epsilon}, {\bf X} \! \right)
\end{multline*}
at the corresponding $\, {\bf X} \in \mathbb{R}^{d}$.

 We can see that the values in parentheses represent again some
smooth $2\pi$-periodic in each $\theta^{\alpha}$ functions of
$\bm{\theta}$ at the same point $\, {\bf X} \in \mathbb{R}^{d}$.
At the same time, the right-hand part of the relation above
represents a linear combination of the flows, generating linear
evolution of the phase shifts $\theta_{0}^{\alpha}$ with the
coefficients $a^{\alpha}_{[{\bf q}, {\bf p}]} ({\bf X})$.
For a regular Hamiltonian submanifold $\Lambda$ with a minimal
set of commuting integrals $(I^{1}, \dots, I^{m+s})$ we can
conclude then that the covector in parentheses is given by some 
linear combination of the functions 
$\bm{\zeta}^{(\gamma)}_{[\bm{\zeta}_{\bf X}, {\bf U}({\bf X})]}
(\bm{\theta} + \bm{\zeta}({\bf X})/\epsilon)$, modulo the kernel
vectors of the operator 
${\hat B}^{kj}_{[0] [\bm{\zeta}]} ({\bf X})$. For
$(\bm{\zeta}_{X^{1}}, \dots, \bm{\zeta}_{X^{d}}) \in {\cal M}$
all the smooth in $\bm{\theta}$ kernel vectors of
${\hat B}^{kj}_{[0] [\bm{\zeta}]} ({\bf X})$ on the space of
$2\pi$-periodic in each $\theta^{\alpha}$ functions are given
by the regular kernel vectors, which are also represented by
linear combinations of
$\bm{\zeta}^{(\gamma)}_{[\bm{\zeta}_{\bf X}, {\bf U}({\bf X})]}  
(\bm{\theta} + \bm{\zeta}({\bf X})/\epsilon)$. Thus, we can 
write for
$(\bm{\zeta}_{\bf X}, {\bf U}({\bf X})) \in {\cal S}^{\prime}$:

\begin{multline}
\label{zetaRepr}
\left. {\delta \{ G^{[\bm{\zeta}]}_{[{\bf q}]}  ,
G^{[\bm{\zeta}]}_{[{\bf p}]} \} \over
\delta \varphi^{k} (\bm{\theta}, {\bf X}) }
\right|_{\bm{\varphi} = \bm{\Phi}_{[{\bf U}, \bm{\zeta}]} \, (1)}
- \,\, q_{\nu} ({\bf X}) \,
{\hat K}^{\nu i}_{k [{\bf U}, \bm{\zeta}]} ({\bf X}) \,
B_{i [{\bf p}]} \! \left( \bm{\theta} +
{\bm{\zeta} ({\bf X}) \over \epsilon}, {\bf X} \right)
\, +    \\
+ \,\, p_{\nu} ({\bf X}) \,
{\hat K}^{\nu i}_{k [{\bf U}, \bm{\zeta}]} ({\bf X}) \,  
B_{i [{\bf q}]} \! \left( \! \bm{\theta} +
{\bm{\zeta} ({\bf X}) \over \epsilon}, {\bf X} \! \right)
\,\, \equiv \, \sum_{\gamma=1}^{m+s}
b_{\gamma [{\bf q}, {\bf p}]} ({\bf X}) \,\,
\zeta^{(\gamma)}_{k [\bm{\zeta}_{\bf X}, {\bf U}({\bf X})]} \!
\left( \! \bm{\theta} + {\bm{\zeta} ({\bf X}) \over \epsilon}, 
{\bf X} \! \right)
\end{multline}
with some coefficients $b_{\gamma [{\bf q}, {\bf p}]} ({\bf X})$.

 Now, let us consider the Jacobi identity of the form
$$\left\{ G^{[\bm{\zeta}]}_{[{\bf q}]} \, , \,
\left\{ G^{[\bm{\zeta}]}_{[{\bf p}]} \, , \,
G^{[\bm{\zeta}]}_{[{\bf r}]} \right\} \right\} \, + \,
\left\{ G^{[\bm{\zeta}]}_{[{\bf p}]} \, , \,
\left\{  G^{[\bm{\zeta}]}_{[{\bf r}]} \, , \,
G^{[\bm{\zeta}]}_{[{\bf q}]} \right\} \right\} \, + \,
\left\{  G^{[\bm{\zeta}]}_{[{\bf r}]} \, , \,
\left\{ G^{[\bm{\zeta}]}_{[{\bf q}]} \, , \,
G^{[\bm{\zeta}]}_{[{\bf p}]} \right\} \right\}
\,\, \equiv \,\, 0 $$
for arbitrary smooth functions $\, {\bf q}({\bf X})$,
$\, {\bf p}({\bf X})$, $\, {\bf r}({\bf X})$.

 In the leading ($\sim \epsilon^{2}$) order of $\epsilon$ at
$\, \bm{\varphi} = \bm{\Phi}_{[{\bf U}, \bm{\zeta}]}$ the above 
identity gives the relations:
\begin{equation}
\label{GGGJacobi}
\int \left. \left\{ G^{[\bm{\zeta}]}_{[{\bf q}]} \, , \,
G^{\nu [\bm{\zeta}]} ({\bf X}) \right\}
\right|_{\bm{\varphi} = \bm{\Phi}_{[{\bf U}, \bm{\zeta}]} \, (1)}
\, \left. {\delta \{ G^{[\bm{\zeta}]}_{[{\bf p}]}  ,
G^{[\bm{\zeta}]}_{[{\bf r}]} \} \over \delta G^{\nu} ({\bf X}) }   
\right|_{\bm{\varphi} = \bm{\Phi}_{[{\bf U}, \bm{\zeta}]} \, (1)}
\, d^{d} X \,\,\,\,\, + \,\,\,\,\, c. p. \,\,\,\,\, +   
\end{equation}
$$+ \int \!\! \left. \left\{ G^{[\bm{\zeta}]}_{[{\bf q}]} \, , \, 
g^{k [\bm{\zeta}]} (\bm{\theta}, {\bf X}) \right\}
\right|_{\bm{\varphi} = \bm{\Phi}_{[{\bf U}, \bm{\zeta}]} \, (1)}
\left. {\delta \{ G^{[\bm{\zeta}]}_{[{\bf p}]} , 
G^{[\bm{\zeta}]}_{[{\bf r}]} \} \over
\delta \varphi^{k} (\bm{\theta}, {\bf X}) }   
\right|_{\bm{\varphi} = \bm{\Phi}_{[{\bf U}, \bm{\zeta}]} \, (1)}
{d^{m} \theta \over (2\pi)^{m}} \, d^{d} X
\,\,\, + \,\,\, c. p. \,\,\, \equiv \,\,\, 0 $$

 Using relations (\ref{zetarhpOrt}) and representations
(\ref{ABBPredstav}), (\ref{zetaRepr}), we can write for
$(\bm{\zeta}_{\bf X}, {\bf U}({\bf X})) \in {\cal S}^{\prime}$:
\begin{equation}
\label{gchastTozhdJ}
\int_{0}^{2\pi}\!\!\!\!\!\!\dots \! \int_{0}^{2\pi} \!\!
\left. \left\{ G^{[\bm{\zeta}]}_{[{\bf q}]} \, , \,  
g^{k [\bm{\zeta}]} (\bm{\theta}, {\bf X}) \right\}
\right|_{\bm{\varphi} = \bm{\Phi}_{[{\bf U}, \bm{\zeta}]} \, (1)}
\left. {\delta \{ G^{[\bm{\zeta}]}_{[{\bf p}]} ,
G^{[\bm{\zeta}]}_{[{\bf r}]} \} \over
\delta \varphi^{k} (\bm{\theta}, {\bf X}) }
\right|_{\bm{\varphi} = \bm{\Phi}_{[{\bf U}, \bm{\zeta}]} \, (1)}
{d^{m} \theta \over (2\pi)^{m}} \,\,\, + \,\,\, c. p. \,\,\, =  
\end{equation}
\begin{multline*}
= \, \int_{0}^{2\pi}\!\!\!\!\!\!\dots \! \int_{0}^{2\pi} 
{d^{m} \theta \over (2\pi)^{m}} \,
\left[ {\hat B}^{kj}_{[0] [\bm{\zeta}]} ({\bf X}) \,\,
B_{j[{\bf q}]} \! \left( \bm{\theta} +
{\bm{\zeta} ({\bf X}) \over \epsilon}, \, {\bf X} \right)
\right] \, \times  \\
\times \left[ p_{\nu} ({\bf X}) \,
{\hat K}^{\nu i}_{k [{\bf U}, \bm{\zeta}]} ({\bf X}) \,
B_{i [{\bf r}]} \! \left( \! \bm{\theta} +
{\bm{\zeta} ({\bf X}) \over \epsilon}, {\bf X} \! \right)
\, - \, r_{\nu} ({\bf X}) \,  
{\hat K}^{\nu i}_{k [{\bf U}, \bm{\zeta}]} ({\bf X}) \,
B_{i [{\bf p}]} \! \left( \! \bm{\theta} +
{\bm{\zeta} ({\bf X}) \over \epsilon}, {\bf X} \! \right)
\! \right] +
\end{multline*}
\begin{multline*}
+ \, \int_{0}^{2\pi}\!\!\!\!\!\!\dots \! \int_{0}^{2\pi} 
{d^{m} \theta \over (2\pi)^{m}} \,
\left[ {\hat B}^{kj}_{[0] [\bm{\zeta}]} ({\bf X}) \,\,
B_{j[{\bf p}]} \! \left( \bm{\theta} +
{\bm{\zeta} ({\bf X}) \over \epsilon}, \, {\bf X} \right)
\right] \, \times  \\
\times \left[ r_{\nu} ({\bf X}) \,
{\hat K}^{\nu i}_{k [{\bf U}, \bm{\zeta}]} ({\bf X}) \,
B_{i [{\bf q}]} \! \left( \! \bm{\theta} +
{\bm{\zeta} ({\bf X}) \over \epsilon}, {\bf X} \! \right)
\, - \, q_{\nu} ({\bf X}) \,
{\hat K}^{\nu i}_{k [{\bf U}, \bm{\zeta}]} ({\bf X}) \,
B_{i [{\bf r}]} \! \left( \! \bm{\theta} +
{\bm{\zeta} ({\bf X}) \over \epsilon}, {\bf X} \! \right)
\! \right] +
\end{multline*}
\begin{multline*}
+ \, \int_{0}^{2\pi}\!\!\!\!\!\!\dots \! \int_{0}^{2\pi}
{d^{m} \theta \over (2\pi)^{m}} \,
\left[ {\hat B}^{kj}_{[0] [\bm{\zeta}]} ({\bf X}) \,\,  
B_{j[{\bf r}]} \! \left( \bm{\theta} + 
{\bm{\zeta} ({\bf X}) \over \epsilon}, \, {\bf X} \right)
\right] \, \times  \\
\times \left[ q_{\nu} ({\bf X}) \,
{\hat K}^{\nu i}_{k [{\bf U}, \bm{\zeta}]} ({\bf X}) \,
B_{i [{\bf p}]} \! \left( \! \bm{\theta} +
{\bm{\zeta} ({\bf X}) \over \epsilon}, {\bf X} \! \right)
\, - \, p_{\nu} ({\bf X}) \,
{\hat K}^{\nu i}_{k [{\bf U}, \bm{\zeta}]} ({\bf X}) \,
B_{i [{\bf q}]} \! \left( \! \bm{\theta} +
{\bm{\zeta} ({\bf X}) \over \epsilon}, {\bf X} \! \right)
\! \right]
\end{multline*}

Making now the substitutions
$${\bf Q} (\bm{\theta}, {\bf X}) \,\, = \,\,
{\bf B}_{[{\bf r}]} (\bm{\theta}, {\bf X}) \,\,\,\,\,\,\,\, ,  
\,\,\,\,\,\,\,\, {\bf P} (\bm{\theta}, {\bf X}) \,\, = \,\,    
{\bf B}_{[{\bf p}]} (\bm{\theta}, {\bf X}) $$
in identity (\ref{ggJRasp}), we get the following identities
\begin{multline*}
\int_{0}^{2\pi}\!\!\!\!\!\!\dots \! \int_{0}^{2\pi} 
{d^{m} \theta \over (2\pi)^{m}} \, 
\left[ {\hat B}^{kj}_{[0] [\bm{\zeta}]} ({\bf X}) \,\,
B_{j [{\bf p}]} \! \left( \bm{\theta} +
{\bm{\zeta} ({\bf X}) \over \epsilon}, \, {\bf X} \right) \right]
\,\, \times   \\
\times \, \left[ q_{\nu} ({\bf X}) \,\,
{\hat K}^{\nu i}_{k [{\bf U}, \bm{\zeta}]} ({\bf X}) \,\,
B_{i [{\bf r}]} \left( \bm{\theta} +
{\bm{\zeta} ({\bf X}) \over \epsilon}, \, {\bf X} \right) \right]
\,\,\, -
\end{multline*}
\begin{multline*}
- \, \int_{0}^{2\pi}\!\!\!\!\!\!\dots \! \int_{0}^{2\pi} 
{d^{m} \theta \over (2\pi)^{m}} \,
\left[ {\hat B}^{kj}_{[0] [\bm{\zeta}]} ({\bf X}) \,\,
B_{j [{\bf r}]} \! \left( \bm{\theta} +
{\bm{\zeta} ({\bf X}) \over \epsilon}, \, {\bf X} \right) \right]
\,\, \times   \\
\times \, \left[ q_{\nu} ({\bf X}) \,\,
{\hat K}^{\nu i}_{k [{\bf U}, \bm{\zeta}]} ({\bf X}) \,\,
B_{i [{\bf p}]} \left( \bm{\theta} +
{\bm{\zeta} ({\bf X}) \over \epsilon}, \, {\bf X} \right) \right]
\,\,\,\,\, \equiv \,\,\,\,\, 0
\end{multline*}
for
$\, (\bm{\zeta}_{\bf X}, {\bf U}({\bf X})) \in {\cal S}^{\prime}$.

  Making the cyclic permutations of the functions 
$\, {\bf q} ({\bf X})$, $\, {\bf p}({\bf X})$, 
and $\, {\bf r}({\bf X})$ in the above identity,
we can see then that the right-hand part of the equality    
(\ref{gchastTozhdJ}) is identically equal to zero for all values
$\, (\bm{\zeta}_{\bf X}, {\bf U}({\bf X})) \in {\cal S}^{\prime}$.
At the same time we can see that
the left-hand side of relation (\ref{gchastTozhdJ})   
represents a smooth regular function of the values
$\, (\bm{\zeta}_{\bf X}, {\bf U}({\bf X}))$ and their derivatives.
From the fact that the set ${\cal S}^{\prime}$ represents an
everywhere dense set in the parameter space, we can then 
conclude that the left-hand side of the relation
(\ref{gchastTozhdJ}) is identically equal to zero under the
conditions of the theorem.

 From the identity (\ref{GGGJacobi}) we get then the relations:
$$\int \left. \left\{ G^{[\bm{\zeta}]}_{[{\bf q}]} \, , \,
G^{\nu [\bm{\zeta}]} ({\bf X}) \right\}
\right|_{\bm{\varphi} = \bm{\Phi}_{[{\bf U}, \bm{\zeta}]} \, (1)}
\, \left. {\delta \{ G^{[\bm{\zeta}]}_{[{\bf p}]}  ,
G^{[\bm{\zeta}]}_{[{\bf r}]} \} \over \delta G^{\nu} ({\bf X}) }
\right|_{\bm{\varphi} = \bm{\Phi}_{[{\bf U}, \bm{\zeta}]} \, (1)}
\, d^{d} X \,\,\,\,\, + \,\,\,\,\, c. p. 
\,\,\,\,\, \equiv \,\,\,\,\, 0 $$

 Using relations (\ref{AvBrRel1}) - (\ref{AvBrRel2}) we get now
the Jacobi identity for the bracket 
$\{ \dots \, , \, \dots \}_{\rm AV}$ at the ``point''
$\, (\bm{\zeta} ({\bf X}), {\bf U}({\bf X}))$ of the
submanifold ${\cal K}$.

{\hfill Theorem 4.1 is proved.}

\vspace{0.2cm}

 According to our remarks above, we can formulate here the following 
theorem about the single-phase case:

\vspace{0.2cm}

{\bf Theorem 4.1$^{\prime}$.}

{\it Let the family $\Lambda$ of single-phase solutions of
(\ref{EvInSyst}) represent a regular Hamiltonian submanifold in the
space of periodic functions equipped with a minimal set of
commuting integrals $\, (I^{1}, \dots, I^{s+1})$. Then the form
$$ \begin{array}{c}
\left\{ S ({\bf X}) \, , \, S ({\bf Y}) \right\}
\, = \, 0 \,\,\, ,    \\ \\
\left\{ S ({\bf X}) \, , \, U^{\gamma} ({\bf Y}) \right\} \,\, = \,\,
\omega^{\gamma} \, \left( S_{\bf X}, \,
U^{1}({\bf X}), \dots, U^{s+1}({\bf X}) \right)
\,\, \delta ({\bf X} - {\bf Y}) \,\, ,  \\   \\
\left\{ U^{\gamma} ({\bf X})\, , \, U^{\rho} ({\bf Y}) \right\}
\,\,\, = \,\,\, \langle A^{\gamma\rho}_{10\dots0} \rangle
\left( S_{\bf X}, \,
U^{1}({\bf X}), \dots, U^{s+1}({\bf X}) \right) \,\,
\delta_{X^{1}} ({\bf X} - {\bf Y}) \,\,\, + \,\, \dots \,\, +  \\ \\  
+ \,\,\, \langle A^{\gamma\rho}_{0\dots01} \rangle
\left( S_{\bf X}, \,
U^{1}({\bf X}), \dots, U^{s+1}({\bf X}) \right) \,\,\,
\delta_{X^{d}} ({\bf X} - {\bf Y}) \,\,\, +  \\   \\
+ \,\,\, \left[ \langle Q^{\gamma\rho \, p} \rangle
\left( S_{\bf X}, \,
U^{1}({\bf X}), \dots, U^{s+1}({\bf X}) \right)
\right]_{X^{p}} \,\,\, \delta ({\bf X} - {\bf Y})
\,\,\,\,\,\,\,\, , \,\,\,\,\,\,\,\,\,\,\,\,\,\,\,
\gamma, \rho \, = \, 1, \dots , s + 1
\end{array}  $$
defines a Poisson bracket on the space of fields
$\, (S ({\bf X}) , \, U^{\gamma} ({\bf X}))$,
$\gamma \, = \, 1, \dots, s + 1$.
}

\vspace{0.2cm}

 The following theorem proves the invariance of the (contravariant)
2-form (\ref{AveragedBracket}) with respect to the choice of the
functionals $\, (I^{1}, \dots, I^{m+s})$.

\vspace{0.2cm}

{\bf Theorem 4.2.}

{\it Let the family $\Lambda$ represent a regular Hamiltonian 
submanifold equipped with a minimal set of commuting integrals  
$\, (I^{1}, \dots, I^{m+s})$. Let the set
$\, (I^{\prime 1}, \dots, I^{\prime m+s})$ represent another set
of commuting integrals, satisfying all the conditions of
Definition 2.2. Then the forms (\ref{AveragedBracket}), obtained
with the aid of the sets $\, (I^{1}, \dots, I^{m+s})$ and
$\, (I^{\prime 1}, \dots, I^{\prime m+s})$ coincide with each other.
}

\vspace{0.2cm}

Proof.

Let us consider two different sets of parameters on $\Lambda$:
$$\left( {\bf k}_{1}, \dots, {\bf k}_{d}, \, U^{1}, \dots,
U^{m+s} \right) \,\,\, , \,\,\,\,\,
\left( {\bf k}_{1}, \dots, {\bf k}_{d}, \, U^{\prime 1}, \dots,
U^{\prime m+s} \right) $$
where $\, U^{\gamma} \, \equiv \, \langle P^{\gamma} \rangle$,
$\, U^{\prime \gamma} \, \equiv \, \langle P^{\prime \gamma} \rangle$.
We can then write on $\Lambda$:
$$U^{\prime \gamma} \,\,\, = \,\,\, U^{\prime \gamma}
\left( {\bf k}_{1}, \dots, {\bf k}_{d}, \, {\bf U} \right) $$
since the values 
$( {\bf k}_{1}, \dots, {\bf k}_{d}, \, {\bf U} )$ give the full set
of parameters on $\Lambda$ excluding the initial phase shifts.

 We have to prove here that the brackets (\ref{AveragedBracket})
$\, \{ \dots , \dots \}_{\rm AV}$ and
$\, \{ \dots , \dots \}^{\prime}_{\rm AV}$, obtained with the aid
of the sets $\, (I^{1}, \dots, I^{m+s})$ and
$\, (I^{\prime 1}, \dots, I^{\prime m+s})$, transform into each other
under the corresponding transformation
$$U^{\prime \gamma} ({\bf X}) \,\,\, = \,\,\, U^{\prime \gamma}
\left( {\bf S}_{X^{1}}, \dots, {\bf S}_{X^{d}}, 
\, {\bf U} ({\bf X}) \right) $$

 We have by definition
$$\left\{ S^{\alpha} ({\bf X}) \, , \, S^{\beta} ({\bf Y}) 
\right\}_{\rm AV} \,\, = \,\, 0 \,\,\,\,\, , \,\,\,\,\,\,\,\,
\left\{ S^{\alpha} ({\bf X}) \, , \, S^{\beta} ({\bf Y})
\right\}^{\prime}_{\rm AV} \,\, = \,\, 0 $$
in both the brackets $\, \{ \dots , \dots \}_{\rm AV}$ and
$\, \{ \dots , \dots \}^{\prime}_{\rm AV}$. Let us consider now
the brackets, containing the functionals ${\bf U} ({\bf X})$ and
${\bf U}^{\prime} ({\bf X})$.

 We can write on $\Lambda$:
\begin{equation}
\label{UPJRel}
U^{\gamma} \,\,\, \equiv \,\,\, \langle P^{\gamma} \rangle
\,\,\, \equiv \,\,\, \left. J^{\gamma} \right|_{\Lambda}
\,\,\,\,\, , \,\,\,\,\,\,\,\,
U^{\prime \gamma} \,\,\, \equiv \,\,\, 
\langle P^{\prime \gamma} \rangle
\,\,\, \equiv \,\,\, \left. J^{\prime \gamma} \right|_{\Lambda}
\end{equation}
where the functionals $J^{\gamma}$ and $J^{\prime \gamma}$ are defined
by formula (\ref{Jnu}).

 By Definition 2.2, both the sets $\, (J^{1}, \dots, J^{m+s})$, 
$\, (J^{\prime 1}, \dots, J^{\prime m+s})$ generate
the linear shifts of $\bm{\theta}_{0}$ on $\Lambda$ according
to bracket (\ref{PBronTorus}), such that we have
$${\rm rk} \,\, || \omega^{\alpha \gamma} ||
\,\, = \,\, m  \,\,\,\,\, , \,\,\,\,\,\,\,\,\,\,
{\rm rk} \,\, || \omega^{\prime \alpha \gamma} ||  
\,\, = \,\, m $$
for the corresponding sets $\, \{ \bm{\omega}^{\gamma} \}$, 
$\, \{ \bm{\omega}^{\prime \gamma} \}$.

 Besides that, the variation derivatives
$$\zeta^{\gamma}_{i} \left( \bm{\theta} + \bm{\theta}_{0}, \,
{\bf k}_{1}, \dots, {\bf k}_{d}, \, {\bf U} \right) \,\,\, = \,\,\,
\left. {\delta J^{\gamma} \over \delta \varphi^{i} (\bm{\theta})}
\right|_{\hat \Lambda} \,\,\,\,\, , \,\,\,\,\,\,\,\,\,\,
\zeta^{\prime \gamma}_{i} \left( \bm{\theta} + \bm{\theta}_{0}, \,
{\bf k}_{1}, \dots, {\bf k}_{d}, \, {\bf U} \right) \,\,\, = \,\,\,
\left. 
{\delta J^{\prime \gamma} \over \delta \varphi^{i} (\bm{\theta})}
\right|_{\hat \Lambda} $$
represent regular covectors on ${\hat \Lambda}$, such that both
the linear spans
$${\rm Span} \, \left\{ \bm{\zeta}^{1}, \dots, \bm{\zeta}^{m+s}
\right\} \,\,\,\,\, , \,\,\,\,\,\,\,\,
{\rm Span} \, \left\{ \bm{\zeta}^{\prime 1}, \dots, 
\bm{\zeta}^{\prime m+s} \right\} $$
contain all the regular kernel vectors of the Hamiltonian operators
(\ref{OperatorB}). As then follows from Definition 2.1, the linear 
spans of the sets 
$\{ \bm{\zeta}^{1}, \dots, \bm{\zeta}^{m+s} \}$ and
$\{ \bm{\zeta}^{\prime 1}, \dots, \bm{\zeta}^{\prime m+s} \}$ 
coincide at every point of the submanifold ${\hat \Lambda}$.
According to (\ref{UPJRel}), we can then write 
$$\left.
{\delta J^{\prime \gamma} \over \delta \varphi^{i} (\bm{\theta})}
\right|_{\hat \Lambda_{{\bf k}_{1}, \dots, {\bf k}_{d}}}
\,\,\, = \,\,\,\,\, 
{\partial U^{\prime \gamma} \over \partial U^{\rho}}
\left( {\bf k}_{1}, \dots, {\bf k}_{d}, {\bf U} \right) \,\,\,
\left. {\delta J^{\rho} \over \delta \varphi^{i} (\bm{\theta})}
\right|_{\hat \Lambda_{{\bf k}_{1}, \dots, {\bf k}_{d}}} $$
for any submanifold $\hat \Lambda_{{\bf k}_{1}, \dots, {\bf k}_{d}}$.

 From the relations above we also immediately get the relations
$$\omega^{\prime \alpha \gamma}
\left( {\bf k}_{1}, \dots, {\bf k}_{d}, {\bf U} \right) 
\,\,\,\,\, = \,\,\,\,\, 
{\partial U^{\prime \gamma} \over \partial U^{\rho}}
\,\,\, \omega^{\alpha \rho}
\left( {\bf k}_{1}, \dots, {\bf k}_{d}, {\bf U} \right) $$
for the corresponding frequencies $\, \{ \bm{\omega}^{\gamma} \}$
and $\, \{ \bm{\omega}^{\prime \gamma} \}$.

 Using the relations
$$\delta U^{\prime \gamma} ({\bf Y}) \,\,\,\,\, = \,\,\,\,\,
{\partial U^{\prime \gamma} \over \partial U^{\rho}} ({\bf Y}) 
\,\,\, \delta U^{\rho} ({\bf Y}) \,\,\,\,\, + \,\,\,\,\,
{\partial U^{\prime \gamma} \over \partial k^{\alpha}_{q}} ({\bf Y}) 
\,\,\, \delta S^{\alpha}_{Y^{q}} $$
and relations (\ref{AveragedBracket}) we get then:
\begin{multline*}
\left\{ S^{\alpha} ({\bf X}) \, , \,
U^{\prime \gamma} ({\bf Y}) \right\}_{\rm AV} \,\,\, = \,\,\,
{\partial U^{\prime \gamma} \over \partial U^{\rho}} ({\bf X}) 
\,\, \omega^{\alpha \rho} ({\bf X}) \,\,\, 
\delta ({\bf X} - {\bf Y})  \,\,\, =   \\
= \,\,\, \omega^{\prime \alpha \gamma} ({\bf X}) \,\,\,
\delta ({\bf X} - {\bf Y})  \,\,\, =  \,\,\,
\left\{ S^{\alpha} ({\bf X}) \, , \,
U^{\prime \gamma} ({\bf Y}) \right\}^{\prime}_{\rm AV}
\end{multline*}

 To finish the proof of the Theorem let us recall, that
we have the relations
$$\left\{ U^{\gamma} ({\bf X}) \, , \, U^{\rho} ({\bf Y}) 
\right\}_{\rm AV} \,\,\, = \,\,\, \left. \left\{ J^{\gamma} ({\bf X}) 
\, , \, J^{\rho} ({\bf Y}) \right\} \right|_{{\cal K} (1)} $$
$$\left\{ U^{\prime \gamma} ({\bf X}) \, , \, 
U^{\prime \rho} ({\bf Y}) \right\}^{\prime}_{\rm AV} 
\,\,\, = \,\,\, \left. \left\{ J^{\prime \gamma} ({\bf X}) \, , \, 
J^{\prime \rho} ({\bf Y}) \right\} \right|_{{\cal K} (1)} $$
where the functionals $\, J^{\gamma} ({\bf X})$,
$\, J^{\prime \gamma} ({\bf X})$ are defined by relations 
(\ref{FuncJnuX}). Consider now the regularized functionals
$$J^{\prime}_{[{\bf a}]} \,\,\, = \,\,\, \int a_{\gamma} ({\bf Z})
\,\, J^{\prime \gamma} ({\bf Z}) \,\, d^{d} Z $$
with some smooth compactly supported functions 
$a_{\gamma} ({\bf Z})$.

 Using relations (\ref{SpecialExp}) we can write on ${\cal K}$:
\begin{multline*}
\delta \, J^{\prime}_{[{\bf a}]} \,\,\, = \,\,\, 
\int a_{\gamma} ({\bf Z}) \,\, \left.
{\delta J^{\prime \gamma} ({\bf Z}) \over \delta G^{\nu} ({\bf X})}
\right|_{\cal K} \,\, \delta G^{\nu [\bm{\zeta}]} ({\bf X}) \,\,\,
d^{d} X \, d^{d} Z \,\, +   \\
+ \int a_{\gamma} ({\bf Z}) \,\, \left.
{\delta J^{\prime \gamma} ({\bf Z}) \over 
\delta \varphi^{i} (\bm{\theta}, {\bf X})} \right|_{\cal K} \,\, 
\delta g^{i [\bm{\zeta}]} (\bm{\theta}, {\bf X}) \,\,\,
{d^{m} \theta \over (2\pi)^{m}} \, d^{d} X \, d^{d} Z
\end{multline*}
where the ``coordinate system''
$({\bf G}^{[\bm{\zeta}]} ({\bf X}), \, 
{\bf g}^{[\bm{\zeta}]} (\bm{\theta}, {\bf X}))$ is introduced
with the aid of the functionals ${\bf J} ({\bf X})$.

 Without loss of generality we will consider all the relations
below at the ``point''  \linebreak
$\, {\bf G} ({\bf X}) \, = \, (\bm{\zeta} ({\bf X}), \,
{\bf U} ({\bf X}))$ of the submanifold ${\cal K}$.

 Expanding the values of the variation derivatives in the
integrands it is not difficult to get the following relations:
$$\delta J^{\prime}_{[{\bf a}]} \, = \,
\int a_{\gamma} ({\bf Z}) \,\, \left.
{\delta U^{\prime \gamma} ({\bf Z}) \over \delta G^{\nu} ({\bf X})}
\right|_{[{\bf U}, \bm{\zeta}]} \,
\delta G^{\nu [\bm{\zeta}]} ({\bf X}) \,\,
d^{d} X \, d^{d} Z \,\,\,
+ \,\,\, \int \epsilon \, U_{\nu [{\bf a}, {\bf G}]}
({\bf X}, \epsilon) \, \delta G^{\nu [\bm{\zeta}]} ({\bf X}) 
\,\, d^{d} X \,\, +  $$
$$+ \int \! \left( \! a_{\gamma} ({\bf X}) \,\,
\zeta^{\prime \gamma}_{i [\bm{\zeta}_{\bf X}, {\bf U}({\bf X})]}
\Big( \bm{\theta} + {\bm{\zeta} ({\bf X}) \over \epsilon} \Big)
\,\, + \,\, \epsilon \, U_{i [{\bf a}, {\bf G}]}
\Big( \bm{\theta} + {\bm{\zeta} ({\bf X}) \over \epsilon}, \,
{\bf X}, \epsilon \Big) \! \right) 
\delta g^{i [\bm{\zeta}]} (\bm{\theta}, {\bf X}) \,\,
{d^{m} \theta \over (2\pi)^{m}} \, d^{d} X $$
where 
$\, U_{\nu [{\bf a}, {\bf G}]} ({\bf X}, \epsilon) \, $ and
$\, U_{i [{\bf a}, {\bf G}]} (\bm{\theta}, {\bf X}, \epsilon) \, $ 
are local functions of 
$\, ({\bf a} ({\bf X}), \, \bm{\zeta}_{\bf X}, \, {\bf U} ({\bf X}))$
and their derivatives, given by regular at
$\, \epsilon \rightarrow 0$ series in $\epsilon$. Easy to see
also, that we have the relations
$${\delta U^{\prime \gamma} ({\bf Z}) \over 
\delta U^{\rho} ({\bf X})}
\,\, = \,\, {\partial U^{\prime \gamma} \over \partial U^{\rho}}
({\bf Z}) \,\,\, \delta ({\bf Z} - {\bf X}) 
\,\,\,\,\, , \,\,\,\,\,\,\,\,
{\delta U^{\prime \gamma} ({\bf Z}) \over 
\delta S^{\alpha} ({\bf X})} \,\, = \,\,
{\partial U^{\prime \gamma} \over \partial k^{\alpha}_{q}} ({\bf Z})
\,\,\, \delta_{Z^{q}} ({\bf Z} - {\bf X}) $$
  
 According to the requirements of the Theorem, the flows, generated
by $\, I^{\prime \gamma}$, leave invariant the submanifold $\Lambda$
and the values of $\, U^{\gamma}$ on it. This property conserves in
the main order of $\epsilon$ for the functionals 
$\, J^{\prime}_{[{\bf a}]}$ and the submanifold ${\cal K}$ with the
coordinates $\, U^{\gamma} ({\bf X})$. So, we can write here
$$\left. \left\{ U^{\gamma [\bm{\zeta}]} ({\bf X}) \, , \,
J^{\prime}_{[{\bf b}]} \right\} 
\right|_{\bm{\varphi} = \bm{\Phi}_{[{\bf U}, \bm{\zeta}]}}
\,\, = \,\,\, O (\epsilon) \,\,\,\,\, , \,\,\,\,\,\,\,\,
\left. \left\{ g^{i [\bm{\zeta}]} (\bm{\theta}, {\bf X}) \, , \,
J^{\prime}_{[{\bf b}]} \right\}
\right|_{\bm{\varphi} = \bm{\Phi}_{[{\bf U}, \bm{\zeta}]}}
\,\, = \,\,\, O (\epsilon) $$

 We have also the relations
$$\left. \left\{ S^{\alpha [\bm{\zeta}]} ({\bf X}) \, , \,
J^{\prime}_{[{\bf b}]} \right\}
\right|_{\bm{\varphi} = \bm{\Phi}_{[{\bf U}, \bm{\zeta}]}}
\,\, = \,\,\, O (\epsilon) $$
according to our definition of the functionals
$\, S^{\alpha [\bm{\zeta}]} ({\bf X})$.

 Besides that, in the full analogy with (\ref{zetarhpOrt}) we can
write the relations
$$\int_{0}^{2\pi}\!\!\!\!\!\dots\int_{0}^{2\pi}
\zeta^{(\rho)}_{i [\bm{\zeta}_{\bf X}, {\bf U}({\bf X})]}
\left( \bm{\theta} + {\bm{\zeta} ({\bf X}) \over \epsilon} \right)
\,\, \left. \left\{ g^{i [\bm{\zeta}]} (\bm{\theta}, {\bf X})
\, , \, J^{\prime}_{[{\bf b}]} \right\}
\right|_{\bm{\varphi} = \bm{\Phi}_{[{\bf U}, \bm{\zeta}]} \, (1)}
\,\, {d^{m} \theta \over (2\pi)^{m}} \,\,\, = \,\,\, 0 $$
for the constraints, defined with the aid of the functionals
${\bf J} ({\bf X})$.

 Using the relations
$$\zeta^{\prime (\gamma)}_{i [\bm{\zeta}_{\bf X}, {\bf U}({\bf X})]} 
( \bm{\theta} ) \,\,\,\,\, = \,\,\,\,\, \left.
{\partial U^{\prime \gamma} \over \partial U^{\rho}}
\right|_{[{\bf U}, \bm{\zeta}]} \,\,\,
\zeta^{(\rho)}_{i [\bm{\zeta}_{\bf X}, {\bf U}({\bf X})]}
( \bm{\theta} ) $$
we can then write
$$\left. \left\{ J^{\prime}_{[{\bf a}]} \, , \, 
J^{\prime}_{[{\bf b}]} \right\}
\right|_{\bm{\varphi} = \bm{\Phi}_{[{\bf U}, \bm{\zeta}]} \, (1)}
\,\,\, =  \,\, \int a_{\gamma} ({\bf Z}) \,\, \left.
{\delta U^{\prime \gamma} ({\bf Z}) \over \delta G^{\nu} ({\bf X})}
\right|_{[{\bf U}, \bm{\zeta}]} \,\,
\left. \left\{ G^{\nu [\bm{\zeta}]} ({\bf X}) \, , \,
J^{\prime}_{[{\bf b}]} \right\}
\right|_{\bm{\varphi} = \bm{\Phi}_{[{\bf U}, \bm{\zeta}]} \, (1)}
\,\,\, d^{d} X \, d^{d} Z $$

 Easy to see, that the same considerations can be repeated now also
for the functional $J^{\prime}_{[{\bf b}]}$. Finally, we can write
$$\left. \left\{ U^{\prime}_{[{\bf a}]} \, , \,
U^{\prime}_{[{\bf b}]} \right\}^{\prime}_{\rm AV}
\right|_{[{\bf U}, \bm{\zeta}]} \,\, = \,\,\,
\left. \left\{ J^{\prime}_{[{\bf a}]} \, , \,
J^{\prime}_{[{\bf b}]} \right\}
\right|_{\bm{\varphi} = \bm{\Phi}_{[{\bf U}, \bm{\zeta}]} \, (1)}
\,\,\, =   $$
\begin{multline*}
= \int a_{\gamma} ({\bf Z}) \, \left.
{\delta U^{\prime \gamma} ({\bf Z}) \over \delta G^{\nu} ({\bf X})}
\right|_{[{\bf U}, \bm{\zeta}]} 
\left. \left\{ G^{\nu [\bm{\zeta}]} ({\bf X}) , 
G^{\mu [\bm{\zeta}]} ({\bf Y}) \right\} 
\right|_{\bm{\varphi} = \bm{\Phi}_{[{\bf U}, \bm{\zeta}]} \, (1)}
\,  \left.
{\delta U^{\prime \rho} ({\bf W}) \over \delta G^{\mu} ({\bf Y})}
\right|_{[{\bf U}, \bm{\zeta}]} \, b_{\rho} ({\bf W}) \,\, \times  \\ 
\times  \,\, d^{d} X \, d^{d} Y \, d^{d} Z \, d^{d} W \,\,\, = 
\end{multline*}  
$$= \,\,\, \left. \left\{ U^{\prime}_{[{\bf a}]} \, , \,
U^{\prime}_{[{\bf b}]} \right\}_{\rm AV}
\right|_{[{\bf U}, \bm{\zeta}]} $$
which completes the proof of the Theorem.

{\hfill Theorem 4.2 is proved.}

\vspace{0.2cm}

 At last, let us consider the regular Whitham system
(\ref{MultDimConsWhithSyst}). We will assume now that the 
family $\Lambda$ represents a complete regular Hamiltonian family
of $m$-phase solutions of system (\ref{EvInSyst}) according to
Definition 3.1.

\vspace{0.2cm}

{\bf Theorem 4.3.}

{\it Let $\Lambda$ represent a complete regular Hamiltonian family
of $m$-phase solutions of system (\ref{EvInSyst}) equipped with
a minimal set of commuting integrals $\, (I^{1}, \dots, I^{m+s})$.
Then the corresponding regular Whitham system
(\ref{MultDimConsWhithSyst}) can be represented in the form
\begin{equation}
\label{WhithSystHamForm}
{S}^{\alpha}_{T} \,\,\, = \,\,\, \left\{ S^{\alpha} ({\bf X}) \, , \,
H_{av} \right\}_{\rm AV} \,\,\,\,\, , \,\,\,\,\,\,\,\,
{U}^{\gamma}_{T} \,\,\, = \,\,\,  \left\{ U^{\gamma} ({\bf X}) \, , \,
H_{av} \right\}_{\rm AV}
\end{equation}
where the functional $\, H_{av}$ is defined by formula
(\ref{HamFunctional}).
}

\vspace{0.2cm}

 Proof.

 According to Theorem 4.2, we can assume without loss of generality
that the Hamiltonian functional $\, H$ is included in the set
$\, (I^{1}, \dots, I^{m+s})$. In this case it can be easily seen
that system (\ref{WhithSystHamForm}) coincides with
(\ref{MultDimConsWhithSyst}).

{\hfill Theorem 4.3 is proved.}

\vspace{0.2cm}

 Let us formulate here also two theorems about the ``canonical forms''
of the bracket (\ref{AveragedBracket}), which were formulated in
\cite{JMP2} with a brief sketch of the proof.

\vspace{0.2cm}

\vspace{0.2cm}

{\bf Theorem 4.4 (\cite{JMP2}).}

{\it Consider the Poisson bracket given by relations
(\ref{AveragedBracket}) with
$\, {\rm rk} \, || \omega^{\alpha\gamma} || \, = \, m$.
There exists locally an invertible coordinate transformation
\begin{multline*}
\left( S^{1}({\bf X}), \dots, S^{m}({\bf X}), \, U^{1}({\bf X}),
\dots, U^{m+s}({\bf X}) \right) \,\, \rightarrow \\
\rightarrow \,\,
\left( S^{1}({\bf X}), \dots, S^{m}({\bf X}),
\, Q_{1}({\bf X}), \dots, Q_{m}({\bf X}),
\, N^{1}({\bf X}), \dots, N^{s}({\bf X}) \right)
\end{multline*}
where
$$Q_{\alpha} \,\, = \,\, Q_{\alpha} \left( {\bf S}_{\bf X}, \,
U^{1}, \dots , U^{m+s} \right)
\,\,\,\,\, , \,\,\,\,\,\,\,\,
N^{l} \,\, = \,\,  N^{l}  \left( {\bf S}_{\bf X}, \,
U^{1}, \dots , U^{m+s} \right) \,\,\, , $$
taking the bracket (\ref{AveragedBracket}) to the form:
$$\left\{ S^{\alpha} ({\bf X}) \, , \, S^{\beta} ({\bf Y}) \right\}
\,\, = \,\, 0 \,\,\, , $$
$$\left\{ S^{\alpha} ({\bf X}) \, , \, Q_{\beta} ({\bf Y}) \right\}
\,\, = \,\, \delta^{\alpha}_{\beta} \,\,
\delta ({\bf X} - {\bf Y}) \,\,\,\,\, , \,\,\,\,\,\,\,\,
\left\{ S^{\alpha} ({\bf X}) \, , \, N^{l}  ({\bf Y}) \right\}
\,\, = \,\, 0 \,\,\, , $$
$$\left\{ Q_{\alpha} ({\bf X}) \, , \, Q_{\beta} ({\bf Y}) \right\}
\,\, = \,\, J_{\alpha\beta} \left[ {\bf S}, {\bf N} \right]
({\bf X}, \, {\bf Y}) \,\,\,\,\, , $$
$$\left\{ Q_{\alpha} ({\bf X}) \, , \, N^{l}  ({\bf Y}) \right\}
\,\, = \,\, J_{\alpha}^{l} \left[ {\bf S}, {\bf N} \right]
({\bf X}, \, {\bf Y}) \,\,\,\,\, , $$
$$\left\{  N^{l}  ({\bf X}) \, , \, N^{q}  ({\bf Y}) \right\}
\,\, = \,\, J^{lq} \left[ {\bf S}, {\bf N} \right]
({\bf X}, \, {\bf Y}) \,\,\,\,\, , $$
where the functionals $J_{\alpha\beta}$, $J_{\alpha}^{l}$, $J^{lq}$
are given by general local distributions of the gradation degree 1.

}

\vspace{0.2cm}

 For the special case $\, s \, = \, 0$ Theorem 4.4 can be formulated
in the stronger form:

\vspace{0.2cm}

{\bf Theorem 4.5 (\cite{JMP2}).}

{\it Consider the Poisson bracket given by relations
(\ref{AveragedBracket}) with
$\, {\rm rk} \, || \omega^{\alpha\gamma} || \, = \, m$ and
$\, s \, = \, 0$. There exists locally an invertible coordinate 
transformation
$$\left( S^{1}({\bf X}), \dots, S^{m}({\bf X}), \, U^{1}({\bf X}),
\dots, U^{m}({\bf X}) \right) \,\,\, \rightarrow \,\,\,
\left( S^{1}({\bf X}), \dots, S^{m}({\bf X}),
\, Q_{1}({\bf X}), \dots, Q_{m}({\bf X}) \right) $$
where 
$Q_{\alpha} \,\, = \,\, Q_{\alpha} ( {\bf S}_{\bf X}, \, {\bf U})$,
which takes the bracket (\ref{AveragedBracket}) to
the following non-degenerate canonical form:
$$ \left\{ S^{\alpha} ({\bf X}) \, , \, S^{\beta} ({\bf Y}) \right\}
\,\, = \,\, 0 \,\,\,\,\, , \,\,\,\,\,\,\,\,
\left\{ Q_{\alpha} ({\bf X}) \, , \, Q_{\beta}  ({\bf Y}) \right\}
\,\, = \,\, 0 \,\,\,\,\, , $$
$$\left\{ S^{\alpha} ({\bf X}) \, , \, Q_{\beta} ({\bf Y}) \right\} 
\,\, = \,\, \delta^{\alpha}_{\beta} \,\,
\delta ({\bf X} - {\bf Y}) \,\,\,\,\, . $$

}

\vspace{0.2cm}

 As we can see, the averaged bracket has the simplest structure
in the absence of additional parameters $\, (n^{1}, \dots, n^{s})$.
Let us say, that the presence of the parameters
$\, (n^{1}, \dots, n^{s})$ can really make the structure of the 
bracket (\ref{AveragedBracket}) more complicated. Let us note also,
that both the theorems above are based just on the form of the
bracket (\ref{AveragedBracket}) and are not connected with the
averaging procedure itself.

\section{Illustrations: KdV and other examples.}
\setcounter{equation}{0}

 Let us continue now with the example of the KdV equation and
consider the construction of the Poisson brackets for the corresponding
Whitham systems according to our scheme. 

 Let us consider again the regular Hamiltonian submanifolds
$\, \Lambda^{(m)} \, $ defined by the sets of equations
(\ref{Extremals}). The corresponding system (\ref{EvPhaseSyst})
can be written here in the form
\begin{equation}
\label{mphaseKdV}
\omega^{\alpha} ({\bf U}) \, \Phi_{\theta^{\alpha}} \,\,\, = \,\,\,
k^{\alpha} ({\bf U}) \, \Phi \, \Phi_{\theta^{\alpha}} \,\, - \,\,   
k^{\alpha} ({\bf U}) \,  k^{\beta} ({\bf U}) \,
k^{\gamma} ({\bf U}) \,\,
\Phi_{\theta^{\alpha}\theta^{\beta}\theta^{\gamma}}
\end{equation}

 Let us say here that the eigenmodes of the linearized operator 
(\ref{mphaseKdV}), as well as the adjoint operator, were studied in 
detail (\cite{KricheverDAN,dobr1,dobr2,krichev1,krichev2}). 
In particular, we can claim here that all the families 
$\, \Lambda^{(m)} \, $ represent complete regular Hamiltonian
families of $m$-phase solutions of KdV equipped with minimal sets
of commuting integrals $\, (I^{1}, \dots, I^{m+1})$. It can be also 
shown, that the minimal set of commuting integrals for
$\, \Lambda^{(m)} \, $ can be given in fact by any 
$\, m \, + \, 1 \, $ independent higher integrals
$\, (I^{j_{1}}, \dots, I^{j_{m+1}}) \, $ of the KdV equation.

 It is well-known that the Whitham systems of the KdV hierarchy
represent a famous object in the theory of integrable systems and
were investigated in different aspect in many details. The Hamiltonian
structures of the Whitham systems for KdV were considered in the frame
of the general Hamiltonian theory of the Whitham equations started
in the works of B.A. Dubrovin and S.P. Novikov 
(\cite{dn1, DubrNovDAN84, dn2, dn3}). Let us say, that the Hamiltonian 
theory of the Whitham equations gives a basis for the integrability
of the Whitham hierarchies corresponding to a wide class of integrable
systems. Usually, the Whitham hierarchies are considered there as
systems of Hydrodynamic Type.

 As we said already, we consider here the Whitham system in the form
(\ref{MultDimConsWhithSyst}) and try to construct the corresponding
Hamiltonian structure, having the form (\ref{AveragedBracket}) - 
(\ref{HamFunctional}).

 The KdV equation gives a very convenient example demonstrating the
applicability of Theorem 4.1 in the multi-phase case. Indeed, we
have to require here the resolvability of the multi-phase systems
(\ref{BgSsystem}) - (\ref{BgUsystem}) on a dense set
$\, {\cal S}^{\prime} \, \subset \, {\cal M}^{\prime} \, $
in the space of parameters on $\, \Lambda^{(m)} , \, $ which
represents in general a nontrivial condition. However, the 
requirements of Theorem 4.1 can be easily established here, using
the same approach which was used in Chapter 2 for the construction
of the finite-dimensional bracket on the family $\, \Lambda^{(m)} $.

 We note first that the operator $\, {\hat B}_{[0]}(X) \, $ 
is given now in the form
$${\hat B}_{[0]}(X) \,\,\, = \,\,\,
k^{1}(X) \, {\partial \over \partial \theta^{1}} \,\, + \,\,
\dots \,\, + \,\,
k^{m}(X) \, {\partial \over \partial \theta^{m}} $$

 The regular zero mode of the operator $\, {\hat B}_{[0]}(X) \, $
is given by the constant function on the torus $\mathbb{T}^{m}$
and is orthogonal to the right-hand parts of systems
(\ref{BgSsystem}) - (\ref{BgUsystem}) according to Lemma 4.1.
The resolvability of systems (\ref{BgSsystem}) - (\ref{BgUsystem})
on a dense set 
$\, {\cal S}^{\prime} \, \subset \, {\cal M}^{\prime} \, $
is based then on the analytic properties of the right-hand
parts of (\ref{BgSsystem}) - (\ref{BgUsystem}).

 Indeed, it is not difficult to get again from the theta-functional
representation of the $m$-phase solutions of KdV that the Fourier
components in $\, \bm{\theta} \, $ of the right-hand parts of systems
(\ref{BgSsystem}) - (\ref{BgUsystem}) decay faster than any power
of $\, | {\bf n} | \,\, $ (${\bf n} \, = \, (n_{1}, \dots , n_{m})$) 
$\, $ for any fixed $\, {\bf X}$. At the same time, the zero
Fourier components $\, $ (${\bf n} \, = \, (0, \dots , 0)$) $\, $
of the right-hand parts of systems
(\ref{BgSsystem}) - (\ref{BgUsystem}) are identically equal to zero
according to Lemma 4.1. As a result, we can claim again that
systems (\ref{BgSsystem}) - (\ref{BgUsystem}) can be resolved on
the space of smooth $2\pi$-periodic in each $\, \theta^{\alpha} \, $
functions for all Diophantine vectors 
$\, {\bf k} \, = \, (k^{1}, \dots \, k^{m}) \, $
with index $\, \nu \, > \, m - 1 \, $. Using again the fact, that
the set of these vectors has the full measure in the space
$\, (k^{1}, \dots , k^{m}) , \, $ we can apply Theorem 4.1 in our
situation.

\vspace{0.3cm}

 As another example, consider now the one-dimensional local 
Poisson bracket
\begin{equation}
\label{ExamplePBr}
\left\{ \varphi (x) \, , \, \varphi (y) \right\} 
\,\,\,\,\, = \,\,\,\,\,
\delta^{\prime\prime\prime} (x - y) \,\,\, + \,\,\, \beta^{2} \,
\delta^{\prime} (x - y)
\end{equation}
and the Hamiltonian functional
$$H \,\,\, = \,\,\, \int \varphi^{3} \,\, d x $$

 The corresponding evolution system has the form
\begin{equation}
\label{ExampleEvSyst}
\varphi_{t} \,\,\, = \,\,\, 6 \, \varphi \, \varphi_{xxx} 
\,\, + \,\, 18 \, \varphi_{x} \, \varphi_{xx} \,\, + \,\,
6 \beta^{2} \, \varphi \, \varphi_{x}
\end{equation}

 The bracket (\ref{ExamplePBr}) has one local translationally
invariant annihilator, having the form
$$N \,\,\, = \,\,\, \int \varphi \,\, d x $$

 The periodic one-phase solutions of (\ref{ExampleEvSyst}) are
defined from the equation
\begin{equation}
\label{ExamplePhaseSyst}
\omega \, \Phi_{\theta} \,\,\, = \,\,\, 6 k^{3} \,
\Phi \, \Phi_{\theta\theta\theta} \,\, + \,\, 18 k^{3} \,
\Phi_{\theta} \, \Phi_{\theta\theta}  \,\, + \,\, 
6 \beta^{2} k \, \Phi \, \Phi_{\theta} 
\end{equation}
which can be integrated to the form
$${\omega \over 3} \,\, \Phi^{3} \,\,\, - \,\,\, 
3  k^{3} \, \Phi^{2} \, \Phi_{\theta}^{2} \,\,\, - \,\,\,
{3 \over 4} \beta^{2} k \, \Phi^{4} \,\, - \,\,
A \, \Phi^{2} \,\, - \,\, B \,\,\, = \,\,\, 0 $$
($A \, = \, {\rm const}$, $\, B \, = \, {\rm const}$).
We have then
$$k \,\,\, {\Phi \, d \Phi \over
\sqrt{\omega \, \Phi^{3} / 9 k \, - \, \beta^{2} \, \Phi^{4} / 4
\, - \, A \, \Phi^{2} / 3 k \, - \, B / 3 k }} \,\,\, = \,\,\,
\pm \, d \theta $$
so, the function $\, \Phi (\theta + \theta_{0})$ can be expressed
in terms of elliptic functions. The corresponding dependence
between $\, \Phi$ and $\, \theta$ is given by the integration with
respect to $\, \Phi$ over the periodic cycles, restricting the 
areas of positive values of the function
$$f \left( \Phi \right) \,\,\, = \,\,\, 
\omega \, \, \Phi^{3} / 9 k \, - \, \beta^{2} \, \Phi^{4} / 4
\, - \, A \, \Phi^{2} / 3 k \, - \, B / 3 k $$

\begin{figure}[t]
\begin{center}
\includegraphics[width=1\linewidth]{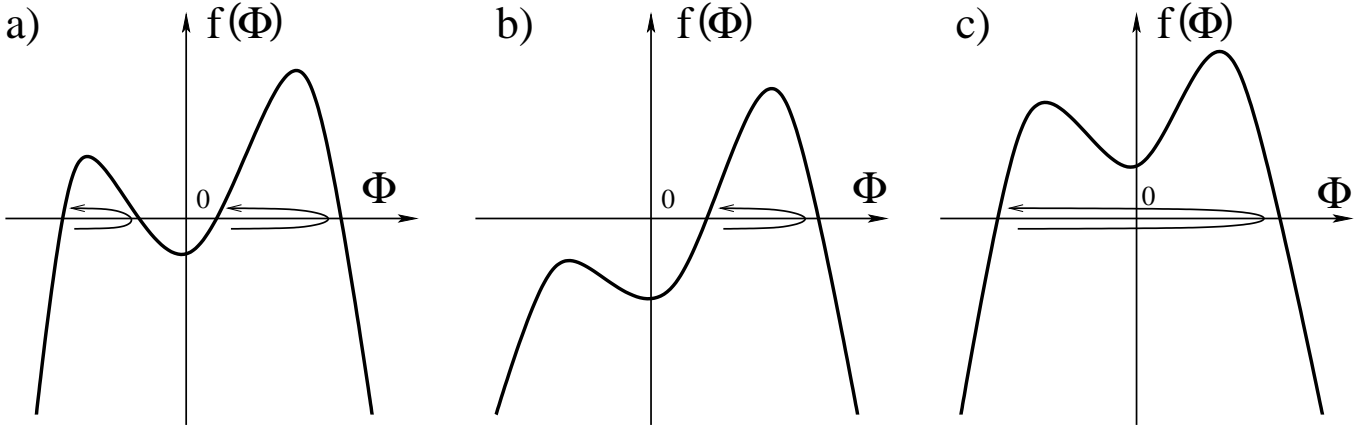}
\end{center}
\caption{Different periodic cycles, corresponding to different
regions in the space of parameters $\, (\omega, \, k, \, A, \, B)$.
}
\label{Figure}
\end{figure}

 Under different choice of the parameters 
$\, (\omega, \, k, \, A, \, B)$ the cycles can have 
different form shown at Fig. \ref{Figure} (a, b, c). 
So, we can consider in fact several families
of the one-phase solutions of system (\ref{ExampleEvSyst}),
corresponding to the cycles of different geometry. Let us say,
however, that the Whitham systems for different families
$\Lambda$ can demonstrate rather different properties, which can 
restrict in fact the applicability of the Whitham theory in
concrete problems. Thus, the regular Whitham system can have
hyperbolic or elliptic nature depending on the type of the roots
of the polynomial $f (\Phi)$ (see e.g. \cite{DriscolloNeil}).
The last case corresponds in fact to the modulation instability
of the solutions of the corresponding type, so, the applicability
of the Whitham approach should be specially studied in every 
concrete case. We will not study here these questions and consider
just the Hamiltonian structure of the regular Whitham system.

 The parameters $\, (\omega, k, A, B)$ are connected by the relation
$$k \,\, \oint \, {\Phi \, d \Phi \over
\sqrt{\omega \, \Phi^{3} / 9 k \, - \, \beta^{2} \, \Phi^{4} / 4
\, - \, A \, \Phi^{2} / 3 k \, - \, B / 3 k }} 
\,\,\,\,\, = \,\,\,\,\, 2 \pi $$
so, the family of one-phase solutions of (\ref{ExampleEvSyst}) is
parametrized by 3 independent parameters excluding the initial
phase shift $\, \theta_{0}$. In particular, the total set of
parameters on the full family $\Lambda$ of one-phase solutions
can be chosen in the form $\, (k, \, U^{1}, U^{2}, \, \theta_{0})$,
where
$$U^{1} \,\, \equiv \,\, \langle \varphi \rangle \,\, = \,\,
\int_{0}^{2\pi} \Phi (\theta) \,\, {d \theta \over 2\pi}
\,\,\,\,\,\,\,\, , \,\,\,\,\,\,\,\,\,\,
U^{2} \,\, \equiv \,\, \langle \varphi^{3} \rangle \,\, = \,\,
\int_{0}^{2\pi} \Phi^{3} (\theta) \,\, {d \theta \over 2\pi} $$ 
are the values of the functionals $\, N$ and $\, H$ on the
corresponding solutions of (\ref{ExampleEvSyst}). It is easy to 
check that every submanifold $\, \Lambda_{k}$ represents a regular
Hamiltonian submanifold in the space of periodic functions with
the period $\, 2 \pi / k$ for all the values of 
$\, k \, \neq \, \pm \beta / n$, $\, n \in \mathbb{N}$. 
It's not difficult to check also, that for the same values 
$\, k \, \neq \, \pm \beta / n$ the total
family $\Lambda$ represents a complete Hamiltonian family of
one-phase solutions of (\ref{ExampleEvSyst}), equipped with the 
minimal set of commuting integrals 
$\, (I^{1}, I^{2}) \, = \, (N, H)$.

 Easy to check that the conservation laws for the functionals
$\, N$ and $\, H$ have the form
$$\varphi_{t} \,\,\, = \,\,\, \left(
6 \, \varphi \, \varphi_{xx} \,\, + \,\, 6 \, \varphi_{x}^{2}
\,\, + \,\, 3 \beta^{2} \, \varphi^{2} \right)_{x} 
\,\,\,\,\, , \,\,\,\,\,\,\,\,
\left( \varphi^{3} \right)_{t} \,\,\, = \,\,\, \Big(
18 \, \varphi^{3} \, \varphi_{xx} \,\, + \,\, 
{9 \over 2}  \, \beta^{2} \, \varphi^{4} \Big)_{x} $$
so the regular Whitham system for the family $\Lambda$ can be
written as
$$S_{T} \,\,\, = \,\,\, \omega \left( S_{X}, U^{1}, U^{2} 
\right)$$
\begin{equation}
\label{ExampleWhithSyst}
U^{1}_{T} \,\,\, = \,\,\, 3 \beta^{2} \, \langle
\Phi^{2} \rangle_{X}
\end{equation}
$$U^{2}_{T} \,\,\, = \,\,\, \Big( - \, 54 \, S_{X}^{2} \,
\langle \Phi^{2} \Phi_{\theta}^{2} \rangle \,\, + \,\,
{9 \over 2}  \, \beta^{2} \, \langle \Phi^{4} \rangle
\Big)_{X} $$
where all the values $\, \langle \dots \rangle$ are represented
as functions of $\, (S_{X}, U^{1}, U^{2})$.

 Calculation of the pairwise Poisson brackets of the densities
of $\, N$ and $\, H$ gives the relations:
$$\left\{ P_{N} (x) \, , \, P_{N} (y) \right\} \,\,\, = \,\,\,
\delta^{\prime\prime\prime} (x - y) \,\, + \,\, \beta^{2} \,
\delta^{\prime} (x - y) \,\,\, ,   $$
\begin{multline*}
\left\{ P_{N} (x) \, , \, P_{H} (y) \right\} \,\,\, = \,\,\,
3 \, \varphi^{2} (x) \,\, \delta^{\prime\prime\prime} (x - y) 
\,\, + \,\, 18 \, \varphi \, \varphi_{x} \,\, 
\delta^{\prime\prime} (x - y) \,\, +  \\
+ \,\, 18 \, (\varphi \, \varphi_{xx} \, + \, \varphi_{x}^{2})
\,\, \delta^{\prime} (x - y) \,\, + \,\, 
3 \beta^{2} \, \varphi^{2} (x) \,\, \delta^{\prime} (x - y) 
\,\, +  \\
+ \,\, 6 \, (\varphi \, \varphi_{xxx} \, + \, 
3 \, \varphi_{x} \, \varphi_{xx}) \,\, \delta (x - y) \,\, + \,\,
6 \beta^{2} \, \varphi \, \varphi_{x} \,\, \delta (x - y) \,\,\, ,
\end{multline*}

$$ \left\{ P_{H} (x) \, , \, P_{N} (y) \right\} \,\,\, = \,\,\,
3 \, \varphi^{2} (x) \,\, \delta^{\prime\prime\prime} (x - y)
\,\, + \,\, 
6 \beta^{2} \, \varphi^{2} (x) \,\, \delta^{\prime} (x - y)
\,\,\, ,   $$
\begin{multline*}
\left\{ P_{H} (x) \, , \, P_{H} (y) \right\} \,\,\, = \,\,\,
9 \, \varphi^{4} (x) \,\, \delta^{\prime\prime\prime} (x - y)
\,\, + \,\, 54 \, \varphi^{3} \, \varphi_{x} \,\,
\delta^{\prime\prime} (x - y) \,\, +  \\
+ \,\, 54 \, (\varphi^{3} \, \varphi_{xx} \, + \, 
\varphi^{2} \, \varphi_{x}^{2})
\,\, \delta^{\prime} (x - y) \,\, + \,\,
9 \beta^{2} \, \varphi^{4} (x) \,\, \delta^{\prime} (x - y)
\,\, +  \\
+ \,\, 18 \, (\varphi^{3} \, \varphi_{xxx} \, + \,
3 \, \varphi^{2} \, \varphi_{x} \, \varphi_{xx}) 
\,\, \delta (x - y) \,\, + \,\, 18 \beta^{2} \, 
\varphi^{3} \, \varphi_{x} \,\, \delta (x - y) \,\,\, ,
\end{multline*}

 Using Theorem 4.1$^{\prime}$ for the set 
$\, k \, \neq \, \pm \beta / n$ and Theorem 4.3, 
we can claim now that the Whitham system (\ref{ExampleWhithSyst}) 
is Hamiltonian with respect to the bracket
$$\left\{ S (X) \, , \, S (Y) \right\} \,\,\, = \,\,\, 0
\,\,\, , \,\,\,\,\,
\left\{ S (X) \, , \, U^{1} (Y) \right\} \,\,\, = \,\,\, 0 
\,\,\, , $$
\begin{equation}
\label{ExampleAvBr}
\left\{ S (X) \, , \, U^{2} (Y) \right\} \,\,\, = \,\,\,
\omega \left( S_{X}, U^{1}, U^{2} \right) \,\, \delta (X - Y) 
\,\,\, , 
\end{equation}
$$\left\{  U^{1} (X) \, , \, U^{1} (Y) \right\} \,\,\, = \,\,\,
\beta^{2} \,\, \delta^{\prime} (X - Y) \,\,\, , $$
$$\left\{  U^{1} (X) \, , \, U^{2} (Y) \right\} \,\,\, = \,\,\,
3 \beta^{2} \, \langle \Phi^{2} \rangle \,\, 
\delta^{\prime} (X - Y) \,\, + \,\, 
3 \beta^{2} \, \langle \Phi^{2} \rangle_{X} \,\, \delta (X - Y) 
\,\,\, , $$
$$\left\{  U^{2} (X) \, , \, U^{1} (Y) \right\} \,\,\, = \,\,\,
3 \beta^{2} \, \langle \Phi^{2} \rangle \,\,
\delta^{\prime} (X - Y) \,\,\, , $$
$$\left\{  U^{2} (X) \, , \, U^{2} (Y) \right\} \,\,\, = \,\,\,
\left( - \, 108 \, S_{X}^{2} \, 
\langle \Phi^{2} \Phi_{\theta}^{2} \rangle \,\, + \,\,
9 \beta^{2} \, \langle \Phi^{4} \rangle \right) 
\,\, \delta^{\prime} (X - Y) \,\,\, +   $$
$$+ \,\, \left( - \, 54 \, S_{X}^{2} \, 
\langle \Phi^{2} \Phi_{\theta}^{2} \rangle \,\, + \,\,
9 \beta^{2} \, \langle \Phi^{4} \rangle / 2 \right)_{X}
\,\, \delta (X - Y) $$
with the Hamiltonian functional
$$H_{av} \,\,\, = \,\,\, \int \langle \Phi^{3} \rangle 
\,\, d X $$

 We will not make here further investigation of the example,
let us just say, that the construction above can be easily
generalized to the multi-dimensional case. Thus, using the
Poisson bracket
$$\left\{ \varphi ({\bf x}) \, , \, \varphi ({\bf y}) \right\}
\,\,\,\,\, = \,\,\,\,\, \sum_{q,p,r = 1}^{d} \alpha_{qpr} \,\,
\delta_{x^{q} x^{p} x^{r}} ({\bf x} - {\bf y}) \,\,\, + \,\,\,
\sum_{q = 1}^{d} \, \beta_{q} \,\, 
\delta_{x^{q}} ({\bf x} - {\bf y}) $$
and the same Hamiltonian functional
$$H \,\,\, = \,\,\, \int \varphi^{3} \,\, d^{d} x $$
we get different multi-dimensional analogs of system
(\ref{ExampleEvSyst}).

 The corresponding one-phase solutions are defined in this case
by equation equivalent to (\ref{ExamplePhaseSyst}), which can be
easily integrated in the same way. The full families of the
one-phase solutions are parametrized in this case by the values
$\, (k_{1}, \dots, k_{d}, \, U^{1}, U^{2}, \, \theta_{0})$
and represent complete Hamiltonian families equipped with
a minimal set of commuting integrals $\, (I^{1}, I^{2})$ on the
open set of the full measure in the space of 
$\, (k_{1}, \dots, k_{d})$. As a result, we can suggest then a 
complete analog of the bracket (\ref{ExampleAvBr}) for the
corresponding regular Whitham systems in the $d$-dimensional
space.

 Let us say again that we have chosen the above example as one
of the simplest examples where the averaging of a Poisson
bracket is made in presence of just a minimal set of commuting
integrals in accordance with the idea of the paper.

\vspace{0.2cm}

 The work was supported by the Russian Federation Government
Grant No. 2010-220-01-077, Grant of the President of Russian Federation
NSh-4995.2012.1, and Grant RFBR No. 13-01-12469-ofi-m-2013.

\end{document}